# ChemDyME: Kinetically Steered, Automated Mechanism Generation Through Combined Molecular Dynamics and Master Equation Calculations

Robin J. Shannon[1,2]*, Emilio Martínez-Núñez[3], Dmitrii V. Shalashilin[2], David R. Glowacki[1,4,5]

[1]*School of Chemistry, Center for Computational Chemistry, University of Bristol, Bristol, BS8 1TS, UK*

[2]*School of Chemistry, University of Leeds, LS2 9JT, UK*

[3]*Department of Physical Chemistry, University of Santiago de Compostela, Spain*

[4]*Intangible Realities Laboratory, School of Chemistry, University of Bristol, Bristol, BS8 1TS, UK*

[5]*Department of Computer Science, Merchant Venturers Building, University of Bristol, Bristol BS8 1TS, United Kingdom*

## Abstract

In many scientific fields, there is an interest in understanding the way in which complex chemical networks evolve. The chemical networks which researchers focus upon, have become increasingly complex and this has motivated the development of automated methods for exploring chemical reactivity or conformational change in a "black-box" manner, harnessing modern computing resources to automate mechanism discovery. In this work we present a new approach to automated mechanism generation implemented which couples molecular dynamics and statistical rate theory to automatically find kinetically important reactions and then solve the time evolution of the species in the evolving network. Key to this ChemDyME approach is the novel concept of "kinetic convergence" whereby the search for new reactions is constrained to those species which are kinetically favorable at the conditions of interest. We demonstrate the capability of the new approach for two systems, a well-studied combustion system, and a multiple oxygen addition system relevant to atmospheric aerosol formation.

## 1. Introduction

Predicting how a reactive system of molecules evolves over time is important in several scientific fields. In the gas phase, there is a need to unravel the chemical mechanisms at play in a combustion engine [1–5] or in the earth's atmosphere;[6] in biological systems there is a need to model complex conformational landscapes;[7] and in catalysis [4,8–10] there is a need to screen potential catalytic mechanisms. In all cases an understanding of the important chemical reactions/conformational changes and the rate at which they occur is vital. Given the prevalence of this type of problem, there are a growing number of automated approaches designed to map the network of possible reactive or conformational changes for a given system. [9,11–21]

What we describe here as automatic network generation of chemical networks, usually in practice involves a number of separate steps. First it is necessary to identify the possible reactions a system can undergo and then it is desirable to identify the transition state or the rate coefficient for these reactions to ascertain the kinetic relevance of each channel. Much work has gone into so called double ended methods [22–24] for automatically finding transition state structures, but such methods assume one already knows the reactant and product of a reaction, which is not always the case when exploring a novel reaction mechanism. Several automated mechanism generation approaches then go on to automatically solve the coupled kinetics of the whole system as a final step by employing master equation or related Markov state model calculations. [4,5,7,14,21,25–27]

In this work we introduce a new automated mechanism generation framework ChemDyME, (**Chem**ical Network Mapping though combined **Dy**namics and **M**aster **E**quation simulations) designed to automatically map reaction networks and then determine the evolution of the species in the system over time. The key aspect to the current approach is novel concept of "kinetic convergence" whereby the search for possible reactions is guided by of the kinetics network. In ChemDyME, molecular dynamics and master equation calculations are interleaved to ensure that only the kinetically relevant part of the reaction network is sampled. Our approach tracks the timescale associated with each new step we make in the growing reaction network and the exploration is considered complete once we exceed some specified timescale (we describe this as reaching kinetic convergence). Thus, we aim to confine our search of the full possible reaction network for a system to only include those reactions deemed kinetically important under the conditions of interest. This kinetic convergence approach can be utilized to add new species to the system at the point when a new bimolecular reaction might be kinetically favorable and to our knowledge ChemDyME is the only automated reaction network approach able to add new species and bimolecular reactions to a growing network.

The aim of the paper is to present the details of the ChemDyME approach and to highlight potential applications. To this end sections 2 and 3 are devoted to introducing the reader to our methodology, firstly in the form of a high level, general, summary of the mechanism generation pipeline (Section 2) and then focusing upon more specific implementation and technical details in Section 3. Having laid out the conceptual and technical core of ChemDyME, Section 4 presents some benchmark calculations, where we utilize ChemDyME to study a well understood combustion system (the chemistry of OH + propyne) which has previously been studied by automated methods.[28] With these pieces in place, Section 5 demonstrate the solution of atmospheric oxidation system with a reactive bath gas by harnessing the newly developed "kinetic convergence" approach, key to the ChemDyME workflow. Specifically, we turn to the atmospheric reaction of the 1-methyl-hexene with ozone which is an archetypal, peroxy radical forming[28] reaction, for investigating the formation of highly oxygenated molecules (HOM),[29] key to the chemistry of atmospheric aerosol. Using ChemDyME we explore the network of the 1-methyl-hexene + $nO_2$ system where the n signifies that additional $O_2$ molecules are added to the system whenever ChemDyME detects addition of another $O_2$ is kinetically favorable.

## 2. High level summary of the three main parts of the ChemDyME pipeline

### 2.1 Framework overview

Given a starting molecular geometry, ChemDyME is designed to explore the network of reactions that the atomic system may undergo in an automated, "black-box" way. As the network is traversed (via master equation calculations determining transition times between species or nodes), ChemDyME tracks the cumulative time associated with all the transitions which have occurred, and the network exploration continues until some maximum time of interest, specified by the user for a given application, is reached. In this paper we term this "kinetic-convergence", and the converged network should be representative of the important reactions under some conditions (temperature and pressure) and timescale of interest.

There are two main principles which have guided the development of the ChemDyME and which are emphasized in our chosen methodologies. The first of these principles is simplification: In ChemDyME we wish to focus only upon the subset of possible chemical reactions which are important under the conditions of interest and we have chosen techniques which prioritize exploring reactions in order of kinetic importance rather than exploring reactions quickly. The second principle is generality: ChemDyME is designed to be applicable to a wide range of systems and we have chosen to minimize the use of "chemical-intuition" in the ChemDyME workflow in order to aid flexibility and avoid human bias arising from "perceived-knowledge." These two principles will be discussed further in sections 2.2 and 2.4.

The overall workflow involved in a ChemDyME calculation may be broadly split into 3 main categories:

1) **Molecular Dynamics (MD)**: Within ChemDyME, reactive MD simulations are used to determine the possible reactions that the system may undergo from a given starting reactive state where the reactive state could be either a stable chemical species or some pair of bimolecular reactive moieties.

From the MD simulations, ChemDyME is interested only in the possible product species which may be formed from the reactive state.

2) **Optimisation and Refinement (OR):** Once a set of reactions has been observed through MD, the reactants and products automatically undergo further refinement through geometry optimization calculations and subsequent vibrational frequency and energy calculations. The reaction path is then automatically characterized (see section 2.3) to determine whether there is a defined transition state or saddle point. If present, then the transition state is also subject to automatic refinement.

3) **Master Equation (ME):** Once the reactants, products and transition state (if present) have been refined, the structures, energies and vibrational frequencies of these species are then input into a master equation simulation. From the specified starting state these calculations determine which of the possible products is formed and track the transition time associated with the reaction. The kinetically favorable product is then used as a reactive state to seed new MD simulations. Master equation calculations are also used to automatically lump or equilibrate species based upon the relative kinetics of the different reactions in the network (see section 2.4). Once the ChemDyME run is complete, the resulting master equation input may be used to obtain phenomenological rate coefficients for the entire network.

These three different types of calculations create a feedback loop (Figure 1) with MD simulations providing input for OR calculations which then provide the input to the ME calculations, which in turn determine the starting point for the next set of MD calculations.

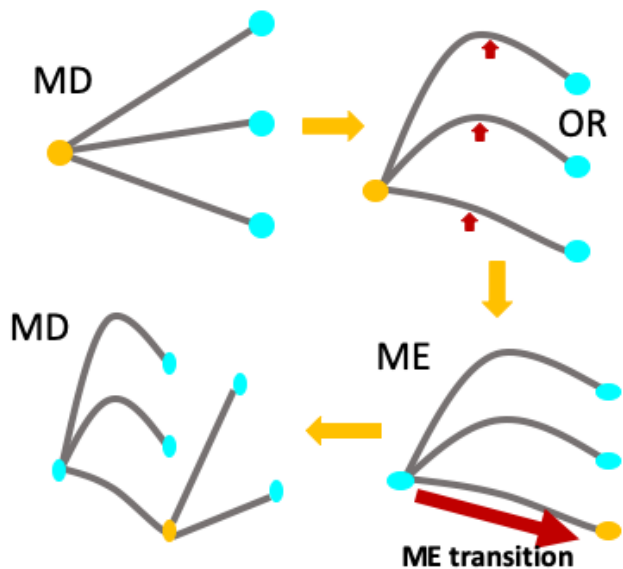

**Figure 1: Outline of the steps involved in a ChemDyME calculation. First molecular dynamics (MD) is used to map the potential products which can be formed from a starting reactant or node. Then optimization and refinement (OR) is used to characterize each reaction path. Master equation calculations (ME)can be used to then identify the kinetically favorable product. New MD calculations are performed from this product and the cycle continues.**

A flow chart showing the overall ChemDyME process in more detail is shown in Figure 2. To summarize, a ChemDyME calculation starts with a set of cartesian co-ordinates for some reactant state and determines whether the reactant state is a single molecule or a pair of bimolecular fragments. ChemDyME then optimizes the geometry of the reactant state and proceeds to run $n_{MD}$ reactive MD simulations from the optimized reactive state to determine the possible reactions this reactant may undergo. Each MD simulation is tracked until a reactive event and product is identified (see 2.2). These products and reaction paths go through the OR process and each of these reactions is then added to a ME calculation. The ME calculation then returns a product species and the time associated with this transition $\tau_{trans}$. This transition time is then added to the cumulative system time $\tau_{tot}$ and the whole process starts again with new MD simulations initialized from the geometry corresponding to the ME reaction product. If at any point the ME product is a state which has already been used to initialize MD, the ME calculation is repeated until an unvisited (a state which has not been used to initialize MD) product is formed. This continues until $\tau_{tot}$ exceeds the specified maximum time of interest $\tau_{max}$, at which point kinetic convergence is considered reached. All reactions in the network are saved in a single ME input file and the ChemDyME process may be repeated multiple times, building upon this input file, to ensure that no channels have been missed. When repeating a previously completed ChemDyME run, $\tau_{tot}$ is reset to zero and the ChemDyME process starts with an ME calculation from the initial reactant state. MD simulations are only run if an unvisited product is encountered in the ME simulations.

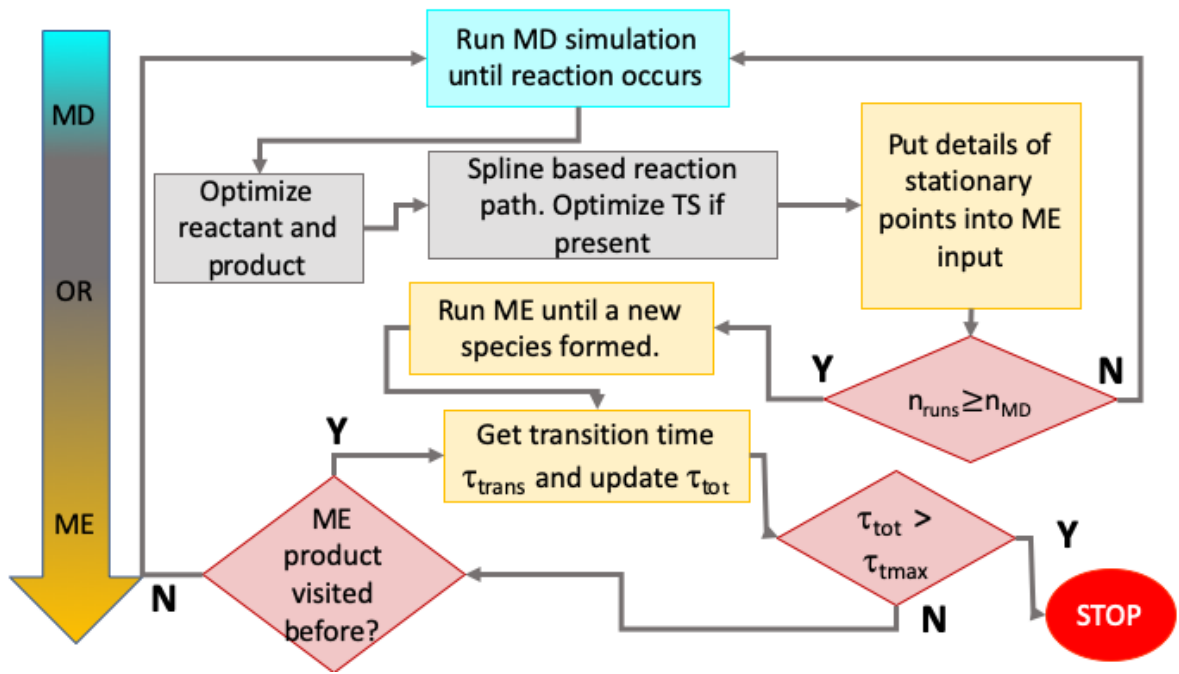

**Figure 2: Flow chart showing the parts of a ChemDyME run in more detail. Cyan represents the MD part, grey represents the OR part and orange represents the ME part of the process.**

## 2.2 Molecular dynamics simulations for discovering reactions

The choice of MD simulations to determine the reactions a given state may undergo is central to the principles of simplification and generality which guide the ChemDyME workflow. Apart from limitations associated with the method used to obtain the forces for the MD simulation and the inherent approximations involved in using a classical equation of motion, an MD simulation replicates the "real" (classical) behaviour of a chemical system. As such MD simulations do not explicitly rely upon any assumptions regarding the types of reaction which may occur and are an ideal tool for investigating novel reactions. With an appropriate forcefield and thermostat/barrostat, unbiased MD samples reactions which are representative of the real reactions the system may undergo under the same conditions and thus the use of MD helps focus upon the subset of reactions which are important under the conditions of interest.

The disadvantage to using MD simulations is that they can be computationally expensive compared to other methods of searching for chemical reactions. In particular, MD suffers from the commonly encountered "rare- event" problem [30–32] whereby the particular process or "event" of interest, in this case chemical reaction, is extremely slow, relative to the fundamental timescale of the MD simulations. While rare event acceleration methods are an active area of research, the majority of these methods require some knowledge of the reaction or process one wishes to accelerate [33–41] and are not suitable for accelerating chemical reactions in general. Zheng et al [2] have successfully used well-tempered metadynamics in conjunction with SPRINT coordinates in order to explore chemical reaction in MD, and recent work from Grimme demonstrated another metadynamics based method for general reaction finding.[42] In other cases, high temperatures or pressures[17,25] have been successfully used to alleviate the rare event problem. However, the product channels observed from such simulations may be different from the products formed at some lower temperature of interest and we recently demonstrated that high temperature MD simulations tend to favor entropically favorable dissociation reactions and may miss the enthalpically favorable but entropically unfavorable reactions which are important at lower temperatures. [6,43]

In this automated mechanism generation framework, we utilize the recently developed "boxed molecular dynamics in energy" (BXDE) [43] method to accelerate the observation of reaction in MD. This BXDE method was shown to accelerate chemical reaction in MD simulations by several orders of magnitude whilst giving product yields in agreement with unbiased simulations at the same temperature. The BXDE approach is part of the boxed molecular dynamics (BXD), [44–47] family of methods which steer molecular dynamics simulations into desired regions of configuration space through the use of reflective boundaries. In the BXDE case, the reflective boundaries are potential energy contours, and this causes the MD simulation to be steered towards configurations of high internal energy which are more likely to promote some reactive event.

For the majority of the MD simulations run in the current framework, BXDE is sufficient to promote the chemical reactions we wish to observe. The exception to this, is when we wish to start a ChemDyME run from two sperate molecular fragments or moieties which undergo a bimolecular reaction. In this case a standard BXD constraint is automatically implemented along the separation of centers of mass (COM) of the two fragments, $r_{com}$. This standard BXD procedure has been documented extensively. In this case, the BXD constraint forces the two reacting fragments towards one another until they reach some specified $r_{com}$ value $r_{cutoff}$. Once $r_{com}$ is less than $r_{cutoff}$, BXDE is used to help surpass any reaction barriers and the BXD constraint is imposed whenever $r_{cutoff}$ is exceeded, thus keeping the two reacting fragments in close proximity. In this cased $r_{com}$ is decreased particularly efficiently by employing a BXD inversion on any MD step at which $r_{com}$ increases. The approach used to accelerate bimolecular reaction is shown schematically in Figure 3.

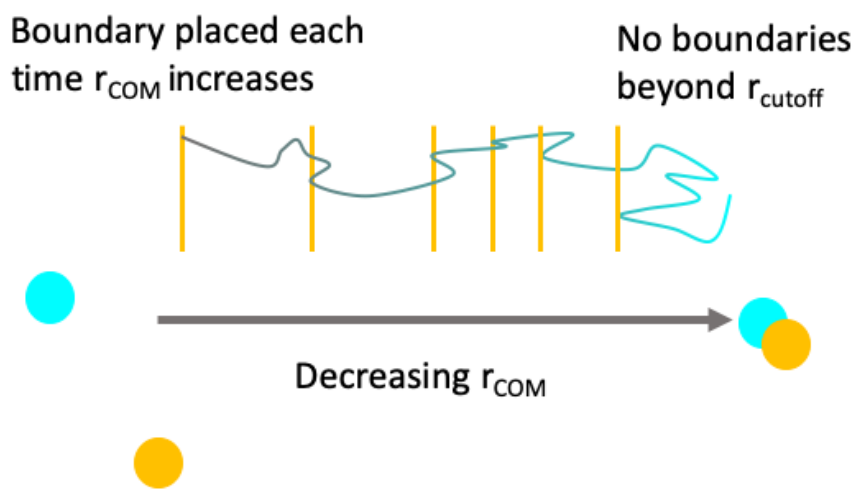

**Figure 3: BXD approach for associating two separate fragments. The two circles represent the centers of mass of two fictitious fragments and r_COM is the distance between them. As the trajectory progresses, a new reflective boundary is placed whenever r_COM would increase until some value r_cutoff is reached.**

For the proposed black-box framework for mechanism generation to be effective, it is necessary to define a criterion which automatically determines when an MD simulation has caused a chemical reaction to occur. In the current work a chemical reaction is defined as a change in the bonding structure of the molecule, i.e at least one chemical bond is formed or broken, however in the future we hope to add a criterion to optionally track conformational changes also. The reaction criterion used here is a simplified form of the TSSCDS algorithm of [25,48] This criterion works as follows: two matrices are defined, $d$ and $d^{REF}$. The first matrix has elements equal to the distance between atoms $i$ and $j$ in the system; the second matrix has elements, consisting of pre-defined ideal bond distances between atoms $i$ and $j$. The ideal bond distances used here are shown in Table 1.

| $ij$ | $d_{ij}^{REF}$/ Angstrom |
|------|------------------------|
| CC   | 1.6 |
| CH   | 1.2 |
| CO   | 1.6 |
| OO   | 1.6 |
| OH   | 1.2 |
| HH   | 0.8 |

**Table 1: Ideal bond distances which make up the elements of the matrix**

Using these two matrices, we form a connectivity matrix with elements C, where:

$$C_{ij} = \begin{cases} 1 \; if \; d_{ij} < d_{ij}^{REF} \\ 0 \; otherwise \end{cases} \quad \text{Eq. 1}$$

For the starting structure, this matrix identifies whether two atoms are bonded ($C_{ij} = 1$) or non-bonded ($C_{ij} = 0$). At each time step, the bonding structure (given by $d$ and $d^{REF}$) of the system is compared with the reactant bonding structure (given by $C$) to monitor for reaction. Specifically, a reaction is then considered to occur if for an atom $i$:

$$\max(\delta_{in}) > \; \min(\delta_{ik}) \; ; \delta_{ij} = \frac{d_{ij}}{d_{ij}^{REF}} \quad \text{Eq. 2}$$

Here index $n$ runs over atoms bonded to $i$ ($C_{ij}$ matrix elements equal to one) and index $k$ runs over atoms which do not have a bond to $i$ ($C_{ij}$ matrix elements equal to zero). In ChemDyME the criteria must be met consistently for a specified number of MD steps to ensure that a reactive event has occurred and that there is no recrossing back to reactants. Assuming the reaction criteria is met consistently, the final molecular configuration is used as a product guess for the OR phase of the ChemDyME workflow.

All BXDE simulations in ChemDyME place the reflective boundaries adaptively as described in our previous work. [43] Typically, the MD simulations are thermostated using a Langevin thermostat and the MD simulations are performed at a specified temperature $T_{MD}$. It should be noted that this temperature is a compromise between the efficiency of reaction sampling (reactions are observed in fewer MD steps at high $T_{MD}$) versus the number of product channels to be considered (higher $T_{MD}$ simulations will produce more product channels which might not be relevant to low T chemistry). This temperature $T_{MD}$ need not be identical to the temperature of interest $T_{ME}$ which is used to define the conditions for the master equation calculation (see section 2.4) .

### 2.3  Optimization and refinement of the stationary points of discovered reaction paths

Once a reaction has been found in MD, it is necessary to find and fully characterize the stationary points involved, to provide input for the master equation part of the framework. This section of the workflow has two distinct parts. First it is necessary to optimize the molecular structures of the correct stationary points. This is trivial in the case of the product(s) of the reaction, but it is much harder to identify (if present) a first order saddle point corresponding to a transition state between reactant and product. Once the stationary points have been optimized, the second part involves a series of refinement steps, to supply all the information necessary to include the reaction in a master equation simulation.

We will consider the optimization step first and Figure 4 presents a flow chart summarizing the key steps. As stated above, finding a stable minimum for the product is usually trivial. The geometry corresponding to the final frame of the reactive trajectory is taken and a geometry optimization is performed. A canonical SMILES [25] string is then obtained for the product and if this string indicates there is more than one molecular fragment then the refinement steps are performed for each product. In the case of a saddle point, getting a good initial guess geometry for the transition state optimization is vital. An initial guess geometry is taken from the geometry at which the TSSCDS reaction criterion is met. To refine this geometry further we partially minimize the structure by freezing all the bonds which change over the course of the reaction and performing a constrained minimization. This partial relaxation is designed to remove excess energy in modes which are unrelated to the reactive event. Finally, a transition state optimization is performed on this partially relaxed structure. This is equivalent to the approach employed by Martínez-Núñez. [22] An intrinsic reaction coordinate (IRC) calculation is then performed to check whether the optimized saddle point (if the TS search was successful)§ connects the expected reactants and products. If the correct transition is not found, this procedure is optionally repeated for several other geometries from snapshots along the reactive MD trajectory.

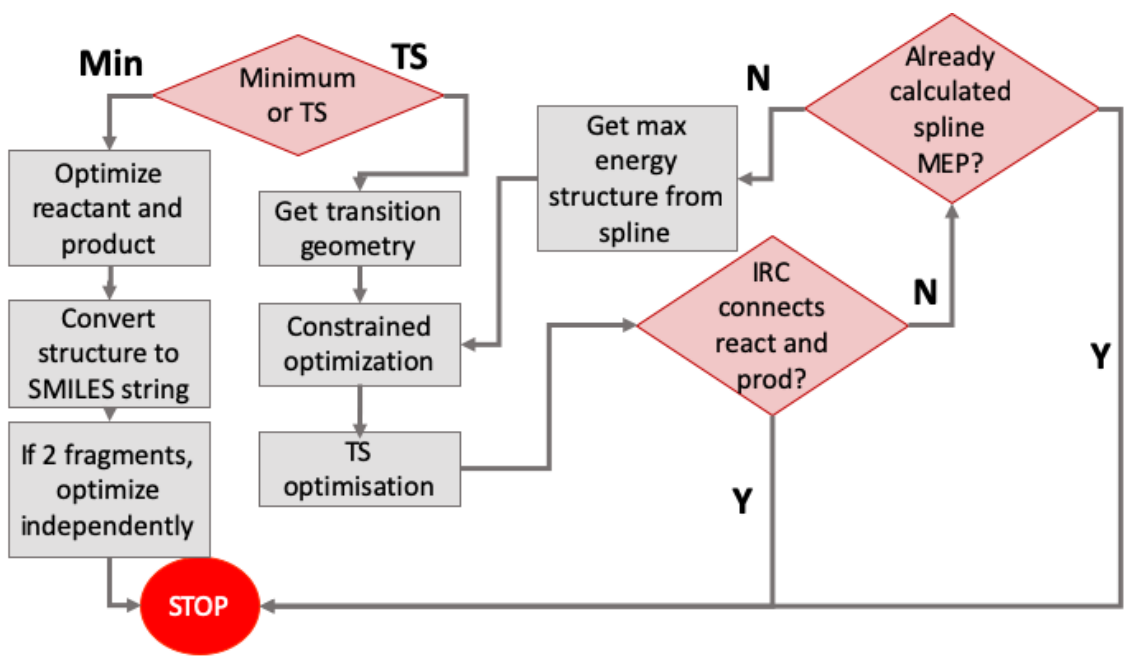

**Figure 4: Flowchart outline the steps in the optimization and refinement (OR) part of the ChemDyME process.**

If none of these transition state searches are successful, we attempt a different approach. Using the spline-based reaction path method [49] in the Scine ReaDuct [50,51] software we optimize the minimum energy path between reactant and product and then attempt another TS search at the geometry corresponding to the maximum in the reaction path. There are of course cases where there is no defined barrier to the reaction and in the particular case where the reaction is either an association (two reactants one product) or a dissociation (one reactant two products), the reaction can be input in the master equation as a barrierless process using the inverse Laplace transform (ILT) method employed in MESMER. [51] This is identified by ChemDyME by looking for a defined barrier in the spline based reaction path. If, however the reaction is an isomerization (one reactant, one product) and no TS can be identified, then this reaction is flagged for further (manual) investigation by ChemDyME and not included in the automated workflow.

Once an optimized structure (either minimum or saddle point) has been found, the refinement procedure begins. A hierarchy of levels of theory are used for this. A defined "initial optimization level" of theory is initially used for the optimizations already described and then an optional "refinement level" of theory is used to optimize the structure at a higher level of theory. Vibrational frequencies are then obtained using this "refinement level." Finally, a single point energy calculation is performed at a "Single Point Level" of theory at the optimized ""Refinement Level"" geometry. The total energy of the species is then stored as the sum of zero-point vibrational energy and the Single Point Level electronic energy.

## 2.4 Master equation calculations for tracking the kinetic evolution of the network and guiding the search

The MD and OR parts of the workflow provide all the necessary information for performing statistical rate theory calculations, and chemical master equation simulations have become a standard tool for determining the kinetic behavior and rate coefficients of a complex chemical network. The methodological details of the chemical master equation have been described in detail in numerous publications[52–56] and there exist multiple open-source codes [51,57,58] designed to solve the chemical master equation for a given network of interconnected intermediates, (wells in potential energy space).

In practice the energy space for each species in the ME network is discretized into energy grains. Microcanonical transition state theory and an appropriate energy transfer model are then used to calculate grain to grain rate coefficients. In ChemDyME we make extensive use of the recently developed Boxed Molecular Kinetics (BXK) algorithm.[59] This BXK methodology has significant conceptual overlap with the BXDE methodology employed for the MD simulations and was recently shown to accelerate Kinetic Monte Carlo, (KMC) ME simulations by several orders of magnitude. Briefly, the KMC approach tracks a stochastic trajectory of the population of the system from one grain to another. All transition probabilities are based upon the microcanonical grain to grain rate coefficients for energy transfer and chemical reaction. The BXK method simply places a boundary in energy space preventing lower energy grains being accessed and helping to alleviate the same rare event problem as encountered in the MD simulations.

A schematic of the BXK KMC approach used here is shown in Figure 5. Usually in a KMC simulation, one would need to run many stochastic trajectories, potentially starting from a range of initial grains, in order to converge the time evolution of the different species. In this work we take a different approach and each ChemDyME run can be viewed as a single KMC trajectory which is paused after each reactive step and for which new reactions are added to the system mid trajectory via the MD and OR procedures. Figure 5 shows such a KMC reactive step. This figure displays a simple two well system and depicts a stochastic trajectory involving a single energy transfer transition followed by reaction from well R to well P. The orange shaded areas signify the inaccessible grains due to the BXK procedure. The total time associated with this particular reactive step $\tau_{trans}$ is equal to the sum of the times for each step leading up to and including the reaction, and ChemDyME stores $\tau_{trans}$, and the product energy grain. Before continuing the stochastic trajectory, ChemDyME examines whether product P has been encountered before and if not, an MD/OR cycle is performed to add additional reaction to the master equation system. This ensures that the MD and OR procedures are only performed for the most kinetically favorable species under the conditions of interest. The KMC simulation finally terminates when $\tau_{tot}$ exceeds $\tau_{max}$, constituting one full KMC trajectory. Additional KMC trajectories may be run to ensure no reaction channels are missed and, in each case, MD/OR calculations are only performed if a previously unencountered product is formed in a particular KMC step.

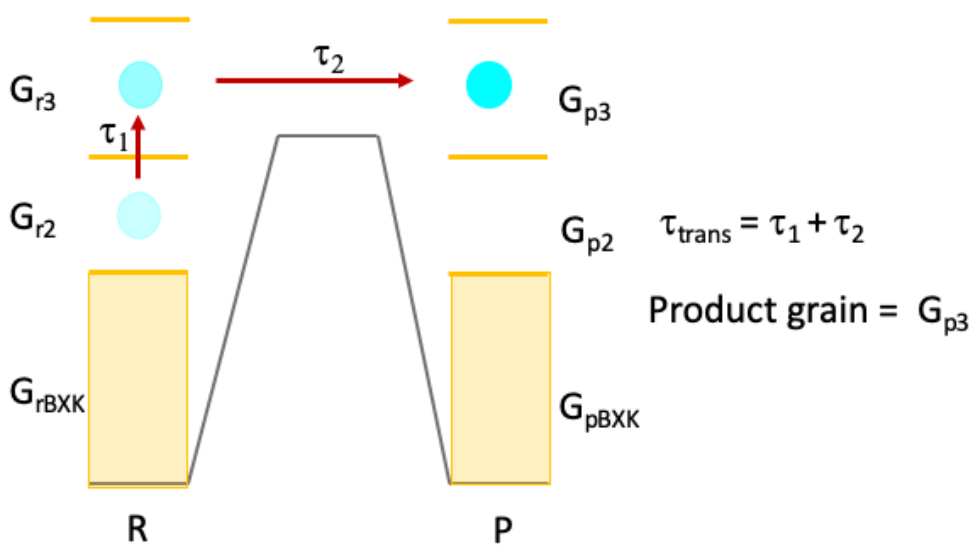

**Figure 5: Stochastic master equation example. Here we have a system of two isomers with two active grains each and an excluded volume of rovibrational energies as per the BXK methodology. In this case the stochastic trajectory involves a single energy transfer step with associated time $\tau_1$ and a reactive step with time $\tau_2$. ChemDyME then terminates the master equation and stores the product grain $G_{p3}$ and the transition time $\tau_{trans} = \tau_1 + \tau_2$.**

## 3. Technical details of the ChemDyME framework

The framework described in Section 2 has been implemented in the object-orientated Python package ChemDyME and utilizes the Atomic Simulation Environment (ASE) package.[59] The ChemDyME framework may be obtained from: https://github.com/RobinShannon/ChemDyME and is distributed under the LGPL license. ASE contains an "atom" class which stores the cartesian coordinates and various chemical properties for a particular chemical configuration. The "atom" class can be readily interfaced with a variety of existing electronic structure theory codes via the ASE "calculator" classes and allows access to the built in ASE optimization methods when required.

### 3.1 External Code interfaces

The ChemDyME source code primarily takes form of a wrapper. The various chemical species in the chemical network are stored as ASE "atom" types and "calculators" are used to run external electronics structure theory calculations which are then gathered and analyzed automatically by ChemDyME. ChemDyME also interfaces with the MESMER[51] code for all master equation calculations. The interaction between ChemDyME and external codes may be broken down into the three parts of the mechanism generation framework as follows:

1)   MD: All MD calculations are carried out using a bespoke MD integrator class within ChemDyME. The starting structure for the MD is turned into an ASE "atoms" object and at each MD timestep forces are obtained by an external code and read back into ChemDyME. BXD classes in ChemDyME check whether a velocity inversion is required at each timestep and reset positions / update velocities accordingly.

2)  By preference optimizations (both minimizations and saddle point searches) and frequency calculations are carried out by external codes, but minimizations and frequency calculations can be performed using built in ASE routines as a backup. If the open source, Scine readuct[49] package is found then the spline-based reaction path [22]methods are utilized as part of the transition state optimization strategy although built in nudged elastic band methods in ASE are an alternate option for minimization the reaction path. Table 2 below lists the currently available codes which may be interfaced with ChemDyME and the capabilities of the interface.

| Code | Capabilities | Appropriate ChemDyME  level |
|------|-------------|----------------------------|
| SCINE (Sparrow and ReaDuct)[49,60] | Energies,Forces, Optimizations,Spline, IRC | Trajectory Level<br><br>Initial Optimization Level<br><br>Refinement Level<br><br>Single Point Level |
| DFTB+[61] | Energies, Forces | Trajectory Level<br><br>Single Point Level |
| XTB[62] | Energies, Forces | Trajectory Level<br><br>Single Point Level |
| Gaussian [63] | Energies,Forces, Optimizations | Refinement Level<br><br>Single Point Level |
| Molpro [64,65] | Energies, Forces | Single Point Level |
| NWChem [66] | Energies, Forces | Single Point Level |

**Table 2: External codes which may be used to generate forces or energies in a ChemDyME run**

3)  ME: At the start of every ChemDyME run a template xml input file is created for MESMER. As each reaction is found and characterized, ChemDyME writes the properties for reactant, transition state and product to the input xml and defines the reaction between them. When required ChemDyME updates the starting species in the xml file and runs the MESMER executable. MESMER runs until a reactive transition occurs and ChemDyME then reads in the time associated with the transition along with the identity and energy grain of the product.

In addition to MESMER and electronic structure theory codes, ChemDyME also utilizes the PyBel interface to the Open Babel [67] software to conveniently convert from one chemical format to another. This simplifies bookkeeping. As the Chemical network grows, ChemDyME needs to track each unique species which has been formed, in a convenient way. This is done through converting each new optimized product geometry into a canonical SMILES string using Open Babel. These SMILES strings are used to label each species in the reaction network and are invariant to the exchange of like atoms. These SMILES labels allow ChemDyME to keep track of which reactions have already been discovered and avoid wasted computational effort re-characterizing reactions which have already been identified. Pybel SMILES strings also indicate whether a given set of cartesian coordinates corresponds to multiple molecular fragments and these strings are used by ChemDyME to determine whether a particular reaction is dissociative. PyBel is additionally used to convert optimized cartesian coordinates into Chemical Mark-up Language (.cml) format to aid the creation of the MESMER xml input.

**3.2 Dynamically changing the number of the atoms in the system**

A ChemDyME run does not need to maintain a static number of atoms. When a dissociation occurs ChemDyME is able to continue exploring the reactive chemistry of one of the fragments formed (by default the fragment with the largest number of atoms). As will be discussed in Section 5, ChemDyME is also capable of adding additional bimolecular channels where it is deemed kinetically appropriate.

To add additional species the ChemDyME input defines a number of SMILES strings corresponding to molecules which might be considered part of the reactive bath for the system of interest and for each molecule the user also specifies a characteristic timescale $\tau_{bi}$ associated with estimated rate constant for reaction between the system and the bath molecule. After the MD and OR cycles have been completed for a particular node or species in the reaction network, ChemDyME runs n (default = 10) ME simulations and stores the slowest transition time $\tau_{slowest}$ from these. If $\tau_{slowest} > \tau_{bi}$ for a given bath molecule then it is considered that bimolecular reaction with this species may be competitvie and an additional MD OR cycle is performed looking for reactions between the two species.

For each run in the bimolecular MD cycle, initial cartesian coordinates are created such that the COM's of the two moieties (reactive intermediate and bath) are separated by $x$ Angstrom (default value is 8). To generate these coordinates randomly the coordinates of the first fragment are translated to the origin of the coordinate system. We then generate a random unit vector $\dot{r}$ and $x\dot{r}$ thus defines a random point upon a sphere of radius $x$ about the origin. Finally, we translate the COM of fragment 2 to $x\dot{r}$ to give the starting geometry for the bimolecular MD run. This is shown in Figure 6. The MD run then uses standard BXD on the separation of COM's (as described in 2.3) to gradually bring the two moieties into proximity with one another.

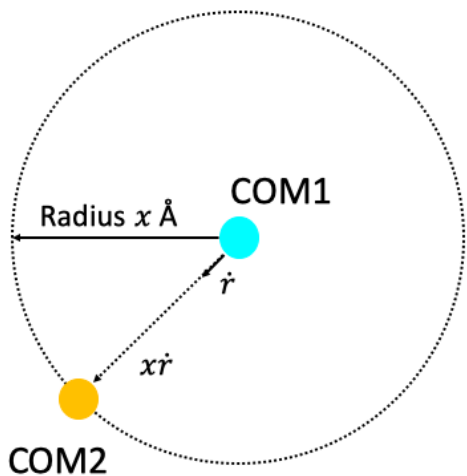

**Figure 6: Procedure for generate coordinates for two separate moieties with centers of mass COM1 and COM2 such that they are randomly separated by x Angstrom.**

## 4. Benchmarking ChemDyME against the well-studied C3H5O atomic system starting from OH + propyne

Propyne and allene are key intermediates in combustion chemistry, originating in both propane flames and through the decomposition of larger alkanes. [68] The reaction of both species with the OH radical can lead to the resonantly stabilized propargyl radical which is of great interest with regards to soot formation mechanisms.[1,69–76] As a result, the $C_3H_5O$ atomic system, particularly initialized from the OH and propyne reaction has been the subject of a great deal of research, both experimentally and theoretically.[1,76]

Theoretically there have been two particularly comprehensive studies studying the OH + propyne mechanism.[1,76] Fitzpatrick[76] performed a manual search for transition states, intermediates and decomposition products formed from the OH + propyne reaction and subsequently Zador and Miller used an automated approach (Kinbot)[1] to thoroughly investigate the $C_3H_5O$ system starting from both OH +

propyne and OH + allene. This automated study focused upon the lowest energy channels and highlighted a small number of channels that were missed by the Fitzpatrick study.

For the purpose of this paper, the OH + propyne system provides an unusually well studied system and thus represents an ideal benchmark for evaluating ChemDyME's use as an automated mechanism generation framework. To this end we have performed ChemDyME runs, starting with separate OH and propyne moieties. Input and output files for these simulations may be found in the examples folder of the ChemDyME github repository https://github.com/RobinShannon/ChemDyME. For the kinetic convergence we have chosen to run ChemDyME with a $T_{ME}$ of 1000K and pressure of 5 bar representative of combustion conditions and a maximum time for kinetic convergence $\tau_{tot}$ of 1s.  With regards to the MD portion of the ChemDyME workflow, we chose a $T_{MD}$ of 1000K also to promote the discovery of lower energy pathways in the MD reaction finding step. A comparison of the various products formed from different $T_{MD}$ can be found in the supporting information section S1.

The levels of theory used in the ChemDyME run are tabulated in Table 3. PM6 calculations are performed using the SCINE Sparrow code,[60] dftb2 calculations are performed using the DFTB+ code,[61]  all DFT calculations are performed using the Gaussian 09 suite of programs [63] and the CCSD(T) calculations are performed using Molpro[64,65] The key requirements of the trajectory level of theory are that is fast (normally necessitates a semi-empirical method) and that it qualitatively orders the possible reaction channels correctly in terms of their kinetic importance. In section S1 of the supporting information we show results from some initial tests where we compared MD runs from two different levels of theory. From these tests we find that with low $T_{MD}$, the dftb2 method as implemented in DFTB+ is more likely to find the lowest energy pathways when compared to PM6 in Scine. For the "low-level" optimizations in the OR part of the work flow the PM6 method in is used, followed by "high-level" optimizations at the B3LYP/3-21G level of theory. Finally, single point energies are calculated at the CCSD(T)-f12/cc-PVDZ level. This heirarchy of levels chosen prioritizes rapid exploration of the reaction network and is of significantly lower accuracy than the calculations employed to generate the Kinbot surface. However, as we shall show, these levels of theory are perfectly sufficient for ChemDyME to find the important chemistry in the OH + propyne system.

| Trajectory Level | DFTB2 |
|---|---|
| Initial Optimization Level | PM6 |
| Refinement Level | B3LYP/3-21G |
| Single Point Level | CCSD(T)-f12/cc-PVDZ |

**Table 3: Levels of theory used for the propyne + OH, ChemDyME calculations**

The network resulting from the ChemDyME run is shown in Figure 7 and those reactions which were also identified as important by Kinbot are colored in orange (the bold reactions from Figure 1 of the Kinbot work[1]) or red (other reactions discovered by Kinbot but not considered kinetically important) and reactions discovered by ChemDyME but not present in the Kinbot network are shown in cyan. This ChemDyME run, tracks the bold reactions found by Kinbot, rapidly forming the wells W4 and W10 before eventually (after many redissociation events back to OH + propyne) forming $CH_3$ via well skipping reactions across the two intermediates W11 and W1. The x axis marks the logarithmic kinetic time at which each species is visited by the stochastic master equation and species or nodes marked at the inf mark were not visited within the ChemDyME run. This nicely illustrates the principal of kinetic convergence: by tracking the kinetic evolution of the growing network, it is possible to cease the exploration once all the kinetically viable species have been visited, simplifying the overall reaction network.

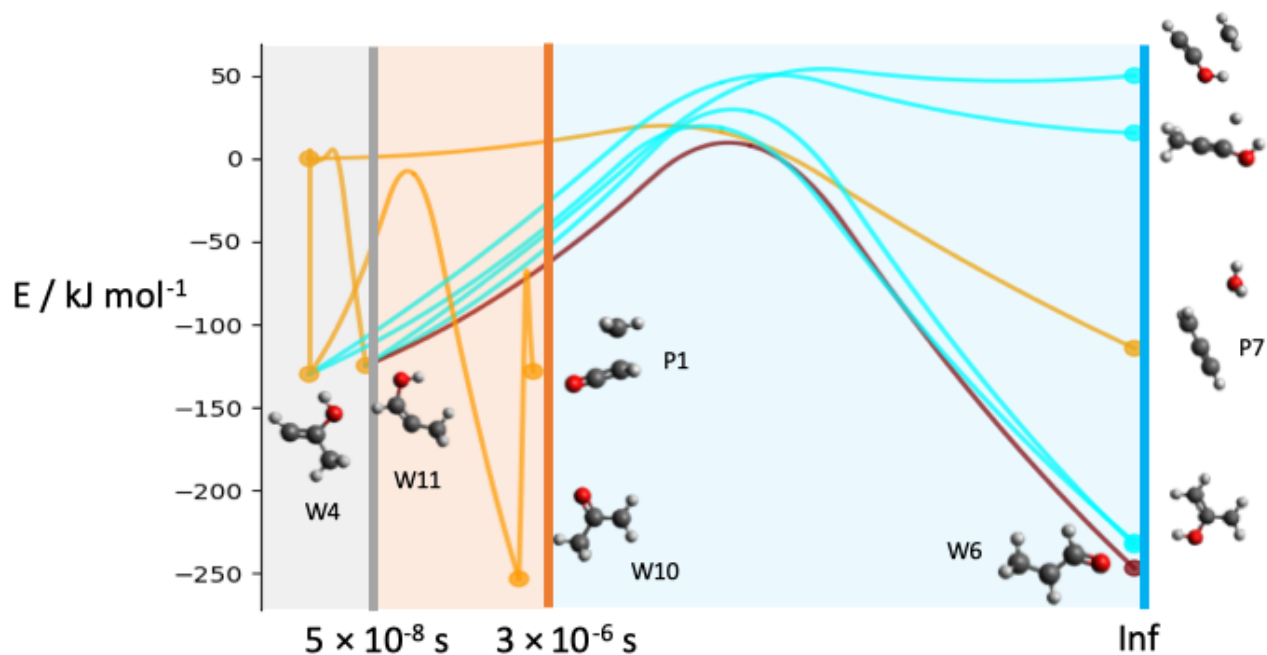

**Figure 7: ChemDyME network for the OH + propyne system. The x axis representing the reaction coordinate additionally reflects the time at which each species is first visited in the stochastic master equation and is spaced logarithmically to better view the full range of timescales. The lines delineate and background shading delineate important timescales and the inf label denotes that this species was not visited observed in the stochastic master equation trajectory. Text labels on the different species give the nomenclature for that species from the Kinbot study[1] where available.**

Figure 7 shows that ChemDyME prioritizes the bold reactions discovered by Kinbot and inspection of the network reveals that only two of these bold reactions originating from OH + propyne are missed by the above ChemDyME network. Thus far we have used the bold reactions from the Kinbot study as a metric for those which are kinetically important under combustion combustions but upon closer inspection these bold reactions may be pruned further and the bold two channels "missed" in the ChemDyME network appear to be uncompetitive.

The bold reactions missing from the ChemDyME network are alternative isomerization product channels from intermediate W10. The MD simulations using DFTB2 do not find these product channels, even when running separate calculation with $T_{MD}$ set to 4000K, however both these channels are higher in energy than the dissociation reaction to form P1 are and entropically "tighter" rendering them uncompetitive with the dissociation channel. Indeed, the Kinbot paper, identifies that "The acetonyl radical (W10) decomposes exclusively to ketene + $CH_3$" [1]. Ignoring the initial initial van der Waals complex found by kinbot, which is discussed below, we argue that ChemDyME does identify all the kinetically important channels from the Kinbot network.

As mentioned above, the other feature missing from the ChemDyME network is the initial van der Waals complex formed between the OH and propyne moieties and the subsequt transition states from this Van der Waals complex. The difficulty of treating van der Waals complexes is a known limitation of ChemDyME currently and this will be addressed in the future. The full master equation xml file resulting from these ChemDyME simulations is in section S3 of the supporting information.

Table 4 compares the energies of the key stationary points in the ChemDyME and Kinbot networks. Here we can see, that despite the much lower level of theory used in the current work, the agreement in energies for the stable species is very good. There are larger discrepancies between the energies of the saddle point structure with the ChemDyME value underestimating the more accurate Kinbot values, although it is noted that the order of the barrier heights is correct in the ChemDyME case. The ChemDyME results could be easily refined through re-optimizing the stationary points with a larger basis, but for the purpose of this work (testing whether ChemDyME finds the important chemistry) the current level of theory is sufficient.

|  | Current work energy / kJ mol$^{-1}$ | Kinbot energy / kJ mol$^{-1}$ |
|---|---|---|
| TS PVDW->W4 | - | 3.51 |
| TS PVDW->W11 | - | 5.59 |
| TS P->P7 | -9.95 | 8.36 |
| P7 | -114.32 | -114.20 |
| W4 | -124.77 | -121.39 |
| W11 | -129.94 | -130.29 |
| TS W11->W10 | -9.37 | 6.02 |
| W10 | -253.32 | -250.42 |
| TS W10 ->P1 | -70.26 | -82.18 |
| P1 | -128.31 | -127.70 |
| TS W10->W9* | -45.05 | -29.09 |
| W9* | -231.13 | -227.52 |
| TS W10->W8* | -32.94 | -20.69 |
| W8* | -256.46 | -251.76 |

Table 4: Comparison of zero point corrected stationary point energies between the current work (CCSD(T)-f12/cc-PVDZ//B3LYP/3-21G) and the Kinbot study (CCSD(T)-F12b/cc-PVQZ-F12//M06-2X/6-311++G(d,p) . The species marked with asterisks were not found in the initial ChemDyME run, but were found from running separate BXDE trajectories using PM6 rather than dftb2 for forces.

There are a few channels ChemDyME discovers, which are not present in the Kinbot network. Mostly these appear (Figure 7) to have barriers in excess of the Kinbot energy cutoff and thus would have been discounted in the Kinbot study. There are other channels observed for which saddle points couldn't be found and hence they are not shown in Figure 6. The initial adduct W4 is found dynamically to dissociate to P7 with the OH moiety pulling off another hydrogen as it dissociates and W4 and W11 are observed to isomerize directly to one another with the OH moiety dissociating and then rapidly transferring carbons. Further studies are needed to ascertain whether these channels really are operable, and it is unlikely they will significantly affect the wider kinetics of this system; however, this does highlight the potential of a dynamical method to observe roaming type channels. [77] Recent work has shown that these dynamical types of reaction, play a larger role in combustion chemistry then previously assumed[78] and the use of MD in ChemDyME means that such channels may be observed, even when there is no defined transition state for the reaction. A videos showing an MD trajectories for one of these dynamical processes is given in the supplementary information.

## 5. Exploring the isobutyl + nO2 surface to for a highly oxygenated molecule (HOM)

The second system we have studied with ChemDyME in the current work, is that of 1-methyl-hexene (1MHE) + ozone (+nO$_2$)[79–81] a prototypical oxidation system for the formation of highly oxygenated organic molecules (HOM)[82–86] in the earth's atmosphere. Such oxidation systems are a key source of atmospheric aerosol.

At the lower temperatures of the earth's atmosphere, bimolecular association reactions are extremely prevalent, and these, typically barrierless, processes are challenging to include in automated approaches. Accurate rate coefficients for a barrierless reaction, usually require variational transition state theory calculations to be performed[87] and such calculations are complicated by the multireference character of the potential energy surface for associations involving radicals or bimolecular oxygen.

The BXD based sampling of associations used in ChemDyME can in principle, be used to generate rate coefficients. This is something we intend to add to future versions of ChemDyME, however, we recognize that the semiempirical methods used here for the dynamics cannot be relied upon for accurate rate estimations. Instead, we take the pragmatic approach of utilizing the inverse Laplace transform method.[50] This relies upon the fact that rate coefficients for barrierless associations are usually relatively independent of temperature and have high pressure limiting values in a relatively small window, when compared to processes which have a defined saddle point.[88] For the purposes of exploring kinetically feasible reaction channels, we simply wish to know for a given species, whether bimolecular association with some bath gas molecule in the system, is competitive with the fastest of its unimolecular loss process. Since the oxygen concentration in the earth's atmosphere is around $1 \times 10^{18}$ and the rate coefficients for bimolecular association with oxygen are typically on the order of $1 \times 10^{-12}$ $cm^3 s^{-1}$ molecule$^{-1}$,[88,89] a typical reaction rate with the bath $\tau_{O2}$ was set to $1 \times 10^6 s^{-1}$. This means that any species in the network which exhibited a ME transition time $\tau_{trans} > \tau_{O2}$ was considered a candidate for $O_2$ addition and BXD association trajectories were performed.

In what follows we discuss results of a ChemDyME run initialized with separated 1-methyl-hexene and ozone moieties. For this run, the "Trajectory Level" and the "Initial Optimization Level" are identical to those used OH + propyne simulations (dftb2 in DFTB+ and PM6 in SCINE Sparrow). The "Refinement Level" is also PM6  however given the open shell singlet structure of the initial part of the network, Gaussian is used for these calculations to ensure the unrestricted, broken symmetry solution to the wavefuntion.[90] CCSD(T) calculations would be extremely time consuming for a system of this potential size so for this run the "Refinement Level" is chosen to be UB3LYP/6-31+G** in Gaussian and again the broken symmetry solution was enforced with the guess=mix keyword for all closed shell species. The master equation temperature and pressure $T_{ME}$, $P_{ME}$ are set to 298K and 1 atm respectively, representative of the earth's lower atmosphere and $t_{max}$ is set to 1000s to represent the longer timescales at play in the earth's atmosphere. $T_{md}$ is set to 1000 K as a compromise between promoting low energy pathways and speed of reaction finding. The system is then considered to be in a bath of $O_2$ as described above and whenever a particular species or node exhibits a lifetime of $> 1 \times 10^6 s^{-1}$ from a KMC ME run, another oxygen molecule is introduced to the system and potential association pathways are sampled. The input and output files may be found in the examples folder of the ChemDyME repository at https://github.com/RobinShannon/ChemDyME

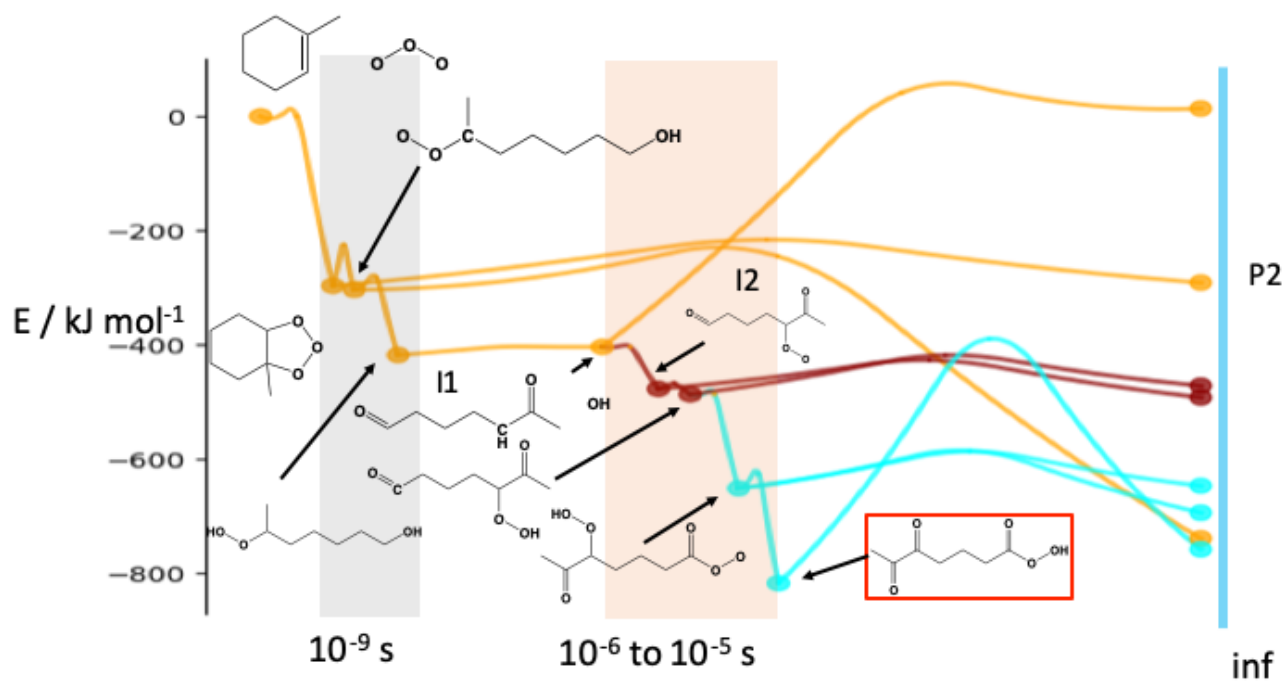

**Figure 8: ChemDyME network for the ozone + 1MHE system. The x axis representing the reaction coordinate additionally reflects the time at which each species is first visited in the stochastic master equation however these have been modified slightly to better separate the different nodes. The background shading shows the approximate timescale for different clusters of nodes and the inf label denotes that this species was not visited/observed in the stochastic master equation trajectory. The different colored edges denote the number of $O_2$ moieties added to the system, with orange = 0, red = 1 and cyan = 2. Labels and the red box highlights the ultimate product species. Key intermediates are labelled. A full figure including all skeletal structures in found in section S2 of the supporting information.**

Figure 8 shows the full network from this ChemDyME simulation with each additional $O_2$ addition colored differently. The ChemDyME run follows much of the expected chemistry. Ozone readily associates with 1MHE forming a 5 membered ring containing the three oxygen atoms. One of the two O-O bonds then cleaves to form a Criegee intermediate species and subsequent hydrogen transfer leads to OH formation. O2 can then bind to the lone pair of the resulting radical co-product to form a peroxy radical. Then we observe the standard peroxy radical cycle of hydrogen transfer to a hydroperoxy followed by either dissociation or O2 addition to the radical center forming a new peroxy radical species.

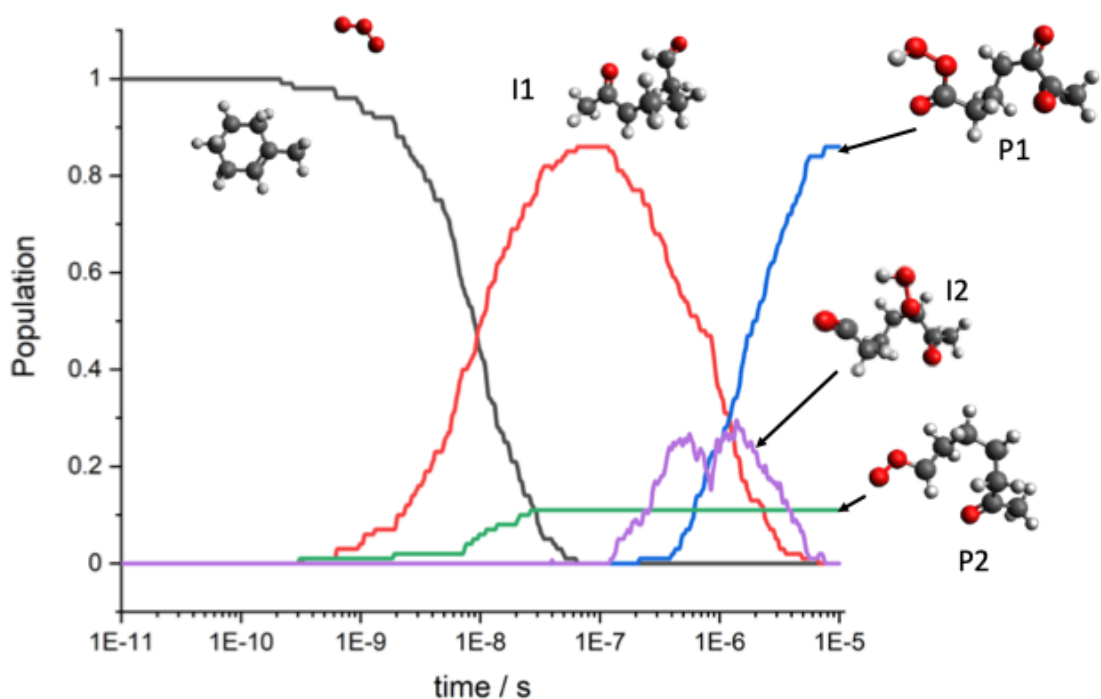

**Figure 9: Evolution of key species populations from stochastic master equation simulations. These master equation simulations consist of 100 stochastic trials and a full input file is given in the supporting information**

In this single ChemDyME run, after an OH elimination and two O₂ additions the run terminates with a second OH elimination forming the species outlined in red In Figure 8 (P1). Interestingly at the PM6 level the transition state for this reaction appears to be concerted with both H transfer and OH loss occurring simultaneously. A video showing the IRC corresponding to this saddle point is given in the online supporting information.

From the MESMER input file resulting from this single ChemDyME run, it is possible to run a full master equation simulation to track the time evolution of the species in the system. Master equations simulations involving multiple bimolecular additions have rarely been performed[89,91–95] although there is now an established method for adding bimolecular reaction to activated species in a master equation in such a way as to satisfy detailed balance.[92,96] Usually, a matrix-based approach would be the preferred method for solving the master equation in MESMER, but given size of the system in terms of species and energy span, we instead use the stochastic BXK algorithm. Figure 9 shows the time evolution of key species in the system from MESMER calculations consisting of 100 stochastic trials. These trials cover a short timescale of up to $10^{-5}$ s, however considering the channels currently found from the ChemDyME run, this timescale covers the important chemistry of the mechanism in Figure 8. The full MESMER xml file for these master equation simulations in given in section S4 of the supporting information.

Figure 9 demonstrates that two intermediates (I1 and I2) form substantial populations before the majority of the population funnels to the final product P1. A minor channel corresponding to 11% of the total population forms product P2 (Highlighted in green in Figure 7) from an alternate cleavage path of the ozone + 1MHE adduct. Additional ChemDyME runs explore this channel further and in particular find the dominant pathways in Figure 10, however this part of the surface has not been explored in any more detail in the current work as quick gaussian calculations by hand identify that the ground electronic state crosses to triplet upon $O_2$ addition and the current ChemDyME simulation is set to treat species as either singlet or doublet depending upon the electron number.

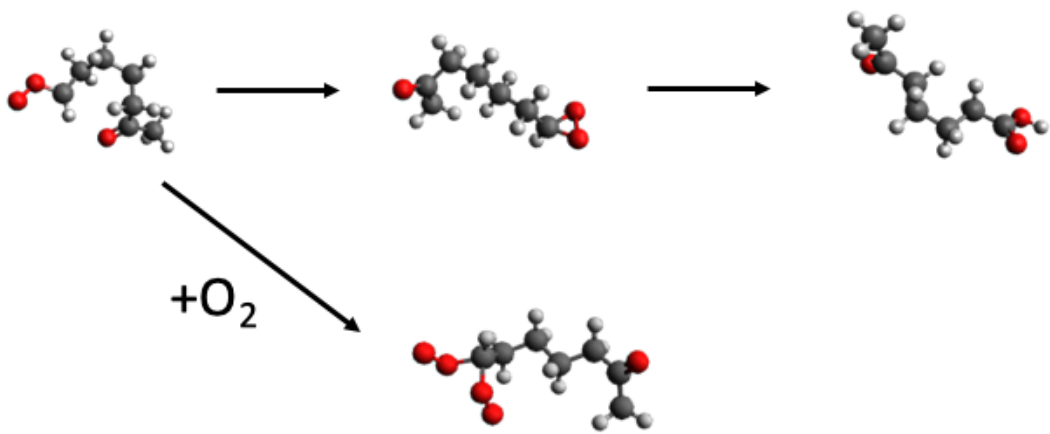

**Figure 10: Schematic of key species in an alternate Criegee intermediate pathway found from additional ChemDyME runs.**

While the current network captures much of the expected chemistry, more MD trials ($n_{MD}$), and higher levels of theory would be required for a comprehensive investigation of the 1MHE network, especially given the complexity of the electronic structure. However, within the scope of the current work, these results demonstrate the novel ability of our ChemDyME approach to map the kinetic evolution of a complex system through both dissociation and association processes. Despite the low level of theory used, the ChemDyME simulation captures the chemistry typical of HOM formation and atmospheric oxidation and despite a combinatorial explosion of possible reactions (given the size of the system and the prospect of multiple $O_2$ additions) the kinetic convergence criterion helps to identify the important chemistry at the level of theory used.

## 6. Conclusions

In this work we demonstrate the capabilities of the ChemDyME approach for exploring the kinetics of reactive mechanism in an automated manner. Through the combined use of molecular dynamics and master equation simulations, ChemDyME narrows the search for chemical reactions, prioritizing those which are kinetically the most important under the conditions of interest. Through benchmark calculations against an OH + propyne combustion network which has already been pruned to neglect high energy pathways, we demonstrate that ChemDyME prunes this network further to find the key reactions.

By utilizing the concept of "kinetic convergence" ChemDyME can add additional molecular moieties to the system, when it is deemed kinetically favorable. We demonstrate this with a prototypical ozonolysis / oxidation system and show that ChemDyME is capable of tracking the network though both the loss of an OH radical and through multiple subsequent $O_2$ additions to form an example highly oxygenated molecule (HOM). The ChemDyME output produces all the input required to run a master equation on the system, and we present the time evolution of an activated oxidation system of unprecedented size giving the timescale of HOM formation.

This work demonstrates the strengths of the ChemDyME approach and of BXDE in general. Previous work using BXDE[6,43] has shown its utility in accelerating reactive MD whilst preserving some sense of the important low energy chemistry. By harnessing both BXDE MD and the energy grained master equation, this work shows that ChemDyME can be used to heavily simplify the reaction of network of interest, prioritizing the discovery of the small subset of reactions which are kinetically important at the conditions of interest. ChemDyME also has the benefit of being a dynamical approach which can give a wider picture of the chemical reactivity of a system when compared to knowledge of the key transition states alone. Finally, the dynamical nature of ChemDyME also makes the approach highly generalizable compared to heuristic based approaches.

Naturally there are also weaknesses to the ChemDyME approach. Currently ChemDyME cannot treat Van der Waals complexes and the reaction finding algorithm does not always identify when multiple reactions have occurred. Advances in both areas have been in the AutoMeKin [25,48,97] software and these improvements are planned for future releases of ChemDyME.

## 7. Acknowledgements

This work was undertaken on ARC3, part of the High Performance Computing facilities at the University of Leeds, UK and using the computational facilities of the Advanced Computing Research Centre, University of Bristol - http://www.bris.ac.uk/acrc/. RJS and DRG acknowledge support of this work through the "CHAMPS" EPSRC program grant EP/P021123/1. D.R.G. acknowledges funding from the Royal Society as a University Research Fellow

## Supporting information

## S1. Comparison of BXDE trajectory products at different temperatures and with different methods.

Figures S1 and S2 show the product species from reactive BXDE trajectories initialized from intermediate W4 in the propyne + OH mechanism. Figure S1 compares products from DFTB2(red writing) and PM6(black writing) trajectories (from SCINE Sparrow and DFTB+ respectively) with a Langevin thermostat set to 1000K whilst Figure S2 compare the same methods but with the thermostat set to 4000K. The species circled in red (intermediate W10) is the reaction channel with the lowest barrier and it can be seen that whilst DFTB2 finds this product channel at both temperatures, PM6 trajectories fail to observe it at all. PM6 trajectories at low temperatures also miss another hydrogen transfer channel, instead favoring dissociative product channels. When comparing different temperatures, it is found that 4000 K MD simulations find a much larger array of exotic dissociation channels, but are less likely to observe the low barriered but entropically tight, hydrogen transfer reactions which dominate the kinetics of the system under combustion conditions.

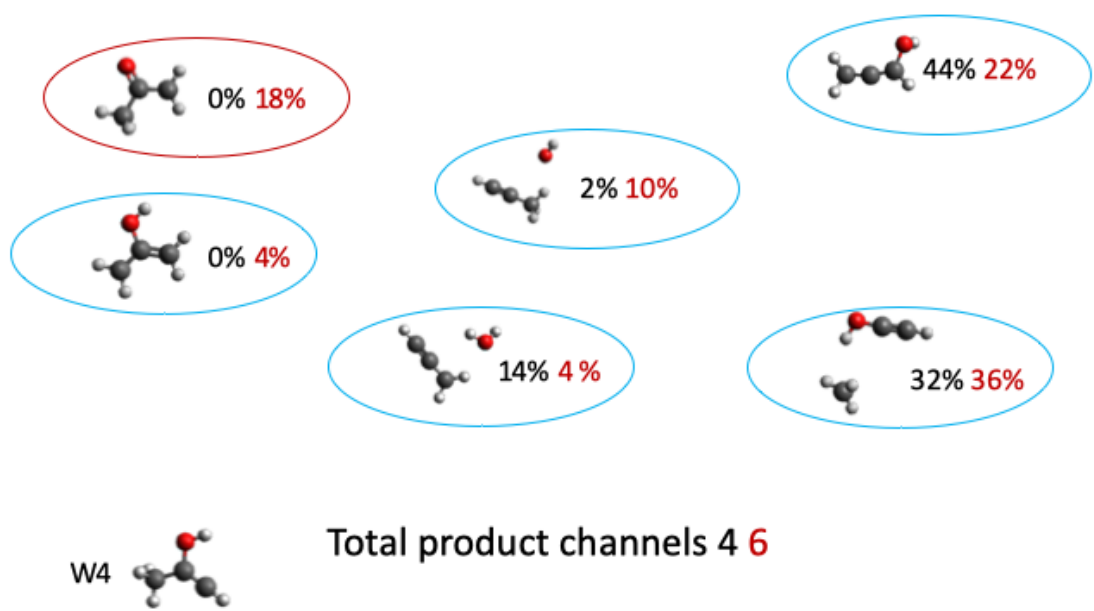

**Figure S1: Products from 50 reactive BXDE trajectories starting from species W4 with a Langevin thermostat set to 1000K. Red writing indicates DFTB2 simulations and black writing indicates PM6. The red circled product corresponds to the lowest energy barrier.**

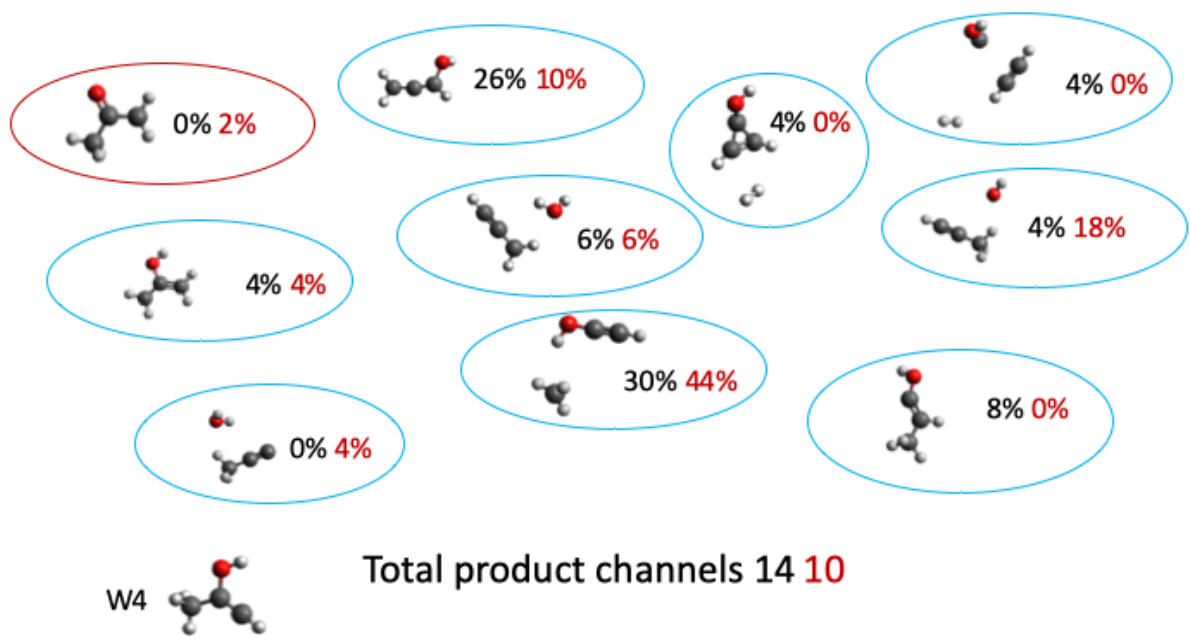

**Figure S2: Products from 50 reactive BXDE trajectories starting from species W4 with a Langevin thermostat set to 4000K. Red writing indicates DFTB2 simulations and black writing indicates PM6. The red circled product corresponds to the lowest energy barrier.**

## S2. More detailed mechanism the 1MHE + ozone network

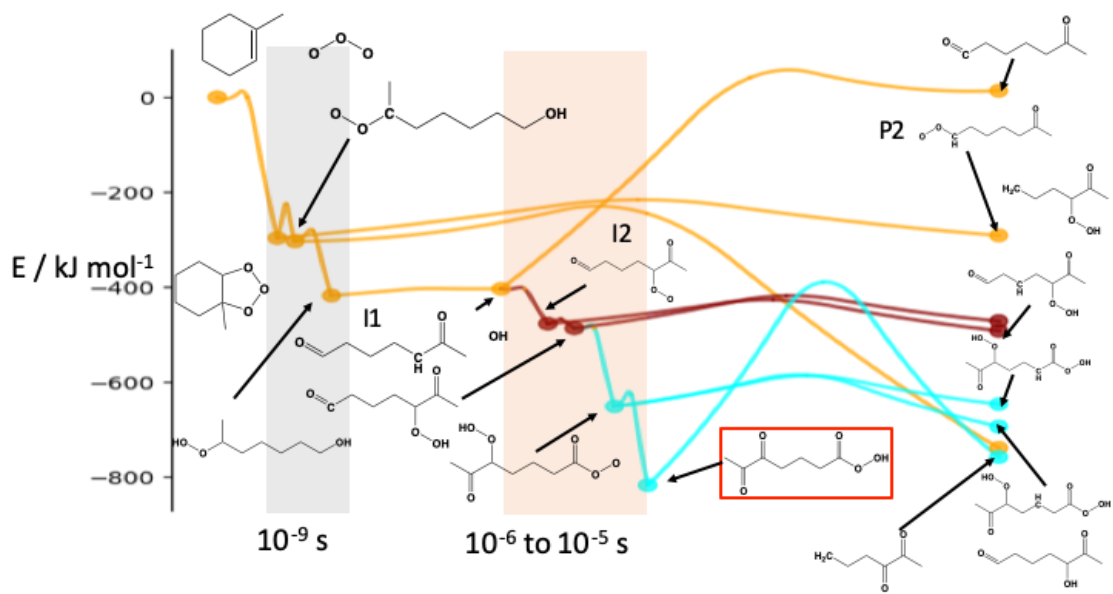

**Figure S2: ChemDyME network for the ozone + 1MHE system. The x axis representing the reaction coordinate additionally reflects the time at which each species is first visited in the stochastic master equation however these have been modified slightly to better separate the different nodes. The background shading shows the approximate timescale for different clusters of nodes and the inf label denotes that this species was not visited/observed in the stochastic master equation trajectory. The different colored edges denote the number of $O_2$ moieties added to the system, with orange = 0, red = 1 and cyan = 2. Labels and the red box highlights the ultimate product species. Key intermediates are labelled. A full figure including all skeletal structures in found in section S2 of the supporting information.**

## S3. OH + propyne MESMER input

<?xml version="1.0" ?>

<me:mesmer xmlns="http://www.xml-cml.org/schema"

xmlns:me="http://www.chem.leeds.ac.uk/mesmer"

xmlns:xsi="http://www.w3.org/2001/XMLSchema-instance">

        `<title>` Glyoxal`</title>`

        `<moleculeList>`

            `<molecule description="Nitrogen" id="N2">`

                `<atom elementType="N2"/>`

                `<propertyList>`

                    `<property dictRef="me:epsilon">`

                        `<scalar>82.0</scalar>`

                    `</property>`

                    `<property dictRef="me:sigma">`

                        `<scalar>3.74</scalar>`

                    `</property>`

                    `<property dictRef="me:MW">`

                        `<scalar units="amu">28.0</scalar>`

                    `</property>`

                `</propertyList>`

            `</molecule>`

            `<molecule id="CC#C">`

                `<atomArray>`

                    `<atom elementType="C" id="a1" x3="0.949211" y3="-0.015143" z3="0.031063"/>`

                    `<atom elementType="C" id="a2" x3="0.461350" y3="-0.027203" z3="-1.348609"/>`

                    `<atom elementType="C" id="a3" x3="0.059521" y3="-0.037142" z3="-2.485403"/>`

                    `<atom elementType="H" id="a4" x3="0.595937" y3="-0.898104" z3="0.577027"/>`

                    `<atom elementType="H" id="a5" x3="0.595938" y3="0.877227" z3="0.561511"/>`

                    `<atom elementType="H" id="a6" x3="2.045500" y3="-0.014926" z3="0.056683"/>`

                    `<atom elementType="H" id="a7" x3="-0.294274" y3="-0.045908" z3="-3.486058"/>`

                `</atomArray>`

                `<bondArray>`

```xml
            <bond atomRefs2="a7 a3" order="1"/>

            <bond atomRefs2="a3 a2" order="3"/>

            <bond atomRefs2="a2 a1" order="1"/>

            <bond atomRefs2="a1 a6" order="1"/>

            <bond atomRefs2="a1 a5" order="1"/>

            <bond atomRefs2="a1 a4" order="1"/>

        </bondArray>

        <propertyList>

            <property dictRef="me:lumpedSpecies">

                    <array> </array>

            </property>

            <property dictRef="me:vibFreqs">

                    <array units="cm-1">351.9936 352.1391 718.542 719.7102 924.5857
1105.2979    1105.4637    1472.9988    1551.0676    1551.2107    2252.0367    3045.1027    3101.5132    3101.5575
3499.0553</array>

            </property>

            <property dictRef="me:ZPE">

                    <scalar units="kJ/mol">0.0</scalar>

            </property>

            <property dictRef="me:spinMultiplicity">

                    <scalar units="cm-1">1</scalar>

            </property>

            <property dictRef="me:epsilon">

                    <scalar>473.17</scalar>

            </property>

            <property dictRef="me:sigma">

                    <scalar>5.09</scalar>

            </property>

            <group>

                    <scalar>1</scalar>

            </group>

        </propertyList>

        <me:energyTransferModel xsi:type="me:ExponentialDown">
```

```xml
                    <scalar units="cm-1">250</scalar>
                </me:energyTransferModel>
            </molecule>
            <molecule id="[OH]" spinMultiplicity="2">
                <atomArray>
                    <atom    elementType="O"    id="a1"    spinMultiplicity="2"    x3="1.027749"
y3="0.080157" z3="0.047252"/>
                    <atom    elementType="H"    id="a2"    x3="2.045522"    y3="0.080157"
z3="0.047252"/>
                </atomArray>
                <bondArray>
                    <bond atomRefs2="a1 a2" order="1"/>
                </bondArray>
                <propertyList>
                    <property dictRef="me:lumpedSpecies">
                        <array> </array>
                    </property>
                    <property dictRef="me:vibFreqs">
                        <array units="cm-1">3251.7758</array>
                    </property>
                    <property dictRef="me:ZPE">
                        <scalar units="kJ/mol">0.0</scalar>
                    </property>
                    <property dictRef="me:spinMultiplicity">
                        <scalar units="cm-1">2</scalar>
                    </property>
                    <property dictRef="me:epsilon">
                        <scalar>473.17</scalar>
                    </property>
                    <property dictRef="me:sigma">
                        <scalar>5.09</scalar>
                    </property>
                </propertyList>
```

```
<me:energyTransferModel xsi:type="me:ExponentialDown">
        <scalar units="cm-1">250</scalar>
    </me:energyTransferModel>
</molecule>
<molecule id="TS_CC#C_[CH2]C#C" spinMultiplicity="2">
    <atomArray>
        <atom elementType="C" id="a1" x3="-7.175975" y3="0.383386" z3="2.037172"/>
        <atom elementType="C" id="a2" x3="-7.502178" y3="0.704736" z3="0.682246"/>
        <atom elementType="H" id="a3" x3="-6.107035" y3="0.436621" z3="2.260837"/>
        <atom elementType="H" id="a4" x3="-7.601217" y3="-0.566247" z3="2.378057"/>
        <atom elementType="H" id="a5" x3="-7.717621" y3="1.189470" z3="2.809769"/>
        <atom elementType="C" id="a6" x3="-7.795809" y3="1.015154" z3="-0.450813"/>
        <atom elementType="H" id="a7" x3="-8.052301" y3="1.274885" z3="-1.447736"/>
        <atom elementType="O" id="a8" spinMultiplicity="2" x3="-7.935072" y3="2.028005" z3="3.776249"/>
        <atom elementType="H" id="a9" x3="-7.266514" y3="2.752672" z3="3.562006"/>
    </atomArray>
    <bondArray>
        <bond atomRefs2="a7 a6" order="1"/>
        <bond atomRefs2="a6 a2" order="3"/>
        <bond atomRefs2="a2 a1" order="1"/>
        <bond atomRefs2="a1 a3" order="1"/>
        <bond atomRefs2="a1 a4" order="1"/>
        <bond atomRefs2="a1 a5" order="1"/>
        <bond atomRefs2="a9 a8" order="1"/>
    </bondArray>
    <propertyList>
        <property dictRef="me:vibFreqs">
```

<array    units="cm-1">132.17294468978673    208.81306042232816
361.35053914984246    377.0879001971972    516.2787979277084    540.2190819728357    783.5874913649636
910.9203454087667    917.1986053798408    937.8176080081103    997.2617366826861    1096.5756612662972
1170.0439214519047    1193.497135288244    1277.6924216064695    2312.7362457132517    2557.0867445970134
2691.070177554716    2745.9204200428535    2800.1601001857857</array>

            </property>

            <property dictRef="me:imFreq">

                    <scalar units="cm-1">1929.229602172262</scalar>

            </property>

            <property dictRef="me:spinMultiplicity">

                    <scalar units="cm-1">2</scalar>

            </property>

            <property dictRef="me:ZPE">

                    <scalar units="kJ/mol">15.80</scalar>

            </property>

            <group>

                    <scalar>1</scalar>

            </group>

        </propertyList>

    </molecule>

    <molecule id="[CH2]C#C" spinMultiplicity="2">

        <atomArray>

                <atom   elementType="C"   id="a1"   spinMultiplicity="2"   x3="1.168308"   y3="-
0.037786" z3="0.012789"/>

                <atom    elementType="H"    id="a2"    x3="0.609528"    y3="0.793684"
z3="0.428252"/>

                <atom   elementType="H"   id="a3"   x3="0.609551"   y3="-0.859211"   z3="-
0.422227"/>

                <atom    elementType="C"    id="a4"    x3="2.536353"    y3="-0.049981"
z3="0.036499"/>

                <atom    elementType="C"    id="a5"    x3="3.762118"    y3="-0.060920"
z3="0.057736"/>

                <atom    elementType="H"    id="a6"    x3="4.823516"    y3="-0.070339"
z3="0.076026"/>

        </atomArray>

        <bondArray>

```xml
                <bond atomRefs2="a3 a1" order="1"/>

                <bond atomRefs2="a1 a4" order="1"/>

                <bond atomRefs2="a1 a2" order="1"/>

                <bond atomRefs2="a4 a5" order="3"/>

                <bond atomRefs2="a5 a6" order="1"/>

        </bondArray>

        <propertyList>

                <property dictRef="me:lumpedSpecies">

                        <array> </array>

                </property>

                <property dictRef="me:vibFreqs">

                        <array units="cm-1">351.5125  409.1058  517.241  686.8534  702.8162
1070.851 1092.0666 1504.1019 2040.4579 3157.2748 3244.8073 3486.1251</array>

                </property>

                <property dictRef="me:ZPE">

                        <scalar units="kJ/mol">-114.32199848723093</scalar>

                </property>

                <property dictRef="me:spinMultiplicity">

                        <scalar units="cm-1">2</scalar>

                </property>

                <property dictRef="me:epsilon">

                        <scalar>473.17</scalar>

                </property>

                <property dictRef="me:sigma">

                        <scalar>5.09</scalar>

                </property>

                <group>

                        <scalar>1</scalar>

                </group>

        </propertyList>

        <me:energyTransferModel xsi:type="me:ExponentialDown">

                <scalar units="cm-1">250</scalar>
```

```
            </me:energyTransferModel>
        </molecule>
        <molecule id="O">
            <atomArray>
                <atom      elementType="O"     id="a1"     x3="0.925269"    y3="0.044414"
z3="0.061097"/>
                <atom     elementType="H"    id="a2"    x3="0.638551"    y3="0.854808"    z3="-
0.443130"/>
                <atom      elementType="H"     id="a3"      x3="1.920710"      y3="0.085023"
z3="0.035830"/>
            </atomArray>
            <bondArray>
                <bond atomRefs2="a2 a1" order="1"/>
                <bond atomRefs2="a3 a1" order="1"/>
            </bondArray>
            <propertyList>
                <property dictRef="me:lumpedSpecies">
                    <array> </array>
                </property>
                <property dictRef="me:vibFreqs">
                    <array units="cm-1">1692.2943 3415.4019 3556.0709</array>
                </property>
                <property dictRef="me:ZPE">
                    <scalar units="kJ/mol">0.0</scalar>
                </property>
                <property dictRef="me:spinMultiplicity">
                    <scalar units="cm-1">1</scalar>
                </property>
                <property dictRef="me:epsilon">
                    <scalar>473.17</scalar>
                </property>
                <property dictRef="me:sigma">
                    <scalar>5.09</scalar>
```

```xml
            </property>

        </propertyList>

        <me:energyTransferModel xsi:type="me:ExponentialDown">

            <scalar units="cm-1">250</scalar>

        </me:energyTransferModel>

    </molecule>

    <molecule id="TS_CC#C_C[C]=CO" spinMultiplicity="2">

        <atomArray>
                            <atom        elementType="C"        id="a1"        x3="-7.899822"        y3="1.247084"
z3="0.669474"/>

                            <atom        elementType="C"        id="a2"        x3="-6.504138"        y3="1.312615"
z3="0.940165"/>

                            <atom        elementType="H"        id="a3"        x3="-8.373126"        y3="1.239873"
z3="1.662485"/>

                            <atom        elementType="H"        id="a4"        x3="-8.217786"        y3="0.335695"
z3="0.133755"/>

                            <atom        elementType="H"        id="a5"        x3="-8.282558"        y3="2.119472"
z3="0.112252"/>

                            <atom        elementType="C"        id="a6"        x3="-5.340011"        y3="1.384989"
z3="1.300266"/>

                            <atom        elementType="H"        id="a7"        x3="-4.314576"        y3="1.420634"
z3="1.329373"/>

                            <atom        elementType="O"        id="a8"        spinMultiplicity="2"        x3="-5.329423"
y3="1.683320" z3="3.659212"/>

                            <atom        elementType="H"        id="a9"        x3="-6.148480"        y3="1.617357"
z3="4.233809"/>

        </atomArray>

        <bondArray>

                <bond atomRefs2="a5 a1" order="1"/>

                <bond atomRefs2="a4 a1" order="1"/>

                <bond atomRefs2="a1 a2" order="1"/>

                <bond atomRefs2="a1 a3" order="1"/>

                <bond atomRefs2="a2 a6" order="3"/>

                <bond atomRefs2="a6 a7" order="1"/>

                <bond atomRefs2="a8 a9" order="1"/>

        </bondArray>
```

```
<propertyList>
    <property dictRef="me:vibFreqs">
        <array      units="cm-1">56.67405370730133       99.9090752685441
385.5475169473353    387.3403841292845    391.43380629053684    510.3845433868839    790.6862854023864
799.0476225355512    1012.0970067272162    1013.8284412827389    1126.5061362683737    1210.8500777530166
1237.3543295923785    1338.4510917250739    2316.277312980587    2605.6452649950197    2680.831283156513
2702.972349573537 2784.844467854648 2888.9414662994595</array>
    </property>
    <property dictRef="me:imFreq">
        <scalar units="cm-1">207.35300052235976</scalar>
    </property>
    <property dictRef="me:spinMultiplicity">
        <scalar units="cm-1">2</scalar>
    </property>
    <property dictRef="me:ZPE">
        <scalar units="kJ/mol">4.126834987442225</scalar>
    </property>
    <group>
        <scalar>1</scalar>
    </group>
</propertyList>
</molecule>
<molecule id="C[C][CH]O" spinMultiplicity="2">
    <atomArray>
        <atom      elementType="C"      id="a1"      x3="-7.074817"      y3="0.655860"
z3="1.058700"/>
        <atom      elementType="C"      id="a2"      spinMultiplicity="2"      x3="-5.695742"
y3="1.061308" z3="1.390010"/>
        <atom      elementType="H"      id="a3"      x3="-7.640264"      y3="0.354363"
z3="1.958510"/>
        <atom      elementType="H"      id="a4"      x3="-7.087114"      y3="-0.197858"
z3="0.368255"/>
        <atom      elementType="H"      id="a5"      x3="-7.631786"      y3="1.476271"
z3="0.587093"/>
        <atom      elementType="C"      id="a6"      x3="-4.843951"      y3="1.208380"
z3="2.377799"/>
```

```xml
                                <atom    elementType="H"    id="a7"    x3="-3.830273"    y3="1.559755"
z3="2.218597"/>

                                <atom    elementType="O"    id="a8"    x3="-5.027502"    y3="0.965869"
z3="3.760903"/>

                                <atom    elementType="H"    id="a9"    x3="-5.955751"    y3="0.642612"
z3="3.915704"/>

                </atomArray>

                <bondArray>

                        <bond atomRefs2="a4 a1" order="1"/>

                        <bond atomRefs2="a5 a1" order="1"/>

                        <bond atomRefs2="a1 a2" order="1"/>

                        <bond atomRefs2="a1 a3" order="1"/>

                        <bond atomRefs2="a2 a6" order="2"/>

                        <bond atomRefs2="a7 a6" order="1"/>

                        <bond atomRefs2="a6 a8" order="1"/>

                        <bond atomRefs2="a8 a9" order="1"/>

                </bondArray>

                <propertyList>

                        <property dictRef="me:lumpedSpecies">

                                <array> </array>

                        </property>

                        <property dictRef="me:vibFreqs">

                                <array units="cm-1">100.6641 160.5634 297.6089 433.8734 596.0979
844.9045  881.9835  1023.2578  1075.5491  1097.8216  1278.9806  1404.4803  1461.9681  1530.9354  1547.1515
1767.578 2967.5919 3050.5056 3075.5969 3190.1399 3485.33</array>

                        </property>

                        <property dictRef="me:ZPE">

                                <scalar units="kJ/mol">-124.77033514945647</scalar>

                        </property>

                        <property dictRef="me:spinMultiplicity">

                                <scalar units="cm-1">2</scalar>

                        </property>

                        <property dictRef="me:epsilon">

                                <scalar>473.17</scalar>
```

```
                </property>
                <property dictRef="me:sigma">
                        <scalar>5.09</scalar>
                </property>
                <group>
                        <scalar>1</scalar>
                </group>
        </propertyList>
        <me:energyTransferModel xsi:type="me:ExponentialDown">
                <scalar units="cm-1">250</scalar>
        </me:energyTransferModel>
</molecule>
<molecule id="TS_CC#C_CC(=[CH])O" spinMultiplicity="2">
        <atomArray>
                <atom elementType="C" id="a1" x3="-1.069585" y3="-0.057111" z3="-0.645718"/>
                <atom elementType="C" id="a2" x3="0.361044" y3="-0.035507" z3="-0.564916"/>
                <atom elementType="H" id="a3" x3="-1.439635" y3="-0.986531" z3="-1.111803"/>
                <atom elementType="H" id="a4" x3="-1.459444" y3="0.789945" z3="-1.236056"/>
                <atom elementType="H" id="a5" x3="-1.527628" y3="0.008276" z3="0.362144"/>
                <atom elementType="C" id="a6" x3="1.564855" y3="-0.016809" z3="-0.489517"/>
                <atom elementType="H" id="a7" x3="2.594684" y3="-0.002220" z3="-0.445112"/>
                <atom elementType="O" id="a8" spinMultiplicity="2" x3="-0.000077" y3="0.141487" z3="2.021862"/>
                <atom elementType="H" id="a9" x3="0.975786" y3="0.158470" z3="2.109115"/>
        </atomArray>
        <bondArray>
                <bond atomRefs2="a4 a1" order="1"/>
                <bond atomRefs2="a3 a1" order="1"/>
```

```xml
            <bond atomRefs2="a1 a2" order="1"/>

            <bond atomRefs2="a1 a5" order="1"/>

            <bond atomRefs2="a2 a6" order="3"/>

            <bond atomRefs2="a6 a7" order="1"/>

            <bond atomRefs2="a8 a9" order="1"/>

        </bondArray>

        <propertyList>

            <property dictRef="me:vibFreqs">

                        <array      units="cm-1">35.55230050838298        78.28997458054556
100.9795682382628    316.4123170378518    366.9776202717837    374.0582311305118    826.1511404119397
830.9786971093747    1017.7136041901801    1021.9556920696441    1107.3601545198612    1219.6134907483956
1225.3873183691705    1335.6333811129975    2387.9501742701023    2630.7765623948735    2668.8713882701054
2683.7618588609616 2773.376525263516 2896.2158417425735</array>

            </property>

            <property dictRef="me:imFreq">

                        <scalar units="cm-1">137.47802242071475</scalar>

            </property>

            <property dictRef="me:spinMultiplicity">

                        <scalar units="cm-1">2</scalar>

            </property>

            <property dictRef="me:ZPE">

                        <scalar units="kJ/mol">-20.07653086909563</scalar>

            </property>

        </propertyList>

    </molecule>

    <molecule id="C[C](O)[CH]" spinMultiplicity="2">

        <atomArray>

                        <atom    elementType="C"    id="a1"    x3="-6.399640"    y3="1.918561"    z3="-
1.309067"/>

                        <atom    elementType="C"    id="a2"    x3="-6.613004"    y3="1.165908"    z3="-
0.019223"/>

                        <atom    elementType="H"    id="a3"    x3="-6.961767"    y3="1.457309"    z3="-
2.122158"/>

                        <atom    elementType="H"    id="a4"    x3="-5.332323"    y3="1.924807"    z3="-
1.557075"/>
```

```
                              <atom elementType="H" id="a5" x3="-6.720932" y3="2.958616" z3="-
1.183512"/>
                              <atom elementType="C" id="a6" spinMultiplicity="2" x3="-7.376321"
y3="0.099977" z3="0.092162"/>
                              <atom elementType="H" id="a7" x3="-7.716586" y3="-0.619107"
z3="0.816982"/>
                              <atom elementType="O" id="a8" x3="-5.876941" y3="1.780011"
z3="1.018696"/>
                              <atom elementType="H" id="a9" x3="-6.023253" y3="1.277637"
z3="1.865033"/>
              </atomArray>
              <bondArray>
                      <bond atomRefs2="a3 a1" order="1"/>
                      <bond atomRefs2="a4 a1" order="1"/>
                      <bond atomRefs2="a1 a5" order="1"/>
                      <bond atomRefs2="a1 a2" order="1"/>
                      <bond atomRefs2="a2 a6" order="2"/>
                      <bond atomRefs2="a2 a8" order="1"/>
                      <bond atomRefs2="a6 a7" order="1"/>
                      <bond atomRefs2="a8 a9" order="1"/>
              </bondArray>
              <propertyList>
                      <property dictRef="me:lumpedSpecies">
                              <array> </array>
                      </property>
                      <property dictRef="me:vibFreqs">
                              <array units="cm-1">161.4588 390.9844 426.3799 433.7199 497.9991
603.9209 755.6762 862.963 1015.0214 1109.4195 1145.702 1338.8846 1455.6831 1539.3294 1540.3064 1697.6358
3059.1306 3109.3996 3162.332 3306.5935 3485.559</array>
                      </property>
                      <property dictRef="me:ZPE">
                              <scalar units="kJ/mol">-129.93697306258343</scalar>
                      </property>
                      <property dictRef="me:spinMultiplicity">
                              <scalar units="cm-1">2</scalar>
```

```xml
          </property>
          <property dictRef="me:epsilon">
                    <scalar>473.17</scalar>
          </property>
          <property dictRef="me:sigma">
                    <scalar>5.09</scalar>
          </property>
          <group>
                    <scalar>1</scalar>
          </group>
      </propertyList>
      <me:energyTransferModel xsi:type="me:ExponentialDown">
                <scalar units="cm-1">250</scalar>
      </me:energyTransferModel>
  </molecule>
  <molecule id="TS_C[C][CH]O_CC(=[CH])O" spinMultiplicity="4">
          <atomArray>
                        <atom elementType="C" id="a1" x3="-1.016240" y3="-0.152491" z3="-0.784219"/>
                        <atom elementType="C" id="a2" spinMultiplicity="3" x3="0.236137" y3="0.235946" z3="-0.199599"/>
                        <atom elementType="H" id="a3" x3="-1.873905" y3="0.137981" z3="-0.152885"/>
                        <atom elementType="H" id="a4" x3="-1.069643" y3="-1.250844" z3="-0.912899"/>
                        <atom elementType="H" id="a5" x3="-1.162962" y3="0.304173" z3="-1.778323"/>
                        <atom elementType="C" id="a6" spinMultiplicity="2" x3="1.323165" y3="0.704805" z3="0.179025"/>
                        <atom elementType="H" id="a7" x3="2.228746" y3="1.146638" z3="0.409805"/>
                        <atom elementType="O" id="a8" x3="0.867845" y3="-0.706260" z3="1.197982"/>
                        <atom elementType="H" id="a9" x3="0.466857" y3="-0.419949" z3="2.041113"/>
          </atomArray>
```

```xml
<bondArray>
        <bond atomRefs2="a5 a1" order="1"/>
        <bond atomRefs2="a4 a1" order="1"/>
        <bond atomRefs2="a1 a2" order="1"/>
        <bond atomRefs2="a1 a3" order="1"/>
        <bond atomRefs2="a2 a6" order="1"/>
        <bond atomRefs2="a6 a7" order="1"/>
        <bond atomRefs2="a6 a8" order="1"/>
        <bond atomRefs2="a8 a9" order="1"/>
    </bondArray>
    <propertyList>
        <property dictRef="me:vibFreqs">
            <array    units="cm-1">50.40036553133097    268.1902182061224
339.7425177625869    418.67367486839663    612.1003245973942    795.0792895810064    834.5093503875759
950.8651233326254    1009.6280243526245    1021.0866870723855    1081.050438022304    1219.436215557928
1222.8660373137398    1328.8188806292499    2226.00242982412    2592.9257660022718    2676.4049560126646
2682.1896779830286 2774.9617574876843 2855.8642962951662</array>
        </property>
        <property dictRef="me:imFreq">
            <scalar units="cm-1">433.2873386030255</scalar>
        </property>
        <property dictRef="me:spinMultiplicity">
            <scalar units="cm-1">2</scalar>
        </property>
        <property dictRef="me:ZPE">
            <scalar units="kJ/mol">86.55539805888725</scalar>
        </property>
        <group>
            <scalar>1</scalar>
        </group>
    </propertyList>
</molecule>
<molecule id="CC(=[CH])O" spinMultiplicity="2">
    <atomArray>
```

```xml
                              <atom     elementType="C"     id="a1"     x3="-3.851699"     y3="-0.799402"
z3="6.286867"/>
                              <atom     elementType="C"     id="a2"     x3="-4.641223"     y3="0.435832"
z3="6.642614"/>
                              <atom     elementType="H"     id="a3"     x3="-3.863620"     y3="-0.941240"
z3="5.200386"/>
                              <atom     elementType="H"     id="a4"     x3="-4.319976"     y3="-1.677784"
z3="6.744979"/>
                              <atom     elementType="H"     id="a5"     x3="-2.821135"     y3="-0.711373"
z3="6.633336"/>
                              <atom     elementType="C"     id="a6"     spinMultiplicity="2"     x3="-4.139579"
y3="1.469413" z3="7.283965"/>
                              <atom     elementType="H"     id="a7"     x3="-4.414677"     y3="2.437768"
z3="7.664540"/>
                              <atom     elementType="O"     id="a8"     x3="-5.970952"     y3="0.300248"
z3="6.184767"/>
                              <atom     elementType="H"     id="a9"     x3="-6.484171"     y3="1.118620"
z3="6.423404"/>
                    </atomArray>
                    <bondArray>
                              <bond atomRefs2="a3 a1" order="1"/>
                              <bond atomRefs2="a8 a9" order="1"/>
                              <bond atomRefs2="a8 a2" order="1"/>
                              <bond atomRefs2="a1 a5" order="1"/>
                              <bond atomRefs2="a1 a2" order="1"/>
                              <bond atomRefs2="a1 a4" order="1"/>
                              <bond atomRefs2="a2 a6" order="2"/>
                              <bond atomRefs2="a6 a7" order="1"/>
                    </bondArray>
                    <propertyList>
                              <property dictRef="me:lumpedSpecies">
                                        <array> </array>
                              </property>
                              <property dictRef="me:vibFreqs">
```

```xml
              <array units="cm-1">161.3752 390.9796 426.3798 433.7177 497.9945
603.9205 755.6782 862.9639 1015.017 1109.412 1145.698 1338.8848 1455.6786 1539.3274 1540.3047 1697.6367
3059.1602 3109.4305 3162.3611 3306.5929 3485.56</array>
          </property>
          <property dictRef="me:ZPE">
                  <scalar units="kJ/mol">-129.93708428864824</scalar>
          </property>
          <property dictRef="me:spinMultiplicity">
                  <scalar units="cm-1">2</scalar>
          </property>
          <property dictRef="me:epsilon">
                  <scalar>473.17</scalar>
          </property>
          <property dictRef="me:sigma">
                  <scalar>5.09</scalar>
          </property>
          <group>
                  <scalar>1</scalar>
          </group>
      </propertyList>
      <me:energyTransferModel xsi:type="me:ExponentialDown">
              <scalar units="cm-1">250</scalar>
      </me:energyTransferModel>
  </molecule>
  <molecule id="TS_C[C][CH]O_CC#[C]" spinMultiplicity="2">
      <atomArray>
                  <atom elementType="C" id="a1" x3="-1.117381" y3="-0.133564" z3="-0.930787"/>
                  <atom elementType="C" id="a2" x3="0.306972" y3="-0.052185" z3="-0.958503"/>
                  <atom elementType="H" id="a3" x3="-1.487823" y3="-0.078732" z3="0.115978"/>
                  <atom elementType="H" id="a4" x3="-1.487605" y3="-1.085720" z3="-1.347428"/>
```

<atom elementType="H" id="a5" x3="-1.592094" y3="0.698661" z3="-1.477877"/>

<atom elementType="C" id="a6" spinMultiplicity="2" x3="1.496839" y3="0.046318" z3="-0.564541"/>

<atom elementType="H" id="a7" x3="1.440517" y3="0.978286" z3="1.683085"/>

<atom elementType="O" id="a8" x3="0.910565" y3="0.176671" z3="1.685118"/>

<atom elementType="H" id="a9" x3="1.530010" y3="-0.549734" z3="1.794955"/>

</atomArray>

<bondArray>

<bond atomRefs2="a5 a1" order="1"/>

<bond atomRefs2="a4 a1" order="1"/>

<bond atomRefs2="a2 a1" order="1"/>

<bond atomRefs2="a2 a6" order="3"/>

<bond atomRefs2="a1 a3" order="1"/>

<bond atomRefs2="a7 a8" order="1"/>

<bond atomRefs2="a8 a9" order="1"/>

</bondArray>

<propertyList>

<property dictRef="me:vibFreqs">

<array units="cm-1">71.21398290434784 121.5143669058337 210.8212173482216 299.7152076819921 352.3670071556001 387.7697498745235 449.37908502364513 937.4191210959578 952.5018789442089 1134.728060286667 1190.5913544510486 1198.6038210861723 1288.6238509342961 1396.334954043544 1890.0697363665236 2532.4398193048364 2625.948977573173 2674.3946134823946 2684.160130859478 2770.462793132719</array>

</property>

<property dictRef="me:imFreq">

<scalar units="cm-1">3877.536214475379</scalar>

</property>

<property dictRef="me:spinMultiplicity">

<scalar units="cm-1">2</scalar>

</property>

<property dictRef="me:ZPE">

<scalar units="kJ/mol">96.03524131376535</scalar>

```xml
                </property>
                <group>
                        <scalar>1</scalar>
                </group>
        </propertyList>
    </molecule>
    <molecule id="CC#[C]" spinMultiplicity="2">
        <atomArray>
                <atom elementType="C" id="a1" x3="0.928991" y3="-0.050946" z3="-0.059534"/>
                <atom elementType="C" id="a2" x3="0.440461" y3="-1.236919" z3="0.649511"/>
                <atom elementType="C" id="a3" spinMultiplicity="2" x3="0.037287" y3="-2.217356" z3="1.236006"/>
                <atom elementType="H" id="a4" x3="0.573774" y3="0.863813" z3="0.428815"/>
                <atom elementType="H" id="a5" x3="0.573837" y3="-0.048199" z3="-1.096508"/>
                <atom elementType="H" id="a6" x3="2.024879" y3="-0.032492" z3="-0.070544"/>
        </atomArray>
        <bondArray>
                <bond atomRefs2="a5 a1" order="1"/>
                <bond atomRefs2="a6 a1" order="1"/>
                <bond atomRefs2="a1 a4" order="1"/>
                <bond atomRefs2="a1 a2" order="1"/>
                <bond atomRefs2="a2 a3" order="3"/>
        </bondArray>
        <propertyList>
                <property dictRef="me:lumpedSpecies">
                        <array> </array>
                </property>
                <property dictRef="me:vibFreqs">
                        <array units="cm-1">204.0608   213.0972   923.2111   1063.0085 1066.0806 1469.7525 1544.888 1545.2852 2264.6709 3049.4891 3108.6759 3108.7372</array>
```

```xml
        </property>
        <property dictRef="me:ZPE">
                <scalar units="kJ/mol">58.184403982761026</scalar>
        </property>
        <property dictRef="me:spinMultiplicity">
                <scalar units="cm-1">2</scalar>
        </property>
        <property dictRef="me:epsilon">
                <scalar>473.17</scalar>
        </property>
        <property dictRef="me:sigma">
                <scalar>5.09</scalar>
        </property>
        <group>
                <scalar>1</scalar>
        </group>
    </propertyList>
    <me:energyTransferModel xsi:type="me:ExponentialDown">
                <scalar units="cm-1">250</scalar>
    </me:energyTransferModel>
  </molecule>
  <molecule id="O">
    <atomArray>
                <atom  elementType="O"  id="a1"  x3="0.951056"  y3="-0.052327"  z3="-0.053915"/>
                <atom  elementType="H"  id="a2"  x3="0.664338"  y3="-0.396681"  z3="-0.944085"/>
                <atom  elementType="H"  id="a3"  x3="1.946496"  y3="-0.069583"  z3="-0.098521"/>
    </atomArray>
    <bondArray>
            <bond atomRefs2="a2 a1" order="1"/>
            <bond atomRefs2="a3 a1" order="1"/>
```

```xml
        </bondArray>
        <propertyList>
                <property dictRef="me:lumpedSpecies">
                        <array> </array>
                </property>
                <property dictRef="me:vibFreqs">
                        <array units="cm-1">1692.2943 3415.4019 3556.0709</array>
                </property>
                <property dictRef="me:ZPE">
                        <scalar units="kJ/mol">0.0</scalar>
                </property>
                <property dictRef="me:spinMultiplicity">
                        <scalar units="cm-1">1</scalar>
                </property>
                <property dictRef="me:epsilon">
                        <scalar>473.17</scalar>
                </property>
                <property dictRef="me:sigma">
                        <scalar>5.09</scalar>
                </property>
        </propertyList>
        <me:energyTransferModel xsi:type="me:ExponentialDown">
                <scalar units="cm-1">250</scalar>
        </me:energyTransferModel>
</molecule>
<molecule id="TS_C[C][CH]O_C[C]=CO" spinMultiplicity="2">
        <atomArray>
                <atom elementType="C" id="a1" x3="-7.075142" y3="0.654958" z3="1.064001"/>
                <atom elementType="C" id="a2" spinMultiplicity="2" x3="-5.695469" y3="1.060818" z3="1.392395"/>
                <atom elementType="H" id="a3" x3="-7.638455" y3="0.352882" z3="1.964485"/>
```

```xml
                <atom    elementType="H"    id="a4"    x3="-7.088696"    y3="-0.198200"
z3="0.373496"/>
                <atom    elementType="H"    id="a5"    x3="-7.633384"    y3="1.474926"
z3="0.594019"/>
                <atom    elementType="C"    id="a6"    x3="-4.841962"    y3="1.208756"
z3="2.378578"/>
                <atom    elementType="H"    id="a7"    x3="-3.828546"    y3="1.559874"
z3="2.217366"/>
                <atom    elementType="O"    id="a8"    x3="-5.023162"    y3="0.967536"
z3="3.762166"/>
                <atom    elementType="H"    id="a9"    x3="-5.951238"    y3="0.644813"
z3="3.918797"/>
        </atomArray>
        <bondArray>
            <bond atomRefs2="a4 a1" order="1"/>
            <bond atomRefs2="a5 a1" order="1"/>
            <bond atomRefs2="a1 a2" order="1"/>
            <bond atomRefs2="a1 a3" order="1"/>
            <bond atomRefs2="a2 a6" order="2"/>
            <bond atomRefs2="a7 a6" order="1"/>
            <bond atomRefs2="a6 a8" order="1"/>
            <bond atomRefs2="a8 a9" order="1"/>
        </bondArray>
        <propertyList>
            <property dictRef="me:vibFreqs">
                        <array    units="cm-1">232.41436457735375    337.1615467439119
418.91285839824343    550.5245372341572    819.7184867863668    917.9258095564156    982.3099883799374
985.0017760466802    1078.609406032168    1177.7131058675168    1217.997421023077    1230.7449240276478
1301.6407464283125    1369.3245976891671    1795.8961978394657    2577.529825371589    2670.7460237672635
2696.2187623735063    2757.128952301289    2782.747385230743</array>
            </property>
            <property dictRef="me:imFreq">
                    <scalar units="cm-1">127.18154494718294</scalar>
            </property>
            <property dictRef="me:spinMultiplicity">
                    <scalar units="cm-1">2</scalar>
```

```xml
            </property>
            <property dictRef="me:ZPE">
                    <scalar units="kJ/mol">-144.26967104781951</scalar>
            </property>
            <group>
                    <scalar>3</scalar>
            </group>
        </propertyList>
    </molecule>
    <molecule id="TS_C[C][CH]O_CC#C" spinMultiplicity="2">
        <atomArray>
                            <atom    elementType="C"    id="a1"    x3="-7.049846"    y3="0.680706"
z3="1.156238"/>
                            <atom    elementType="C"    id="a2"    x3="-5.676533"    y3="1.051419"
z3="1.500217"/>
                            <atom    elementType="H"    id="a3"    x3="-7.595503"    y3="0.558560"
z3="2.096081"/>
                            <atom    elementType="H"    id="a4"    x3="-7.079985"    y3="-0.254921"
z3="0.585892"/>
                            <atom    elementType="H"    id="a5"    x3="-7.529175"    y3="1.466126"
z3="0.560715"/>
                            <atom    elementType="C"    id="a6"    x3="-4.544758"    y3="1.362499"
z3="1.794154"/>
                            <atom    elementType="H"    id="a7"    x3="-3.554152"    y3="1.654663"
z3="2.039384"/>
                            <atom    elementType="O"    id="a8"    spinMultiplicity="2"    x3="-6.145147"
y3="1.030735" z3="3.955625"/>
                            <atom    elementType="H"    id="a9"    x3="-5.600954"    y3="0.176578"
z3="3.976996"/>
        </atomArray>
        <bondArray>
                <bond atomRefs2="a5 a1" order="1"/>
                <bond atomRefs2="a4 a1" order="1"/>
                <bond atomRefs2="a1 a2" order="1"/>
                <bond atomRefs2="a1 a3" order="1"/>
                <bond atomRefs2="a2 a6" order="3"/>
```

```
                    <bond atomRefs2="a6 a7" order="1"/>

                    <bond atomRefs2="a8 a9" order="1"/>

               </bondArray>

          <propertyList>

               <property dictRef="me:vibFreqs">

                    <array    units="cm-1">111.16980917052545    124.10976306728078
372.5324536070673    386.3330916109452    496.05302465843835    623.0336560424662    918.0188916605947
922.8947410853016    1009.5662644217307    1010.893172479569    1041.7920163871954    1218.5554918511066
1229.3554335711392    1304.0526782251968    2332.832555531182    2591.0430536458766    2703.5650904085805
2714.369887243828 2785.2130649993624 2817.045239098539</array>

               </property>

               <property dictRef="me:imFreq">

                    <scalar units="cm-1">191.06818825545528</scalar>

               </property>

               <property dictRef="me:spinMultiplicity">

                    <scalar units="cm-1">2</scalar>

               </property>

               <property dictRef="me:ZPE">

                    <scalar units="kJ/mol">-16.841491435177478</scalar>

               </property>

          </propertyList>

     </molecule>

     <molecule id="TS_C[C][CH]O_[CH2]C=CO" spinMultiplicity="4">

          <atomArray>

                    <atom    elementType="C"    id="a1"    spinMultiplicity="2"    x3="-0.504435"
y3="1.090785" z3="-0.935849"/>

                    <atom    elementType="C"    id="a2"    spinMultiplicity="2"    x3="0.263384"
y3="0.206812" z3="-0.089887"/>

                    <atom    elementType="H"    id="a3"    x3="-1.532946"    y3="0.873827"    z3="-
1.222339"/>

                    <atom    elementType="H"    id="a4"    x3="0.024135"    y3="1.854043"    z3="-
1.498729"/>

                    <atom    elementType="H"    id="a5"    x3="-0.188611"    y3="1.363287"    z3="
0.342961"/>

                    <atom    elementType="C"    id="a6"    spinMultiplicity="2"    x3="0.244168"    y3="-
1.070908" z3="0.275435"/>
```

```xml
                    <atom elementType="H" id="a7" x3="-0.086623" y3="-1.873917" z3="-0.379125"/>
                    <atom elementType="O" id="a8" x3="0.722258" y3="-1.598906" z3="1.474359"/>
                    <atom elementType="H" id="a9" x3="1.058670" y3="-0.845023" z3="2.033174"/>
        </atomArray>
        <bondArray>
            <bond atomRefs2="a4 a1" order="1"/>
            <bond atomRefs2="a3 a1" order="1"/>
            <bond atomRefs2="a1 a2" order="1"/>
            <bond atomRefs2="a7 a6" order="1"/>
            <bond atomRefs2="a2 a6" order="1"/>
            <bond atomRefs2="a2 a5" order="1"/>
            <bond atomRefs2="a6 a8" order="1"/>
            <bond atomRefs2="a8 a9" order="1"/>
        </bondArray>
        <propertyList>
            <property dictRef="me:vibFreqs">
                    <array units="cm-1">254.34531885244397 269.082744651459 347.2203049346683 494.4839654438206 537.0013857876403 813.1637822155385 856.3241801233448 949.401884343447 1008.6743323418297 1028.1369112504347 1158.1377602164282 1226.5268802730898 1316.1710652130173 1377.1375084781073 1794.3594764896202 2515.5225598391594 2584.826952161911 2691.1767377434753 2730.663604076424 2738.410589042136</array>
            </property>
            <property dictRef="me:imFreq">
                    <scalar units="cm-1">2300.055427149821</scalar>
            </property>
            <property dictRef="me:spinMultiplicity">
                    <scalar units="cm-1">2</scalar>
            </property>
            <property dictRef="me:ZPE">
                    <scalar units="kJ/mol">28.32279575507182</scalar>
            </property>
        </propertyList>
```

```xml
        </molecule>
        <molecule id="[CH2]C=CO" spinMultiplicity="2">
                <atomArray>
                        <atom elementType="C" id="a1" spinMultiplicity="2" x3="-3.025113" y3="3.443468" z3="0.150748"/>
                        <atom elementType="C" id="a2" x3="-3.358411" y3="2.172208" z3="0.583650"/>
                        <atom elementType="H" id="a3" x3="-3.574380" y3="3.927605" z3="-0.645672"/>
                        <atom elementType="H" id="a4" x3="-2.206131" y3="4.013180" z3="0.582158"/>
                        <atom elementType="H" id="a5" x3="-4.190179" y3="1.665480" z3="0.103439"/>
                        <atom elementType="C" id="a6" x3="-2.730675" y3="1.453841" z3="1.590471"/>
                        <atom elementType="H" id="a7" x3="-3.044884" y3="0.460931" z3="1.878269"/>
                        <atom elementType="O" id="a8" x3="-1.653676" y3="1.876847" z3="2.348719"/>
                        <atom elementType="H" id="a9" x3="-1.375157" y3="2.790535" z3="2.073933"/>
                </atomArray>
                <bondArray>
                        <bond atomRefs2="a3 a1" order="1"/>
                        <bond atomRefs2="a5 a2" order="1"/>
                        <bond atomRefs2="a1 a4" order="1"/>
                        <bond atomRefs2="a1 a2" order="1"/>
                        <bond atomRefs2="a2 a6" order="2"/>
                        <bond atomRefs2="a6 a7" order="1"/>
                        <bond atomRefs2="a6 a8" order="1"/>
                        <bond atomRefs2="a9 a8" order="1"/>
                </bondArray>
                <propertyList>
                        <property dictRef="me:lumpedSpecies">
                                <array> </array>
                        </property>
```

```
<property dictRef="me:vibFreqs">
                    <array units="cm-1">202.7645 284.0556 448.7911 550.9416 653.1868
732.5476  743.2004  961.9957  969.2038  1068.4569  1193.8409  1268.9026  1376.5879  1467.8237  1505.6289
1567.8247 3150.8331 3181.5087 3246.3629 3256.2508 3501.9003</array>
        </property>

        <property dictRef="me:ZPE">
                    <scalar units="kJ/mol">-232.93939314055623</scalar>
        </property>

        <property dictRef="me:spinMultiplicity">
                    <scalar units="cm-1">2</scalar>
        </property>

        <property dictRef="me:epsilon">
                    <scalar>473.17</scalar>
        </property>

        <property dictRef="me:sigma">
                    <scalar>5.09</scalar>
        </property>

        <group>
                    <scalar>3</scalar>
        </group>

    </propertyList>

    <me:energyTransferModel xsi:type="me:ExponentialDown">
                    <scalar units="cm-1">250</scalar>
    </me:energyTransferModel>

</molecule>

<molecule id="TS_C[C][CH]O_C[CH]C=O" spinMultiplicity="4">

    <atomArray>
                    <atom     elementType="C"     id="a1"     x3="-0.863323"     y3="0.967527"
z3="0.268105"/>
                    <atom  elementType="C"  id="a2"  spinMultiplicity="3"  x3="-0.227577"  y3="-
0.287159" z3="-0.229679"/>
                    <atom     elementType="H"     id="a3"     x3="-0.405850"     y3="1.373726"
z3="1.181087"/>
```

```xml
                        <atom     elementType="H"     id="a4"     x3="-1.935458"     y3="0.814037"
z3="0.438390"/>
                        <atom     elementType="H"     id="a5"     x3="-0.785459"     y3="1.739889"     z3="-
0.513751"/>
                        <atom     elementType="C"     id="a6"     spinMultiplicity="2"     x3="1.011288"     y3="-
0.804376"     z3="0.096580"/>
                        <atom     elementType="H"     id="a7"     x3="1.726678"     y3="-0.632412"
z3="0.889912"/>
                        <atom     elementType="O"     id="a8"     x3="1.299086"     y3="-1.730685"     z3="-
0.859786"/>
                        <atom     elementType="H"     id="a9"     x3="0.180615"     y3="-1.440547"     z3="-
1.270859"/>
                </atomArray>
                <bondArray>
                        <bond atomRefs2="a9 a8" order="1"/>
                        <bond atomRefs2="a8 a6" order="1"/>
                        <bond atomRefs2="a5 a1" order="1"/>
                        <bond atomRefs2="a2 a6" order="1"/>
                        <bond atomRefs2="a2 a1" order="1"/>
                        <bond atomRefs2="a6 a7" order="1"/>
                        <bond atomRefs2="a1 a4" order="1"/>
                        <bond atomRefs2="a1 a3" order="1"/>
                </bondArray>
                <propertyList>
                        <property dictRef="me:vibFreqs">
                                <array     units="cm-1">153.28991942160718     375.0312952130748
536.4365266448415     753.1712148621019     776.1008079659374     924.0629574395135     953.8934819040064
1018.8699452918532     1131.6956136783374     1206.737891824023     1229.758059567231     1251.3774290731797
1298.3378519985395     1318.9328877797132     1505.2671649700746     2376.4772332521984     2676.355741139525
2691.6598463164937     2717.627255808547     2779.939722599298</array>
                        </property>
                        <property dictRef="me:imFreq">
                                <scalar units="cm-1">364.7556034136078</scalar>
                        </property>
                        <property dictRef="me:spinMultiplicity">
                                <scalar units="cm-1">2</scalar>
```

```xml
            </property>
            <property dictRef="me:ZPE">
                    <scalar units="kJ/mol">7.876088014139941</scalar>
            </property>
            <group>
                    <scalar>2</scalar>
            </group>
        </propertyList>
    </molecule>
    <molecule id="C[CH]C=O" spinMultiplicity="2">
        <atomArray>
                        <atom    elementType="C"    id="a1"    x3="-5.113564"    y3="3.364407"
z3="3.829451"/>
                        <atom    elementType="C"    id="a2"    spinMultiplicity="2"    x3="-5.148302"
y3="1.883041" z3="3.655048"/>
                        <atom    elementType="H"    id="a3"    x3="-4.148161"    y3="3.696909"
z3="4.224929"/>
                        <atom    elementType="H"    id="a4"    x3="-5.905451"    y3="3.697600"
z3="4.516079"/>
                        <atom    elementType="H"    id="a5"    x3="-5.293978"    y3="3.876642"
z3="2.873062"/>
                        <atom    elementType="C"    id="a6"    x3="-4.057198"    y3="1.038747"
z3="3.967190"/>
                        <atom    elementType="H"    id="a7"    x3="-3.164888"    y3="1.562078"
z3="4.355003"/>
                        <atom    elementType="O"    id="a8"    x3="-4.069407"    y3="-0.221094"
z3="3.825066"/>
                        <atom    elementType="H"    id="a9"    x3="-6.047777"    y3="1.411894"
z3="3.270291"/>
        </atomArray>
        <bondArray>
            <bond atomRefs2="a5 a1" order="1"/>
            <bond atomRefs2="a9 a2" order="1"/>
            <bond atomRefs2="a2 a1" order="1"/>
            <bond atomRefs2="a2 a6" order="1"/>
            <bond atomRefs2="a8 a6" order="2"/>
```

```xml
                <bond atomRefs2="a1 a3" order="1"/>

                <bond atomRefs2="a1 a4" order="1"/>

                <bond atomRefs2="a6 a7" order="1"/>

        </bondArray>

        <propertyList>

                <property dictRef="me:lumpedSpecies">

                        <array> </array>

                </property>

                <property dictRef="me:vibFreqs">

                        <array units="cm-1">153.8415  227.3141  308.8449  550.9265  776.0854
923.9827  966.0921  1064.444  1141.6243  1191.0508  1361.7099  1393.456  1457.6491  1485.7886  1535.3116
1553.6566 2945.3263 3019.9059 3054.4639 3113.0958 3194.4559</array>

                </property>

                <property dictRef="me:ZPE">

                        <scalar units="kJ/mol">-246.89788444670899</scalar>

                </property>

                <property dictRef="me:spinMultiplicity">

                        <scalar units="cm-1">2</scalar>

                </property>

                <property dictRef="me:epsilon">

                        <scalar>473.17</scalar>

                </property>

                <property dictRef="me:sigma">

                        <scalar>5.09</scalar>

                </property>

                <group>

                        <scalar>2</scalar>

                </group>

        </propertyList>

        <me:energyTransferModel xsi:type="me:ExponentialDown">

                <scalar units="cm-1">250</scalar>

        </me:energyTransferModel>

    </molecule>
```

```xml
<molecule id="TS_C[C][CH]O_CC#CO" spinMultiplicity="2">
    <atomArray>
        <atom elementType="C" id="a1" x3="-1.095078" y3="-0.264544" z3="-1.138453"/>
        <atom elementType="C" id="a2" x3="-0.281258" y3="0.144139" z3="-0.035895"/>
        <atom elementType="H" id="a3" x3="-1.356531" y3="-1.335646" z3="-1.079181"/>
        <atom elementType="H" id="a4" x3="-0.589087" y3="-0.095798" z3="-2.105215"/>
        <atom elementType="H" id="a5" x3="-2.043016" y3="0.304100" z3="-1.160345"/>
        <atom elementType="C" id="a6" x3="0.481233" y3="0.421629" z3="0.869624"/>
        <atom elementType="H" id="a7" spinMultiplicity="2" x3="1.884498" y3="-0.644170" z3="0.629458"/>
        <atom elementType="O" id="a8" x3="1.050272" y3="0.955207" z3="1.940267"/>
        <atom elementType="H" id="a9" x3="1.948967" y3="0.515082" z3="2.079740"/>
    </atomArray>
    <bondArray>
        <bond atomRefs2="a4 a1" order="1"/>
        <bond atomRefs2="a5 a1" order="1"/>
        <bond atomRefs2="a1 a3" order="1"/>
        <bond atomRefs2="a1 a2" order="1"/>
        <bond atomRefs2="a2 a6" order="3"/>
        <bond atomRefs2="a6 a8" order="1"/>
        <bond atomRefs2="a8 a9" order="1"/>
    </bondArray>
    <propertyList>
        <property dictRef="me:vibFreqs">
            <array units="cm-1">13.32273992116023 151.5092104141068 199.43913398441245 240.42779620062592 259.31491383087524 363.0012212532987 432.92853467769413 867.5189368731693 1004.2717501539537 1009.9243330546212 1202.474939631859 1228.168742624596 1232.0201151239237 1314.864737788541 1421.7729758224352 2411.977991123494 2551.3367865849486 2678.9429909807213 2681.563436674905 2775.829849756846</array>
```

```xml
            </property>
            <property dictRef="me:imFreq">
                    <scalar units="cm-1">785.5317281613985</scalar>
            </property>
            <property dictRef="me:spinMultiplicity">
                    <scalar units="cm-1">2</scalar>
            </property>
            <property dictRef="me:ZPE">
                    <scalar units="kJ/mol">45.80723734603867</scalar>
            </property>
            <group>
                    <scalar>3</scalar>
            </group>
        </propertyList>
    </molecule>
    <molecule id="CC#CO">
        <atomArray>
                            <atom     elementType="C"    id="a1"    x3="0.893616"    y3="-0.079524"
z3="0.091534"/>
                            <atom     elementType="C"    id="a2"    x3="0.389398"    y3="-0.922443"
z3="1.175907"/>
                            <atom     elementType="C"    id="a3"    x3="-0.002588"    y3="-1.622260"
z3="2.074296"/>
                            <atom     elementType="O"    id="a4"    x3="-0.407577"    y3="-2.405232"
z3="3.076820"/>
                            <atom     elementType="H"    id="a5"    x3="0.568548"    y3="0.961683"
z3="0.210741"/>
                            <atom    elementType="H"    id="a6"   x3="0.545210"   y3="-0.434256"   z3="-
0.886603"/>
                            <atom     elementType="H"    id="a7"    x3="1.990767"    y3="-0.086937"
z3="0.078233"/>
                            <atom     elementType="H"    id="a8"    x3="-1.400663"    y3="-2.404180"
z3="3.161761"/>
        </atomArray>
        <bondArray>
```

```xml
                    <bond atomRefs2="a6 a1" order="1"/>

                    <bond atomRefs2="a7 a1" order="1"/>

                    <bond atomRefs2="a1 a5" order="1"/>

                    <bond atomRefs2="a1 a2" order="1"/>

                    <bond atomRefs2="a2 a3" order="3"/>

                    <bond atomRefs2="a3 a4" order="1"/>

                    <bond atomRefs2="a4 a8" order="1"/>

            </bondArray>

            <propertyList>

                    <property dictRef="me:lumpedSpecies">

                            <array> </array>

                    </property>

                    <property dictRef="me:vibFreqs">

                            <array units="cm-1">18.3677 229.0649 240.6195 545.536 562.9581
749.1289 1107.2134 1110.3528 1192.4502 1333.4675 1476.3374 1553.1612 1556.7301 2443.2839 3036.254
3086.7825 3091.0246 3436.2808</array>

                    </property>

                    <property dictRef="me:ZPE">

                            <scalar units="kJ/mol">49.9862611906463</scalar>

                    </property>

                    <property dictRef="me:spinMultiplicity">

                            <scalar units="cm-1">1</scalar>

                    </property>

                    <property dictRef="me:epsilon">

                            <scalar>473.17</scalar>

                    </property>

                    <property dictRef="me:sigma">

                            <scalar>5.09</scalar>

                    </property>

                    <group>

                            <scalar>3</scalar>

                    </group>

            </propertyList>
```

```xml
            <me:energyTransferModel xsi:type="me:ExponentialDown">
                    <scalar units="cm-1">250</scalar>
            </me:energyTransferModel>
    </molecule>
    <molecule id="[H]" spinMultiplicity="2">
            <atomArray>
                    <atom     elementType="H"     id="a1"     spinMultiplicity="2"     x3="0.946941"
y3="0.070217" z3="-0.047481"/>
            </atomArray>
            <propertyList>
                    <property dictRef="me:lumpedSpecies">
                            <array> </array>
                    </property>
                    <property dictRef="me:vibFreqs">
                            <array units="cm-1"/>
                    </property>
                    <property dictRef="me:ZPE">
                            <scalar units="kJ/mol">0.0</scalar>
                    </property>
                    <property dictRef="me:spinMultiplicity">
                            <scalar units="cm-1">2</scalar>
                    </property>
                    <property dictRef="me:epsilon">
                            <scalar>473.17</scalar>
                    </property>
                    <property dictRef="me:sigma">
                            <scalar>5.09</scalar>
                    </property>
            </propertyList>
            <me:energyTransferModel xsi:type="me:ExponentialDown">
                    <scalar units="cm-1">250</scalar>
            </me:energyTransferModel>
```

```
        </molecule>
        <molecule id="TS_C[C](O)[CH]_[CH2]C(=C)O" spinMultiplicity="2">
                <atomArray>
                        <atom    elementType="C"    id="a1"    x3="0.453905"    y3="-0.678090"
z3="0.672403"/>
                        <atom    elementType="C"    id="a2"    x3="-0.290874"    y3="0.149963"    z3="-
0.359072"/>
                        <atom    elementType="H"    id="a3"    x3="1.250108"    y3="-1.335239"
z3="0.322244"/>
                        <atom    elementType="H"    id="a4"    x3="-0.099446"    y3="-1.026359"
z3="1.544703"/>
                        <atom    elementType="H"    id="a5"    x3="0.954507"    y3="0.634898"
z3="0.893431"/>
                        <atom    elementType="C"    id="a6"    spinMultiplicity="2"    x3="0.214588"
y3="1.358395" z3="-0.106275"/>
                        <atom    elementType="H"    id="a7"    x3="0.121202"    y3="2.372667"    z3="-
0.465928"/>
                        <atom    elementType="O"    id="a8"    x3="-1.223731"    y3="-0.249847"    z3="-
1.288224"/>
                        <atom    elementType="H"    id="a9"    x3="-1.380259"    y3="-1.226389"    z3="-
1.213282"/>
                </atomArray>
                <bondArray>
                        <bond atomRefs2="a8 a9" order="1"/>
                        <bond atomRefs2="a8 a2" order="1"/>
                        <bond atomRefs2="a7 a6" order="1"/>
                        <bond atomRefs2="a2 a6" order="2"/>
                        <bond atomRefs2="a2 a1" order="1"/>
                        <bond atomRefs2="a3 a1" order="1"/>
                        <bond atomRefs2="a1 a5" order="1"/>
                        <bond atomRefs2="a1 a4" order="1"/>
                </bondArray>
                <propertyList>
                        <property dictRef="me:vibFreqs">
                                <array    units="cm-1">326.3562874269104    408.7888926820271
430.4793422586303    516.0420840154303    553.5650096891037    806.0174182962111    917.0303708653427
```

924.8787892374581   993.6488571897237   1026.7286320504868   1071.6794093046008   1215.2160067359052
1278.1056824766079   1294.9440084280131   1685.1814241877198   2172.66647635055   2558.5797060920413
2675.9581151355137 2690.655914438867 2717.1897457141235</array>

                    </property>

                    <property dictRef="me:imFreq">

                            <scalar units="cm-1">2665.106895868166</scalar>

                    </property>

                    <property dictRef="me:spinMultiplicity">

                            <scalar units="cm-1">2</scalar>

                    </property>

                    <property dictRef="me:ZPE">

                            <scalar units="kJ/mol">18.51630083896057</scalar>

                    </property>

                    <group>

                            <scalar>3</scalar>

                    </group>

            </propertyList>

        </molecule>

        <molecule id="[CH2]C(=C)O" spinMultiplicity="2">

                <atomArray>

                        <atom   elementType="C"   id="a1"   x3="-6.060003"   y3="6.666519"   z3="-4.422005"/>

                        <atom   elementType="C"   id="a2"   x3="-6.498541"   y3="7.506354"   z3="-5.429832"/>

                        <atom   elementType="H"   id="a3"   x3="-6.349024"   y3="5.622706"   z3="-4.371056"/>

                        <atom   elementType="H"   id="a4"   x3="-5.405957"   y3="7.043997"   z3="-3.648722"/>

                        <atom   elementType="H"   id="a5"   x3="-5.485612"   y3="9.286398"   z3="-4.766695"/>

                        <atom   elementType="C"   id="a6"   spinMultiplicity="2"   x3="-6.136945" y3="8.848595" z3="-5.509560"/>

                        <atom   elementType="H"   id="a7"   x3="-6.509159"   y3="9.459814"   z3="-6.319708"/>

                        <atom   elementType="O"   id="a8"   x3="-7.345363"   y3="7.062489"   z3="-6.451369"/>

```xml
                    <atom elementType="H" id="a9" x3="-7.549932" y3="6.101022" z3="-6.318033"/>
        </atomArray>
        <bondArray>
            <bond atomRefs2="a8 a9" order="1"/>
            <bond atomRefs2="a8 a2" order="1"/>
            <bond atomRefs2="a7 a6" order="1"/>
            <bond atomRefs2="a6 a2" order="1"/>
            <bond atomRefs2="a6 a5" order="1"/>
            <bond atomRefs2="a2 a1" order="2"/>
            <bond atomRefs2="a1 a3" order="1"/>
            <bond atomRefs2="a1 a4" order="1"/>
        </bondArray>
        <propertyList>
            <property dictRef="me:lumpedSpecies">
                <array> </array>
            </property>
            <property dictRef="me:vibFreqs">
                <array units="cm-1">325.2567 437.8659 472.4316 491.6812 560.2839 618.9998 729.3939 774.995 898.5765 1000.2331 1042.7613 1182.7614 1389.2008 1445.2873 1516.9382 1557.4363 3177.5813 3198.9231 3276.4602 3302.6831 3524.1273</array>
            </property>
            <property dictRef="me:ZPE">
                <scalar units="kJ/mol">-231.1580956367015</scalar>
            </property>
            <property dictRef="me:spinMultiplicity">
                <scalar units="cm-1">2</scalar>
            </property>
            <property dictRef="me:epsilon">
                <scalar>473.17</scalar>
            </property>
            <property dictRef="me:sigma">
                <scalar>5.09</scalar>
```

```xml
            </property>
            <group>
                    <scalar>3</scalar>
            </group>
        </propertyList>
        <me:energyTransferModel xsi:type="me:ExponentialDown">
                <scalar units="cm-1">250</scalar>
        </me:energyTransferModel>
    </molecule>
    <molecule id="TS_C[C](O)[CH]_OC#C" spinMultiplicity="2">
        <atomArray>
                    <atom    elementType="C"    id="a1"    spinMultiplicity="2"    x3="0.102890"
y3="0.561175" z3="-1.198322"/>
                    <atom    elementType="C"    id="a2"    x3="-0.024479"    y3="-0.278087"
z3="0.736569"/>
                    <atom    elementType="H"    id="a3"    x3="-0.713492"    y3="0.101305"    z3="-
1.701778"/>
                    <atom    elementType="H"    id="a4"    x3="1.087203"    y3="0.204483"    z3="-
1.392615"/>
                    <atom    elementType="H"    id="a5"    x3="-0.016495"    y3="1.560585"    z3="-
0.845250"/>
                    <atom    elementType="C"    id="a6"    x3="-0.728302"    y3="-1.269773"
z3="0.603738"/>
                    <atom    elementType="H"    id="a7"    x3="-1.327316"    y3="-2.054699"
z3="0.310849"/>
                    <atom    elementType="O"    id="a8"    x3="0.737998"    y3="0.675229"
z3="1.253191"/>
                    <atom    elementType="H"    id="a9"    x3="0.881993"    y3="0.499783"
z3="2.233619"/>
        </atomArray>
        <bondArray>
                <bond atomRefs2="a3 a1" order="1"/>
                <bond atomRefs2="a4 a1" order="1"/>
                <bond atomRefs2="a1 a5" order="1"/>
                <bond atomRefs2="a7 a6" order="1"/>
                <bond atomRefs2="a6 a2" order="3"/>
```

```xml
                        <bond atomRefs2="a2 a8" order="1"/>

                        <bond atomRefs2="a8 a9" order="1"/>

                </bondArray>

                <propertyList>

                        <property dictRef="me:vibFreqs">

                                <array    units="cm-1">12.913904880445045    194.69981839493403
    196.63940382934905    338.8857351645034    402.4029767492693    592.2937798457657    610.0105091626947
    741.1069586835702    763.9267136116574    1019.0155781271725    1095.516579155151    1251.68722990812
    1253.090512257065    1254.6746893456186    2182.669184827119    2549.1057398620333    2740.387555391615
    2745.3499210536406 2785.960512390446 2862.355900555297</array>

                        </property>

                        <property dictRef="me:imFreq">

                                <scalar units="cm-1">669.6227255404566</scalar>

                        </property>

                        <property dictRef="me:spinMultiplicity">

                                <scalar units="cm-1">2</scalar>

                        </property>

                        <property dictRef="me:ZPE">

                                <scalar units="kJ/mol">45.89527165605323</scalar>

                        </property>

                        <group>

                                <scalar>3</scalar>

                        </group>

                </propertyList>

        </molecule>

        <molecule id="OC#C">

                <atomArray>

                                <atom    elementType="O"    id="a1"    x3="0.923063"    y3="-0.026606"
    z3="0.120218"/>

                                <atom    elementType="C"    id="a2"    x3="0.503645"    y3="0.688961"
    z3="1.157405"/>

                                <atom    elementType="C"    id="a3"    x3="0.095992"    y3="1.331995"
    z3="2.089476"/>

                                <atom    elementType="H"    id="a4"    x3="1.917747"    y3="-0.062133"
    z3="0.069338"/>
```

```xml
                              <atom     elementType="H"     id="a5"     x3="-0.280159"     y3="1.894292"
z3="2.904453"/>
               </atomArray>
               <bondArray>
                       <bond atomRefs2="a4 a1" order="1"/>
                       <bond atomRefs2="a1 a2" order="1"/>
                       <bond atomRefs2="a2 a3" order="3"/>
                       <bond atomRefs2="a3 a5" order="1"/>
               </bondArray>
               <propertyList>
                       <property dictRef="me:lumpedSpecies">
                               <array> </array>
                       </property>
                       <property dictRef="me:vibFreqs">
                               <array     units="cm-1">481.1291     529.0111     713.4735     804.1704
1055.8237 1273.393 2314.5103 3444.4906 3526.5169</array>
                       </property>
                       <property dictRef="me:ZPE">
                               <scalar units="kJ/mol">15.415531057421084</scalar>
                       </property>
                       <property dictRef="me:spinMultiplicity">
                               <scalar units="cm-1">1</scalar>
                       </property>
                       <property dictRef="me:epsilon">
                               <scalar>473.17</scalar>
                       </property>
                       <property dictRef="me:sigma">
                               <scalar>5.09</scalar>
                       </property>
                       <group>
                               <scalar>3</scalar>
                       </group>
               </propertyList>
```

```
            <me:energyTransferModel xsi:type="me:ExponentialDown">
                    <scalar units="cm-1">250</scalar>
            </me:energyTransferModel>
        </molecule>
        <molecule id="[CH3]" spinMultiplicity="2">
            <atomArray>
                    <atom   elementType="C"   id="a1"   spinMultiplicity="2"   x3="1.127687"
y3="0.174923" z3="0.029150"/>
                    <atom   elementType="H"   id="a2"   x3="0.617669"   y3="0.439596"
z3="0.945898"/>
                    <atom   elementType="H"   id="a3"   x3="0.617696"   y3="0.269482"   z3="-
0.920361"/>
                    <atom   elementType="H"   id="a4"   x3="2.147768"   y3="-0.184201"
z3="0.061851"/>
            </atomArray>
            <bondArray>
                    <bond atomRefs2="a3 a1" order="1"/>
                    <bond atomRefs2="a1 a4" order="1"/>
                    <bond atomRefs2="a1 a2" order="1"/>
            </bondArray>
            <propertyList>
                    <property dictRef="me:lumpedSpecies">
                            <array> </array>
                    </property>
                    <property dictRef="me:vibFreqs">
                            <array   units="cm-1">493.1447   1451.0848   1451.1138   3124.5656
3302.1608 3302.1879</array>
                    </property>
                    <property dictRef="me:ZPE">
                            <scalar units="kJ/mol">0.0</scalar>
                    </property>
                    <property dictRef="me:spinMultiplicity">
                            <scalar units="cm-1">2</scalar>
                    </property>
```

```xml
            <property dictRef="me:epsilon">
                    <scalar>473.17</scalar>
            </property>
            <property dictRef="me:sigma">
                    <scalar>5.09</scalar>
            </property>
        </propertyList>
        <me:energyTransferModel xsi:type="me:ExponentialDown">
                <scalar units="cm-1">250</scalar>
        </me:energyTransferModel>
    </molecule>
    <molecule id="TS_C[C](O)[CH]_CC#C" spinMultiplicity="2">
        <atomArray>
                <atom    elementType="C"   id="a1"    x3="0.069986"    y3="0.579037"    z3="-1.106369"/>
                <atom    elementType="C"   id="a2"    x3="-0.533534"    y3="-0.354169"    z3="-0.201613"/>
                <atom    elementType="H"   id="a3"    x3="0.440415"    y3="0.082906"    z3="-2.019951"/>
                <atom    elementType="H"   id="a4"    x3="0.936026"    y3="1.087967"    z3="-0.636443"/>
                <atom    elementType="H"   id="a5"    x3="-0.639499"    y3="1.364808"    z3="-1.418113"/>
                <atom    elementType="C"    id="a6"    x3="-1.036545"    y3="-1.137918"    z3="0.565142"/>
                <atom    elementType="H"    id="a7"    x3="-1.479943"    y3="-1.812475"    z3="1.206317"/>
                <atom    elementType="O"   id="a8"    spinMultiplicity="2"    x3="1.316987"    y3="0.409862" z3="1.485096"/>
                <atom    elementType="H"    id="a9"    x3="0.926107"    y3="-0.220018"    z3="2.125935"/>
        </atomArray>
        <bondArray>
                <bond atomRefs2="a3 a1" order="1"/>
                <bond atomRefs2="a5 a1" order="1"/>
                <bond atomRefs2="a1 a4" order="1"/>
```

```xml
        <bond atomRefs2="a1 a2" order="1"/>

        <bond atomRefs2="a2 a6" order="3"/>

        <bond atomRefs2="a6 a7" order="1"/>

        <bond atomRefs2="a8 a9" order="1"/>

      </bondArray>

      <propertyList>

        <property dictRef="me:vibFreqs">

          <array      units="cm-1">34.07882004073068      75.51211460811304
99.94923250422882   315.53097558731133   366.3426099017207   373.89180741144355   825.7496675429453
830.6681822953637   1017.6227740860062   1021.8853836336792   1107.2579190956844   1219.5226467421644
1225.429067380171   1335.569043763414   2387.5337086069903   2630.7701683237497   2668.853331902999
2683.764878686289 2773.3476622820094 2896.002092720524</array>

        </property>

        <property dictRef="me:imFreq">

          <scalar units="cm-1">137.10202497727704</scalar>

        </property>

        <property dictRef="me:spinMultiplicity">

          <scalar units="cm-1">2</scalar>

        </property>

        <property dictRef="me:ZPE">

          <scalar units="kJ/mol">-20.130727365375378</scalar>

        </property>

      </propertyList>

    </molecule>

    <molecule id="TS_C[C](O)[CH]_C[C]=CO" spinMultiplicity="4">

      <atomArray>

        <atom   elementType="C"   id="a1"   x3="0.111064"   y3="0.625976"   z3="-
1.125529"/>

        <atom   elementType="C"   id="a2"   spinMultiplicity="3"   x3="-0.228517"   y3="-
0.303033" z3="-0.084982"/>

        <atom   elementType="H"   id="a3"   x3="-0.356632"   y3="0.352313"   z3="-
2.087133"/>

        <atom   elementType="H"   id="a4"   x3="1.205982"   y3="0.657534"   z3="-
1.286618"/>

        <atom   elementType="H"   id="a5"   x3="-0.206207"   y3="1.654379"   z3="-
0.880026"/>
```

```xml
                    <atom elementType="C" id="a6" spinMultiplicity="2" x3="-0.655668" y3="-1.164507" z3="0.702559"/>
                    <atom elementType="H" id="a7" x3="-1.063238" y3="-1.917324" z3="1.281961"/>
                    <atom elementType="O" id="a8" x3="0.744304" y3="-0.299447" z3="1.429498"/>
                    <atom elementType="H" id="a9" x3="0.448911" y3="0.394108" z3="2.050271"/>
            </atomArray>
            <bondArray>
                    <bond atomRefs2="a3 a1" order="1"/>
                    <bond atomRefs2="a4 a1" order="1"/>
                    <bond atomRefs2="a1 a5" order="1"/>
                    <bond atomRefs2="a1 a2" order="1"/>
                    <bond atomRefs2="a2 a6" order="1"/>
                    <bond atomRefs2="a6 a7" order="1"/>
                    <bond atomRefs2="a6 a8" order="1"/>
                    <bond atomRefs2="a8 a9" order="1"/>
            </bondArray>
            <propertyList>
                    <property dictRef="me:vibFreqs">
                            <array units="cm-1">50.875028839757505    268.0726345636456
339.67167136375684    418.66661321368076    611.9277247972842    795.0681570893986    834.3111950740998
950.8703395008797    1009.6021140354819    1021.1206585360286    1081.0070234926156    1219.479045563087
1222.8530589088093    1328.8212885951423    2225.8572953702583    2592.9292049338574    2676.4145195613128
2682.20592834257    2774.9719291893784    2855.806365475667</array>
                    </property>
                    <property dictRef="me:imFreq">
                            <scalar units="cm-1">433.46351852410413</scalar>
                    </property>
                    <property dictRef="me:spinMultiplicity">
                            <scalar units="cm-1">2</scalar>
                    </property>
                    <property dictRef="me:ZPE">
                            <scalar units="kJ/mol">86.55713133951558</scalar>
```

```xml
                    </property>
                    <group>
                            <scalar>3</scalar>
                    </group>
                </propertyList>
            </molecule>
            <molecule id="C[C]=CO" spinMultiplicity="2">
                <atomArray>
                            <atom    elementType="C"    id="a1"    x3="-2.944266"    y3="-2.485408"    z3="-3.370526"/>
                            <atom    elementType="C"    id="a2"    spinMultiplicity="2"    x3="-3.987747"    y3="-1.444196"    z3="-3.324232"/>
                            <atom    elementType="H"    id="a3"    x3="-2.143092"    y3="-2.233175"    z3="-4.076148"/>
                            <atom    elementType="H"    id="a4"    x3="-3.354100"    y3="-3.461467"    z3="-3.659650"/>
                            <atom    elementType="H"    id="a5"    x3="-2.500471"    y3="-2.587130"    z3="-2.366965"/>
                            <atom    elementType="C"    id="a6"    x3="-4.567396"    y3="-0.707779"    z3="-2.404043"/>
                            <atom    elementType="H"    id="a7"    x3="-5.365002"    y3="-0.009295"    z3="-2.649823"/>
                            <atom    elementType="O"    id="a8"    x3="-4.226570"    y3="-0.826731"    z3="-1.027738"/>
                            <atom    elementType="H"    id="a9"    x3="-4.407591"    y3="0.044126"    z3="-0.583892"/>
                </atomArray>
                <bondArray>
                            <bond atomRefs2="a3 a1" order="1"/>
                            <bond atomRefs2="a4 a1" order="1"/>
                            <bond atomRefs2="a1 a2" order="1"/>
                            <bond atomRefs2="a1 a5" order="1"/>
                            <bond atomRefs2="a2 a6" order="2"/>
                            <bond atomRefs2="a7 a6" order="1"/>
                            <bond atomRefs2="a6 a8" order="1"/>
                            <bond atomRefs2="a8 a9" order="1"/>
```

```xml
        </bondArray>
        <propertyList>
                <property dictRef="me:lumpedSpecies">
                        <array> </array>
                </property>
                <property dictRef="me:vibFreqs">
                        <array units="cm-1">126.5642 143.1921 191.2416 350.8836 630.14
855.6127 887.573 1010.6129 1082.0165 1098.7118 1244.0042 1390.9419 1443.3808 1530.8824 1543.8514
1779.8199 3002.6783 3067.9887 3084.5372 3138.9076 3479.7145</array>
                </property>
                <property dictRef="me:ZPE">
                        <scalar units="kJ/mol">-122.84193448622456</scalar>
                </property>
                <property dictRef="me:spinMultiplicity">
                        <scalar units="cm-1">2</scalar>
                </property>
                <property dictRef="me:epsilon">
                        <scalar>473.17</scalar>
                </property>
                <property dictRef="me:sigma">
                        <scalar>5.09</scalar>
                </property>
                <group>
                        <scalar>3</scalar>
                </group>
        </propertyList>
        <me:energyTransferModel xsi:type="me:ExponentialDown">
                <scalar units="cm-1">250</scalar>
        </me:energyTransferModel>
</molecule>
<molecule id="TS_C[C](O)[CH]_CC(=O)[CH2]" spinMultiplicity="4">
        <atomArray>
```

```xml
                              <atom elementType="C" id="a1" x3="0.168977" y3="0.757948" z3="-
0.891574"/>
                              <atom elementType="C" id="a2" spinMultiplicity="2" x3="0.081041" y3="-
0.098879" z3="0.325005"/>
                              <atom elementType="H" id="a3" x3="-0.647715" y3="0.540095" z3="-
1.582088"/>
                              <atom elementType="H" id="a4" x3="1.126757" y3="0.583731" z3="-
1.397157"/>
                              <atom elementType="H" id="a5" x3="0.133798" y3="1.817870" z3="-
0.614242"/>
                              <atom elementType="C" id="a6" spinMultiplicity="3" x3="-0.770981" y3="-
1.079873" z3="0.782417"/>
                              <atom elementType="H" id="a7" x3="-1.399206" y3="-1.722851"
z3="0.166891"/>
                              <atom elementType="O" id="a8" x3="1.019187" y3="0.013286"
z3="1.320973"/>
                              <atom elementType="H" id="a9" x3="0.288142" y3="-0.811327"
z3="1.889775"/>
                    </atomArray>
                    <bondArray>
                              <bond atomRefs2="a3 a1" order="1"/>
                              <bond atomRefs2="a4 a1" order="1"/>
                              <bond atomRefs2="a1 a5" order="1"/>
                              <bond atomRefs2="a1 a2" order="1"/>
                              <bond atomRefs2="a7 a6" order="1"/>
                              <bond atomRefs2="a2 a6" order="1"/>
                              <bond atomRefs2="a2 a8" order="1"/>
                              <bond atomRefs2="a8 a9" order="1"/>
                    </bondArray>
                    <propertyList>
                              <property dictRef="me:vibFreqs">
                                        <array units="cm-1">112.58658852550651    343.423347815558
386.6417187273103    496.9564165393998    745.99148381268    798.8939449178897    868.4377151726121
961.4081239372495    1002.5350139083008    1022.5955896590792    1208.5268385115346    1219.4513074024842
1297.6967642787952    1305.4597666115576    1530.8373275211966    2372.300934554525    2661.006162730493
2697.699056384478    2715.0066877621684    2797.76774368756</array>
                              </property>
```

```xml
<property dictRef="me:imFreq">
        <scalar units="cm-1">2614.7442402567326</scalar>
</property>
<property dictRef="me:spinMultiplicity">
        <scalar units="cm-1">2</scalar>
</property>
<property dictRef="me:ZPE">
        <scalar units="kJ/mol">-9.370536989602442</scalar>
</property>
<group>
        <scalar>1</scalar>
</group>
</propertyList>
</molecule>
<molecule id="CC(=O)[CH2]" spinMultiplicity="2">
        <atomArray>
                <atom elementType="C" id="a1" x3="-5.986158" y3="0.110076" z3="2.213133"/>
                <atom elementType="C" id="a2" x3="-6.185807" y3="-0.707875" z3="3.488542"/>
                <atom elementType="H" id="a3" x3="-6.921668" y3="0.600747" z3="1.919555"/>
                <atom elementType="H" id="a4" x3="-5.667551" y3="-0.534977" z3="1.385968"/>
                <atom elementType="H" id="a5" x3="-5.222599" y3="0.868338" z3="2.396186"/>
                <atom elementType="C" id="a6" spinMultiplicity="2" x3="-7.141796" y3="-1.765039" z3="3.439485"/>
                <atom elementType="H" id="a7" x3="-7.703270" y3="-1.982229" z3="2.538829"/>
                <atom elementType="O" id="a8" x3="-5.533493" y3="-0.460483" z3="4.544576"/>
                <atom elementType="H" id="a9" x3="-7.308304" y3="-2.363708" z3="4.326396"/>
        </atomArray>
        <bondArray>
```

```xml
                    <bond atomRefs2="a4 a1" order="1"/>

                    <bond atomRefs2="a3 a1" order="1"/>

                    <bond atomRefs2="a1 a5" order="1"/>

                    <bond atomRefs2="a1 a2" order="1"/>

                    <bond atomRefs2="a7 a6" order="1"/>

                    <bond atomRefs2="a6 a2" order="1"/>

                    <bond atomRefs2="a6 a9" order="1"/>

                    <bond atomRefs2="a2 a8" order="2"/>

            </bondArray>

            <propertyList>

                    <property dictRef="me:lumpedSpecies">

                            <array> </array>

                    </property>

                    <property dictRef="me:vibFreqs">

                            <array units="cm-1">58.4303 396.3258 432.5256 504.4372 530.2006
803.9659 805.7888 950.6176 1071.0276 1075.9865 1259.1608 1432.3522 1452.619 1516.8278 1537.3198
1549.7219 3054.6874 3109.2343 3160.1011 3172.1526 3284.316</array>

                    </property>

                    <property dictRef="me:ZPE">

                            <scalar units="kJ/mol">-253.32880552020177</scalar>

                    </property>

                    <property dictRef="me:spinMultiplicity">

                            <scalar units="cm-1">2</scalar>

                    </property>

                    <property dictRef="me:epsilon">

                            <scalar>473.17</scalar>

                    </property>

                    <property dictRef="me:sigma">

                            <scalar>5.09</scalar>

                    </property>

                    <group>

                            <scalar>1</scalar>

                    </group>
```

```xml
        </propertyList>

        <me:energyTransferModel xsi:type="me:ExponentialDown">

                <scalar units="cm-1">250</scalar>

        </me:energyTransferModel>

    </molecule>

    <molecule id="TS_CC(=O)[CH2]_C=C=O" spinMultiplicity="2">

        <atomArray>

                        <atom    elementType="C"   id="a1"   spinMultiplicity="2"   x3="0.143125"
y3="1.175279" z3="0.323718"/>

                        <atom    elementType="C"    id="a2"    x3="-0.224020"    y3="-0.858827"
z3="0.524965"/>

                        <atom    elementType="O"    id="a3"    x3="-0.433693"    y3="-1.048596"
z3="1.662263"/>

                        <atom   elementType="C"   id="a4"   x3="-0.055145"   y3="-1.079479"   z3="-
0.783920"/>

                        <atom   elementType="H"   id="a5"   x3="0.909884"   y3="-1.267644"   z3="-
1.224523"/>

                        <atom   elementType="H"   id="a6"   x3="-0.852736"   y3="-0.976884"   z3="-
1.500955"/>

                        <atom    elementType="H"    id="a7"    x3="-0.805129"    y3="1.475581"
z3="0.709577"/>

                        <atom   elementType="H"   id="a8"   x3="0.340548"   y3="1.399033"   z3="-
0.700117"/>

                        <atom    elementType="H"    id="a9"    x3="0.977166"    y3="1.181538"
z3="0.988991"/>

        </atomArray>

        <bondArray>

                <bond atomRefs2="a6 a4" order="1"/>

                <bond atomRefs2="a5 a4" order="1"/>

                <bond atomRefs2="a4 a2" order="2"/>

                <bond atomRefs2="a8 a1" order="1"/>

                <bond atomRefs2="a1 a7" order="1"/>

                <bond atomRefs2="a1 a9" order="1"/>

                <bond atomRefs2="a2 a3" order="2"/>

        </bondArray>

        <propertyList>
```

```xml
            <property dictRef="me:vibFreqs">
                              <array    units="cm-1">43.53383695127201    223.46437201006412
244.13117077184697    376.8747725995906    432.5408222085705    646.60267552218    677.3818647429399
880.5791677694465    974.420899810055    1051.7093769237404    1181.4124956535857    1245.73345124743
1254.7316437109375    1266.5881277442281    2138.3825998149755    2728.7491052026817    2736.8409703234584
2738.805125974537   2784.5377768325607   2797.334761522665</array>
            </property>
            <property dictRef="me:imFreq">
                    <scalar units="cm-1">670.8472319303758</scalar>
            </property>
            <property dictRef="me:spinMultiplicity">
                    <scalar units="cm-1">2</scalar>
            </property>
            <property dictRef="me:ZPE">
                    <scalar units="kJ/mol">-70.26064437976538</scalar>
            </property>
            <group>
                    <scalar>1</scalar>
            </group>
        </propertyList>
    </molecule>
    <molecule id="C=C=O">
        <atomArray>
                        <atom    elementType="C"    id="a1"    x3="1.012457"    y3="-0.033580"
z3="0.064922"/>
                        <atom    elementType="C"    id="a2"    x3="0.358743"    y3="1.093125"
z3="0.177009"/>
                        <atom    elementType="O"    id="a3"    x3="-0.234259"    y3="2.115189"
z3="0.278686"/>
                        <atom   elementType="H"   id="a4"   x3="0.479545"   y3="-0.970476"   z3="-
0.028283"/>
                        <atom    elementType="H"    id="a5"    x3="2.094294"    y3="-0.042780"
z3="0.064006"/>
        </atomArray>
        <bondArray>
                <bond atomRefs2="a4 a1" order="1"/>
```

```xml
                    <bond atomRefs2="a5 a1" order="1"/>

                    <bond atomRefs2="a1 a2" order="2"/>

                    <bond atomRefs2="a2 a3" order="2"/>

            </bondArray>

            <propertyList>

                    <property dictRef="me:lumpedSpecies">

                            <array>  </array>

                    </property>

                    <property dictRef="me:vibFreqs">

                            <array    units="cm-1">488.8962    595.5193    724.3075    1049.2654
1186.4846 1463.3583 2206.0246 3200.894 3281.1398</array>

                    </property>

                    <property dictRef="me:ZPE">

                            <scalar units="kJ/mol">-128.31340641696676</scalar>

                    </property>

                    <property dictRef="me:spinMultiplicity">

                            <scalar units="cm-1">1</scalar>

                    </property>

                    <property dictRef="me:epsilon">

                            <scalar>473.17</scalar>

                    </property>

                    <property dictRef="me:sigma">

                            <scalar>5.09</scalar>

                    </property>

                    <group>

                            <scalar>1</scalar>

                    </group>

            </propertyList>

            <me:energyTransferModel xsi:type="me:ExponentialDown">

                    <scalar units="cm-1">250</scalar>

            </me:energyTransferModel>

    </molecule>
```

```xml
<molecule id="[CH3]" spinMultiplicity="2">
    <atomArray>
        <atom elementType="C" id="a1" spinMultiplicity="2" x3="1.109812" y3="-0.023564" z3="0.154798"/>
        <atom elementType="H" id="a2" x3="0.600615" y3="-0.973792" z3="0.062502"/>
        <atom elementType="H" id="a3" x3="0.600626" y3="0.822334" z3="0.597442"/>
        <atom elementType="H" id="a4" x3="2.130801" y3="0.078704" z3="-0.188523"/>
    </atomArray>
    <bondArray>
        <bond atomRefs2="a4 a1" order="1"/>
        <bond atomRefs2="a2 a1" order="1"/>
        <bond atomRefs2="a1 a3" order="1"/>
    </bondArray>
    <propertyList>
        <property dictRef="me:lumpedSpecies">
            <array> </array>
        </property>
        <property dictRef="me:vibFreqs">
            <array units="cm-1">493.7794 1451.1913 1451.2012 3123.9667 3301.5144 3301.5564</array>
        </property>
        <property dictRef="me:ZPE">
            <scalar units="kJ/mol">0.0</scalar>
        </property>
        <property dictRef="me:spinMultiplicity">
            <scalar units="cm-1">2</scalar>
        </property>
        <property dictRef="me:epsilon">
            <scalar>473.17</scalar>
        </property>
        <property dictRef="me:sigma">
```

```xml
                    <scalar>5.09</scalar>

                </property>

            </propertyList>

            <me:energyTransferModel xsi:type="me:ExponentialDown">

                    <scalar units="cm-1">250</scalar>

            </me:energyTransferModel>

        </molecule>

        <molecule id="TS_C=C=O_[C]=O" spinMultiplicity="5">

            <atomArray>

                    <atom     elementType="C"    id="a1"    spinMultiplicity="2"    x3="0.962559"
y3="0.043737" z3="0.159587"/>

                    <atom     elementType="C"    id="a2"    spinMultiplicity="3"    x3="0.305550"
y3="1.123960" z3="0.799537"/>

                    <atom     elementType="O"    id="a3"    spinMultiplicity="2"    x3="0.031514"
y3="1.702080" z3="-0.218640"/>

                    <atom    elementType="H"    id="a4"    x3="0.392290"    y3="-0.821495"    z3="-
0.138544"/>

                    <atom    elementType="H"    id="a5"    x3="2.018868"    y3="0.113196"    z3="-
0.045600"/>

            </atomArray>

            <bondArray>

                    <bond atomRefs2="a3 a2" order="1"/>

                    <bond atomRefs2="a4 a1" order="1"/>

                    <bond atomRefs2="a5 a1" order="1"/>

                    <bond atomRefs2="a1 a2" order="1"/>

            </bondArray>

            <propertyList>

                <property dictRef="me:vibFreqs">

                            <array    units="cm-1">685.8937249385536      749.2639637260459
1217.9279191240007   1295.9533412727633   1980.534495119045   2648.548437097153   2737.0098366483826
4756.65681887489</array>

                </property>

                <property dictRef="me:imFreq">

                    <scalar units="cm-1">5965.107286011783</scalar>

                </property>
```

```xml
                    <property dictRef="me:spinMultiplicity">
                            <scalar units="cm-1">2</scalar>
                    </property>
                    <property dictRef="me:ZPE">
                            <scalar units="kJ/mol">389.6554521052303</scalar>
                    </property>
                    <group>
                            <scalar>3</scalar>
                    </group>
            </propertyList>
    </molecule>
    <molecule id="[C]=O" spinMultiplicity="3">
            <atomArray>
                    <atom  elementType="C"  id="a1"  spinMultiplicity="3"  x3="1.017656"  y3="-0.088845" z3="0.020507"/>
                    <atom     elementType="O"     id="a2"     x3="2.171120"     y3="-0.088845" z3="0.020507"/>
            </atomArray>
            <bondArray>
                    <bond atomRefs2="a1 a2" order="2"/>
            </bondArray>
            <propertyList>
                    <property dictRef="me:lumpedSpecies">
                            <array> </array>
                    </property>
                    <property dictRef="me:vibFreqs">
                            <array units="cm-1">2087.6529</array>
                    </property>
                    <property dictRef="me:ZPE">
                            <scalar units="kJ/mol">480.4276222939575</scalar>
                    </property>
                    <property dictRef="me:spinMultiplicity">
                            <scalar units="cm-1">1</scalar>
```

```xml
                        </property>
                        <property dictRef="me:epsilon">
                                <scalar>473.17</scalar>
                        </property>
                        <property dictRef="me:sigma">
                                <scalar>5.09</scalar>
                        </property>
                        <group>
                                <scalar>3</scalar>
                        </group>
                </propertyList>
                <me:energyTransferModel xsi:type="me:ExponentialDown">
                        <scalar units="cm-1">250</scalar>
                </me:energyTransferModel>
        </molecule>
        <molecule id="[CH2]" spinMultiplicity="3">
                <atomArray>
                        <atom    elementType="C"    id="a1"    spinMultiplicity="3"    x3="1.298784"
y3="0.065739" z3="0.297053"/>
                        <atom    elementType="H"    id="a2"    x3="0.427056"    y3="0.330993"
z3="0.854103"/>
                        <atom    elementType="H"    id="a3"    x3="2.171053"    y3="-0.199386"    z3="-
0.259213"/>
                </atomArray>
                <bondArray>
                        <bond atomRefs2="a3 a1" order="1"/>
                        <bond atomRefs2="a1 a2" order="1"/>
                </bondArray>
                <propertyList>
                        <property dictRef="me:lumpedSpecies">
                                <array> </array>
                        </property>
                        <property dictRef="me:vibFreqs">
```

                    <array units="cm-1">1214.0809 3228.1343 3583.5784</array>
                </property>
                <property dictRef="me:ZPE">
                    <scalar units="kJ/mol">0.0</scalar>
                </property>
                <property dictRef="me:spinMultiplicity">
                    <scalar units="cm-1">1</scalar>
                </property>
                <property dictRef="me:epsilon">
                    <scalar>473.17</scalar>
                </property>
                <property dictRef="me:sigma">
                    <scalar>5.09</scalar>
                </property>
            </propertyList>
            <me:energyTransferModel xsi:type="me:ExponentialDown">
                <scalar units="cm-1">250</scalar>
            </me:energyTransferModel>
        </molecule>
        <molecule id="TS_C[CH]C=O_CC=C=O" spinMultiplicity="2">
            <atomArray>
                            <atom    elementType="C"    id="a1"    x3="0.073018"    y3="1.225015"
z3="0.046821"/>
                            <atom    elementType="C"    id="a2"    x3="-0.563927"    y3="-0.118035"
z3="0.169336"/>
                            <atom    elementType="H"    id="a3"    x3="-1.601360"    y3="-0.136828"
z3="0.492122"/>
                            <atom  elementType="C"  id="a4"  spinMultiplicity="2"  x3="0.085577"  y3="-
1.238168" z3="-0.097076"/>
                            <atom    elementType="O"    id="a5"    x3="1.160172"    y3="-1.859839"    z3="-
0.467856"/>
                            <atom    elementType="H"    id="a6"    x3="0.062152"    y3="1.751114"
z3="1.011850"/>
                            <atom    elementType="H"    id="a7"    x3="-0.466297"    y3="1.847282"    z3="-
0.681261"/>

```xml
                              <atom elementType="H" id="a8" x3="1.119686" y3="1.167229" z3="-0.283145"/>
                              <atom elementType="H" id="a9" x3="0.130980" y3="-2.637771" z3="-0.190790"/>
                 </atomArray>
                 <bondArray>
                        <bond atomRefs2="a7 a1" order="1"/>
                        <bond atomRefs2="a5 a9" order="1"/>
                        <bond atomRefs2="a5 a4" order="1"/>
                        <bond atomRefs2="a8 a1" order="1"/>
                        <bond atomRefs2="a4 a2" order="2"/>
                        <bond atomRefs2="a1 a2" order="1"/>
                        <bond atomRefs2="a1 a6" order="1"/>
                        <bond atomRefs2="a2 a3" order="1"/>
                 </bondArray>
                 <propertyList>
                        <property dictRef="me:vibFreqs">
                              <array units="cm-1">72.18627647383583    202.2784057614637
269.7410997907628    409.2389569153034    603.2352505835829    768.4556231534973    969.981875262907
1006.4738514522224  1087.838369555904   1190.387229704533   1246.1443435136725  1251.3812943574355
1332.087021701302   1338.6203795428926  1765.8878622998056  2503.8116121921603  2689.3560589113017
2690.1869165287794  2763.282530269368   2783.2328721818067</array>
                        </property>
                        <property dictRef="me:imFreq">
                              <scalar units="cm-1">2097.5479441638863</scalar>
                        </property>
                        <property dictRef="me:spinMultiplicity">
                              <scalar units="cm-1">2</scalar>
                        </property>
                        <property dictRef="me:ZPE">
                              <scalar units="kJ/mol">14.946732222533365</scalar>
                        </property>
                 </propertyList>
          </molecule>
```

```xml
<molecule id="CC=C=O">
    <atomArray>
        <atom elementType="C" id="a1" x3="0.936392" y3="-0.050765" z3="-0.074897"/>
        <atom elementType="C" id="a2" x3="0.404373" y3="-0.368566" z3="-1.465557"/>
        <atom elementType="C" id="a3" x3="-0.349485" y3="-1.402507" z3="-1.737427"/>
        <atom elementType="O" id="a4" x3="-1.034879" y3="-2.344721" z3="-1.978285"/>
        <atom elementType="H" id="a5" x3="0.606298" y3="-0.795266" z3="0.655249"/>
        <atom elementType="H" id="a6" x3="0.581372" y3="0.931944" z3="0.257334"/>
        <atom elementType="H" id="a7" x3="2.032742" y3="-0.038584" z3="-0.076120"/>
        <atom elementType="H" id="a8" x3="0.659738" y3="0.294955" z3="-2.285266"/>
    </atomArray>
    <bondArray>
        <bond atomRefs2="a8 a2" order="1"/>
        <bond atomRefs2="a4 a3" order="2"/>
        <bond atomRefs2="a3 a2" order="2"/>
        <bond atomRefs2="a2 a1" order="1"/>
        <bond atomRefs2="a7 a1" order="1"/>
        <bond atomRefs2="a1 a6" order="1"/>
        <bond atomRefs2="a1 a5" order="1"/>
    </bondArray>
    <propertyList>
        <property dictRef="me:lumpedSpecies">
            <array> </array>
        </property>
        <property dictRef="me:vibFreqs">
            <array units="cm-1">141.9181 228.596 541.9935 632.952 676.3183 900.1678 1092.2419 1110.7066 1190.2559 1416.0245 1472.5531 1554.9434 1571.8748 2195.4382 3044.6408 3092.7643 3125.9167 3197.6519</array>
```

```xml
            </property>
            <property dictRef="me:ZPE">
                    <scalar units="kJ/mol">-72.98140111101776</scalar>
            </property>
            <property dictRef="me:spinMultiplicity">
                    <scalar units="cm-1">1</scalar>
            </property>
            <property dictRef="me:epsilon">
                    <scalar>473.17</scalar>
            </property>
            <property dictRef="me:sigma">
                    <scalar>5.09</scalar>
            </property>
        </propertyList>
        <me:energyTransferModel xsi:type="me:ExponentialDown">
                <scalar units="cm-1">250</scalar>
        </me:energyTransferModel>
    </molecule>
    <molecule id="[H]" spinMultiplicity="2">
            <atomArray>
                    <atom    elementType="H"    id="a1"    spinMultiplicity="2"    x3="0.936577"
y3="0.071655" z3="-0.029037"/>
            </atomArray>
            <propertyList>
                    <property dictRef="me:lumpedSpecies">
                            <array> </array>
                    </property>
                    <property dictRef="me:vibFreqs">
                            <array units="cm-1"/>
                    </property>
                    <property dictRef="me:ZPE">
                            <scalar units="kJ/mol">0.0</scalar>
```

```xml
                              </property>
                              <property dictRef="me:spinMultiplicity">
                                        <scalar units="cm-1">2</scalar>
                              </property>
                              <property dictRef="me:epsilon">
                                        <scalar>473.17</scalar>
                              </property>
                              <property dictRef="me:sigma">
                                        <scalar>5.09</scalar>
                              </property>
                    </propertyList>
                    <me:energyTransferModel xsi:type="me:ExponentialDown">
                              <scalar units="cm-1">250</scalar>
                    </me:energyTransferModel>
          </molecule>
          <molecule id="TS_C[CH]C=O_[CH2]CC=O" spinMultiplicity="2">
                    <atomArray>
                              <atom    elementType="C"    id="a1"    spinMultiplicity="2"    x3="-0.009356"
y3="1.192452" z3="0.209601"/>
                              <atom    elementType="C"    id="a2"    x3="-0.489000"    y3="-0.123482"    z3="-
0.147599"/>
                              <atom    elementType="H"    id="a3"    x3="-1.557084"    y3="-0.286409"    z3="-
0.257646"/>
                              <atom    elementType="C"    id="a4"    x3="0.355087"    y3="-1.311751"
z3="0.065539"/>
                              <atom    elementType="O"    id="a5"    x3="1.565959"    y3="-1.295740"    z3="-
0.046730"/>
                              <atom    elementType="H"    id="a6"    x3="0.980066"    y3="1.324170"
z3="0.620624"/>
                              <atom    elementType="H"    id="a7"    x3="-0.686870"    y3="2.013683"
z3="0.355104"/>
                              <atom    elementType="H"    id="a8"    x3="0.025623"    y3="0.724329"    z3="-
1.124365"/>
                              <atom    elementType="H"    id="a9"    x3="-0.184425"    y3="-2.237252"
z3="0.325471"/>
                    </atomArray>
```

```xml
<bondArray>
        <bond atomRefs2="a8 a2" order="1"/>
        <bond atomRefs2="a3 a2" order="1"/>
        <bond atomRefs2="a2 a4" order="1"/>
        <bond atomRefs2="a2 a1" order="1"/>
        <bond atomRefs2="a5 a4" order="2"/>
        <bond atomRefs2="a4 a9" order="1"/>
        <bond atomRefs2="a1 a7" order="1"/>
        <bond atomRefs2="a1 a6" order="1"/>
    </bondArray>
    <propertyList>
        <property dictRef="me:vibFreqs">
            <array     units="cm-1">116.13755744868739      267.4055284124708
417.7639331998863    637.1447404267433     840.2674055483908     916.7198772432943     965.0931591623482
1002.8411181867331    1017.200777014368    1099.4730971553624    1214.5868495386085    1284.2064030883578
1309.6563297751495    1412.299859775307    1759.1240153321974    2313.7411976070443    2651.528224265043
2725.3728221063684 2757.1614081480643 2776.7215623307197</array>
        </property>
        <property dictRef="me:imFreq">
            <scalar units="cm-1">1985.8112890657078</scalar>
        </property>
        <property dictRef="me:spinMultiplicity">
            <scalar units="cm-1">2</scalar>
        </property>
        <property dictRef="me:ZPE">
            <scalar units="kJ/mol">-70.68122321210944</scalar>
        </property>
    </propertyList>
</molecule>
<molecule id="[CH2]CC=O" spinMultiplicity="2">
    <atomArray>
        <atom    elementType="C"   id="a1"   spinMultiplicity="2"   x3="-2.375094"   y3="-4.292843" z3="7.404252"/>
```

```xml
                        <atom    elementType="C"    id="a2"    x3="-3.223858"    y3="-4.070516"
z3="6.199714"/>
                        <atom    elementType="H"    id="a3"    x3="-3.493820"    y3="-5.016868"
z3="5.695017"/>
                        <atom    elementType="C"    id="a4"    x3="-2.542395"    y3="-3.183660"
z3="5.166830"/>
                        <atom    elementType="O"    id="a5"    x3="-1.420365"    y3="-2.698279"
z3="5.296299"/>
                        <atom    elementType="H"    id="a6"    x3="-1.397534"    y3="-3.829804"
z3="7.443634"/>
                        <atom    elementType="H"    id="a7"    x3="-2.722337"    y3="-4.903696"
z3="8.227129"/>
                        <atom    elementType="H"    id="a8"    x3="-4.193430"    y3="-3.600598"
z3="6.449379"/>
                        <atom    elementType="H"    id="a9"    x3="-3.156020"    y3="-3.005353"
z3="4.263002"/>
            </atomArray>
            <bondArray>
                    <bond atomRefs2="a9 a4" order="1"/>
                    <bond atomRefs2="a4 a5" order="2"/>
                    <bond atomRefs2="a4 a2" order="1"/>
                    <bond atomRefs2="a3 a2" order="1"/>
                    <bond atomRefs2="a2 a8" order="1"/>
                    <bond atomRefs2="a2 a1" order="1"/>
                    <bond atomRefs2="a1 a6" order="1"/>
                    <bond atomRefs2="a1 a7" order="1"/>
            </bondArray>
            <propertyList>
                    <property dictRef="me:lumpedSpecies">
                            <array> </array>
                    </property>
                    <property dictRef="me:vibFreqs">
                            <array units="cm-1">181.8528 227.8894 301.4778 463.1241 673.0739
721.2343 853.9788 1019.9529 1026.5853 1115.0294 1241.3555 1353.8068 1439.1631 1472.5816 1484.1905
1744.4016 2916.612 2971.284 2973.1368 3179.3526 3293.0998</array>
                    </property>
```

```xml
                    <property dictRef="me:ZPE">
                            <scalar units="kJ/mol">-205.6810779428848</scalar>
                    </property>
                    <property dictRef="me:spinMultiplicity">
                            <scalar units="cm-1">2</scalar>
                    </property>
                    <property dictRef="me:epsilon">
                            <scalar>473.17</scalar>
                    </property>
                    <property dictRef="me:sigma">
                            <scalar>5.09</scalar>
                    </property>
            </propertyList>
            <me:energyTransferModel xsi:type="me:ExponentialDown">
                    <scalar units="cm-1">250</scalar>
            </me:energyTransferModel>
        </molecule>
        <molecule id="TS_C[CH]C=O_CC[C]=O" spinMultiplicity="2">
            <atomArray>
                    <atom    elementType="C"    id="a1"       x3="0.071563"    y3="1.067740"    z3="0.124156"/>
                    <atom   elementType="C"   id="a2"   spinMultiplicity="2"   x3="-0.498600"   y3="-0.299454"   z3="0.003246"/>
                    <atom    elementType="H"    id="a3"       x3="-1.579359"    y3="-0.397735"    z3="0.056906"/>
                    <atom   elementType="C"   id="a4"   x3="0.268546"   y3="-1.388909"   z3="-0.440115"/>
                    <atom   elementType="O"   id="a5"   x3="1.173901"   y3="-1.837549"   z3="-1.060555"/>
                    <atom    elementType="H"    id="a6"   x3="-0.272288"    y3="1.570411"    z3="1.038354"/>
                    <atom   elementType="H"   id="a7"   x3="-0.238334"   y3="1.690666"   z3="-0.731163"/>
                    <atom    elementType="H"    id="a8"    x3="1.170507"    y3="1.072883"    z3="0.135187"/>
```

```xml
                    <atom     elementType="H"     id="a9"     x3="-0.095935"     y3="-1.478054"
z3="0.873983"/>
        </atomArray>
        <bondArray>
            <bond atomRefs2="a5 a4" order="2"/>
            <bond atomRefs2="a7 a1" order="1"/>
            <bond atomRefs2="a4 a2" order="1"/>
            <bond atomRefs2="a4 a9" order="1"/>
            <bond atomRefs2="a2 a3" order="1"/>
            <bond atomRefs2="a2 a1" order="1"/>
            <bond atomRefs2="a1 a8" order="1"/>
            <bond atomRefs2="a1 a6" order="1"/>
        </bondArray>
        <propertyList>
            <property dictRef="me:vibFreqs">
                        <array     units="cm-1">72.6652227118511        204.2884170060438
407.90783198169703    556.8859424690622    629.7500144818786    938.178259206811    954.3363314882394
1010.2862999156665   1087.1221393178587   1189.7178399839183   1241.378060425396   1245.8604135810317
1333.4311725011157   1342.306669553698   1929.5183984307798   2206.7483381529382   2682.0945665957593
2691.008616734564 2748.189379061061 2782.579571106665</array>
            </property>
            <property dictRef="me:imFreq">
                    <scalar units="cm-1">2397.5872356912278</scalar>
            </property>
            <property dictRef="me:spinMultiplicity">
                    <scalar units="cm-1">2</scalar>
            </property>
            <property dictRef="me:ZPE">
                    <scalar units="kJ/mol">-61.88005925905986</scalar>
            </property>
        </propertyList>
    </molecule>
    <molecule id="CC[C]=O" spinMultiplicity="2">
        <atomArray>
```

```xml
                        <atom  elementType="C"  id="a1"  x3="-1.042393"  y3="3.058704"  z3="-
1.204380"/>
                        <atom  elementType="C"  id="a2"  x3="-1.526576"  y3="1.701664"  z3="-
0.656848"/>
                        <atom  elementType="H"  id="a3"  x3="-2.401899"  y3="1.359045"  z3="-
1.226513"/>
                        <atom  elementType="C"  id="a4"  spinMultiplicity="2"  x3="-0.453134"
y3="0.628850" z3="-0.847706"/>
                        <atom  elementType="O"  id="a5"  x3="-0.186896"  y3="-0.340018"  z3="-
0.178736"/>
                        <atom  elementType="H"  id="a6"  x3="-0.749202"  y3="2.957206"  z3="-
2.253324"/>
                        <atom  elementType="H"  id="a7"  x3="-0.177167"  y3="3.417084"  z3="-
0.638359"/>
                        <atom  elementType="H"  id="a8"  x3="-1.838814"  y3="3.805042"  z3="-
1.129585"/>
                        <atom  elementType="H"  id="a9"  x3="-1.810329"  y3="1.760653"
z3="0.402094"/>
                </atomArray>
                <bondArray>
                        <bond atomRefs2="a6 a1" order="1"/>
                        <bond atomRefs2="a3 a2" order="1"/>
                        <bond atomRefs2="a1 a8" order="1"/>
                        <bond atomRefs2="a1 a2" order="1"/>
                        <bond atomRefs2="a1 a7" order="1"/>
                        <bond atomRefs2="a4 a2" order="1"/>
                        <bond atomRefs2="a4 a5" order="2"/>
                        <bond atomRefs2="a2 a9" order="1"/>
                </bondArray>
                <propertyList>
                        <property dictRef="me:lumpedSpecies">
                                <array> </array>
                        </property>
                        <property dictRef="me:vibFreqs">
```

```xml
                        <array units="cm-1">70.2123 221.8977 310.0937 475.6451 735.6191
805.6244 991.274 1062.3115 1115.8632 1303.6974 1337.0866 1468.3124 1500.5331 1563.5091 1570.8063
1837.9206 3034.7733 3066.4632 3079.1214 3132.5208 3137.8561</array>
                    </property>
                    <property dictRef="me:ZPE">
                            <scalar units="kJ/mol">-253.8989628640034</scalar>
                    </property>
                    <property dictRef="me:spinMultiplicity">
                            <scalar units="cm-1">2</scalar>
                    </property>
                    <property dictRef="me:epsilon">
                            <scalar>473.17</scalar>
                    </property>
                    <property dictRef="me:sigma">
                            <scalar>5.09</scalar>
                    </property>
                </propertyList>
                <me:energyTransferModel xsi:type="me:ExponentialDown">
                        <scalar units="cm-1">250</scalar>
                </me:energyTransferModel>
            </molecule>
            <molecule id="TS_C[CH]C=O_[CH2]C=CO" spinMultiplicity="4">
                    <atomArray>
                            <atom elementType="C" id="a1" spinMultiplicity="2" x3="0.407438" y3="-
0.488081" z3="0.908124"/>
                            <atom elementType="C" id="a2" spinMultiplicity="2" x3="0.557302"
y3="0.510401" z3="-0.124022"/>
                            <atom elementType="H" id="a3" x3="1.455990" y3="1.080695" z3="-
0.318846"/>
                            <atom elementType="C" id="a4" spinMultiplicity="2" x3="-0.548026"
y3="0.510849" z3="-0.968925"/>
                            <atom elementType="O" id="a5" x3="-1.387683" y3="-0.514289" z3="-
0.807488"/>
                            <atom elementType="H" id="a6" x3="-0.641381" y3="-1.093347"
z3="0.059912"/>
```

```xml
                              <atom      elementType="H"      id="a7"      x3="-0.372418"      y3="-0.310592"
z3="1.657143"/>

                              <atom      elementType="H"      id="a8"      x3="1.301983"      y3="-0.934761"
z3="1.332204"/>

                              <atom   elementType="H"   id="a9"   x3="-0.773205"   y3="1.239124"   z3="-
1.738103"/>

                    </atomArray>

                    <bondArray>

                              <bond atomRefs2="a9 a4" order="1"/>

                              <bond atomRefs2="a4 a5" order="1"/>

                              <bond atomRefs2="a4 a2" order="1"/>

                              <bond atomRefs2="a5 a6" order="1"/>

                              <bond atomRefs2="a3 a2" order="1"/>

                              <bond atomRefs2="a2 a1" order="1"/>

                              <bond atomRefs2="a1 a8" order="1"/>

                              <bond atomRefs2="a1 a7" order="1"/>

                    </bondArray>

                    <propertyList>

                              <property dictRef="me:vibFreqs">

                                        <array      units="cm-1">387.91439577831915      486.7588942395599
500.6328932388724    714.6053860569169    818.8610608588384    843.6609538367111    910.1579403076456
977.3132895531262   1014.8866262494092   1138.8406805359448   1162.608762405642   1240.6684208447687
1299.161904513013   1371.5699206657323   1504.1156493950593   1966.3836283428584   2674.9065914604307
2728.4444050080588 2741.398873549646 2756.7107823841866</array>

                              </property>

                              <property dictRef="me:imFreq">

                                        <scalar units="cm-1">2199.1947184534956</scalar>

                              </property>

                              <property dictRef="me:spinMultiplicity">

                                        <scalar units="cm-1">2</scalar>

                              </property>

                              <property dictRef="me:ZPE">

                                        <scalar units="kJ/mol">-119.32596146628535</scalar>

                              </property>

                    </propertyList>
```

```
        </molecule>
        <molecule id="[CH2]C=CO" spinMultiplicity="2">
                <atomArray>
                                <atom elementType="C" id="a1" spinMultiplicity="2" x3="8.240427"
y3="3.473715" z3="1.994384"/>
                                <atom elementType="C" id="a2" x3="8.169663" y3="3.872480"
z3="0.671291"/>
                                <atom elementType="H" id="a3" x3="9.028918" y3="4.375266"
z3="0.237581"/>
                                <atom elementType="C" id="a4" x3="7.093437" y3="3.696352" z3="-
0.185743"/>
                                <atom elementType="O" id="a5" x3="5.890179" y3="3.088610"
z3="0.125092"/>
                                <atom elementType="H" id="a6" x3="5.891456" y3="2.783145"
z3="1.070946"/>
                                <atom elementType="H" id="a7" x3="7.421931" y3="2.965425"
z3="2.497522"/>
                                <atom elementType="H" id="a8" x3="9.125953" y3="3.654629"
z3="2.588936"/>
                                <atom elementType="H" id="a9" x3="7.106648" y3="4.034153" z3="-
1.211971"/>
                </atomArray>
                <bondArray>
                                <bond atomRefs2="a9 a4" order="1"/>
                                <bond atomRefs2="a4 a5" order="1"/>
                                <bond atomRefs2="a4 a2" order="2"/>
                                <bond atomRefs2="a5 a6" order="1"/>
                                <bond atomRefs2="a3 a2" order="1"/>
                                <bond atomRefs2="a2 a1" order="1"/>
                                <bond atomRefs2="a1 a7" order="1"/>
                                <bond atomRefs2="a1 a8" order="1"/>
                </bondArray>
                <propertyList>
                                <property dictRef="me:lumpedSpecies">
                                        <array> </array>
                                </property>
```

```xml
<property dictRef="me:vibFreqs">
    <array units="cm-1">203.0814 284.0746 448.9446 550.9639 653.2084
732.5866 743.2102 962.0162 969.2106 1068.4704 1193.8555 1268.9194 1376.6082 1467.8257 1505.6338
1567.8338 3150.7364 3181.4921 3246.3482 3256.2004 3501.6458</array>
</property>
<property dictRef="me:ZPE">
    <scalar units="kJ/mol">-232.93487229782465</scalar>
</property>
<property dictRef="me:spinMultiplicity">
    <scalar units="cm-1">2</scalar>
</property>
<property dictRef="me:epsilon">
    <scalar>473.17</scalar>
</property>
<property dictRef="me:sigma">
    <scalar>5.09</scalar>
</property>
</propertyList>
<me:energyTransferModel xsi:type="me:ExponentialDown">
    <scalar units="cm-1">250</scalar>
</me:energyTransferModel>
</molecule>
</moleculeList>
<reactionList>
    <reaction active="true" id="CC#C_C[C]=CO">
        <reactant>
            <molecule me:type="modelled" ref="CC#C"/>
        </reactant>
        <reactant>
            <molecule me:type="excessReactant" ref="[OH]"/>
        </reactant>
        <product>
            <molecule me:type="modelled" ref="C[C][CH]O"/>
```

```xml
            </product>
            <group>
                    <scalar>1</scalar>
            </group>
            <me:MCRCMethod xsi:type="MesmerILT">
                    <me:preExponential units="cm3 molecule-1 s-1">1E-10</me:preExponential>
                    <me:activationEnergy units="kJ/mol">0</me:activationEnergy>
                    <me:TInfinity>298.0</me:TInfinity>
                    <me:nInfinity>0.0</me:nInfinity>
            </me:MCRCMethod>
            <me:excessReactantConc>1E18</me:excessReactantConc>
    </reaction>
    <reaction active="true" id="CC#C_CC(=[CH])O">
            <reactant>
                    <molecule me:type="modelled" ref="CC#C"/>
            </reactant>
            <reactant>
                    <molecule me:type="excessReactant" ref="[OH]"/>
            </reactant>
            <product>
                    <molecule me:type="modelled" ref="C[C](O)[CH]"/>
            </product>
            <group>
                    <scalar>1</scalar>
            </group>
            <me:MCRCMethod xsi:type="MesmerILT">
                    <me:preExponential units="cm3 molecule-1 s-1">1E-10</me:preExponential>
                    <me:activationEnergy units="kJ/mol">0</me:activationEnergy>
                    <me:TInfinity>298.0</me:TInfinity>
                    <me:nInfinity>0.0</me:nInfinity>
            </me:MCRCMethod>
            <me:excessReactantConc>1E18</me:excessReactantConc>
```

```xml
        </reaction>
        <reaction active="true" id="CC#C_[CH2]C#C">
                <reactant>
                        <molecule me:type="modelled" ref="CC#C"/>
                </reactant>
                <reactant>
                        <molecule me:type="excessReactant" ref="[OH]"/>
                </reactant>
                <product>
                        <molecule me:type="modelled" ref="[CH2]C#C"/>
                </product>
                <group>
                        <scalar>1</scalar>
                </group>
                <me:transitionState>
                        <molecule me:type="transitionState" ref="TS_CC#C_[CH2]C#C"/>
                </me:transitionState>
                <me:MCRCMethod name="SimpleRRKM"/>
        </reaction>
        <reaction active="true" id="C[C][CH]O_[CH2]C=CO">
                <reactant>
                        <molecule me:type="modelled" ref="C[C][CH]O"/>
                </reactant>
                <product>
                        <molecule me:type="modelled" ref="[CH2]C=CO"/>
                </product>
                <group>
                        <scalar>3</scalar>
                </group>
                <me:transitionState>
                        <molecule me:type="transitionState" ref="TS_C[C][CH]O_[CH2]C=CO"/>
                </me:transitionState>
```

```xml
                <me:MCRCMethod name="SimpleRRKM"/>
    </reaction>
    <reaction active="true" id="C[C][CH]O_C[CH]C=O">
            <reactant>
                    <molecule me:type="modelled" ref="C[C][CH]O"/>
            </reactant>
            <product>
                    <molecule me:type="modelled" ref="C[CH]C=O"/>
            </product>
            <group>
                    <scalar>2</scalar>
            </group>
            <me:transitionState>
                    <molecule me:type="transitionState" ref="TS_C[C][CH]O_C[CH]C=O"/>
            </me:transitionState>
            <me:MCRCMethod name="SimpleRRKM"/>
    </reaction>
    <reaction active="true" id="C[C](O)[CH]_[CH2]C(=C)O">
            <reactant>
                    <molecule me:type="modelled" ref="C[C](O)[CH]"/>
            </reactant>
            <product>
                    <molecule me:type="modelled" ref="[CH2]C(=C)O"/>
            </product>
            <group>
                    <scalar>3</scalar>
            </group>
            <me:transitionState>
                    <molecule me:type="transitionState" ref="TS_C[C](O)[CH]_[CH2]C(=C)O"/>
            </me:transitionState>
            <me:MCRCMethod name="SimpleRRKM"/>
    </reaction>
```

```xml
<reaction active="true" id="C[C](O)[CH]_CC(=O)[CH2]">
        <reactant>
                <molecule me:type="modelled" ref="C[C](O)[CH]"/>
        </reactant>
        <product>
                <molecule me:type="modelled" ref="CC(=O)[CH2]"/>
        </product>
        <group>
                <scalar>1</scalar>
        </group>
        <me:transitionState>
                <molecule me:type="transitionState" ref="TS_C[C](O)[CH]_CC(=O)[CH2]"/>
        </me:transitionState>
        <me:MCRCMethod name="SimpleRRKM"/>
</reaction>
<reaction active="true" id="C[CH]C=O_[CH2]CC=O">
        <reactant>
                <molecule me:type="modelled" ref="C[CH]C=O"/>
        </reactant>
        <product>
                <molecule me:type="modelled" ref="[CH2]CC=O"/>
        </product>
        <me:transitionState>
                <molecule me:type="transitionState" ref="TS_C[CH]C=O_[CH2]CC=O"/>
        </me:transitionState>
        <me:MCRCMethod name="SimpleRRKM"/>
</reaction>
<reaction active="true" id="C[CH]C=O_CC[C]=O">
        <reactant>
                <molecule me:type="modelled" ref="C[CH]C=O"/>
        </reactant>
        <product>
```

```xml
                        <molecule me:type="modelled" ref="CC[C]=O"/>
            </product>
            <me:transitionState>
                        <molecule me:type="transitionState" ref="TS_C[CH]C=O_CC[C]=O"/>
            </me:transitionState>
            <me:MCRCMethod name="SimpleRRKM"/>
    </reaction>
    <reaction active="true" id="C[CH]C=O_[CH2]C=CO">
            <reactant>
                        <molecule me:type="modelled" ref="C[CH]C=O"/>
            </reactant>
            <product>
                        <molecule me:type="modelled" ref="[CH2]C=CO"/>
            </product>
            <me:transitionState>
                        <molecule me:type="transitionState" ref="TS_C[CH]C=O_[CH2]C=CO"/>
            </me:transitionState>
            <me:MCRCMethod name="SimpleRRKM"/>
    </reaction>
    <reaction active="true" id="CC(=O)[CH2]_C=C=O">
            <reactant>
                        <molecule me:type="modelled" ref="CC(=O)[CH2]"/>
            </reactant>
            <product>
                        <molecule me:type="sink" ref="C=C=O"/>
            </product>
            <group>
                        <scalar>1</scalar>
            </group>
            <me:transitionState>
                        <molecule me:type="transitionState" ref="TS_CC(=O)[CH2]_C=C=O"/>
            </me:transitionState>
```

```
                        <me:MCRCMethod name="SimpleRRKM"/>

                </reaction>

        </reactionList>

        <me:conditions>

                <me:bathGas>N2</me:bathGas>

                <me:PTs>

                        <me:PTpair    me:P="3750"    me:T="1000"    me:precision="d"    me:units="Torr"
refReactionExcess="R2">        </me:PTpair>

                </me:PTs>

                <me:modelParameters>

                        <me:grainSize units="cm-1">100</me:grainSize>

                        <me:energyAboveTheTopHill>10.0</me:energyAboveTheTopHill>

                </me:modelParameters>

                <me:InitialPopulation>

                        <me:molecule grain="0.0" population="1.0" ref="C=C=O"/>

                </me:InitialPopulation>

        </me:conditions>

        <me:modelParameters>

                <me:numberStochasticTrials>1</me:numberStochasticTrials>

                <me:stochasticStartTime>1E-11</me:stochasticStartTime>

                <me:stochasticEndTime>1E10</me:stochasticEndTime>

                <me:StochasticThermalThreshold>5000</me:StochasticThermalThreshold>

                <me:StochasticEqilThreshold>100000000</me:StochasticEqilThreshold>

                <me:StochasticAxdLimit>10</me:StochasticAxdLimit>

                <me:grainSize units="cm-1">100</me:grainSize>

                <me:energyAboveTheTopHill>10.0</me:energyAboveTheTopHill>

        </me:modelParameters>

        <me:control>

                <me:printSpeciesProfile/>

                <me:stochasticOnePass/>

                <me:stochasticSimulation/>

        </me:control>
```

</me:mesmer>

# S4 1MHE + Ozone MESMER input

```
<?xml version="1.0" ?>

<?xml-stylesheet type='text/xsl' href='../../mesmer2.xsl' media='other'?>

<?xml-stylesheet type='text/xsl' href='../../mesmer1.xsl' media='screen'?>

<me:mesmer      xmlns="http://www.xml-cml.org/schema"      xmlns:me="http://www.chem.leeds.ac.uk/mesmer"
xmlns:xsi="http://www.w3.org/2001/XMLSchema-instance">

        <title> Glyoxal</title>

        <moleculeList>

                <molecule description="Nitrogen" id="N2">

                        <atom elementType="N2"/>

                        <propertyList>

                                <property dictRef="me:epsilon">

                                        <scalar>82.0</scalar>

                                </property>

                                <property dictRef="me:sigma">

                                        <scalar>3.74</scalar>

                                </property>

                                <property dictRef="me:MW">

                                        <scalar units="amu">28.0</scalar>

                                </property>

                        </propertyList>

                </molecule>

                <molecule id="CC1=CCCCC1">

                        <atomArray>

                                <atom   elementType="C"   id="a1"   x3="1.040783"   y3="0.023213"   z3="-
0.048069"/>

                                <atom   elementType="C"   id="a2"   x3="0.584003"   y3="-1.403730"   z3="-
0.040902"/>

                                <atom   elementType="C"   id="a3"   x3="1.442342"   y3="-2.427240"   z3="-
0.157845"/>

                                <atom   elementType="C"   id="a4"   x3="1.019753"   y3="-3.866305"   z3="-
0.173240"/>
```

```
                              <atom     elementType="C"     id="a5"     x3="-0.417421"     y3="-4.069778"
z3="0.331406"/>
                              <atom     elementType="C"     id="a6"   x3="-1.355511"   y3="-3.012147"   z3="-
0.263940"/>
                              <atom     elementType="C"     id="a7"     x3="-0.901082"     y3="-1.598014"
z3="0.127192"/>
                              <atom     elementType="H"     id="a8"     x3="0.651474"     y3="0.574409"
z3="0.817737"/>
                              <atom     elementType="H"     id="a9"   x3="0.693326"   y3="0.541697"   z3="-
0.952058"/>
                              <atom     elementType="H"     id="a10"   x3="2.133883"   y3="0.116003"   z3="-
0.023602"/>
                              <atom     elementType="H"     id="a11"   x3="2.512029"   y3="-2.266953"   z3="-
0.267347"/>
                              <atom     elementType="H"     id="a12"     x3="1.718451"     y3="-4.470219"
z3="0.439211"/>
                              <atom     elementType="H"     id="a13"   x3="1.112061"   y3="-4.251503"   z3="-
1.210698"/>
                              <atom     elementType="H"     id="a14"     x3="-0.435864"     y3="-4.004879"
z3="1.436275"/>
                              <atom     elementType="H"     id="a15"     x3="-0.770765"     y3="-5.084333"
z3="0.074344"/>
                              <atom     elementType="H"     id="a16"   x3="-1.369800"   y3="-3.105956"   z3="-
1.367077"/>
                              <atom     elementType="H"     id="a17"     x3="-2.392670"     y3="-3.183796"
z3="0.076047"/>
                              <atom     elementType="H"     id="a18"   x3="-1.457308"   y3="-0.849898"   z3="-
0.470483"/>
                              <atom     elementType="H"     id="a19"     x3="-1.165132"     y3="-1.397709"
z3="1.186489"/>
               </atomArray>
               <bondArray>
                              <bond atomRefs2="a16 a6" order="1"/>
                              <bond atomRefs2="a13 a4" order="1"/>
                              <bond atomRefs2="a9 a1" order="1"/>
                              <bond atomRefs2="a18 a7" order="1"/>
                              <bond atomRefs2="a11 a3" order="1"/>
                              <bond atomRefs2="a6 a17" order="1"/>
```

```xml
                    <bond atomRefs2="a6 a7" order="1"/>

                    <bond atomRefs2="a6 a5" order="1"/>

                    <bond atomRefs2="a4 a3" order="1"/>

                    <bond atomRefs2="a4 a5" order="1"/>

                    <bond atomRefs2="a4 a12" order="1"/>

                    <bond atomRefs2="a3 a2" order="2"/>

                    <bond atomRefs2="a1 a2" order="1"/>

                    <bond atomRefs2="a1 a10" order="1"/>

                    <bond atomRefs2="a1 a8" order="1"/>

                    <bond atomRefs2="a2 a7" order="1"/>

                    <bond atomRefs2="a15 a5" order="1"/>

                    <bond atomRefs2="a7 a19" order="1"/>

                    <bond atomRefs2="a5 a14" order="1"/>

            </bondArray>

            <propertyList>

                <property dictRef="me:lumpedSpecies">

                        <array> </array>

                </property>

                <property dictRef="me:vibFreqs">

                        <array  units="cm-1">73.0106  95.4525  208.4203  257.2198  342.7305
421.6592 448.9588 497.42 631.2393 797.8003 824.6365 869.9617 927.7906 940.8585 969.2288 990.1898 1003.566
1046.1146  1075.4865  1117.0681  1126.6359  1165.4946  1179.3836  1194.3502  1215.1995  1240.39  1241.9422
1250.2303  1262.8491  1267.6545  1274.4901  1289.404  1294.2255  1310.9239  1317.5845  1338.5299  1340.2835
1372.1616  1812.6884  2663.8642  2666.424  2678.3377  2686.6316  2693.8272  2695.8669  2738.5534  2742.0071
2743.8627 2747.8841 2758.515 2788.9372</array>

                </property>

                <property dictRef="me:ZPE">

                        <scalar units="kJ/mol">0.0</scalar>

                </property>

<group>1</group>

                <property dictRef="me:spinMultiplicity">

                        <scalar units="cm-1">1</scalar>

                </property>

                <property dictRef="me:epsilon">
```

```xml
                        <scalar>473.17</scalar>
                    </property>
                    <property dictRef="me:sigma">
                            <scalar>5.09</scalar>
                    </property>
                </propertyList>
                <me:energyTransferModel xsi:type="me:ExponentialDown">
                        <scalar units="cm-1">250</scalar>
                </me:energyTransferModel>
        </molecule>
        <molecule id="[O]O[O]" spinMultiplicity="3">
                <atomArray>
                        <atom elementType="O" id="a1" spinMultiplicity="2" x3="1.074646" y3="0.017911" z3="0.157278"/>
                        <atom elementType="O" id="a2" x3="2.225518" y3="-0.000266" z3="-0.042897"/>
                        <atom elementType="O" id="a3" spinMultiplicity="2" x3="2.798639" y3="-0.092332" z3="-1.056784"/>
                </atomArray>
                <bondArray>
                        <bond atomRefs2="a3 a2" order="1"/>
                        <bond atomRefs2="a2 a1" order="1"/>
                </bondArray>
                <propertyList>
                        <property dictRef="me:lumpedSpecies">
                                <array> </array>
                        </property>
                        <property dictRef="me:vibFreqs">
                                <array units="cm-1">664.0727 1202.6833 1225.2525</array>
                        </property>
<group>1</group>
                        <property dictRef="me:ZPE">
                                <scalar units="kJ/mol">0.0</scalar>
```

```
                    </property>
                    <property dictRef="me:spinMultiplicity">
                            <scalar units="cm-1">1</scalar>
                    </property>
                    <property dictRef="me:epsilon">
                            <scalar>473.17</scalar>
                    </property>
                    <property dictRef="me:sigma">
                            <scalar>5.09</scalar>
                    </property>
            </propertyList>
            <me:energyTransferModel xsi:type="me:ExponentialDown">
                    <scalar units="cm-1">250</scalar>
            </me:energyTransferModel>
    </molecule>
    <molecule id="TS_CC1=CCCCC1_C[C]12CCCC[CH]2OOO1">
            <atomArray>
                            <atom        elementType="C"        id="a1"        x3="4.599201"        y3="-0.779737"
z3="1.738264"/>
                            <atom        elementType="C"        id="a2"        x3="3.400800"        y3="-0.592401"
z3="0.803886"/>
                            <atom        elementType="C"        id="a3"        x3="2.699806"        y3="-1.924481"
z3="0.528376"/>
                            <atom        elementType="C"        id="a4"        x3="2.491438"        y3="-2.808885"
z3="1.760098"/>
                            <atom        elementType="C"        id="a5"        x3="3.231719"        y3="-2.488003"
z3="3.014362"/>
                            <atom        elementType="C"        id="a6"        x3="4.128166"        y3="-1.281822"
z3="3.105694"/>
                            <atom        elementType="H"        id="a7"        x3="5.314512"        y3="-1.506542"
z3="1.300026"/>
                            <atom        elementType="H"        id="a8"        x3="5.153252"        y3="0.170356"
z3="1.851105"/>
                            <atom        elementType="H"        id="a9"        x3="2.686304"        y3="0.132675"
z3="1.237338"/>
```

<atom elementType="H" id="a10" x3="3.736846" y3="-0.151032" z3="-0.154742"/>

<atom elementType="H" id="a11" x3="1.734087" y3="-1.736753" z3="0.019313"/>

<atom elementType="H" id="a12" x3="3.302896" y3="-2.514545" z3="-0.202173"/>

<atom elementType="C" id="a13" x3="1.043156" y3="-3.201054" z3="1.995927"/>

<atom elementType="H" id="a14" x3="2.687270" y3="-2.698945" z3="3.949124"/>

<atom elementType="H" id="a15" x3="5.008817" y3="-1.533196" z3="3.737275"/>

<atom elementType="H" id="a16" x3="3.597573" y3="-0.468312" z3="3.643113"/>

<atom elementType="H" id="a17" x3="0.588115" y3="-3.606941" z3="1.081777"/>

<atom elementType="H" id="a18" x3="0.433526" y3="-2.355160" z3="2.327790"/>

<atom elementType="H" id="a19" x3="0.963880" y3="-4.002026" z3="2.746611"/>

<atom elementType="O" id="a20" x3="4.286792" y3="-3.869183" z3="3.107036"/>

<atom elementType="O" id="a21" x3="4.415134" y3="-4.145958" z3="1.793105"/>

<atom elementType="O" id="a22" x3="3.105058" y3="-4.212034" z3="1.327697"/>

</atomArray>

<bondArray>

<bond atomRefs2="a12 a3" order="1"/>

<bond atomRefs2="a10 a2" order="1"/>

<bond atomRefs2="a11 a3" order="1"/>

<bond atomRefs2="a3 a2" order="1"/>

<bond atomRefs2="a3 a4" order="1"/>

<bond atomRefs2="a2 a9" order="1"/>

<bond atomRefs2="a2 a1" order="1"/>

<bond atomRefs2="a17 a13" order="1"/>

<bond atomRefs2="a7 a1" order="1"/>

```xml
            <bond atomRefs2="a22 a4" order="1"/>

            <bond atomRefs2="a22 a21" order="1"/>

            <bond atomRefs2="a1 a8" order="1"/>

            <bond atomRefs2="a1 a6" order="1"/>

            <bond atomRefs2="a4 a13" order="1"/>

            <bond atomRefs2="a4 a5" order="1"/>

            <bond atomRefs2="a21 a20" order="1"/>

            <bond atomRefs2="a13 a18" order="1"/>

            <bond atomRefs2="a13 a19" order="1"/>

            <bond atomRefs2="a5 a6" order="1"/>

            <bond atomRefs2="a5 a20" order="1"/>

            <bond atomRefs2="a5 a14" order="1"/>

            <bond atomRefs2="a6 a16" order="1"/>

            <bond atomRefs2="a6 a15" order="1"/>

        </bondArray>

        <propertyList>

            <property dictRef="me:vibFreqs">

                    <array     units="cm-1">117.54888850699412     130.76531082122264
228.54479027388513    255.59535073221167    274.55068677759476     303.1817782303052    332.1181023305691
406.3934147676966    435.5660063790524    458.75128843289895    513.5195417469403    551.6507922223664
639.8322185338416    659.8817997038757    743.392837202982    786.5930228299931    802.5686503763551
864.5990684847445    905.290901509595    929.3010930258779    955.1369009392503    964.5811952644094
973.7015092051806    1027.747380490401    1052.4735400374957    1074.5642542492735    1105.7877276166737
1123.6188589551584    1129.709139000201    1158.907156771512    1179.7967341914837    1205.0347624797214
1213.495010534761    1234.668657549782    1239.6637579000605    1245.1562681803987    1253.9757143330885
1257.2259206861452    1274.347200989638    1282.7435995835751    1287.58684358688    1310.491865085394
1318.9679209377232    1325.5294779712317    1332.5392746918683    1343.9671314318948    1428.6504950192038
2652.413792829269    2659.218387139831    2671.900173833504    2681.1118401416848    2684.5169004802137
2698.8270359973703    2699.8680569857797    2732.6504991663382    2735.7787459356114    2742.006011580537
2745.9243814817964 2789.1813051207073</array>

            </property>

            <property dictRef="me:imFreq">

                    <scalar units="cm-1">39.08628656635503</scalar>

            </property>

            <property dictRef="me:spinMultiplicity">

                    <scalar units="cm-1">1</scalar>

            </property>
```

```
                    <property dictRef="me:ZPE">
                                <scalar units="kJ/mol">-215.702461258948</scalar>
                    </property>
            </propertyList>
        </molecule>
        <molecule id="C[C]12CCCC[CH]2OOO1">
            <atomArray>
                                    <atom    elementType="C"    id="a1"    x3="4.484559"    y3="-0.638656"
z3="1.756743"/>
                                    <atom    elementType="C"    id="a2"    x3="3.268916"    y3="-0.582532"
z3="0.827075"/>
                                    <atom    elementType="C"    id="a3"    x3="2.823364"    y3="-1.999553"
z3="0.467472"/>
                                    <atom    elementType="C"    id="a4"    x3="2.508712"    y3="-2.869683"
z3="1.688037"/>
                                    <atom    elementType="C"    id="a5"    x3="3.360699"    y3="-2.600536"
z3="2.976828"/>
                                    <atom    elementType="C"    id="a6"    x3="4.106130"    y3="-1.274594"
z3="3.094960"/>
                                    <atom    elementType="H"    id="a7"    x3="5.298573"    y3="-1.222847"
z3="1.280927"/>
                                    <atom    elementType="H"    id="a8"    x3="4.892859"    y3="0.375686"
z3="1.922799"/>
                                    <atom    elementType="H"    id="a9"    x3="2.441059"    y3="-0.026349"
z3="1.306520"/>
                                    <atom    elementType="H"    id="a10"    x3="3.517593"    y3="-0.023117"    z3="-
0.095177"/>
                                    <atom    elementType="H"    id="a11"    x3="1.953231"    y3="-1.966526"    z3="-
0.215007"/>
                                    <atom    elementType="H"    id="a12"    x3="3.629240"    y3="-2.507367"    z3="-
0.110199"/>
                                    <atom    elementType="C"    id="a13"    x3="1.023751"    y3="-3.000816"
z3="1.978681"/>
                                    <atom    elementType="H"    id="a14"    x3="2.755056"    y3="-2.778002"
z3="3.895872"/>
                                    <atom    elementType="H"    id="a15"    x3="5.024888"    y3="-1.446733"
z3="3.696428"/>
```

<atom elementType="H" id="a16" x3="3.487032" y3="-0.565903" z3="3.681077"/>

<atom elementType="H" id="a17" x3="0.475997" y3="-3.371940" z3="1.101371"/>

<atom elementType="H" id="a18" x3="0.585767" y3="-2.036937" z3="2.261829"/>

<atom elementType="H" id="a19" x3="0.830087" y3="-3.719806" z3="2.787096"/>

<atom elementType="O" id="a20" x3="4.305973" y3="-3.720519" z3="3.019133"/>

<atom elementType="O" id="a21" x3="4.291881" y3="-4.249454" z3="1.698461"/>

<atom elementType="O" id="a22" x3="2.935721" y3="-4.258054" z3="1.367464"/>

</atomArray>

<bondArray>

<bond atomRefs2="a11 a3" order="1"/>

<bond atomRefs2="a12 a3" order="1"/>

<bond atomRefs2="a10 a2" order="1"/>

<bond atomRefs2="a3 a2" order="1"/>

<bond atomRefs2="a3 a4" order="1"/>

<bond atomRefs2="a2 a9" order="1"/>

<bond atomRefs2="a2 a1" order="1"/>

<bond atomRefs2="a17 a13" order="1"/>

<bond atomRefs2="a7 a1" order="1"/>

<bond atomRefs2="a22 a4" order="1"/>

<bond atomRefs2="a22 a21" order="1"/>

<bond atomRefs2="a4 a13" order="1"/>

<bond atomRefs2="a4 a5" order="1"/>

<bond atomRefs2="a21 a20" order="1"/>

<bond atomRefs2="a1 a8" order="1"/>

<bond atomRefs2="a1 a6" order="1"/>

<bond atomRefs2="a13 a18" order="1"/>

<bond atomRefs2="a13 a19" order="1"/>

<bond atomRefs2="a5 a20" order="1"/>

```
                <bond atomRefs2="a5 a6" order="1"/>

                <bond atomRefs2="a5 a14" order="1"/>

                <bond atomRefs2="a6 a16" order="1"/>

                <bond atomRefs2="a6 a15" order="1"/>

        </bondArray>

        <propertyList>

                <property dictRef="me:lumpedSpecies">

                        <array> </array>

                </property>

                <property dictRef="me:vibFreqs">

                        <array  units="cm-1">65.0954  86.4121  127.7438  236.2826  246.7335
264.5431  331.507  373.1912  433.0633  467.6885  472.871  578.02  639.0643  712.4661  737.1936  791.5275  849.6194
886.8949  893.9222  922.9051  940.9752  963.1078  980.775  998.1681  1043.8445  1060.9926  1110.9605  1120.8336
1144.0116  1145.1783  1184.696  1198.0778  1209.1477  1219.1032  1235.9689  1243.4946  1246.9358  1264.002
1265.9554  1281.2643  1285.6048  1290.2791  1299.6373  1310.3435  1316.6637  1324.2783  1330.4444  1344.6441
2660.949  2661.9906  2673.0307  2676.723  2682.271  2687.7443  2698.0042  2737.0101  2739.3369  2743.0258
2747.5085  2788.4159</array>

                </property>

                <property dictRef="me:ZPE">

                        <scalar units="kJ/mol">-297.02560234249734</scalar>

                </property>

                <property dictRef="me:spinMultiplicity">

                        <scalar units="cm-1">1</scalar>

                </property>

                <property dictRef="me:epsilon">

                        <scalar>473.17</scalar>

                </property>

<group>1</group>

                <property dictRef="me:sigma">

                        <scalar>5.09</scalar>

                </property>

        </propertyList>

        <me:energyTransferModel xsi:type="me:ExponentialDown">

                <scalar units="cm-1">250</scalar>
```

```
                    </me:energyTransferModel>
              </molecule>
        <molecule id="TS2_C[C]12CCCC[CH]2OOO1_[O]O[CH]CCCCC(=O)C" spinMultiplicity="3">
              <atomArray>
                              <atom   elementType="C"   id="a1"   x3="1.262344"   y3="1.432355"   z3="-
0.033194"/>
                              <atom   elementType="C"   id="a2"   x3="0.096484"   y3="1.349492"   z3="-
1.025012"/>
                              <atom   elementType="C"   id="a3"   x3="-0.161828"   y3="-0.087557"   z3="-
1.472227"/>
                              <atom   elementType="C"   id="a4"   x3="-0.631867"   y3="-1.039361"   z3="-
0.365914"/>
                              <atom   elementType="C"   id="a5"   x3="0.490555"   y3="-0.687168"
z3="1.113849"/>
                              <atom   elementType="C"   id="a6"   x3="0.924617"   y3="0.732929"
z3="1.289489"/>
                              <atom   elementType="H"   id="a7"   x3="2.168379"   y3="0.975670"   z3="-
0.479306"/>
                              <atom   elementType="H"   id="a8"   x3="1.519161"   y3="2.491012"
z3="0.162458"/>
                              <atom   elementType="H"   id="a9"   x3="-0.818899"   y3="1.782402"   z3="-
0.577211"/>
                              <atom   elementType="H"   id="a10"   x3="0.321231"   y3="1.973950"   z3="-
1.912703"/>
                              <atom   elementType="H"   id="a11"   x3="-0.921474"   y3="-0.104984"   z3="-
2.279429"/>
                              <atom   elementType="H"   id="a12"   x3="0.753879"   y3="-0.519764"   z3="-
1.928511"/>
                              <atom   elementType="C"   id="a13"   x3="-1.999625"   y3="-0.741552"
z3="0.229637"/>
                              <atom   elementType="H"   id="a14"   x3="-0.139722"   y3="-1.139937"
z3="1.880350"/>
                              <atom   elementType="H"   id="a15"   x3="1.802482"   y3="0.775230"
z3="1.970468"/>
                              <atom   elementType="H"   id="a16"   x3="0.118219"   y3="1.294135"
z3="1.810440"/>
                              <atom   elementType="H"   id="a17"   x3="-2.760245"   y3="-0.791209"   z3="-
0.562091"/>
```

```xml
                              <atom    elementType="H"    id="a18"    x3="-2.060792"    y3="0.248999"
z3="0.686555"/>
                              <atom    elementType="H"    id="a19"    x3="-2.281795"    y3="-1.497471"
z3="0.973486"/>
                              <atom    elementType="O"    id="a20"    x3="1.522692"    y3="-1.499532"
z3="0.605151"/>
                              <atom  elementType="O"  id="a21"  spinMultiplicity="2"  x3="1.152884"  y3="-
2.664210" z3="0.490825"/>
                              <atom  elementType="O"  id="a22"  spinMultiplicity="2"  x3="-0.356679"  y3="-
2.283429" z3="-0.577107"/>
                    </atomArray>
                    <bondArray>
                              <bond atomRefs2="a11 a3" order="1"/>
                              <bond atomRefs2="a12 a3" order="1"/>
                              <bond atomRefs2="a10 a2" order="1"/>
                              <bond atomRefs2="a3 a2" order="1"/>
                              <bond atomRefs2="a3 a4" order="1"/>
                              <bond atomRefs2="a2 a9" order="1"/>
                              <bond atomRefs2="a2 a1" order="1"/>
                              <bond atomRefs2="a22 a4" order="1"/>
                              <bond atomRefs2="a17 a13" order="1"/>
                              <bond atomRefs2="a7 a1" order="1"/>
                              <bond atomRefs2="a4 a13" order="1"/>
                              <bond atomRefs2="a4 a5" order="1"/>
                              <bond atomRefs2="a1 a8" order="1"/>
                              <bond atomRefs2="a1 a6" order="1"/>
                              <bond atomRefs2="a13 a18" order="1"/>
                              <bond atomRefs2="a13 a19" order="1"/>
                              <bond atomRefs2="a21 a20" order="1"/>
                              <bond atomRefs2="a20 a5" order="1"/>
                              <bond atomRefs2="a5 a6" order="1"/>
                              <bond atomRefs2="a5 a14" order="1"/>
                              <bond atomRefs2="a6 a16" order="1"/>
                              <bond atomRefs2="a6 a15" order="1"/>
```

```xml
          </bondArray>

          <propertyList>

               <property dictRef="me:vibFreqs">

                              <array      units="cm-1">95.7658369349441      122.83325320295587
151.11722507837123   206.47489090487073   247.56844454072893   291.0871264733645   315.5751657500706
364.5910590882535   373.66564820411355   448.4139868827974   466.4907969210786   500.2668032212167
560.2470314107927   595.3496755580154   646.3986703183577   804.8340415846452   847.930753963756
884.1041771562008   913.3428277529799   933.643271727125   951.1525167694837   994.4021979379344
1011.593758895479   1046.130578909843   1056.4427929841934   1067.8962052585741   1102.7564534522057
1126.8765091763394   1158.1108853338978   1170.6749118246419   1189.402794544728   1206.575287328422
1223.7317011954074   1239.256441394532   1246.5888009316864   1256.0029239298929   1257.2633805003434
1269.0890349070587   1281.165617824477   1282.246551728036   1290.1530353208273   1306.8492992553975
1314.6199133021144   1320.937236080375   1328.0788735820227   1344.1714758311991   1440.2405941784464
2658.1143216488304   2664.3756718480604   2670.618723037822   2678.0178380103325   2688.3613515447264
2699.3116567152924   2704.1377037741368   2733.0996764401452   2740.5946876474486   2741.87169676947
2745.868880716369 2789.055805948768</array>

               </property>

               <property dictRef="me:imFreq">

                              <scalar units="cm-1">608.8864896956935</scalar>

               </property>

<group>1</group>

               <property dictRef="me:spinMultiplicity">

                              <scalar units="cm-1">1</scalar>

               </property>

               <property dictRef="me:ZPE">

                              <scalar units="kJ/mol">-216.521047936143</scalar>

               </property>

          </propertyList>

     </molecule>

     <molecule id="[O]O[CH]CCCCC(=O)C" spinMultiplicity="3">

          <atomArray>

                              <atom      elementType="C"      id="a1"      x3="3.355564"      y3="0.258985"
z3="2.526429"/>

                              <atom      elementType="C"      id="a2"      x3="2.111173"      y3="0.286813"
z3="1.629281"/>

                              <atom      elementType="C"      id="a3"      x3="2.263668"      y3="-0.517034"
z3="0.331218"/>
```

```xml
          <atom     elementType="C"     id="a4"     x3="1.625367"     y3="-1.888881"
z3="0.438068"/>
          <atom elementType="C" id="a5" spinMultiplicity="2" x3="4.515219" y3="-
1.951865" z3="2.388226"/>
          <atom     elementType="C"     id="a6"     x3="3.620712"     y3="-1.102362"
z3="3.195571"/>
          <atom     elementType="H"     id="a7"     x3="4.247336"     y3="0.571601"
z3="1.946042"/>
          <atom     elementType="H"     id="a8"     x3="3.237956"     y3="1.025996"
z3="3.318913"/>
          <atom     elementType="H"     id="a9"     x3="1.228878"     y3="-0.061964"
z3="2.201929"/>
          <atom     elementType="H"     id="a10"     x3="1.890784"     y3="1.343024"
z3="1.368181"/>
          <atom     elementType="H"     id="a11"     x3="1.821199"     y3="0.039681"     z3="-
0.517220"/>
          <atom     elementType="H"     id="a12"     x3="3.339595"     y3="-0.634393"
z3="0.076773"/>
          <atom     elementType="C"     id="a13"     x3="0.128531"     y3="-1.940905"
z3="0.319670"/>
          <atom     elementType="H"     id="a14"     x3="4.252583"     y3="-2.522549"
z3="1.518415"/>
          <atom     elementType="H"     id="a15"     x3="4.055979"     y3="-0.935557"
z3="4.206080"/>
          <atom     elementType="H"     id="a16"     x3="2.664258"     y3="-1.643904"
z3="3.371044"/>
          <atom     elementType="H"     id="a17"     x3="-0.177226"     y3="-1.933920"     z3="-
0.736473"/>
          <atom     elementType="H"     id="a18"     x3="-0.372358"     y3="-1.098437"
z3="0.808813"/>
          <atom     elementType="H"     id="a19"     x3="-0.271195"     y3="-2.870575"
z3="0.752736"/>
          <atom     elementType="O"     id="a20"     x3="5.817882"     y3="-1.960584"
z3="2.890653"/>
          <atom elementType="O" id="a21" spinMultiplicity="2" x3="6.637982" y3="-
2.624858" z3="2.277392"/>
          <atom     elementType="O"     id="a22"     x3="2.287574"     y3="-2.891467"
z3="0.587630"/>
       </atomArray>
```

```
<bondArray>
        <bond atomRefs2="a17 a13" order="1"/>
        <bond atomRefs2="a11 a3" order="1"/>
        <bond atomRefs2="a12 a3" order="1"/>
        <bond atomRefs2="a13 a4" order="1"/>
        <bond atomRefs2="a13 a19" order="1"/>
        <bond atomRefs2="a13 a18" order="1"/>
        <bond atomRefs2="a3 a4" order="1"/>
        <bond atomRefs2="a3 a2" order="1"/>
        <bond atomRefs2="a4 a22" order="2"/>
        <bond atomRefs2="a10 a2" order="1"/>
        <bond atomRefs2="a14 a5" order="1"/>
        <bond atomRefs2="a2 a9" order="1"/>
        <bond atomRefs2="a2 a1" order="1"/>
        <bond atomRefs2="a7 a1" order="1"/>
        <bond atomRefs2="a21 a20" order="1"/>
        <bond atomRefs2="a5 a20" order="1"/>
        <bond atomRefs2="a5 a6" order="1"/>
        <bond atomRefs2="a1 a6" order="1"/>
        <bond atomRefs2="a1 a8" order="1"/>
        <bond atomRefs2="a6 a16" order="1"/>
        <bond atomRefs2="a6 a15" order="1"/>
</bondArray>
<propertyList>
        <property dictRef="me:lumpedSpecies">
                <array> </array>
        </property>
        <property dictRef="me:vibFreqs">
                <array   units="cm-1">13.901   36.2829   49.4159   60.2522   86.4812
100.0598  157.8254  231.6843  248.9137  302.9637  340.5114  378.7578  453.2157  492.6611  565.8494  594.0313
754.6165  822.2193  846.3869  865.9125  907.1243  941.9912  992.2804  1015.0922  1028.2491  1059.9917  1071.8328
1093.6825  1111.9856  1161.1312  1177.6937  1180.4776  1194.1854  1213.4608  1223.4801  1226.0161  1254.0995
1258.0772  1265.6216  1274.8381  1279.6495  1286.6899  1297.6868  1308.5137  1319.7887  1325.435  1351.2296
```

1802.7419 2658.9921 2661.1445 2667.4509 2667.9594 2685.5057 2696.3259 2711.6218 2733.4355 2735.8947 2739.9809 2744.4768 2789.9251</array>

              </property>

              <property dictRef="me:ZPE">

                      <scalar units="kJ/mol">-291.5686490017663</scalar>

              </property>

<group>1</group>

              <property dictRef="me:spinMultiplicity">

                      <scalar units="cm-1">1</scalar>

              </property>

              <property dictRef="me:epsilon">

                      <scalar>473.17</scalar>

              </property>

              <property dictRef="me:sigma">

                      <scalar>5.09</scalar>

              </property>

           </propertyList>

           <me:energyTransferModel xsi:type="me:ExponentialDown">

              <scalar units="cm-1">250</scalar>

           </me:energyTransferModel>

        </molecule>

        <molecule id="TS_C[C]12CCCC[CH]2OOO1_O=CCCCC[C](O[O])C" spinMultiplicity="3">

           <atomArray>

              <atom elementType="C" id="a1" x3="1.266320" y3="1.452803" z3="-0.000045"/>

              <atom elementType="C" id="a2" x3="0.165511" y3="1.343178" z3="-1.059311"/>

              <atom elementType="C" id="a3" x3="-0.108938" y3="-0.123851" z3="-1.394456"/>

              <atom elementType="C" id="a4" x3="-0.626557" y3="-0.942572" z3="-0.247606"/>

              <atom elementType="C" id="a5" x3="0.621499" y3="-0.716203" z3="1.205007"/>

              <atom elementType="C" id="a6" x3="0.830786" y3="0.792694" z3="1.306343"/>

<div align="right">145</div>

                    <atom   elementType="H"   id="a7"   x3="2.194920"   y3="0.962425"   z3="-0.363903"/>
                    <atom    elementType="H"    id="a8"    x3="1.524138"    y3="2.513433"   z3="0.176846"/>
                    <atom   elementType="H"   id="a9"   x3="-0.759578"   y3="1.839470"   z3="-0.709896"/>
                    <atom   elementType="H"   id="a10"   x3="0.471445"   y3="1.879889"   z3="-1.978722"/>
                    <atom   elementType="H"   id="a11"   x3="-0.827413"   y3="-0.192633"   z3="-2.237816"/>
                    <atom   elementType="H"   id="a12"   x3="0.839999"   y3="-0.600490"   z3="-1.753690"/>
                    <atom    elementType="C"    id="a13"    x3="-2.031293"    y3="-0.789192"   z3="0.226756"/>
                    <atom    elementType="H"    id="a14"    x3="-0.019892"    y3="-1.123850"   z3="2.006912"/>
                    <atom    elementType="H"    id="a15"    x3="1.611968"    y3="0.955771"   z3="2.080421"/>
                    <atom    elementType="H"    id="a16"    x3="-0.088453"    y3="1.272466"   z3="1.691123"/>
                    <atom   elementType="H"   id="a17"   x3="-2.750040"   y3="-0.772167"   z3="-0.605999"/>
                    <atom    elementType="H"    id="a18"    x3="-2.157376"    y3="0.157041"   z3="0.775598"/>
                    <atom    elementType="H"    id="a19"    x3="-2.334552"    y3="-1.602473"   z3="0.903185"/>
                    <atom   elementType="O"   id="a20"   spinMultiplicity="2"   x3="1.596195"   y3="-1.462653"   z3="0.828024"/>
                    <atom   elementType="O"   id="a21"   spinMultiplicity="2"   x3="0.879350"   y3="-2.520604"   z3="-0.575661"/>
                    <atom   elementType="O"   id="a22"   x3="-0.298039"   y3="-2.322481"   z3="-0.273109"/>
                </atomArray>
                <bondArray>
                    <bond atomRefs2="a11 a3" order="1"/>
                    <bond atomRefs2="a10 a2" order="1"/>
                    <bond atomRefs2="a12 a3" order="1"/>
                    <bond atomRefs2="a3 a2" order="1"/>

```
                        <bond atomRefs2="a3 a4" order="1"/>

                        <bond atomRefs2="a2 a9" order="1"/>

                        <bond atomRefs2="a2 a1" order="1"/>

                        <bond atomRefs2="a17 a13" order="1"/>

                        <bond atomRefs2="a21 a22" order="1"/>

                        <bond atomRefs2="a7 a1" order="1"/>

                        <bond atomRefs2="a22 a4" order="1"/>

                        <bond atomRefs2="a4 a13" order="1"/>

                        <bond atomRefs2="a4 a5" order="1"/>

                        <bond atomRefs2="a1 a8" order="1"/>

                        <bond atomRefs2="a1 a6" order="1"/>

                        <bond atomRefs2="a13 a18" order="1"/>

                        <bond atomRefs2="a13 a19" order="1"/>

                        <bond atomRefs2="a20 a5" order="1"/>

                        <bond atomRefs2="a5 a6" order="1"/>

                        <bond atomRefs2="a5 a14" order="1"/>

                        <bond atomRefs2="a6 a16" order="1"/>

                        <bond atomRefs2="a6 a15" order="1"/>

                </bondArray>

                <propertyList>

                        <property dictRef="me:vibFreqs">

                                      <array     units="cm-1">99.20013088990835       105.63407521805078
130.75685736193947   201.40491186411225   241.23221118352075   267.9163198560149    299.48353201739275
340.813254142569     371.11509001801875   413.0567020583921    486.5538375437101    495.3874170081798
563.5849902936138    621.0021380376588    665.2267629003444    814.3231873690355    843.8106894040087
900.1929756715776    903.7714318379506    934.0052351831063    953.9035621179031    966.641045764062
1014.7055813033094   1032.3541551477067   1063.9735707541688   1073.2894423249322   1092.9690008505306
1122.4278809533494   1152.3821903483044   1186.0590055087994   1193.3102787911193   1209.8380335248187
1227.5202367822833   1232.341702935477    1253.6008975800014   1256.421999173375    1265.8373111210265
1270.789270157949    1281.9562346634843   1291.7337515001748   1292.5580029192138   1314.1322950850338
1323.1506200300978   1323.9568041068758   1336.0556988244614   1381.1404483212848   1454.6949078476803
2643.7974833292524   2645.812544099057    2664.2677080941185   2671.5143134883147   2678.9948199595424
2682.1200856414466   2686.0600499661946   2718.4253641832042   2737.8242735702493   2740.7741433215715
2744.8535556632946   2781.3262791117904</array>

                        </property>

                        <property dictRef="me:imFreq">

                                <scalar units="cm-1">626.1905623489981</scalar>
```

```xml
            </property>
<group>1</group>
                <property dictRef="me:spinMultiplicity">
                        <scalar units="cm-1">2</scalar>
                </property>
                <property dictRef="me:ZPE">
                        <scalar units="kJ/mol">-225.3498220403596</scalar>
                </property>
        </propertyList>
    </molecule>
    <molecule id="O=CCCCC[C](O[O])C" spinMultiplicity="3">
        <atomArray>
                <atom      elementType="C"      id="a1"     x3="3.447950"     y3="-1.180080"
z3="2.011923"/>
                <atom      elementType="C"      id="a2"     x3="2.321705"     y3="-0.683561"
z3="1.092803"/>
                <atom    elementType="C"   id="a3"   x3="2.471369"   y3="-1.173783"   z3="-
0.358871"/>
                <atom   elementType="C"   id="a4"   spinMultiplicity="2"   x3="2.363002"   y3="-
2.634709" z3="-0.482199"/>
                <atom      elementType="C"      id="a5"     x3="4.221455"     y3="-3.028482"
z3="3.556445"/>
                <atom      elementType="C"      id="a6"     x3="3.076538"     y3="-2.471076"
z3="2.744124"/>
                <atom      elementType="H"      id="a7"     x3="4.371011"     y3="-1.342216"
z3="1.412111"/>
                <atom      elementType="H"      id="a8"     x3="3.695789"     y3="-0.392723"
z3="2.750220"/>
                <atom      elementType="H"      id="a9"     x3="1.333641"     y3="-0.983149"
z3="1.491451"/>
                <atom      elementType="H"      id="a10"     x3="2.316784"     y3="0.425029"
z3="1.087915"/>
                <atom   elementType="H"   id="a11"   x3="1.713680"   y3="-0.675532"   z3="-
0.999898"/>
                <atom   elementType="H"   id="a12"   x3="3.467541"   y3="-0.854766"   z3="-
0.755624"/>
```

```
                              <atom  elementType="C"  id="a13"   x3="1.144528"  y3="-3.446420"  z3="-
0.572894"/>
                              <atom     elementType="H"     id="a14"    x3="5.236326"     y3="-2.758475"
z3="3.223465"/>
                              <atom     elementType="H"     id="a15"    x3="2.192305"     y3="-2.317494"
z3="3.396230"/>
                              <atom     elementType="H"     id="a16"    x3="2.759554"     y3="-3.249448"
z3="2.014475"/>
                              <atom  elementType="H"  id="a17"  x3="0.578866"  y3="-3.218029"  z3="-
1.491686"/>
                              <atom     elementType="H"     id="a18"    x3="0.471925"     y3="-3.253526"
z3="0.279281"/>
                              <atom  elementType="H"  id="a19"  x3="1.356959"  y3="-4.527829"  z3="-
0.585275"/>
                              <atom     elementType="O"     id="a20"    x3="4.035000"     y3="-3.746310"
z3="4.514107"/>
                              <atom  elementType="O"  id="a21"  spinMultiplicity="2"  x3="4.595893"  y3="-
2.913591" z3="-0.460125"/>
                              <atom  elementType="O"  id="a22"  x3="3.507467"  y3="-3.455935"  z3="-
0.531917"/>
               </atomArray>
               <bondArray>
                              <bond atomRefs2="a17 a13" order="1"/>
                              <bond atomRefs2="a11 a3" order="1"/>
                              <bond atomRefs2="a12 a3" order="1"/>
                              <bond atomRefs2="a19 a13" order="1"/>
                              <bond atomRefs2="a13 a4" order="1"/>
                              <bond atomRefs2="a13 a18" order="1"/>
                              <bond atomRefs2="a22 a4" order="1"/>
                              <bond atomRefs2="a22 a21" order="1"/>
                              <bond atomRefs2="a4 a3" order="1"/>
                              <bond atomRefs2="a3 a2" order="1"/>
                              <bond atomRefs2="a10 a2" order="1"/>
                              <bond atomRefs2="a2 a9" order="1"/>
                              <bond atomRefs2="a2 a1" order="1"/>
                              <bond atomRefs2="a7 a1" order="1"/>
```

```xml
				<bond atomRefs2="a1 a6" order="1"/>

				<bond atomRefs2="a1 a8" order="1"/>

				<bond atomRefs2="a16 a6" order="1"/>

				<bond atomRefs2="a6 a15" order="1"/>

				<bond atomRefs2="a6 a5" order="1"/>

				<bond atomRefs2="a14 a5" order="1"/>

				<bond atomRefs2="a5 a20" order="2"/>

			</bondArray>

			<propertyList>

				<property dictRef="me:lumpedSpecies">

						<array> </array>

				</property>

				<property dictRef="me:vibFreqs">

						<array   units="cm-1">29.7502   31.9487   44.6624   68.1024   82.5887
91.7115 161.6621 202.2067 234.9025 241.762 304.8349 343.5114 421.4287 498.6219 514.4676 617.6612 728.7632
793.9084 856.009 886.0988 910.758 958.4471 973.5699 1013.4121 1037.6275 1088.4427 1100.1223 1118.2889
1134.4386 1148.3565 1174.6087 1195.4347 1204.1546 1222.1674 1226.2998 1231.2943 1243.3918 1253.3152
1268.4932 1278.7735 1284.9756 1291.3289 1302.9754 1314.2262 1328.4144 1333.2515 1467.9156 1786.7189
2645.473 2650.6921 2654.4639 2670.3103 2671.4898 2677.4041 2683.2931 2727.839 2729.3874 2741.0489
2745.5239 2776.7273</array>

				</property>

				<property dictRef="me:ZPE">

						<scalar units="kJ/mol">-303.9626712204279</scalar>

				</property>

<group>1</group>

				<property dictRef="me:spinMultiplicity">

						<scalar units="cm-1">1</scalar>

				</property>

				<property dictRef="me:epsilon">

						<scalar>473.17</scalar>

				</property>

				<property dictRef="me:sigma">

						<scalar>5.09</scalar>

				</property>
```

```
            </propertyList>

            <me:energyTransferModel xsi:type="me:ExponentialDown">

                    <scalar units="cm-1">250</scalar>

            </me:energyTransferModel>

        </molecule>

        <molecule id="TS_O=CCCCC[C](O[O])C_CC(=CCCCC=O)OO" spinMultiplicity="3">

            <atomArray>
                            <atom     elementType="C"     id="a1"     x3="5.461667"     y3="-4.672230"
z3="0.789698"/>
                            <atom     elementType="C"   id="a2"   x3="6.772155"   y3="-4.579526"   z3="-
0.005018"/>
                            <atom     elementType="C"   id="a3"   x3="6.795985"   y3="-5.534227"   z3="-
1.164779"/>
                            <atom   elementType="C"   id="a4"   spinMultiplicity="2"   x3="5.786265"   y3="-
5.483131" z3="-2.167119"/>
                            <atom     elementType="C"     id="a5"     x3="4.087012"     y3="-3.476522"
z3="2.539082"/>
                            <atom     elementType="C"     id="a6"     x3="5.480339"     y3="-3.777807"
z3="2.033709"/>
                            <atom     elementType="H"     id="a7"     x3="4.600420"     y3="-4.387880"
z3="0.147075"/>
                            <atom     elementType="H"     id="a8"     x3="5.270218"     y3="-5.722782"
z3="1.085291"/>
                            <atom     elementType="H"   id="a9"   x3="6.927151"   y3="-3.540337"   z3="-
0.360927"/>
                            <atom     elementType="H"     id="a10"     x3="7.630303"     y3="-4.800446"
z3="0.663110"/>
                            <atom   elementType="H"   id="a11"   x3="7.807281"   y3="-5.794572"   z3="-
1.519334"/>
                            <atom   elementType="H"   id="a12"   x3="6.626114"   y3="-6.875586"   z3="-
0.962608"/>
                            <atom   elementType="C"   id="a13"   x3="4.954004"   y3="-4.380215"   z3="-
2.666567"/>
                            <atom     elementType="H"     id="a14"     x3="4.022254"     y3="-3.211502"
z3="3.607261"/>
                            <atom     elementType="H"     id="a15"     x3="6.093565"     y3="-4.242072"
z3="2.830489"/>
```

```
                              <atom    elementType="H"    id="a16"    x3="5.979497"    y3="-2.810472"
z3="1.807449"/>
                              <atom    elementType="H"    id="a17"    x3="4.380583"    y3="-3.915536"    z3="-
1.838964"/>
                              <atom    elementType="H"    id="a18"    x3="4.222120"    y3="-4.698469"    z3="-
3.427397"/>
                              <atom    elementType="H"    id="a19"    x3="5.569328"    y3="-3.577805"    z3="-
3.107981"/>
                              <atom    elementType="O"    id="a20"    x3="3.103928"    y3="-3.499364"
z3="1.833313"/>
                              <atom elementType="O" id="a21" spinMultiplicity="2" x3="6.182814" y3="-
7.656283" z3="-2.084372"/>
                              <atom    elementType="O"    id="a22"    x3="5.442890"    y3="-6.691713"    z3="-
2.668613"/>
                    </atomArray>
                    <bondArray>
                              <bond atomRefs2="a18 a13" order="1"/>
                              <bond atomRefs2="a19 a13" order="1"/>
                              <bond atomRefs2="a22 a4" order="1"/>
                              <bond atomRefs2="a22 a21" order="1"/>
                              <bond atomRefs2="a13 a4" order="1"/>
                              <bond atomRefs2="a13 a17" order="1"/>
                              <bond atomRefs2="a4 a3" order="1"/>
                              <bond atomRefs2="a11 a3" order="1"/>
                              <bond atomRefs2="a3 a12" order="1"/>
                              <bond atomRefs2="a3 a2" order="1"/>
                              <bond atomRefs2="a9 a2" order="1"/>
                              <bond atomRefs2="a2 a10" order="1"/>
                              <bond atomRefs2="a2 a1" order="1"/>
                              <bond atomRefs2="a7 a1" order="1"/>
                              <bond atomRefs2="a1 a8" order="1"/>
                              <bond atomRefs2="a1 a6" order="1"/>
                              <bond atomRefs2="a16 a6" order="1"/>
                              <bond atomRefs2="a20 a5" order="2"/>
                              <bond atomRefs2="a6 a5" order="1"/>
```

```
                        <bond atomRefs2="a6 a15" order="1"/>

                        <bond atomRefs2="a5 a14" order="1"/>

                </bondArray>

                <propertyList>

                        <property dictRef="me:vibFreqs">

                                <array    units="cm-1">8.352833210495808    43.474484374314066
49.762928104531596    72.22096238440494    79.09606586241556    98.81392811314056    188.671638584588
198.81221103224627    261.70477543091204    299.6623515813628    356.051902434508    375.8512995762834
497.03190727162587    597.3021520450689    605.9943857036379    681.4317181724504    730.9112083880217
804.9721943364204    824.7978735027932    909.7172395060809    935.8377301387466    967.7569356088552
988.9807191395653    1019.4252614951495    1036.4631939580954    1055.8769505239175    1065.0659888288899
1100.484895800605    1123.6021044175936    1159.7043433908113    1164.3029827687253    1183.2127504421128
1199.583971433675    1205.5123073385591    1211.0702708713554    1237.3860699103368    1261.4658418195424
1272.0439908976548    1280.3192886446432    1285.4954188568777    1300.854046981803    1314.1747322644223
1317.7442872480297    1334.3850310504893    1392.6970590751575    1569.5880469270135    1791.2557124929003
1985.4137691305634    2643.2891872022838    2646.3115755766657    2654.726627829677    2665.459621003975
2676.92944295259    2679.4761269572373    2715.6582338019816    2728.6534147343336    2738.7910717855016
2747.18513377038 2770.114401870106</array>

                        </property>

                        <property dictRef="me:imFreq">

                                <scalar units="cm-1">1375.4540419003886</scalar>

                        </property>

<group>1</group>

                        <property dictRef="me:spinMultiplicity">

                                <scalar units="cm-1">1</scalar>

                        </property>

                        <property dictRef="me:ZPE">

                                <scalar units="kJ/mol">-299.8726276128774</scalar>

                        </property>

                </propertyList>

        </molecule>

        <molecule id="CC(=CCCCC=O)OO">

                <atomArray>

                        <atom    elementType="C"    id="a1"    x3="5.650525"    y3="-4.493320"
z3="0.956750"/>

                        <atom    elementType="C"    id="a2"    x3="6.729865"    y3="-4.372900"    z3="-
0.129981"/>
```

```xml
<atom elementType="C" id="a3" x3="6.623386" y3="-5.502045" z3="-1.105082"/>
<atom elementType="C" id="a4" x3="5.966052" y3="-5.427545" z3="-2.266408"/>
<atom elementType="C" id="a5" x3="4.630657" y3="-3.388499" z3="2.990948"/>
<atom elementType="C" id="a6" x3="5.783132" y3="-3.392550" z3="2.013553"/>
<atom elementType="H" id="a7" x3="4.640044" y3="-4.453736" z3="0.500108"/>
<atom elementType="H" id="a8" x3="5.706279" y3="-5.487906" z3="1.442062"/>
<atom elementType="H" id="a9" x3="6.645729" y3="-3.393655" z3="0.642595"/>
<atom elementType="H" id="a10" x3="7.739161" y3="-4.380035" z3="0.331318"/>
<atom elementType="H" id="a11" x3="7.121677" y3="-6.428483" z3="-0.818049"/>
<atom elementType="H" id="a12" x3="5.777568" y3="-8.215505" z3="-2.472139"/>
<atom elementType="C" id="a13" x3="5.226621" y3="-4.297051" z3="-2.891824"/>
<atom elementType="H" id="a14" x3="4.773886" y3="-2.741055" z3="3.872832"/>
<atom elementType="H" id="a15" x3="6.741415" y3="-3.500868" z3="2.560726"/>
<atom elementType="H" id="a16" x3="5.837168" y3="-2.396313" z3="1.525304"/>
<atom elementType="H" id="a17" x3="4.421139" y3="-3.938950" z3="-2.233452"/>
<atom elementType="H" id="a18" x3="4.764532" y3="-4.587592" z3="-3.850318"/>
<atom elementType="H" id="a19" x3="5.899643" y3="-3.452314" z3="-3.101125"/>
<atom elementType="O" id="a20" x3="3.614728" y3="-4.027418" z3="2.838079"/>
<atom elementType="O" id="a21" x3="6.547770" y3="-7.629693" z3="-2.742898"/>
<atom elementType="O" id="a22" x3="5.836267" y3="-6.482692" z3="-3.188340"/>
```

```
</atomArray>
<bondArray>
        <bond atomRefs2="a18 a13" order="1"/>
        <bond atomRefs2="a22 a21" order="1"/>
        <bond atomRefs2="a22 a4" order="1"/>
        <bond atomRefs2="a19 a13" order="1"/>
        <bond atomRefs2="a13 a4" order="1"/>
        <bond atomRefs2="a13 a17" order="1"/>
        <bond atomRefs2="a21 a12" order="1"/>
        <bond atomRefs2="a4 a3" order="2"/>
        <bond atomRefs2="a3 a11" order="1"/>
        <bond atomRefs2="a3 a2" order="1"/>
        <bond atomRefs2="a9 a2" order="1"/>
        <bond atomRefs2="a2 a10" order="1"/>
        <bond atomRefs2="a2 a1" order="1"/>
        <bond atomRefs2="a7 a1" order="1"/>
        <bond atomRefs2="a1 a8" order="1"/>
        <bond atomRefs2="a1 a6" order="1"/>
        <bond atomRefs2="a16 a6" order="1"/>
        <bond atomRefs2="a6 a15" order="1"/>
        <bond atomRefs2="a6 a5" order="1"/>
        <bond atomRefs2="a20 a5" order="2"/>
        <bond atomRefs2="a5 a14" order="1"/>
</bondArray>
<propertyList>
        <property dictRef="me:lumpedSpecies">
                <array> </array>
        </property>
        <property dictRef="me:vibFreqs">
                <array units="cm-1">11.5333 34.7823 56.2132 64.384 88.8749 96.4234
129.1752  179.0074  192.0345  211.5295  238.2397  322.5291  386.6111  456.3518  473.8314  602.991  632.3385
716.8557 796.6764 872.2471 894.3927 934.3117 982.0436 1003.3105 1013.6198 1039.4417 1063.0043 1111.4759
1136.5952  1156.8477  1164.6978  1176.4366  1195.3239  1226.8467  1229.5563  1232.7471  1240.6809  1260.0381
1269.8426  1284.7121  1296.7639  1314.0751  1319.6963  1328.6434  1338.738  1393.0002  1794.3745  1836.3111
```

2641.9207  2653.5672  2654.2219  2666.7958  2681.4376  2683.5982  2686.9307  2734.3596  2739.4047  2752.047
2756.1288  2778.8865</array>

```
				</property>

				<property dictRef="me:ZPE">

						<scalar units="kJ/mol">-448.8518386285451</scalar>

				</property>

				<property dictRef="me:spinMultiplicity">

						<scalar units="cm-1">1</scalar>

				</property>
```

<group>1</group>

```
				<property dictRef="me:epsilon">

						<scalar>473.17</scalar>

				</property>

				<property dictRef="me:sigma">

						<scalar>5.09</scalar>

				</property>

			</propertyList>

			<me:energyTransferModel xsi:type="me:ExponentialDown">

					<scalar units="cm-1">250</scalar>

			</me:energyTransferModel>

		</molecule>

		<molecule id="TS_O=CCCCC[C](O[O])C_CC(=CCCCC=O)OO" spinMultiplicity="3">

			<atomArray>

					<atom    elementType="C"     id="a1"     x3="0.762854"     y3="-0.290747"
z3="1.207557"/>

					<atom    elementType="C"     id="a2"     x3="-0.349592"     y3="-0.141565"
z3="0.158253"/>

					<atom    elementType="C"     id="a3"     spinMultiplicity="2"    x3="0.039069"
y3="0.863797" z3="-0.909247"/>

					<atom    elementType="C"     id="a4"     spinMultiplicity="2"    x3="-0.642768"
y3="0.828659" z3="-2.159923"/>

					<atom    elementType="C"     id="a5"     x3="1.433414"     y3="-1.459922"
z3="3.349153"/>

					<atom    elementType="C"     id="a6"     x3="0.359135"     y3="-1.277584"
z3="2.303928"/>
```

```xml
            <atom    elementType="H"    id="a7"    x3="1.695894"    y3="-0.628886"
z3="0.712896"/>
            <atom    elementType="H"    id="a8"    x3="0.993859"    y3="0.697264"
z3="1.652193"/>
            <atom elementType="H" id="a9" x3="-0.562564" y3="-1.128375" z3="-
0.303894"/>
            <atom    elementType="H"    id="a10"    x3="-1.289171"    y3="0.186636"
z3="0.646467"/>
            <atom elementType="H" id="a11" x3="0.265790" y3="1.849697" z3="-
0.506765"/>
            <atom elementType="H" id="a12" x3="1.137667" y3="0.153857" z3="-
1.476668"/>
            <atom elementType="C" id="a13" x3="-1.733520" y3="1.654324" z3="-
2.698542"/>
            <atom    elementType="H"    id="a14"    x3="2.457009"    y3="-1.177975"
z3="3.054948"/>
            <atom    elementType="H"    id="a15"    x3="-0.585864"    y3="-0.960134"
z3="2.793815"/>
            <atom    elementType="H"    id="a16"    x3="0.120535"    y3="-2.272214"
z3="1.867125"/>
            <atom    elementType="H"    id="a17"    x3="-1.478365"    y3="2.729061"    z3="-
2.658386"/>
            <atom    elementType="H"    id="a18"    x3="-2.654598"    y3="1.533641"    z3="-
2.099979"/>
            <atom    elementType="H"    id="a19"    x3="-1.986167"    y3="1.419211"    z3="-
3.746072"/>
            <atom    elementType="O"    id="a20"    x3="1.187955"    y3="-1.901907"
z3="4.449520"/>
            <atom    elementType="O"    id="a21"    x3="0.951906"    y3="-0.660982"    z3="-
2.561661"/>
            <atom    elementType="O"    id="a22"    x3="-0.122477"    y3="-0.015854"    z3="-
3.074718"/>
        </atomArray>
        <bondArray>
            <bond atomRefs2="a19 a13" order="1"/>
            <bond atomRefs2="a22 a21" order="1"/>
            <bond atomRefs2="a22 a4" order="1"/>
            <bond atomRefs2="a13 a17" order="1"/>
```

```xml
            <bond atomRefs2="a13 a4" order="1"/>

            <bond atomRefs2="a13 a18" order="1"/>

            <bond atomRefs2="a21 a12" order="1"/>

            <bond atomRefs2="a4 a3" order="1"/>

            <bond atomRefs2="a3 a11" order="1"/>

            <bond atomRefs2="a3 a2" order="1"/>

            <bond atomRefs2="a9 a2" order="1"/>

            <bond atomRefs2="a2 a10" order="1"/>

            <bond atomRefs2="a2 a1" order="1"/>

            <bond atomRefs2="a7 a1" order="1"/>

            <bond atomRefs2="a1 a8" order="1"/>

            <bond atomRefs2="a1 a6" order="1"/>

            <bond atomRefs2="a16 a6" order="1"/>

            <bond atomRefs2="a6 a15" order="1"/>

            <bond atomRefs2="a6 a5" order="1"/>

            <bond atomRefs2="a14 a5" order="1"/>

            <bond atomRefs2="a5 a20" order="2"/>

        </bondArray>

        <propertyList>

            <property dictRef="me:vibFreqs">

                        <array     units="cm-1">26.66767468296296        34.17064086556612
55.511308002808384      79.04209804628034      88.43062554976746      91.448174725042      137.89016991321088
191.10321731262286      265.2335904469886      272.9950872505833      353.72936845183585      387.00662306514135
529.3246617388709      585.2272055028317      593.7428710119082      684.3147000945121      718.7710573939934
787.0107484150495      850.8815635370667      929.2412971995097      958.4641896308003      972.4958548264635
1005.7317898282344      1036.7120169281661      1071.1321362894669      1103.9292321220669      1113.8372254958717
1130.8186768296202      1135.473287439981      1161.6212748514672      1175.5254693576937      1185.3763167897996
1191.2171772504782      1198.1040574468072      1212.031761560985      1223.5459998870033      1252.0205157728324
1264.3540725151213      1277.7896032315512      1286.2797711495284      1301.7928530506822      1313.6240594296044
1318.4572652393538      1325.1046584037936      1388.2830622074443      1552.4182044407949      1787.7046968699815
1949.8343000750283      2645.1597523264327      2652.0540024785805      2661.1911907999533      2667.1734280481314
2677.1441033663987      2682.073847025421      2726.5985661289055      2735.998048354507      2748.4450248951703
2755.0443181201153 2774.30062944853</array>

            </property>

            <property dictRef="me:imFreq">

                        <scalar units="cm-1">1333.8682052519675</scalar>

            </property>
```

```xml
                    <property dictRef="me:spinMultiplicity">
                                <scalar units="cm-1">1</scalar>
                    </property>
<group>1</group>
                    <property dictRef="me:ZPE">
                                <scalar units="kJ/mol">-282.3483709347917</scalar>
                    </property>
            </propertyList>
        </molecule>
        <molecule id="CC(=CCCCC=O)OO">
            <atomArray>
                    <atom     elementType="C"     id="a1"     x3="4.431979"     y3="-4.369807"     z3="0.341742"/>
                    <atom     elementType="C"     id="a2"     x3="3.342352"     y3="-4.507618"     z3="-0.734923"/>
                    <atom     elementType="C"     id="a3"     x3="2.607781"     y3="-3.226718"     z3="-0.967672"/>
                    <atom     elementType="C"     id="a4"     x3="2.803662"     y3="-2.352400"     z3="-1.959181"/>
                    <atom     elementType="C"     id="a5"     x3="4.894414"     y3="-4.084206"     z3="2.819046"/>
                    <atom     elementType="C"     id="a6"     x3="3.843924"     y3="-4.193312"     z3="1.742522"/>
                    <atom     elementType="H"     id="a7"     x3="5.079457"     y3="-5.268637"     z3="0.315290"/>
                    <atom     elementType="H"     id="a8"     x3="5.084105"     y3="-3.509041"     z3="0.095704"/>
                    <atom     elementType="H"     id="a9"     x3="3.803742"     y3="-4.872114"     z3="-1.680152"/>
                    <atom     elementType="H"     id="a10"     x3="2.617836"     y3="-5.296759"     z3="-0.443936"/>
                    <atom     elementType="H"     id="a11"     x3="1.828111"     y3="-3.018185"     z3="-0.232419"/>
                    <atom     elementType="H"     id="a12"     x3="5.418461"     y3="-3.066111"     z3="2.669240"/>
                    <atom     elementType="C"     id="a13"     x3="2.035944"     y3="-1.090472"     z3="-2.197485"/>
```

                          <atom    elementType="H"    id="a14"    x3="5.942668"    y3="-4.201899"
z3="2.500415"/>

                          <atom    elementType="H"    id="a15"    x3="3.189963"    y3="-3.294421"
z3="1.777249"/>

                          <atom    elementType="H"    id="a16"    x3="3.156532"    y3="-5.030996"
z3="1.989202"/>

                          <atom   elementType="H"   id="a17"   x3="2.140443"   y3="-0.393940"   z3="-
1.353933"/>

                          <atom   elementType="H"   id="a18"   x3="0.965977"   y3="-1.300535"   z3="-
2.342116"/>

                          <atom   elementType="H"   id="a19"   x3="2.382988"   y3="-0.562174"   z3="-
3.102437"/>

                          <atom    elementType="O"    id="a20"    x3="4.606513"    y3="-3.878942"
z3="3.977784"/>

                          <atom   elementType="O"   id="a21"   x3="4.575215"   y3="-3.477938"   z3="-
3.027272"/>

                          <atom   elementType="O"   id="a22"   x3="3.720239"   y3="-2.347386"   z3="-
3.018777"/>

              </atomArray>

              <bondArray>

                          <bond atomRefs2="a19 a13" order="1"/>

                          <bond atomRefs2="a21 a22" order="1"/>

                          <bond atomRefs2="a21 a12" order="1"/>

                          <bond atomRefs2="a22 a4" order="1"/>

                          <bond atomRefs2="a18 a13" order="1"/>

                          <bond atomRefs2="a13 a4" order="1"/>

                          <bond atomRefs2="a13 a17" order="1"/>

                          <bond atomRefs2="a4 a3" order="2"/>

                          <bond atomRefs2="a9 a2" order="1"/>

                          <bond atomRefs2="a3 a2" order="1"/>

                          <bond atomRefs2="a3 a11" order="1"/>

                          <bond atomRefs2="a2 a10" order="1"/>

                          <bond atomRefs2="a2 a1" order="1"/>

                          <bond atomRefs2="a8 a1" order="1"/>

                          <bond atomRefs2="a7 a1" order="1"/>

                          <bond atomRefs2="a1 a6" order="1"/>

```xml
                        <bond atomRefs2="a6 a15" order="1"/>

                        <bond atomRefs2="a6 a16" order="1"/>

                        <bond atomRefs2="a6 a5" order="1"/>

                        <bond atomRefs2="a14 a5" order="1"/>

                        <bond atomRefs2="a5 a20" order="2"/>

                </bondArray>

                <propertyList>

                        <property dictRef="me:lumpedSpecies">

                                <array> </array>

                        </property>

                        <property dictRef="me:vibFreqs">

                                <array  units="cm-1">17.1094   25.8733   49.2416   57.0165   77.399
108.6058 137.5053 171.0302 196.0663 260.9154 290.4941 298.3987 410.797 460.6515 506.199 517.3868 615.0838
702.1278 828.2541 848.7278 899.6551 930.7738 973.4576 1000.6821 1014.6234 1031.4955 1088.2325 1115.9611
1131.574  1146.5909  1167.7284  1180.397  1206.6695  1215.13  1228.7442  1230.6142  1245.7023  1263.9764
1280.6098  1291.3554  1302.0351  1311.9131  1319.9678  1326.9541  1336.5351  1391.5903  1786.8264  1838.968
2645.4567 2651.7663 2653.7952 2654.9354 2677.9424 2678.8214 2688.0085 2730.7344 2735.2094 2748.8142
2754.1085 2779.2934</array>

                        </property>

                        <property dictRef="me:ZPE">

                                <scalar units="kJ/mol">-418.04590699289173</scalar>

                        </property>

                        <property dictRef="me:spinMultiplicity">

                                <scalar units="cm-1">1</scalar>

                        </property>

<group>1</group>

                        <property dictRef="me:epsilon">

                                <scalar>473.17</scalar>

                        </property>

                        <property dictRef="me:sigma">

                                <scalar>5.09</scalar>

                        </property>

                </propertyList>

                <me:energyTransferModel xsi:type="me:ExponentialDown">
```

```
                    <scalar units="cm-1">250</scalar>
              </me:energyTransferModel>
        </molecule>
    <molecule id="TS_O=CCCC[C](O[O])C_O=CCCC[CH](C(=O)C)O" spinMultiplicity="3">
           <atomArray>
                    <atom    elementType="C"    id="a1"    x3="4.234252"    y3="-3.105857"
z3="0.853135"/>
                    <atom   elementType="C"   id="a2"   x3="4.951356"   y3="-4.021023"   z3="-
0.150313"/>
                    <atom   elementType="C"   id="a3"   x3="5.369468"   y3="-3.259059"   z3="-
1.396586"/>
                    <atom   elementType="C"   id="a4"   x3="5.909419"   y3="-4.109826"   z3="-
2.513833"/>
                    <atom    elementType="C"    id="a5"    x3="5.076527"    y3="-4.212681"
z3="2.962527"/>
                    <atom    elementType="C"    id="a6"    x3="3.863392"    y3="-3.853757"
z3="2.134253"/>
                    <atom    elementType="H"    id="a7"    x3="4.862581"    y3="-2.224300"
z3="1.089087"/>
                    <atom    elementType="H"    id="a8"    x3="3.311519"    y3="-2.696003"
z3="0.390630"/>
                    <atom    elementType="H"    id="a9"    x3="5.842391"    y3="-4.482487"
z3="0.322936"/>
                    <atom   elementType="H"   id="a10"   x3="4.283966"   y3="-4.865360"   z3="-
0.424283"/>
                    <atom   elementType="H"   id="a11"   x3="4.386779"   y3="-2.708225"   z3="-
1.841959"/>
                    <atom   elementType="H"   id="a12"   x3="6.039238"   y3="-2.407333"   z3="-
1.185251"/>
                    <atom   elementType="C"   id="a13"   x3="6.985993"   y3="-5.103708"   z3="-
2.239064"/>
                    <atom    elementType="H"    id="a14"    x3="5.856376"    y3="-3.437674"
z3="3.034352"/>
                    <atom    elementType="H"    id="a15"    x3="3.178387"    y3="-3.237248"
z3="2.755063"/>
                    <atom    elementType="H"    id="a16"    x3="3.277944"    y3="-4.769304"
z3="1.901146"/>
                    <atom   elementType="H"   id="a17"   x3="7.794886"   y3="-4.687379"   z3="-
1.622796"/>
```

```xml
<atom elementType="H" id="a18" x3="7.439763" y3="-5.480226" z3="-3.171028"/>
<atom elementType="H" id="a19" x3="6.592502" y3="-5.982809" z3="-1.704221"/>
<atom elementType="O" id="a20" x3="5.184468" y3="-5.281424" z3="3.519050"/>
<atom elementType="O" id="a21" spinMultiplicity="3" x3="4.169385" y3="-2.578754" z3="-3.277931"/>
<atom elementType="O" id="a22" x3="5.451230" y3="-3.954564" z3="-3.646406"/>
</atomArray>
<bondArray>
<bond atomRefs2="a22 a4" order="2"/>
<bond atomRefs2="a18 a13" order="1"/>
<bond atomRefs2="a4 a13" order="1"/>
<bond atomRefs2="a4 a3" order="1"/>
<bond atomRefs2="a13 a19" order="1"/>
<bond atomRefs2="a13 a17" order="1"/>
<bond atomRefs2="a11 a3" order="1"/>
<bond atomRefs2="a3 a12" order="1"/>
<bond atomRefs2="a3 a2" order="1"/>
<bond atomRefs2="a10 a2" order="1"/>
<bond atomRefs2="a2 a9" order="1"/>
<bond atomRefs2="a2 a1" order="1"/>
<bond atomRefs2="a8 a1" order="1"/>
<bond atomRefs2="a1 a7" order="1"/>
<bond atomRefs2="a1 a6" order="1"/>
<bond atomRefs2="a16 a6" order="1"/>
<bond atomRefs2="a6 a15" order="1"/>
<bond atomRefs2="a6 a5" order="1"/>
<bond atomRefs2="a5 a14" order="1"/>
<bond atomRefs2="a5 a20" order="2"/>
</bondArray>
<propertyList>
```

```
<property dictRef="me:vibFreqs">
                          <array    units="cm-1">0.12997323200476377    10.611936470983219
25.37505967659918    29.35531504916968    42.368379870549774    71.25071381140158    102.9223984977037
152.57415896574005    262.18851863072575    270.9371634352591    294.55224480509764    384.99543902747604
424.41206623253606    509.1997557073725    528.34835049626    553.4166209335257    681.8239867317437
752.5512939179042    791.5542956070551    840.2283458968652    929.4832276004112    969.6687438516702
988.9345287478033    999.3266233534439    1032.220155539663    1064.8416582388995    1085.8094907506957
1097.0675159727643    1153.709868308275    1156.8091255947647    1170.5122912344893    1187.9367227073137
1194.6233882136155    1201.0522666135257    1212.654985670432    1237.2201159243377    1241.4675436576872
1274.336540431527    1274.8492642647789    1278.1197752157477    1284.227233636882    1297.8786139888896
1306.958598326873    1314.1088779485965    1330.9608906979875    1667.7291453109779    1792.4405869118682
2024.5625619920975    2644.2880503705355    2652.4539245857036    2662.549088007921    2674.0081122611878
2675.620137933847    2683.8715562475672    2704.8594530766622    2728.6931848304066    2741.2584570453105
2744.8961600858265 2783.605131393014</array>
                    </property>
                    <property dictRef="me:imFreq">
                              <scalar units="cm-1">1239.581184812474</scalar>
                    </property>
                    <property dictRef="me:spinMultiplicity">
                              <scalar units="cm-1">1</scalar>
                    </property>
          <group>1</group>

                    <property dictRef="me:ZPE">
                              <scalar units="kJ/mol">-245.2645228428667</scalar>
                    </property>
               </propertyList>
          </molecule>
          <molecule id="O=CCCC[CH](C(=O)C)O">
               <atomArray>
                              <atom    elementType="C"    id="a1"    x3="4.467279"    y3="-2.978858"
z3="0.856914"/>
                              <atom    elementType="C"    id="a2"    x3="4.989674"    y3="-3.965012"    z3="-
0.195263"/>
                              <atom    elementType="C"    id="a3"    x3="5.520355"    y3="-3.210594"    z3="-
1.423379"/>
                              <atom    elementType="C"    id="a4"    x3="5.996952"    y3="-4.216551"    z3="-
2.493131"/>
                              <atom    elementType="C"    id="a5"    x3="4.820776"    y3="-4.429279"
z3="2.901516"/>
```

```xml
                              <atom     elementType="C"     id="a6"     x3="3.820126"     y3="-3.691893"
z3="2.042968"/>
                              <atom     elementType="H"     id="a7"     x3="5.284046"     y3="-2.314778"
z3="1.201158"/>
                              <atom     elementType="H"     id="a8"     x3="3.724737"     y3="-2.302847"
z3="0.375959"/>
                              <atom     elementType="H"     id="a9"     x3="5.781579"     y3="-4.608360"
z3="0.231535"/>
                         <atom   elementType="H"   id="a10"   x3="4.173360"   y3="-4.652455"   z3="-
0.500310"/>
                         <atom   elementType="H"   id="a11"   x3="3.860638"   y3="-2.939850"   z3="-
2.505529"/>
                         <atom   elementType="H"   id="a12"   x3="6.312586"   y3="-2.475233"   z3="-
1.154643"/>
                         <atom   elementType="C"   id="a13"   x3="7.416949"   y3="-4.669868"   z3="-
2.397281"/>
                              <atom     elementType="H"     id="a14"     x3="5.827487"     y3="-3.985242"
z3="2.960371"/>
                              <atom     elementType="H"     id="a15"     x3="3.278908"     y3="-2.960044"
z3="2.681382"/>
                              <atom     elementType="H"     id="a16"     x3="3.027839"     y3="-4.391848"
z3="1.700009"/>
                         <atom   elementType="H"   id="a17"   x3="8.104211"   y3="-3.894835"   z3="-
2.770566"/>
                         <atom   elementType="H"   id="a18"   x3="7.593157"   y3="-5.564951"   z3="-
3.016371"/>
                         <atom   elementType="H"   id="a19"   x3="7.723387"   y3="-4.911705"   z3="-
1.372303"/>
                              <atom     elementType="O"     id="a20"     x3="4.535878"     y3="-5.442152"
z3="3.499460"/>
                         <atom   elementType="O"   id="a21"   x3="4.508862"   y3="-2.387247"   z3="-
1.996750"/>
                         <atom   elementType="O"   id="a22"   x3="5.226436"   y3="-4.583709"   z3="-
3.349214"/>
               </atomArray>
               <bondArray>
                    <bond atomRefs2="a22 a4" order="2"/>
                    <bond atomRefs2="a18 a13" order="1"/>
                    <bond atomRefs2="a17 a13" order="1"/>
```

```xml
                        <bond atomRefs2="a11 a21" order="1"/>

                        <bond atomRefs2="a4 a13" order="1"/>

                        <bond atomRefs2="a4 a3" order="1"/>

                        <bond atomRefs2="a13 a19" order="1"/>

                        <bond atomRefs2="a21 a3" order="1"/>

                        <bond atomRefs2="a3 a12" order="1"/>

                        <bond atomRefs2="a3 a2" order="1"/>

                        <bond atomRefs2="a10 a2" order="1"/>

                        <bond atomRefs2="a2 a9" order="1"/>

                        <bond atomRefs2="a2 a1" order="1"/>

                        <bond atomRefs2="a8 a1" order="1"/>

                        <bond atomRefs2="a1 a7" order="1"/>

                        <bond atomRefs2="a1 a6" order="1"/>

                        <bond atomRefs2="a16 a6" order="1"/>

                        <bond atomRefs2="a6 a15" order="1"/>

                        <bond atomRefs2="a6 a5" order="1"/>

                        <bond atomRefs2="a5 a14" order="1"/>

                        <bond atomRefs2="a5 a20" order="2"/>

                </bondArray>

                <propertyList>

                        <property dictRef="me:lumpedSpecies">

                                <array> </array>

                        </property>

                        <property dictRef="me:vibFreqs">

                                <array units="cm-1">25.8439 27.922 46.4542 61.4664 65.2023 98.1771
135.0672 209.9234 254.2339 289.818 333.2801 377.6777 383.5444 465.0354 533.5121 588.5702 609.2314
759.5133 795.2653 898.6603 943.0774 969.2746 1004.2116 1014.5223 1050.9426 1083.3972 1093.4526 1138.224
1148.7042 1152.6673 1167.7893 1194.1819 1202.0759 1208.5304 1220.3932 1228.903 1233.6928 1247.1674
1265.4382 1274.6838 1278.5166 1294.4639 1307.7524 1317.2139 1325.281 1373.9309 1791.0842 1808.3226
2536.8426 2644.6305 2653.0374 2661.5183 2674.9571 2680.4473 2691.0348 2698.5038 2732.6514 2742.5582
2747.3693 2786.6108</array>

                        </property>

                        <property dictRef="me:ZPE">

                                <scalar units="kJ/mol">-739.6147471997558</scalar>
```

```
                              </property>
                              <property dictRef="me:spinMultiplicity">
                                      <scalar units="cm-1">1</scalar>
                              </property>
                              <property dictRef="me:epsilon">
                                      <scalar>473.17</scalar>
                              </property>
<group>1</group>

                              <property dictRef="me:sigma">
                                      <scalar>5.09</scalar>
                              </property>
                      </propertyList>
                      <me:energyTransferModel xsi:type="me:ExponentialDown">
                              <scalar units="cm-1">250</scalar>
                      </me:energyTransferModel>
              </molecule>
              <molecule id="TS_CC(=CCCCC=O)OO_O=CCCC[CH]1O[C]1(C)O" spinMultiplicity="3">
                      <atomArray>
                              <atom     elementType="C"     id="a1"     x3="-0.283810"     y3="0.245649"
z3="1.526376"/>
                              <atom     elementType="C"     id="a2"     x3="-0.317091"     y3="-0.098064"
z3="0.031198"/>
                              <atom   elementType="C"   id="a3"   spinMultiplicity="2"   x3="0.645214"
y3="0.691830" z3="-0.753197"/>
                              <atom   elementType="C"   id="a4"   x3="0.763986"   y3="0.600646"   z3="-
2.212734"/>
                              <atom   elementType="C"   id="a5"   x3="-1.298966"   y3="-0.232630"
z3="3.791140"/>
                              <atom   elementType="C"   id="a6"   x3="-1.271226"   y3="-0.609006"
z3="2.325795"/>
                              <atom     elementType="H"     id="a7"     x3="0.740061"     y3="0.108571"
z3="1.931899"/>
                              <atom     elementType="H"     id="a8"     x3="-0.512177"     y3="1.321019"
z3="1.681791"/>
                              <atom   elementType="H"   id="a9"   x3="-0.104472"   y3="-1.183810"   z3="-
0.113896"/>
```

```xml
                        <atom   elementType="H"   id="a10"   x3="-1.346157"   y3="0.048889"   z3="-
0.369367"/>
                        <atom   elementType="H"   id="a11"   x3="1.324952"   y3="1.373053"   z3="-
0.251881"/>
                        <atom   elementType="H"   id="a12"   x3="3.647362"   y3="0.975528"   z3="-
2.843065"/>
                        <atom   elementType="C"   id="a13"   x3="-0.075494"   y3="-0.400266"   z3="-
2.952382"/>
                        <atom   elementType="H"   id="a14"   x3="-1.946041"   y3="-0.868677"
z3="4.418319"/>
                        <atom   elementType="H"   id="a15"   x3="-2.295385"   y3="-0.508159"
z3="1.909651"/>
                        <atom   elementType="H"   id="a16"   x3="-1.018470"   y3="-1.684734"
z3="2.226913"/>
                        <atom   elementType="H"   id="a17"   x3="0.178053"   y3="-1.428418"   z3="-
2.660371"/>
                        <atom   elementType="H"   id="a18"   x3="0.094590"   y3="-0.325977"   z3="-
4.038387"/>
                        <atom   elementType="H"   id="a19"   x3="-1.147704"   y3="-0.254523"   z3="-
2.778059"/>
                        <atom   elementType="O"   id="a20"   x3="-0.672991"   y3="0.690109"
z3="4.258375"/>
                        <atom   elementType="O"   id="a21"   spinMultiplicity="2"   x3="3.360380"
y3="0.200124" z3="-2.313656"/>
                        <atom   elementType="O"   id="a22"   x3="1.535388"   y3="1.338843"   z3="-
2.814463"/>
                </atomArray>
                <bondArray>
                        <bond atomRefs2="a18 a13" order="1"/>
                        <bond atomRefs2="a13 a19" order="1"/>
                        <bond atomRefs2="a13 a17" order="1"/>
                        <bond atomRefs2="a13 a4" order="1"/>
                        <bond atomRefs2="a12 a21" order="1"/>
                        <bond atomRefs2="a22 a4" order="2"/>
                        <bond atomRefs2="a4 a3" order="1"/>
                        <bond atomRefs2="a3 a11" order="1"/>
                        <bond atomRefs2="a3 a2" order="1"/>
```

```xml
        <bond atomRefs2="a10 a2" order="1"/>

        <bond atomRefs2="a9 a2" order="1"/>

        <bond atomRefs2="a2 a1" order="1"/>

        <bond atomRefs2="a1 a8" order="1"/>

        <bond atomRefs2="a1 a7" order="1"/>

        <bond atomRefs2="a1 a6" order="1"/>

        <bond atomRefs2="a15 a6" order="1"/>

        <bond atomRefs2="a16 a6" order="1"/>

        <bond atomRefs2="a6 a5" order="1"/>

        <bond atomRefs2="a5 a20" order="2"/>

        <bond atomRefs2="a5 a14" order="1"/>

    </bondArray>

    <propertyList>

        <property dictRef="me:vibFreqs">

            <array    units="cm-1">15.967055960823197    23.496133274746686
57.709250699152996    62.19133550905053    78.02155521571092    95.36800366266498    115.4958979996861
134.9663779489954    165.4384412217635    180.80717019086916    307.5307663416405    328.2514709759217
459.4256134984474    471.4917773045559    491.5406235499775    507.89982048007795    642.1420301900458
646.9676422927561    717.0584558825849    727.4504670309806    825.4699401764915    925.1222240633508
942.5112079448934    1004.6391008954905    1012.8499706631959    1036.7313551130744    1044.6122284419432
1095.9390199880388    1140.7802985513904    1149.276650630745    1156.6723501473716    1159.5881784226344
1171.0704382085175    1176.7400892436747    1212.1648663413694    1224.0622812756253    1229.0598134210045
1236.1252185892058    1252.9761962963942    1258.9827579627909    1270.459373015645    1302.282199877263
1318.2578361880633    1322.5417843977064    1347.8883475059706    1632.8074588619997    1794.20041123107
2524.7884822670626    2633.9888766099975    2642.419943634926    2654.7490050329175    2676.586272661116
2686.0833557903525    2692.3503163200585    2709.502946576403    2729.1154482026122    2735.1591216903043
2748.354878570176 2786.749481041312</array>

        </property>

        <property dictRef="me:imFreq">

            <scalar units="cm-1">380.60574261784615</scalar>

        </property>

        <property dictRef="me:spinMultiplicity">

            <scalar units="cm-1">1</scalar>

        </property>

<group>1</group>

        <property dictRef="me:ZPE">

            <scalar units="kJ/mol">-418.64495922189326</scalar>
```

            </property>

          </propertyList>

        </molecule>

      <molecule id="O=CCCC[CH]1O[C]1(C)O">

          <atomArray>

                      <atom     elementType="C"     id="a1"      x3="5.747315"      y3="-6.340087"   z3="0.796952"/>

                      <atom    elementType="C"    id="a2"    x3="5.250898"    y3="-5.705334"    z3="-0.505555"/>

                      <atom    elementType="C"    id="a3"    x3="6.408045"    y3="-5.205468"    z3="-1.331611"/>

                      <atom    elementType="C"    id="a4"    x3="7.049153"    y3="-6.056613"    z3="-2.413219"/>

                      <atom     elementType="C"     id="a5"     x3="5.055226"     y3="-7.527152"   z3="2.919311"/>

                      <atom     elementType="C"     id="a6"     x3="4.583472"     y3="-6.838730"   z3="1.658066"/>

                      <atom     elementType="H"     id="a7"     x3="6.438506"     y3="-7.181584"   z3="0.581055"/>

                      <atom     elementType="H"     id="a8"     x3="6.354316"     y3="-5.615127"   z3="1.377463"/>

                      <atom    elementType="H"    id="a9"    x3="4.643186"    y3="-6.427417"    z3="-1.089145"/>

                      <atom    elementType="H"    id="a10"    x3="4.564171"    y3="-4.858660"    z3="-0.293634"/>

                      <atom    elementType="H"    id="a11"    x3="6.974125"    y3="-4.375831"    z3="-0.897666"/>

                      <atom    elementType="H"    id="a12"    x3="8.458264"    y3="-4.775833"    z3="-2.974794"/>

                      <atom    elementType="C"    id="a13"    x3="6.746202"    y3="-7.483239"    z3="-2.723314"/>

                      <atom     elementType="H"     id="a14"     x3="4.248570"     y3="-7.847674"   z3="3.599967"/>

                      <atom     elementType="H"     id="a15"     x3="3.913913"     y3="-5.995155"   z3="1.925892"/>

                      <atom     elementType="H"     id="a16"     x3="3.947788"     y3="-7.541021"   z3="1.078795"/>

                      <atom    elementType="H"    id="a17"    x3="6.636334"    y3="-8.084711"    z3="-1.810582"/>

<atom elementType="H" id="a18" x3="7.559124" y3="-7.934678" z3="-3.318982"/>

<atom elementType="H" id="a19" x3="5.820447" y3="-7.572614" z3="-3.311785"/>

<atom elementType="O" id="a20" x3="6.219432" y3="-7.724589" z3="3.180035"/>

<atom elementType="O" id="a21" x3="8.347505" y3="-5.742967" z3="-2.814087"/>

<atom elementType="O" id="a22" x3="6.106776" y3="-5.006068" z3="-2.767463"/>

</atomArray>

<bondArray>

<bond atomRefs2="a18 a13" order="1"/>

<bond atomRefs2="a19 a13" order="1"/>

<bond atomRefs2="a12 a21" order="1"/>

<bond atomRefs2="a21 a4" order="1"/>

<bond atomRefs2="a22 a4" order="1"/>

<bond atomRefs2="a22 a3" order="1"/>

<bond atomRefs2="a13 a4" order="1"/>

<bond atomRefs2="a13 a17" order="1"/>

<bond atomRefs2="a4 a3" order="1"/>

<bond atomRefs2="a3 a11" order="1"/>

<bond atomRefs2="a3 a2" order="1"/>

<bond atomRefs2="a9 a2" order="1"/>

<bond atomRefs2="a2 a10" order="1"/>

<bond atomRefs2="a2 a1" order="1"/>

<bond atomRefs2="a7 a1" order="1"/>

<bond atomRefs2="a1 a8" order="1"/>

<bond atomRefs2="a1 a6" order="1"/>

<bond atomRefs2="a16 a6" order="1"/>

<bond atomRefs2="a6 a15" order="1"/>

<bond atomRefs2="a6 a5" order="1"/>

<bond atomRefs2="a5 a20" order="2"/>

<bond atomRefs2="a5 a14" order="1"/>

```
</bondArray>

<propertyList>

    <property dictRef="me:lumpedSpecies">

        <array> </array>

    </property>

    <property dictRef="me:vibFreqs">

        <array units="cm-1">12.1438 38.22 63.8126 66.6144 83.4399 98.4684
174.3221 203.9744 274.8783 283.1181 348.0964 387.383 406.409 475.1007 506.5719 642.6907 712.129 757.5324
769.8601 822.5325 925.3554 957.3596 992.0564 1013.8108 1039.5558 1075.2519 1090.7837 1113.7147 1147.8181
1160.1155 1162.0971 1172.5455 1175.0577 1201.1623 1225.658 1227.176 1230.9094 1246.9348 1260.0016
1268.2472 1274.1943 1294.6116 1314.9205 1323.3889 1327.1307 1358.0894 1540.3214 1795.3881 2542.1109
2642.6993 2651.357 2663.9588 2680.1601 2682.1838 2689.6964 2722.1113 2734.0144 2740.9513 2753.1001
2781.5459</array>

    </property>

    <property dictRef="me:ZPE">

        <scalar units="kJ/mol">-655.6205729989728</scalar>

    </property>

    <property dictRef="me:spinMultiplicity">

        <scalar units="cm-1">1</scalar>

    </property>

    <property dictRef="me:epsilon">

        <scalar>473.17</scalar>

    </property>

    <property dictRef="me:sigma">

        <scalar>5.09</scalar>

    </property>

<group>1</group>

    </propertyList>

    <me:energyTransferModel xsi:type="me:ExponentialDown">

        <scalar units="cm-1">250</scalar>

    </me:energyTransferModel>

</molecule>

<molecule id="TS_CC(=CCCCC=O)OO_O=CCCC[C]C(=O)C" spinMultiplicity="3">

    <atomArray>
```

```xml
<atom elementType="C" id="a1" x3="5.376315" y3="-8.080470" z3="-1.096602"/>
<atom elementType="C" id="a2" x3="6.595338" y3="-7.213553" z3="-1.463114"/>
<atom elementType="C" id="a3" spinMultiplicity="2" x3="6.968884" y3="-7.094372" z3="-2.849304"/>
<atom elementType="C" id="a4" x3="6.159574" y3="-7.216752" z3="-4.157393"/>
<atom elementType="C" id="a5" x3="6.589572" y3="-9.199329" z3="0.827980"/>
<atom elementType="C" id="a6" x3="5.758828" y3="-9.399264" z3="-0.421755"/>
<atom elementType="H" id="a7" x3="4.706586" y3="-7.514316" z3="-0.410989"/>
<atom elementType="H" id="a8" x3="4.755588" y3="-8.287982" z3="-1.994278"/>
<atom elementType="H" id="a9" x3="6.440741" y3="-6.170084" z3="-1.058751"/>
<atom elementType="H" id="a10" x3="7.494740" y3="-7.531935" z3="-0.861227"/>
<atom elementType="H" id="a11" x3="7.968619" y3="-6.810482" z3="-3.224198"/>
<atom elementType="H" id="a12" x3="9.262935" y3="-9.163338" z3="-4.626435"/>
<atom elementType="C" id="a13" x3="5.161487" y3="-6.083007" z3="-4.185777"/>
<atom elementType="H" id="a14" x3="7.281592" y3="-8.339218" z3="0.795858"/>
<atom elementType="H" id="a15" x3="4.842117" y3="-9.985224" z3="-0.176296"/>
<atom elementType="H" id="a16" x3="6.317629" y3="-10.066942" z3="-1.112840"/>
<atom elementType="H" id="a17" x3="5.055617" y3="-5.498558" z3="-3.270240"/>
<atom elementType="H" id="a18" x3="4.168628" y3="-6.471388" z3="-4.461536"/>
<atom elementType="H" id="a19" x3="5.446656" y3="-5.387632" z3="-4.998751"/>
<atom elementType="O" id="a20" x3="6.504320" y3="-9.926211" z3="1.788393"/>
```

```xml
                    <atom elementType="O" id="a21" spinMultiplicity="2" x3="8.940877" y3="-8.166474" z3="-4.986280"/>
                    <atom elementType="O" id="a22" x3="6.330532" y3="-7.925203" z3="-5.122022"/>
            </atomArray>
            <bondArray>
                    <bond atomRefs2="a22 a4" order="2"/>
                    <bond atomRefs2="a19 a13" order="1"/>
                    <bond atomRefs2="a21 a12" order="1"/>
                    <bond atomRefs2="a18 a13" order="1"/>
                    <bond atomRefs2="a13 a4" order="1"/>
                    <bond atomRefs2="a13 a17" order="1"/>
                    <bond atomRefs2="a4 a3" order="1"/>
                    <bond atomRefs2="a11 a3" order="1"/>
                    <bond atomRefs2="a3 a2" order="1"/>
                    <bond atomRefs2="a8 a1" order="1"/>
                    <bond atomRefs2="a2 a1" order="1"/>
                    <bond atomRefs2="a2 a9" order="1"/>
                    <bond atomRefs2="a2 a10" order="1"/>
                    <bond atomRefs2="a16 a6" order="1"/>
                    <bond atomRefs2="a1 a6" order="1"/>
                    <bond atomRefs2="a1 a7" order="1"/>
                    <bond atomRefs2="a6 a15" order="1"/>
                    <bond atomRefs2="a6 a5" order="1"/>
                    <bond atomRefs2="a14 a5" order="1"/>
                    <bond atomRefs2="a5 a20" order="2"/>
            </bondArray>
            <propertyList>
                    <property dictRef="me:vibFreqs">
                            <array units="cm-1">0.018096619822066742  22.957462694713893
30.236756427758174  49.400102891011734  62.524860462152354  67.2659610424418  84.48118438277353
93.74727850491273  149.81972518795672  233.92159419768373  270.20356845062634  390.40985695415264
439.72917296529846  457.39731801584463  505.9640813360781  579.2001773960383  665.697561575255
762.477832212507  792.3852818225199  844.867013653103  883.0530222561075  967.6877237439242
992.8112525466115  1010.172281402178  1018.304110923591  1028.1423699991506  1038.8992692491327
```

1056.944312920204   1091.7293732766657   1109.6440162193305   1124.2097724086964   1140.1784170602734
1171.1936943313826   1176.6216933983635   1199.398430483081   1204.3557301253795   1221.0769718818863
1224.2856027487305   1249.04672314084   1273.204189976623   1278.446390885718   1294.9494730565743
1320.7761599532043   1322.0896406293207   1390.7896959562472   1742.4151808630952   1800.5698837612092
2316.920264843209   2571.805249948933   2608.276187611612   2620.8758067652516   2645.533317746872
2655.5251844250565   2666.05941943975   2681.6659267311948   2694.7232169235904   2729.924851165746
2736.8519016423816 2781.9659918426455</array>

                    </property>

                    <property dictRef="me:imFreq">

                            <scalar units="cm-1">201.6410693839681</scalar>

                    </property>

                    <property dictRef="me:spinMultiplicity">

                            <scalar units="cm-1">1</scalar>

                    </property>

<group>1</group>

                    <property dictRef="me:ZPE">

                            <scalar units="kJ/mol">-281.46244980442435</scalar>

                    </property>

            </propertyList>

        </molecule>

        <molecule id="O=CCCC[C]C(=O)C" spinMultiplicity="3">

                <atomArray>

                            <atom   elementType="O"   id="a1"   x3="2.078588"   y3="1.128698"   z3="-1.354926"/>

                            <atom   elementType="C"   id="a2"   x3="1.694103"   y3="0.141507"   z3="-0.765945"/>

                            <atom   elementType="C"   id="a3"   x3="2.526624"   y3="-1.100723"   z3="-0.573003"/>

                            <atom   elementType="C"   id="a4"   x3="3.839587"   y3="-1.096330"   z3="-1.359426"/>

                            <atom   elementType="C"   id="a5"   x3="4.812035"   y3="-0.006412"   z3="-0.856820"/>

                            <atom   elementType="C"   id="a6"   spinMultiplicity="3"   x3="5.216676"   y3="-0.173963" z3="0.481231"/>

                            <atom   elementType="C"   id="a7"   x3="5.152571"   y3="-0.783977"   z3="1.748726"/>

                            <atom   elementType="O"   id="a8"   x3="4.067352"   y3="-1.008476"   z3="2.282668"/>

```xml
                        <atom    elementType="C"    id="a9"    x3="6.459932"    y3="-1.080541"
z3="2.426375"/>
                        <atom    elementType="H"    id="a10"    x3="0.687388"    y3="0.100544"    z3="-
0.316749"/>
                        <atom    elementType="H"    id="a11"    x3="2.740608"    y3="-1.210270"
z3="0.525040"/>
                        <atom    elementType="H"    id="a12"    x3="1.931607"    y3="-1.999119"    z3="-
0.831788"/>
                        <atom    elementType="H"    id="a13"    x3="4.318368"    y3="-2.091606"    z3="-
1.290553"/>
                        <atom    elementType="H"    id="a14"    x3="3.631144"    y3="-0.915735"    z3="-
2.433612"/>
                        <atom    elementType="H"    id="a15"    x3="5.700842"    y3="0.066704"    z3="-
1.530639"/>
                        <atom    elementType="H"    id="a16"    x3="4.310437"    y3="1.003763"    z3="-
0.970365"/>
                        <atom    elementType="H"    id="a17"    x3="7.122453"    y3="-1.692101"
z3="1.801993"/>
                        <atom    elementType="H"    id="a18"    x3="7.000586"    y3="-0.154307"
z3="2.668388"/>
                        <atom    elementType="H"    id="a19"    x3="6.295771"    y3="-1.619433"
z3="3.371890"/>
                </atomArray>
                <bondArray>
                        <bond atomRefs2="a14 a4" order="1"/>
                        <bond atomRefs2="a15 a5" order="1"/>
                        <bond atomRefs2="a4 a13" order="1"/>
                        <bond atomRefs2="a4 a5" order="1"/>
                        <bond atomRefs2="a4 a3" order="1"/>
                        <bond atomRefs2="a1 a2" order="2"/>
                        <bond atomRefs2="a16 a5" order="1"/>
                        <bond atomRefs2="a5 a6" order="1"/>
                        <bond atomRefs2="a12 a3" order="1"/>
                        <bond atomRefs2="a2 a3" order="1"/>
                        <bond atomRefs2="a2 a10" order="1"/>
                        <bond atomRefs2="a3 a11" order="1"/>
```

<bond atomRefs2="a6 a7" order="1"/>

<bond atomRefs2="a7 a8" order="2"/>

<bond atomRefs2="a7 a9" order="1"/>

<bond atomRefs2="a17 a9" order="1"/>

<bond atomRefs2="a9 a18" order="1"/>

<bond atomRefs2="a9 a19" order="1"/>

</bondArray>

<propertyList>

<property dictRef="me:lumpedSpecies">

<array> </array>

</property>

<property dictRef="me:vibFreqs">

<array units="cm-1">25.8171  45.7199  54.0707  79.518  96.5227
105.9833 192.5877 234.3675 314.9725 373.5494 450.7001 497.682 523.8576 623.3313 743.8016 801.556 825.4967
922.0475 955.5289 996.3351 1010.2105 1024.9669 1049.798 1081.861 1129.9347 1167.4002 1168.6547 1181.8789
1198.7564 1215.7004 1229.0811 1231.8076 1258.7409 1275.8639 1286.3883 1313.0567 1317.6673 1328.2527
1636.475  1686.4188  1784.3186  2596.0311  2613.4926  2645.0143  2671.9152  2688.1156  2693.4322  2702.9203
2725.8355 2738.4735 2787.5705</array>

</property>

<property dictRef="me:ZPE">

<scalar units="kJ/mol">-433.1037259647566</scalar>

</property>

<property dictRef="me:spinMultiplicity">

<scalar units="cm-1">1</scalar>

</property>

<group>1</group>

<property dictRef="me:epsilon">

<scalar>473.17</scalar>

</property>

<property dictRef="me:sigma">

<scalar>5.09</scalar>

</property>

</propertyList>

<me:energyTransferModel xsi:type="me:ExponentialDown">

```xml
                        <scalar units="cm-1">250</scalar>
                </me:energyTransferModel>
        </molecule>
        <molecule id="O">
                <atomArray>
                        <atom   elementType="O"   id="a1"   x3="0.978241"   y3="-0.028820"   z3="-0.087158"/>
                        <atom   elementType="H"   id="a2"   x3="0.677023"   y3="-0.677926"   z3="0.536334"/>
                        <atom   elementType="H"   id="a3"   x3="1.927217"   y3="-0.040374"   z3="-0.076060"/>
                </atomArray>
                <bondArray>
                        <bond atomRefs2="a1 a3" order="1"/>
                        <bond atomRefs2="a1 a2" order="1"/>
                </bondArray>
                <propertyList>
                        <property dictRef="me:lumpedSpecies">
                                <array> </array>
                        </property>
                        <property dictRef="me:vibFreqs">
                                <array units="cm-1">1336.6341 2528.7149 2614.885</array>
                        </property>
                        <property dictRef="me:ZPE">
                                <scalar units="kJ/mol">0.0</scalar>
                        </property>
                        <property dictRef="me:spinMultiplicity">
                                <scalar units="cm-1">1</scalar>
                        </property>
                        <property dictRef="me:epsilon">
                                <scalar>473.17</scalar>
                        </property>
<group>1</group>
```

```xml
                        <property dictRef="me:sigma">
                                <scalar>5.09</scalar>
                        </property>
                </propertyList>
                <me:energyTransferModel xsi:type="me:ExponentialDown">
                        <scalar units="cm-1">250</scalar>
                </me:energyTransferModel>
        </molecule>
        <molecule id="TS_CC(=CCCCC=O)OO_O=CCCC[CH]C(=O)C" spinMultiplicity="3">
                <atomArray>
                        <atom    elementType="C"    id="a1"    x3="-0.363095"    y3="-0.004529"
z3="1.413982"/>
                        <atom    elementType="C"    id="a2"    x3="0.037745"    y3="0.257518"    z3="-
0.042666"/>
                        <atom    elementType="C"    id="a3"    spinMultiplicity="2"    x3="-0.291058"    y3="-
0.884623" z3="-0.937217"/>
                        <atom    elementType="C"    id="a4"    spinMultiplicity="2"    x3="-0.148481"    y3="-
0.804304" z3="-2.342852"/>
                        <atom    elementType="C"    id="a5"    x3="-0.462062"    y3="0.969111"
z3="3.744502"/>
                        <atom    elementType="C"    id="a6"    x3="0.034794"    y3="1.157947"
z3="2.328024"/>
                        <atom    elementType="H"    id="a7"    x3="-1.457071"    y3="-0.178938"
z3="1.491008"/>
                        <atom    elementType="H"    id="a8"    x3="0.105334"    y3="-0.941778"
z3="1.779995"/>
                        <atom    elementType="H"    id="a9"    x3="-0.450359"    y3="1.186058"    z3="-
0.412245"/>
                        <atom    elementType="H"    id="a10"    x3="1.131900"    y3="0.461370"    z3="-
0.104414"/>
                        <atom    elementType="H"    id="a11"    x3="-0.710705"    y3="-1.790117"    z3="-
0.487089"/>
                        <atom    elementType="H"    id="a12"    x3="1.619559"    y3="-3.079269"    z3="-
1.546204"/>
                        <atom    elementType="C"    id="a13"    x3="0.060505"    y3="0.436109"    z3="-
3.137465"/>
                        <atom    elementType="H"    id="a14"    x3="-0.097901"    y3="1.720239"
z3="4.465443"/>
```

```xml
                                    <atom    elementType="H"    id="a15"    x3="1.137061"    y3="1.282773"
z3="2.334028"/>
                                    <atom    elementType="H"    id="a16"    x3="-0.366614"    y3="2.115180"
z3="1.933809"/>
                                    <atom   elementType="H"   id="a17"   x3="-0.789400"   y3="1.124711"   z3="-
3.045938"/>
                                    <atom   elementType="H"   id="a18"   x3="0.184012"   y3="0.202602"   z3="-
4.206853"/>
                                    <atom   elementType="H"   id="a19"   x3="0.967537"   y3="0.972764"   z3="-
2.823832"/>
                                    <atom    elementType="O"    id="a20"    x3="-1.205416"    y3="0.076713"
z3="4.080666"/>
                                    <atom   elementType="O"   id="a21"   x3="1.029619"   y3="-2.303444"   z3="-
1.590843"/>
                                    <atom   elementType="O"   id="a22"   x3="0.034094"   y3="-1.976091"   z3="-
2.893839"/>
                    </atomArray>
                    <bondArray>
                                    <bond atomRefs2="a18 a13" order="1"/>
                                    <bond atomRefs2="a13 a17" order="1"/>
                                    <bond atomRefs2="a13 a19" order="1"/>
                                    <bond atomRefs2="a13 a4" order="1"/>
                                    <bond atomRefs2="a22 a4" order="1"/>
                                    <bond atomRefs2="a22 a21" order="1"/>
                                    <bond atomRefs2="a4 a3" order="1"/>
                                    <bond atomRefs2="a21 a12" order="1"/>
                                    <bond atomRefs2="a3 a11" order="1"/>
                                    <bond atomRefs2="a3 a2" order="1"/>
                                    <bond atomRefs2="a9 a2" order="1"/>
                                    <bond atomRefs2="a10 a2" order="1"/>
                                    <bond atomRefs2="a2 a1" order="1"/>
                                    <bond atomRefs2="a1 a7" order="1"/>
                                    <bond atomRefs2="a1 a8" order="1"/>
                                    <bond atomRefs2="a1 a6" order="1"/>
                                    <bond atomRefs2="a16 a6" order="1"/>
```

```xml
            <bond atomRefs2="a6 a15" order="1"/>

            <bond atomRefs2="a6 a5" order="1"/>

            <bond atomRefs2="a5 a20" order="2"/>

            <bond atomRefs2="a5 a14" order="1"/>

        </bondArray>

        <propertyList>

            <property dictRef="me:vibFreqs">

                                <array     units="cm-1">14.48471820090016      34.638247521404715
58.8933434453765     77.58533914623469     88.07117876360067     96.60700328055856     159.33193301054294
170.26221625338772   227.14533592862273    310.7785647941167    320.00533682610217    396.88536815822266
421.2804707434764    477.05480195012194    544.341952159953    580.1220486208701     645.5280053822147
714.5078498273515    764.651270558891     848.494604152173    932.2000205237745    972.7259350111262
990.720154494529     1013.9953637447542   1039.0014973898549   1051.2873261779625   1110.9394807170263
1135.1655862994355   1145.6521760112232   1158.4002085858879   1171.082723968269   1179.4929323776294
1203.4069869890018   1222.3193598241673   1227.5307437560407   1234.2892018651282   1241.9220292812195
1251.8462074060026   1263.9356890154083   1277.2781434626913   1302.9072809607305   1315.5223193982092
1323.2760595600764   1328.785907068344    1397.8862586112268   1554.0994826968963   1795.2760521088578
2597.840882774681    2642.276194031044    2646.759341956519   2658.855061719387    2678.4045974166543
2686.803144137047    2693.077810712295    2711.970211211421   2729.863361149177    2736.1861526583916
2749.138771510489 2785.3835845521107</array>

            </property>

            <property dictRef="me:imFreq">

                    <scalar units="cm-1">732.3626876991656</scalar>

            </property>

            <property dictRef="me:spinMultiplicity">

                    <scalar units="cm-1">1</scalar>

            </property>

<group>1</group>

            <property dictRef="me:ZPE">

                    <scalar units="kJ/mol">-258.5805006183524</scalar>

            </property>

        </propertyList>

    </molecule>

    <molecule id="O=CCCC[CH]C(=O)C" spinMultiplicity="2">

        <atomArray>

                    <atom    elementType="O"    id="a1"    x3="0.703292"    y3="-0.485620"    z3="-0.225957"/>
```

```
                              <atom    elementType="C"    id="a2"    x3="1.882794"    y3="-0.343060"
z3="0.003393"/>
                              <atom    elementType="C"    id="a3"    x3="2.771734"    y3="-1.524253"
z3="0.347815"/>
                              <atom    elementType="C"    id="a4"    x3="3.905793"    y3="-1.710568"    z3="-
0.663003"/>
                              <atom    elementType="C"    id="a5"    x3="4.840692"    y3="-0.492139"    z3="-
0.780561"/>
                              <atom    elementType="C"    id="a6"    spinMultiplicity="2"    x3="5.473534"    y3="-
0.156154" z3="0.509160"/>
                              <atom    elementType="H"    id="a7"    x3="6.191273"    y3="-0.833863"
z3="0.943138"/>
                              <atom    elementType="C"    id="a8"    x3="5.098156"    y3="1.075148"
z3="1.221185"/>
                              <atom    elementType="O"    id="a9"    x3="4.092474"    y3="1.703182"
z3="0.938083"/>
                              <atom    elementType="C"    id="a10"    x3="6.017893"    y3="1.510677"
z3="2.325120"/>
                              <atom    elementType="H"    id="a11"    x3="2.378048"    y3="0.645058"    z3="-
0.008792"/>
                              <atom    elementType="H"    id="a12"    x3="3.177801"    y3="-1.379963"
z3="1.368974"/>
                              <atom    elementType="H"    id="a13"    x3="2.160352"    y3="-2.450058"
z3="0.396439"/>
                              <atom    elementType="H"    id="a14"    x3="4.494040"    y3="-2.607282"    z3="-
0.382240"/>
                              <atom    elementType="H"    id="a15"    x3="3.478082"    y3="-1.929997"    z3="-
1.663217"/>
                              <atom    elementType="H"    id="a16"    x3="5.627881"    y3="-0.708352"    z3="-
1.537656"/>
                              <atom    elementType="H"    id="a17"    x3="4.273705"    y3="0.378082"    z3="-
1.184233"/>
                              <atom    elementType="H"    id="a18"    x3="6.041390"    y3="0.784218"
z3="3.147831"/>
                              <atom    elementType="H"    id="a19"    x3="7.049459"    y3="1.641996"
z3="1.973423"/>
                              <atom    elementType="H"    id="a20"    x3="5.693613"    y3="2.472642"
z3="2.752781"/>
               </atomArray>
```

```xml
<bondArray>
        <bond atomRefs2="a15 a4" order="1"/>
        <bond atomRefs2="a16 a5" order="1"/>
        <bond atomRefs2="a17 a5" order="1"/>
        <bond atomRefs2="a5 a4" order="1"/>
        <bond atomRefs2="a5 a6" order="1"/>
        <bond atomRefs2="a4 a14" order="1"/>
        <bond atomRefs2="a4 a3" order="1"/>
        <bond atomRefs2="a1 a2" order="2"/>
        <bond atomRefs2="a11 a2" order="1"/>
        <bond atomRefs2="a2 a3" order="1"/>
        <bond atomRefs2="a3 a13" order="1"/>
        <bond atomRefs2="a3 a12" order="1"/>
        <bond atomRefs2="a6 a7" order="1"/>
        <bond atomRefs2="a6 a8" order="1"/>
        <bond atomRefs2="a9 a8" order="2"/>
        <bond atomRefs2="a8 a10" order="1"/>
        <bond atomRefs2="a19 a10" order="1"/>
        <bond atomRefs2="a10 a20" order="1"/>
        <bond atomRefs2="a10 a18" order="1"/>
</bondArray>
<propertyList>
        <property dictRef="me:lumpedSpecies">
                <array> </array>
        </property>
        <property dictRef="me:vibFreqs">
                <array units="cm-1">26.7553 40.3879 56.7 81.1813 90.819 126.1154
191.3789 273.4314 322.4071 389.2903 447.2133 477.8143 493.9236 572.4351 727.4703 805.7205 865.0221
895.7077 953.3173 983.8287 1012.2171 1019.526 1035.7814 1094.5356 1101.3296 1125.8154 1170.992 1195.3298
1206.7476 1218.6595 1228.231 1238.9813 1252.8927 1260.097 1263.4231 1277.6931 1280.0828 1304.0581
1316.3807 1326.2049 1389.8978 1741.0479 1789.2675 2616.5773 2652.5315 2664.9542 2675.7469 2688.1977
2691.3781 2726.3562 2737.7211 2743.3667 2756.5034 2788.701</array>
        </property>
        <property dictRef="me:ZPE">
```

```xml
                    <scalar units="kJ/mol">-403.96693688981384</scalar>
                </property>
                <property dictRef="me:spinMultiplicity">
                        <scalar units="cm-1">2</scalar>
                </property>
<group>1</group>
                <property dictRef="me:epsilon">
                        <scalar>473.17</scalar>
                </property>
                <property dictRef="me:sigma">
                        <scalar>5.09</scalar>
                </property>
            </propertyList>
            <me:energyTransferModel xsi:type="me:ExponentialDown">
                    <scalar units="cm-1">250</scalar>
            </me:energyTransferModel>
        </molecule>
        <molecule id="[OH]" spinMultiplicity="2">
            <atomArray>
                    <atom elementType="O" id="a1" spinMultiplicity="2" x3="0.913827" y3="-0.023061" z3="-0.010259"/>
                    <atom elementType="H" id="a2" x3="1.892649" y3="-0.023061" z3="-0.010259"/>
            </atomArray>
            <bondArray>
                    <bond atomRefs2="a1 a2" order="1"/>
            </bondArray>
            <propertyList>
                <property dictRef="me:lumpedSpecies">
                        <array> </array>
                </property>
                <property dictRef="me:vibFreqs">
                        <array units="cm-1">2641.8918</array>
```

```xml
            </property>
            <property dictRef="me:ZPE">
                    <scalar units="kJ/mol">0.0</scalar>
            </property>
            <property dictRef="me:spinMultiplicity">
                    <scalar units="cm-1">2</scalar>
            </property>
<group>1</group>
            <property dictRef="me:epsilon">
                    <scalar>473.17</scalar>
            </property>
            <property dictRef="me:sigma">
                    <scalar>5.09</scalar>
            </property>
        </propertyList>
        <me:energyTransferModel xsi:type="me:ExponentialDown">
                <scalar units="cm-1">250</scalar>
        </me:energyTransferModel>
    </molecule>
    <molecule id="TS_O=CCCC[CH]C(=O)C_O=[C]CCCCC(=O)C" spinMultiplicity="2">
        <atomArray>
            <atom    elementType="O"    id="a1"    x3="-2.499699"    y3="-0.649098"
z3="1.527824"/>
            <atom    elementType="C"    id="a2"    x3="-1.704096"    y3="-0.682199"
z3="0.643362"/>
            <atom    elementType="C"    id="a3"    x3="-1.789187"    y3="-1.273136"    z3="-
0.724056"/>
            <atom    elementType="C"    id="a4"    x3="-0.396196"    y3="-1.573777"    z3="-
1.290593"/>
            <atom    elementType="C"    id="a5"    x3="0.538827"    y3="-0.354729"    z3="-
1.223309"/>
            <atom    elementType="C"    id="a6"    spinMultiplicity="2"    x3="0.762799"
y3="0.115871" z3="0.184858"/>
            <atom    elementType="H"    id="a7"    x3="1.335665"    y3="-0.545257"
z3="0.829717"/>
```

                            <atom     elementType="C"     id="a8"     x3="0.920842"     y3="1.575440"
z3="0.393901"/>

                            <atom     elementType="O"   id="a9"   x3="0.617046"   y3="2.395291"   z3="-
0.450617"/>

                            <atom     elementType="C"     id="a10"     x3="1.462667"     y3="1.991241"
z3="1.734523"/>

                            <atom     elementType="H"     id="a11"     x3="-0.536773"     y3="-0.122308"
z3="0.817720"/>

                            <atom   elementType="H"   id="a12"   x3="-2.415329"   y3="-2.189815"   z3="-
0.713147"/>

                            <atom   elementType="H"   id="a13"   x3="-2.327527"   y3="-0.554768"   z3="-
1.384274"/>

                            <atom   elementType="H"   id="a14"   x3="0.054226"   y3="-2.426220"   z3="-
0.744627"/>

                            <atom   elementType="H"   id="a15"   x3="-0.492208"   y3="-1.905027"   z3="-
2.343619"/>

                            <atom   elementType="H"   id="a16"   x3="1.513029"   y3="-0.609160"   z3="-
1.691740"/>

                            <atom   elementType="H"   id="a17"   x3="0.125045"   y3="0.476766"   z3="-
1.839555"/>

                            <atom     elementType="H"     id="a18"     x3="0.871861"     y3="1.586125"
z3="2.565047"/>

                            <atom     elementType="H"     id="a19"     x3="2.498641"     y3="1.657312"
z3="1.872790"/>

                            <atom     elementType="H"     id="a20"     x3="1.460368"     y3="3.087448"
z3="1.835796"/>

                    </atomArray>

                    <bondArray>

                            <bond atomRefs2="a15 a4" order="1"/>

                            <bond atomRefs2="a17 a5" order="1"/>

                            <bond atomRefs2="a16 a5" order="1"/>

                            <bond atomRefs2="a13 a3" order="1"/>

                            <bond atomRefs2="a4 a5" order="1"/>

                            <bond atomRefs2="a4 a14" order="1"/>

                            <bond atomRefs2="a4 a3" order="1"/>

                            <bond atomRefs2="a5 a6" order="1"/>

                            <bond atomRefs2="a3 a12" order="1"/>

```xml
                <bond atomRefs2="a3 a2" order="1"/>

                <bond atomRefs2="a9 a8" order="2"/>

                <bond atomRefs2="a6 a8" order="1"/>

                <bond atomRefs2="a6 a7" order="1"/>

                <bond atomRefs2="a8 a10" order="1"/>

                <bond atomRefs2="a2 a11" order="1"/>

                <bond atomRefs2="a2 a1" order="2"/>

                <bond atomRefs2="a10 a20" order="1"/>

                <bond atomRefs2="a10 a19" order="1"/>

                <bond atomRefs2="a10 a18" order="1"/>

        </bondArray>

        <propertyList>

            <property dictRef="me:vibFreqs">

                                    <array     units="cm-1">37.15833920583505       64.44628215234935
67.8104761710448    76.46694144141347    184.3058549453589    251.15916066304663    290.07335921300734
327.5749611565306    379.95972166947547    438.8860352462389    455.0010670508147    534.9883231288363
568.3850914042389    597.6764136460966    771.7853625130832    835.44469391561    876.8585207280718
929.0791105363403    957.865943002568    982.9559171754719    1005.9350788521713    1035.342881198563
1058.3646009991846    1080.3179949414505    1096.5704422229865    1149.854879671323    1162.9153891440471
1180.2311647647105    1201.5115103132403    1218.705785124888    1226.9523400160087    1230.5241269458909
1236.8246589647827    1252.5619804645175    1268.5033722377461    1283.5079400045486    1299.0062825478935
1314.9528664853592    1319.163108842599    1351.4931874023405    1454.0202365725725    1772.227539430247
1885.8478310762953    2652.528673669451    2655.421586122581    2673.9834818935697    2690.180134538133
2693.5323903423896    2727.307196861528    2733.8141743503993    2740.2666166611284    2743.4959607616947
2789.8089034044806</array>

            </property>

            <property dictRef="me:imFreq">

                        <scalar units="cm-1">2369.086171737593</scalar>

            </property>

            <property dictRef="me:spinMultiplicity">

                        <scalar units="cm-1">2</scalar>

            </property>

<group>1</group>

            <property dictRef="me:ZPE">

                        <scalar units="kJ/mol">41.00241605527717</scalar>

            </property>
```

```
                    </propertyList>
            </molecule>
            <molecule id="O=[C]CCCCC(=O)C" spinMultiplicity="2">
                    <atomArray>
                                    <atom     elementType="O"     id="a1"     x3="2.576865"     y3="-0.682217"
z3="1.859890"/>
                                    <atom  elementType="C"  id="a2"  spinMultiplicity="2"  x3="3.164313"  y3="-
0.656495" z3="2.869662"/>
                                    <atom     elementType="C"     id="a3"     x3="4.306699"     y3="-0.065131"
z3="3.562982"/>
                                    <atom     elementType="C"     id="a4"     x3="5.062596"     y3="1.005720"
z3="2.761081"/>
                                    <atom     elementType="C"     id="a5"     x3="5.953783"     y3="0.430479"
z3="1.654197"/>
                                    <atom     elementType="C"     id="a6"     x3="5.221568"     y3="0.263525"
z3="0.319601"/>
                                    <atom  elementType="H"  id="a7"  x3="4.915633"  y3="1.250041"  z3="-
0.080872"/>
                                    <atom  elementType="C"  id="a8"  x3="6.065257"  y3="-0.472980"  z3="-
0.698706"/>
                                    <atom  elementType="O"  id="a9"  x3="7.140378"  y3="-0.961891"  z3="-
0.433910"/>
                                    <atom  elementType="C"  id="a10"  x3="5.470678"  y3="-0.563793"  z3="-
2.077125"/>
                                    <atom     elementType="H"     id="a11"     x3="4.265627"     y3="-0.288748"
z3="0.467024"/>
                                    <atom     elementType="H"     id="a12"     x3="3.947191"     y3="0.364421"
z3="4.527005"/>
                                    <atom     elementType="H"     id="a13"     x3="4.998903"     y3="-0.886645"
z3="3.864874"/>
                                    <atom     elementType="H"     id="a14"     x3="4.344683"     y3="1.736241"
z3="2.338692"/>
                                    <atom     elementType="H"     id="a15"     x3="5.694064"     y3="1.586165"
z3="3.465214"/>
                                    <atom     elementType="H"     id="a16"     x3="6.833814"     y3="1.089754"
z3="1.504540"/>
                                    <atom     elementType="H"     id="a17"     x3="6.385986"     y3="-0.544866"
z3="1.962922"/>
```

```xml
                              <atom   elementType="H"   id="a18"   x3="4.451578"   y3="-0.969843"   z3="-2.064580"/>
                              <atom   elementType="H"   id="a19"   x3="5.424648"   y3="0.418756"   z3="-2.563333"/>
                              <atom   elementType="H"   id="a20"   x3="6.075826"   y3="-1.215839"   z3="-2.725349"/>
                    </atomArray>
                    <bondArray>
                              <bond atomRefs2="a20 a10" order="1"/>
                              <bond atomRefs2="a19 a10" order="1"/>
                              <bond atomRefs2="a10 a18" order="1"/>
                              <bond atomRefs2="a10 a8" order="1"/>
                              <bond atomRefs2="a8 a9" order="2"/>
                              <bond atomRefs2="a8 a6" order="1"/>
                              <bond atomRefs2="a7 a6" order="1"/>
                              <bond atomRefs2="a6 a11" order="1"/>
                              <bond atomRefs2="a6 a5" order="1"/>
                              <bond atomRefs2="a16 a5" order="1"/>
                              <bond atomRefs2="a5 a17" order="1"/>
                              <bond atomRefs2="a5 a4" order="1"/>
                              <bond atomRefs2="a1 a2" order="2"/>
                              <bond atomRefs2="a14 a4" order="1"/>
                              <bond atomRefs2="a4 a15" order="1"/>
                              <bond atomRefs2="a4 a3" order="1"/>
                              <bond atomRefs2="a2 a3" order="1"/>
                              <bond atomRefs2="a3 a13" order="1"/>
                              <bond atomRefs2="a3 a12" order="1"/>
                    </bondArray>
                    <propertyList>
                              <property dictRef="me:lumpedSpecies">
                                        <array> </array>
                              </property>
                              <property dictRef="me:vibFreqs">
```

```xml
                              <array   units="cm-1">18.0934   37.9132   40.5454   59.322   109.2807
144.2281  161.0022  212.3629  298.2717  392.9727  402.6786  487.8352  561.9521  576.1865  766.4383  836.3641
855.7699  925.3648  945.7354  992.3852  1016.288  1030.055  1044.1886  1105.0839  1125.2582  1156.3413  1168.7343
1191.086  1208.2818  1215.1156  1223.256  1228.5429  1230.4313  1242.4041  1266.4477  1272.6415  1289.4101
1303.7032  1317.4309  1319.6153  1330.6349  1805.9897  2008.7459  2647.1436  2655.0955  2666.9575  2669.0287
2690.5778  2693.6409  2734.2926  2735.9959  2739.9811  2748.1631  2790.9882</array>
                              </property>
                              <property dictRef="me:ZPE">
                                      <scalar units="kJ/mol">13.71977730214985</scalar>
                              </property>
                              <property dictRef="me:spinMultiplicity">
                                      <scalar units="cm-1">2</scalar>
                              </property>
                              <property dictRef="me:epsilon">
                                      <scalar>473.17</scalar>
                              </property>
<group>1</group>
                              <property dictRef="me:sigma">
                                      <scalar>5.09</scalar>
                              </property>
                      </propertyList>
                      <me:energyTransferModel xsi:type="me:ExponentialDown">
                              <scalar units="cm-1">250</scalar>
                      </me:energyTransferModel>
              </molecule>
              <molecule id="TS_O=CCCC[CH]C(=O)C_O=CCCCC=C=O" spinMultiplicity="2">
                      <atomArray>
                                      <atom     elementType="O"    id="a1"    x3="1.163869"     y3="-3.046589"
z3="2.040525"/>
                                      <atom     elementType="C"    id="a2"    x3="2.284376"     y3="-2.634081"
z3="2.244063"/>
                                      <atom     elementType="C"    id="a3"    x3="2.712779"     y3="-1.263179"
z3="1.776149"/>
                                      <atom     elementType="C"    id="a4"    x3="4.059321"     y3="-0.815872"
z3="2.345602"/>
```

```xml
                        <atom     elementType="C"    id="a5"     x3="4.515254"    y3="0.550801"
z3="1.794462"/>
                        <atom     elementType="C"    id="a6"     x3="5.490121"    y3="0.449259"
z3="0.662155"/>
                        <atom   elementType="H"   id="a7"   x3="5.174382"   y3="0.857063"   z3="-
0.295461"/>
                        <atom     elementType="C"    id="a8"     x3="6.703332"    y3="-0.042332"
z3="0.815810"/>
                        <atom     elementType="O"    id="a9"     x3="7.740303"    y3="-0.560459"
z3="0.934571"/>
                        <atom   elementType="C"   id="a10"   spinMultiplicity="2"    x3="7.693058"
y3="2.251950" z3="0.746128"/>
                        <atom     elementType="H"    id="a11"    x3="3.034732"    y3="-3.249357"
z3="2.765018"/>
                        <atom     elementType="H"    id="a12"    x3="2.743905"    y3="-1.269983"
z3="0.664026"/>
                        <atom     elementType="H"    id="a13"    x3="1.910281"    y3="-0.531047"
z3="2.022770"/>
                        <atom     elementType="H"    id="a14"    x3="3.977862"    y3="-0.745713"
z3="3.448707"/>
                        <atom     elementType="H"    id="a15"    x3="4.837831"    y3="-1.581828"
z3="2.144662"/>
                        <atom     elementType="H"    id="a16"    x3="3.639859"    y3="1.148920"
z3="1.462051"/>
                        <atom     elementType="H"    id="a17"    x3="4.971321"    y3="1.144469"
z3="2.616260"/>
                        <atom     elementType="H"    id="a18"    x3="6.940324"    y3="2.802243"
z3="0.241172"/>
                        <atom     elementType="H"    id="a19"    x3="7.704031"    y3="2.233915"
z3="1.806536"/>
                        <atom     elementType="H"    id="a20"    x3="8.489426"    y3="1.815626"
z3="0.199713"/>
                </atomArray>
                <bondArray>
                        <bond atomRefs2="a7 a6" order="1"/>
                        <bond atomRefs2="a20 a10" order="1"/>
                        <bond atomRefs2="a18 a10" order="1"/>
                        <bond atomRefs2="a6 a8" order="2"/>
```

```
                    <bond atomRefs2="a6 a5" order="1"/>

                    <bond atomRefs2="a12 a3" order="1"/>

                    <bond atomRefs2="a10 a19" order="1"/>

                    <bond atomRefs2="a8 a9" order="2"/>

                    <bond atomRefs2="a16 a5" order="1"/>

                    <bond atomRefs2="a3 a13" order="1"/>

                    <bond atomRefs2="a3 a2" order="1"/>

                    <bond atomRefs2="a3 a4" order="1"/>

                    <bond atomRefs2="a5 a4" order="1"/>

                    <bond atomRefs2="a5 a17" order="1"/>

                    <bond atomRefs2="a1 a2" order="2"/>

                    <bond atomRefs2="a15 a4" order="1"/>

                    <bond atomRefs2="a2 a11" order="1"/>

                    <bond atomRefs2="a4 a14" order="1"/>

            </bondArray>

            <propertyList>

                <property dictRef="me:vibFreqs">

                        <array    units="cm-1">0.061100554329539404    17.30941683914933
22.2871048586058    37.161675860651194    74.3702255726438    94.88161276662605    165.7747528619866
207.74931718093018    223.6457151254805    312.8869423059929    336.58105731229847    343.09648070441693
429.72868164462847    436.06940721769183    505.8055338906333    623.0027471749232    705.4794750721594
742.6520343726789    863.0624423787241    893.1797842486051    957.5600982086837    996.3822899114708
1056.2673609244462    1102.9519376331377    1117.1473833634052    1136.7690391235726    1139.7276660692294
1179.9221163208822    1188.562152053263    1209.6102291984232    1221.4557083414197    1250.8773406123719
1254.734335961059    1268.0468776773741    1274.949472840056    1288.5685204399063    1306.5018942696897
1309.723296979897    1329.7358704731218    1435.5989435406834    1785.6222559254952    2235.1959406852084
2645.383115546048    2649.2002477709057    2657.967255559535    2671.1988739754133    2727.7473024848896
2734.1536770106927    2743.0617147347252    2749.392987955489    2751.057161008031    2758.947477593203
2793.9775151855793</array>

                </property>

                <property dictRef="me:imFreq">

                        <scalar units="cm-1">67.05595802149526</scalar>

                </property>

                <property dictRef="me:spinMultiplicity">

                        <scalar units="cm-1">2</scalar>

                </property>
```

```xml
<group>1</group>
                                <property dictRef="me:ZPE">
                                        <scalar units="kJ/mol">196.58112527658577</scalar>
                                </property>
                        </propertyList>
                </molecule>
                <molecule id="O=CCCCC=O">
                        <atomArray>
                                <atom    elementType="O"    id="a1"    x3="1.430648"    y3="-0.192244"    z3="0.814652"/>
                                <atom    elementType="C"    id="a2"    x3="2.094422"    y3="-0.094857"    z3="-0.193641"/>
                                <atom    elementType="C"    id="a3"    x3="3.081226"    y3="-1.159489"    z3="-0.607515"/>
                                <atom    elementType="C"    id="a4"    x3="4.085101"    y3="-0.685678"    z3="-1.657646"/>
                                <atom    elementType="C"    id="a5"    x3="5.074585"    y3="0.362349"    z3="-1.118113"/>
                                <atom    elementType="C"    id="a6"    x3="6.017188"    y3="-0.210893"    z3="-0.106070"/>
                                <atom    elementType="C"    id="a7"    x3="7.221225"    y3="-0.638733"    z3="-0.412354"/>
                                <atom    elementType="O"    id="a8"    x3="8.293416"    y3="-1.012027"    z3="-0.673207"/>
                                <atom    elementType="H"    id="a9"    x3="2.001730"    y3="0.774846"    z3="-0.863626"/>
                                <atom    elementType="H"    id="a10"    x3="3.614485"    y3="-1.540250"    z3="0.291494"/>
                                <atom    elementType="H"    id="a11"    x3="2.516588"    y3="-2.040667"    z3="-0.981008"/>
                                <atom    elementType="H"    id="a12"    x3="4.649384"    y3="-1.561815"    z3="-2.037353"/>
                                <atom    elementType="H"    id="a13"    x3="3.552162"    y3="-0.270470"    z3="-2.535774"/>
                                <atom    elementType="H"    id="a14"    x3="5.633266"    y3="0.806710"    z3="-1.967874"/>
                                <atom    elementType="H"    id="a15"    x3="4.524205"    y3="1.207877"    z3="-0.654107"/>
```

```xml
                    <atom    elementType="H"    id="a16"    x3="5.648796"    y3="-0.275341"
z3="0.916599"/>
          </atomArray>
          <bondArray>
                    <bond atomRefs2="a13 a4" order="1"/>
                    <bond atomRefs2="a12 a4" order="1"/>
                    <bond atomRefs2="a14 a5" order="1"/>
                    <bond atomRefs2="a4 a5" order="1"/>
                    <bond atomRefs2="a4 a3" order="1"/>
                    <bond atomRefs2="a5 a15" order="1"/>
                    <bond atomRefs2="a5 a6" order="1"/>
                    <bond atomRefs2="a11 a3" order="1"/>
                    <bond atomRefs2="a9 a2" order="1"/>
                    <bond atomRefs2="a8 a7" order="2"/>
                    <bond atomRefs2="a3 a2" order="1"/>
                    <bond atomRefs2="a3 a10" order="1"/>
                    <bond atomRefs2="a7 a6" order="2"/>
                    <bond atomRefs2="a2 a1" order="2"/>
                    <bond atomRefs2="a6 a16" order="1"/>
          </bondArray>
          <propertyList>
                    <property dictRef="me:lumpedSpecies">
                              <array> </array>
                    </property>
                    <property dictRef="me:vibFreqs">
                              <array   units="cm-1">18.369   22.1961   50.126   115.5291   199.8143
256.2064  344.2322  417.7428  442.9197  520.1808  620.7642  731.3914  738.6108  840.0216  946.4525  975.2682
1008.4666  1082.4298  1109.3203  1134.1538  1149.8114  1172.701  1190.6615  1218.2435  1226.307  1270.8741
1279.4873  1285.2176  1297.8937  1313.1434  1327.9191  1446.5108  1787.815  2253.5667  2645.3671  2658.2854
2663.4866  2671.4664  2736.0804  2741.4119  2743.2964  2756.5116</array>
                    </property>
                    <property dictRef="me:ZPE">
                              <scalar units="kJ/mol">145.28162029093735</scalar>
                    </property>
```

```xml
                        <property dictRef="me:spinMultiplicity">
                                    <scalar units="cm-1">1</scalar>
                        </property>
                        <property dictRef="me:epsilon">
                                    <scalar>473.17</scalar>
                        </property>
<group>1</group>
                        <property dictRef="me:sigma">
                                    <scalar>5.09</scalar>
                        </property>
                    </propertyList>
                    <me:energyTransferModel xsi:type="me:ExponentialDown">
                                    <scalar units="cm-1">250</scalar>
                    </me:energyTransferModel>
                </molecule>
                <molecule id="[CH3]" spinMultiplicity="2">
                        <atomArray>
                                    <atom    elementType="C"    id="a1"    spinMultiplicity="2"    x3="1.152573"
y3="0.309211" z3="-0.035298"/>
                                    <atom    elementType="H"    id="a2"    x3="0.653708"    y3="0.741585"
z3="0.792048"/>
                                    <atom    elementType="H"    id="a3"    x3="0.653714"    y3="0.215309"    z3="-
0.964073"/>
                                    <atom    elementType="H"    id="a4"    x3="2.150575"    y3="-0.028505"
z3="0.065907"/>
                        </atomArray>
                        <bondArray>
                                    <bond atomRefs2="a3 a1" order="1"/>
                                    <bond atomRefs2="a1 a4" order="1"/>
                                    <bond atomRefs2="a1 a2" order="1"/>
                        </bondArray>
                        <propertyList>
                                    <property dictRef="me:lumpedSpecies">
                                                <array> </array>
```

```
                </property>
                <property dictRef="me:vibFreqs">
                                <array   units="cm-1">929.1196   1256.3748   1256.3807   2756.1273
2756.1357 2799.8185</array>
                </property>
                <property dictRef="me:ZPE">
                                <scalar units="kJ/mol">0.0</scalar>
                </property>
                <property dictRef="me:spinMultiplicity">
                                <scalar units="cm-1">2</scalar>
                </property>
                <property dictRef="me:epsilon">
                                <scalar>473.17</scalar>
                </property>
<group>1</group>
                <property dictRef="me:sigma">
                                <scalar>5.09</scalar>
                </property>
            </propertyList>
            <me:energyTransferModel xsi:type="me:ExponentialDown">
                                <scalar units="cm-1">250</scalar>
            </me:energyTransferModel>
        </molecule>
        <molecule id="TS_O=CCCC[CH]C(=O)C_CC(=O)C=C" spinMultiplicity="2">
                <atomArray>
                                <atom     elementType="O"     id="a1"     x3="3.636780"     y3="1.402564"
z3="3.840005"/>
                                <atom     elementType="C"     id="a2"     x3="4.744226"     y3="1.007233"
z3="3.557870"/>
                                <atom     elementType="C"     id="a3"     x3="5.198879"     y3="0.684531"
z3="2.149371"/>
                                <atom     elementType="C"     id="a4"     spinMultiplicity="2"     x3="4.229804"
y3="1.026723" z3="1.096978"/>
```

```
                              <atom   elementType="C"   id="a5"   x3="3.863367"   y3="-1.143755"   z3="-
0.192348"/>
                              <atom   elementType="C"   id="a6"   x3="4.022717"   y3="-2.194404"
z3="0.620801"/>
                              <atom   elementType="H"   id="a7"   x3="3.184959"   y3="-2.681554"
z3="1.120503"/>
                              <atom   elementType="C"   id="a8"   x3="5.376114"   y3="-2.764445"
z3="0.873910"/>
                              <atom   elementType="O"   id="a9"   x3="6.383951"   y3="-2.300726"
z3="0.373580"/>
                              <atom   elementType="C"   id="a10"   x3="5.459851"   y3="-3.916965"
z3="1.837512"/>
                              <atom   elementType="H"   id="a11"   x3="5.512993"   y3="0.858095"
z3="4.334746"/>
                              <atom   elementType="H"   id="a12"   x3="6.188240"   y3="1.152126"
z3="1.958994"/>
                              <atom   elementType="H"   id="a13"   x3="5.414059"   y3="-0.412229"
z3="2.090216"/>
                              <atom   elementType="H"   id="a14"   x3="3.192292"   y3="1.138799"
z3="1.348623"/>
                              <atom   elementType="H"   id="a15"   x3="4.552050"   y3="1.316110"
z3="0.119516"/>
                              <atom   elementType="H"   id="a16"   x3="2.905767"   y3="-0.708531"   z3="-
0.424892"/>
                              <atom   elementType="H"   id="a17"   x3="4.708504"   y3="-0.705468"   z3="-
0.722098"/>
                              <atom   elementType="H"   id="a18"   x3="4.501624"   y3="-4.409517"
z3="2.038744"/>
                              <atom   elementType="H"   id="a19"   x3="6.167786"   y3="-4.687818"
z3="1.486790"/>
                              <atom   elementType="H"   id="a20"   x3="5.854100"   y3="-3.566484"
z3="2.805749"/>
                    </atomArray>
                    <bondArray>
                              <bond atomRefs2="a17 a5" order="1"/>
                              <bond atomRefs2="a16 a5" order="1"/>
                              <bond atomRefs2="a5 a6" order="2"/>
                              <bond atomRefs2="a15 a4" order="1"/>
```

```xml
                    <bond atomRefs2="a9 a8" order="2"/>

                    <bond atomRefs2="a6 a8" order="1"/>

                    <bond atomRefs2="a6 a7" order="1"/>

                    <bond atomRefs2="a8 a10" order="1"/>

                    <bond atomRefs2="a4 a14" order="1"/>

                    <bond atomRefs2="a4 a3" order="1"/>

                    <bond atomRefs2="a19 a10" order="1"/>

                    <bond atomRefs2="a10 a18" order="1"/>

                    <bond atomRefs2="a10 a20" order="1"/>

                    <bond atomRefs2="a12 a3" order="1"/>

                    <bond atomRefs2="a13 a3" order="1"/>

                    <bond atomRefs2="a3 a2" order="1"/>

                    <bond atomRefs2="a2 a1" order="2"/>

                    <bond atomRefs2="a2 a11" order="1"/>

            </bondArray>

            <propertyList>

                <property dictRef="me:vibFreqs">

                        <array       units="cm-1">24.19826678829911       36.31564902038544
47.93984317257393    51.37725844495845    86.9257731446819    110.25730211957355    125.43612458563801
191.3237992080306    254.89762786726592    285.9267666419308    349.3384912674128    411.03897141452796
473.2148085734514    579.6927867707722    624.0146516262008    685.1598856992462    695.4449420689649
845.669825777908    861.5624570440068    924.0167964160466    939.8119024594072    994.4553669814131
1007.2897826192585    1038.0761178313726    1058.9437109270198    1071.0679796899105    1072.8199211227056
1108.954887856739    1188.839817020772    1199.8239652378365    1209.1237236389818    1220.7594100087674
1231.10194513538    1270.4406435276917    1294.4690803313285    1314.3919588208985    1323.005979187458
1332.5999757785792    1353.9988106459207    1721.8556057816259    1778.7849233053516    1795.0102140082388
2641.7653257054335    2647.7566816249123    2673.332870917575    2690.0524099645904    2710.197403774069
2726.608765090135    2742.8111954302813    2752.4524678535277    2770.178715317746    2774.6548336966025
2782.9619390404036</array>

                </property>

                <property dictRef="me:imFreq">

                        <scalar units="cm-1">58.77699491430155</scalar>

                </property>

                <property dictRef="me:spinMultiplicity">

                        <scalar units="cm-1">2</scalar>

                </property>
```

```xml
<group>1</group>
                              <property dictRef="me:ZPE">
                                      <scalar units="kJ/mol">114.65776192642889</scalar>
                              </property>
                      </propertyList>
              </molecule>
      <molecule id="CC(=O)C=C">
              <atomArray>
                                      <atom     elementType="C"     id="a1"     x3="1.022201"     y3="0.073670"
z3="0.235911"/>
                                      <atom     elementType="C"     id="a2"     x3="0.317316"     y3="-0.960226"
z3="1.066529"/>
                                      <atom     elementType="O"     id="a3"     x3="-0.746836"    y3="-0.748702"
z3="1.602466"/>
                                      <atom     elementType="C"     id="a4"     x3="1.024585"     y3="-2.268035"
z3="1.223600"/>
                                      <atom     elementType="C"     id="a5"     x3="1.349643"     y3="-3.057877"
z3="0.205775"/>
                                      <atom     elementType="H"     id="a6"     x3="0.481780"     y3="1.031895"
z3="0.249745"/>
                                      <atom     elementType="H"     id="a7"     x3="1.105790"     y3="-0.240575"    z3="-
0.812591"/>
                                      <atom     elementType="H"     id="a8"     x3="2.041190"     y3="0.261141"
z3="0.598712"/>
                                      <atom     elementType="H"     id="a9"     x3="1.217931"     y3="-2.537236"
z3="2.265321"/>
                                      <atom     elementType="H"     id="a10"    x3="1.835367"     y3="-4.017532"
z3="0.337054"/>
                                      <atom     elementType="H"     id="a11"    x3="1.150061"     y3="-2.820991"    z3="-
0.831524"/>
              </atomArray>
              <bondArray>
                      <bond atomRefs2="a11 a5" order="1"/>
                      <bond atomRefs2="a7 a1" order="1"/>
                      <bond atomRefs2="a5 a10" order="1"/>
                      <bond atomRefs2="a5 a4" order="2"/>
```

```xml
                              <bond atomRefs2="a1 a6" order="1"/>

                              <bond atomRefs2="a1 a8" order="1"/>

                              <bond atomRefs2="a1 a2" order="1"/>

                              <bond atomRefs2="a2 a4" order="1"/>

                              <bond atomRefs2="a2 a3" order="2"/>

                              <bond atomRefs2="a4 a9" order="1"/>

                      </bondArray>

                      <propertyList>

                              <property dictRef="me:lumpedSpecies">

                                      <array> </array>

                              </property>

                              <property dictRef="me:vibFreqs">

                                      <array   units="cm-1">32.365   69.3622   258.1539   335.3972   460.506
581.2063 657.3665 867.7598 944.2815 986.9893 1011.8303 1056.1937 1059.0437 1212.4736 1229.0052 1243.7502
1266.9702 1315.7522 1340.833 1793.3761 1805.1199 2690.9739 2692.3251 2714.6675 2746.602 2778.8929
2789.585</array>

                              </property>

                              <property dictRef="me:ZPE">

                                      <scalar units="kJ/mol">115.03694618131432</scalar>

                              </property>

                              <property dictRef="me:spinMultiplicity">

                                      <scalar units="cm-1">1</scalar>

                              </property>

<group>1</group>

                              <property dictRef="me:epsilon">

                                      <scalar>473.17</scalar>

                              </property>

                              <property dictRef="me:sigma">

                                      <scalar>5.09</scalar>

                              </property>

                      </propertyList>

                      <me:energyTransferModel xsi:type="me:ExponentialDown">

                              <scalar units="cm-1">250</scalar>
```

```xml
            </me:energyTransferModel>
        </molecule>
        <molecule id="[CH2]CC=O" spinMultiplicity="2">
            <atomArray>
                <atom elementType="C" id="a1" spinMultiplicity="2" x3="1.149616" y3="0.145833" z3="-0.440571"/>
                <atom elementType="H" id="a2" x3="0.566211" y3="0.995876" z3="-0.161807"/>
                <atom elementType="H" id="a3" x3="0.682271" y3="-0.596495" z3="-1.051968"/>
                <atom elementType="C" id="a4" x3="2.540993" y3="-0.018605" z3="-0.012881"/>
                <atom elementType="C" id="a5" x3="3.074107" y3="1.106728" z3="0.848073"/>
                <atom elementType="O" id="a6" x3="4.207704" y3="1.099549" z3="1.271481"/>
                <atom elementType="H" id="a7" x3="2.668885" y3="-0.985946" z3="0.533764"/>
                <atom elementType="H" id="a8" x3="3.209312" y3="-0.144371" z3="-0.900835"/>
                <atom elementType="H" id="a9" x3="2.374799" y3="1.928891" z3="1.066351"/>
            </atomArray>
            <bondArray>
                <bond atomRefs2="a3 a1" order="1"/>
                <bond atomRefs2="a8 a4" order="1"/>
                <bond atomRefs2="a1 a2" order="1"/>
                <bond atomRefs2="a1 a4" order="1"/>
                <bond atomRefs2="a4 a7" order="1"/>
                <bond atomRefs2="a4 a5" order="1"/>
                <bond atomRefs2="a5 a9" order="1"/>
                <bond atomRefs2="a5 a6" order="2"/>
            </bondArray>
            <propertyList>
                <property dictRef="me:lumpedSpecies">
                    <array> </array>
```

```xml
            </property>
            <property dictRef="me:vibFreqs">
                        <array units="cm-1">36.0546  105.9062  322.4084  523.8085  674.3924
817.9363  911.6714  975.7754  1115.9277  1128.3744  1170.679  1198.9244  1275.1532  1323.8802  1361.0519
1789.9254 2638.8589 2645.5513 2723.1514 2744.015 2785.1219</array>
            </property>
            <property dictRef="me:ZPE">
                        <scalar units="kJ/mol">0.0</scalar>
            </property>
            <property dictRef="me:spinMultiplicity">
                        <scalar units="cm-1">2</scalar>
            </property>
            <property dictRef="me:epsilon">
                        <scalar>473.17</scalar>
            </property>
<group>1</group>
            <property dictRef="me:sigma">
                        <scalar>5.09</scalar>
            </property>
        </propertyList>
        <me:energyTransferModel xsi:type="me:ExponentialDown">
                    <scalar units="cm-1">250</scalar>
        </me:energyTransferModel>
    </molecule>
    <molecule id="O=CCCC[CH](C(=O)C)O[O]" spinMultiplicity="2">
            <atomArray>
                                    <atom      elementType="O"      id="a1"      x3="0.753247"      y3="1.955068"
z3="2.535040"/>
                                    <atom      elementType="C"      id="a2"      x3="0.124995"      y3="2.290910"
z3="1.555966"/>
                                    <atom      elementType="C"      id="a3"      x3="0.593418"      y3="2.051628"
z3="0.139727"/>
                                    <atom      elementType="C"      id="a4"      x3="1.753020"      y3="1.056364"
z3="0.051924"/>
```

```xml
                            <atom    elementType="C"    id="a5"    x3="1.321971"    y3="-0.390548"
z3="0.325801"/>
                            <atom    elementType="C"    id="a6"    x3="0.339888"    y3="-0.979830"    z3="-
0.669209"/>
                            <atom    elementType="H"    id="a7"    x3="0.304543"    y3="-2.082522"    z3="-
0.613154"/>
                            <atom    elementType="C"    id="a8"    x3="0.483590"    y3="-0.471359"    z3="-
2.091761"/>
                            <atom    elementType="O"    id="a9"    x3="-0.232140"    y3="0.398789"    z3="-
2.530536"/>
                            <atom    elementType="C"    id="a10"    x3="1.556708"    y3="-1.134754"    z3="-
2.900936"/>
                            <atom    elementType="H"    id="a11"    x3="-0.842932"    y3="2.811653"
z3="1.644554"/>
                            <atom    elementType="H"    id="a12"    x3="-0.258175"    y3="1.703263"    z3="-
0.491319"/>
                            <atom    elementType="H"    id="a13"    x3="0.892586"    y3="3.023225"    z3="-
0.306987"/>
                            <atom    elementType="H"    id="a14"    x3="2.239641"    y3="1.138552"    z3="-
0.938230"/>
                            <atom    elementType="H"    id="a15"    x3="2.526399"    y3="1.338698"
z3="0.801267"/>
                            <atom    elementType="H"    id="a16"    x3="2.221216"    y3="-1.035093"
z3="0.391818"/>
                            <atom    elementType="H"    id="a17"    x3="0.849272"    y3="-0.439469"
z3="1.340200"/>
                            <atom    elementType="H"    id="a18"    x3="1.261570"    y3="-2.149902"    z3="-
3.202777"/>
                            <atom    elementType="H"    id="a19"    x3="1.757180"    y3="-0.575844"    z3="-
3.828843"/>
                            <atom    elementType="H"    id="a20"    x3="2.504453"    y3="-1.220553"    z3="-
2.353656"/>
                            <atom elementType="O"  id="a21"  spinMultiplicity="2"  x3="-1.421786"  y3="-
0.719539" z3="0.818503"/>
                            <atom    elementType="O"    id="a22"    x3="-1.106042"    y3="-0.532481"    z3="-
0.300264"/>
                </atomArray>
                <bondArray>
                            <bond atomRefs2="a19 a10" order="1"/>
```

```xml
                    <bond atomRefs2="a18 a10" order="1"/>

                    <bond atomRefs2="a10 a20" order="1"/>

                    <bond atomRefs2="a10 a8" order="1"/>

                    <bond atomRefs2="a9 a8" order="2"/>

                    <bond atomRefs2="a8 a6" order="1"/>

                    <bond atomRefs2="a14 a4" order="1"/>

                    <bond atomRefs2="a6 a7" order="1"/>

                    <bond atomRefs2="a6 a22" order="1"/>

                    <bond atomRefs2="a6 a5" order="1"/>

                    <bond atomRefs2="a12 a3" order="1"/>

                    <bond atomRefs2="a13 a3" order="1"/>

                    <bond atomRefs2="a22 a21" order="1"/>

                    <bond atomRefs2="a4 a3" order="1"/>

                    <bond atomRefs2="a4 a5" order="1"/>

                    <bond atomRefs2="a4 a15" order="1"/>

                    <bond atomRefs2="a3 a2" order="1"/>

                    <bond atomRefs2="a5 a16" order="1"/>

                    <bond atomRefs2="a5 a17" order="1"/>

                    <bond atomRefs2="a2 a11" order="1"/>

                    <bond atomRefs2="a2 a1" order="2"/>

        </bondArray>

        <propertyList>

                    <property dictRef="me:lumpedSpecies">

                            <array> </array>

                    </property>

                    <property dictRef="me:vibFreqs">

                            <array units="cm-1">30.4913 43.6484 48.0859 54.8101 74.046 77.7232
127.1747  143.253  160.3526  251.2789  273.3556  334.6881  351.9927  445.2601  522.6213  557.0752  615.4268
653.5872  744.4523  805.7959  867.8319  895.8072  951.0298  972.6208  1007.1981  1013.8963  1023.9042  1077.3591
1103.5636  1120.2926  1138.4735  1149.5752  1175.2975  1185.4437  1210.0533  1214.7504  1223.3583  1235.3751
1237.739  1258.0415  1274.1136  1281.5071  1292.7581  1314.6968  1316.9228  1325.7975  1338.828  1789.2695
1804.4377  2640.3951  2643.1614  2646.7465  2666.8178  2683.8095  2690.5528  2702.1001  2727.2021  2731.0479
2735.3294 2787.2142</array>

                    </property>
```

```xml
<group>2</group>
                        <property dictRef="me:ZPE">
                                <scalar units="kJ/mol">-477.2069</scalar>
                        </property>
                        <property dictRef="me:spinMultiplicity">
                                <scalar units="cm-1">2</scalar>
                        </property>
                        <property dictRef="me:epsilon">
                                <scalar>473.17</scalar>
                        </property>
                        <property dictRef="me:sigma">
                                <scalar>5.09</scalar>
                        </property>
                </propertyList>
                <me:energyTransferModel xsi:type="me:ExponentialDown">
                        <scalar units="cm-1">250</scalar>
                </me:energyTransferModel>
        </molecule>
        <molecule id="O=O">
                <atomArray>
                        <atom    elementType="O"    id="a1"    x3="1.093843"    y3="0.048467"
z3="0.067410"/>
                        <atom    elementType="O"    id="a2"    x3="2.199035"    y3="0.048467"
z3="0.067410"/>
                </atomArray>
                <bondArray>
                        <bond atomRefs2="a1 a2" order="2"/>
                </bondArray>
                <propertyList>
                        <property dictRef="me:lumpedSpecies">
                                <array> </array>
                        </property>
                        <property dictRef="me:vibFreqs">
```

```xml
                    <array units="cm-1">1602.4124</array>
            </property>
            <property dictRef="me:ZPE">
                    <scalar units="kJ/mol">0.0</scalar>
            </property>
            <property dictRef="me:spinMultiplicity">
                    <scalar units="cm-1">1</scalar>
            </property>
            <property dictRef="me:epsilon">
                    <scalar>473.17</scalar>
            </property>
            <property dictRef="me:sigma">
                    <scalar>5.09</scalar>
            </property>
        </propertyList>
        <me:energyTransferModel xsi:type="me:ExponentialDown">
                <scalar units="cm-1">250</scalar>
        </me:energyTransferModel>
    </molecule>
    <molecule id="TS_O=CCCC[CH](C(=O)C)O[O]_O=[C]CCC[CH](C(=O)C)OO" spinMultiplicity="2">
        <atomArray>
                <atom    elementType="O"    id="a1"    x3="-1.155237"    y3="1.814529"
z3="3.272074"/>
                <atom    elementType="C"    id="a2"    spinMultiplicity="2"    x3="-1.041595"
y3="1.240109" z3="2.239362"/>
                <atom    elementType="C"    id="a3"    x3="-0.608009"    y3="1.692488"
z3="0.888322"/>
                <atom    elementType="C"    id="a4"    x3="0.793670"    y3="1.179334"
z3="0.538369"/>
                <atom    elementType="C"    id="a5"    x3="0.941661"    y3="-0.346334"
z3="0.607385"/>
                <atom    elementType="C"    id="a6"    x3="-0.072746"    y3="-1.147333"    z3="-
0.204559"/>
                <atom    elementType="H"    id="a7"    x3="0.188852"    y3="-2.226874"    z3="-
0.202227"/>
```

```xml
                    <atom   elementType="C"   id="a8"   x3="-0.268698"   y3="-0.603530"   z3="-1.623639"/>
                    <atom   elementType="O"   id="a9"   x3="-1.058515"   y3="0.273379"   z3="-1.877283"/>
                    <atom   elementType="C"   id="a10"   x3="0.601173"   y3="-1.244790"   z3="-2.662608"/>
                    <atom   elementType="H"   id="a11"   x3="-1.353722"   y3="-0.048212"   z3="2.206570"/>
                    <atom   elementType="H"   id="a12"   x3="-1.337959"   y3="1.316415"   z3="0.119482"/>
                    <atom   elementType="H"   id="a13"   x3="-0.641881"   y3="2.799968"   z3="0.819793"/>
                    <atom   elementType="H"   id="a14"   x3="1.046993"   y3="1.538037"   z3="-0.481612"/>
                    <atom   elementType="H"   id="a15"   x3="1.540750"   y3="1.643700"   z3="1.215103"/>
                    <atom   elementType="H"   id="a16"   x3="1.967501"   y3="-0.626613"   z3="0.293686"/>
                    <atom   elementType="H"   id="a17"   x3="0.853588"   y3="-0.682960"   z3="1.667341"/>
                    <atom   elementType="H"   id="a18"   x3="0.262682"   y3="-2.266582"   z3="-2.890389"/>
                    <atom   elementType="H"   id="a19"   x3="0.565063"   y3="-0.684646"   z3="-3.610808"/>
                    <atom   elementType="H"   id="a20"   x3="1.652412"   y3="-1.309161"   z3="-2.356580"/>
                    <atom   elementType="O"   id="a21"   x3="-1.430778"   y3="-1.239385"   z3="1.667552"/>
                    <atom   elementType="O"   id="a22"   x3="-1.445205"   y3="-1.071538"   z3="0.374666"/>
            </atomArray>
            <bondArray>
                    <bond atomRefs2="a19 a10" order="1"/>
                    <bond atomRefs2="a18 a10" order="1"/>
                    <bond atomRefs2="a10 a20" order="1"/>
                    <bond atomRefs2="a10 a8" order="1"/>
                    <bond atomRefs2="a9 a8" order="2"/>
                    <bond atomRefs2="a8 a6" order="1"/>
```

```xml
            <bond atomRefs2="a14 a4" order="1"/>

            <bond atomRefs2="a6 a7" order="1"/>

            <bond atomRefs2="a6 a22" order="1"/>

            <bond atomRefs2="a6 a5" order="1"/>

            <bond atomRefs2="a12 a3" order="1"/>

            <bond atomRefs2="a16 a5" order="1"/>

            <bond atomRefs2="a22 a21" order="1"/>

            <bond atomRefs2="a4 a5" order="1"/>

            <bond atomRefs2="a4 a3" order="1"/>

            <bond atomRefs2="a4 a15" order="1"/>

            <bond atomRefs2="a5 a17" order="1"/>

            <bond atomRefs2="a13 a3" order="1"/>

            <bond atomRefs2="a3 a2" order="1"/>

            <bond atomRefs2="a21 a11" order="1"/>

            <bond atomRefs2="a2 a1" order="2"/>

        </bondArray>

        <propertyList>

            <property dictRef="me:vibFreqs">

                        <array    units="cm-1">34.70226330671719      46.37501214147475
84.69448951592042    98.03506231980222    152.08808871107098    173.5934223095347    185.49959840493935
227.27234066558074    292.7491714257415    315.8875016000293    349.9461294469529    372.0402061359201
414.25400028541287    469.44448925998483    541.5287419307543    553.8971302816622    569.3149588721868
655.9552752883696    799.3189456865811    807.0589722288864    879.5579902288439    887.0673272981502
965.4662315132944    979.8539961429576    1003.2644807583397    1013.582653450335    1032.7849020299689
1048.7731480576526    1087.3726713122246    1115.0754452470337    1135.3498594777584    1163.3154980535382
1173.7186374579692    1176.8741463920887    1210.5835952159246    1218.0741850618795    1223.0238504787837
1228.9504173175728    1250.581619774325    1259.9591040277564    1267.4274168254324    1274.6981156870067
1293.0243053805325    1314.1963389901507    1317.191783227851    1334.3665481747573    1341.7786765976784
1810.0152483320253    1905.1174869450278    2619.3621835237554    2650.860458303254    2667.588491746626
2680.894386895566    2686.019017461484    2691.313619677388    2716.4416107978723    2731.211349435521
2734.664183430492 2786.9020593080045</array>

            </property>

            <property dictRef="me:imFreq">

                        <scalar units="cm-1">2655.3506647725117</scalar>

            </property>

    <group>2</group>

            <property dictRef="me:spinMultiplicity">
```

```xml
                    <scalar units="cm-1">2</scalar>
                </property>
                <property dictRef="me:ZPE">
                    <scalar units="kJ/mol">-467.603</scalar>
                </property>
            </propertyList>
        </molecule>
        <molecule id="O=[C]CCC[CH](C(=O)C)OO" spinMultiplicity="2">
            <atomArray>
                <atom elementType="O" id="a1" x3="-1.718862" y3="2.646161" z3="2.310165"/>
                <atom elementType="C" id="a2" spinMultiplicity="2" x3="-0.903960" y3="2.439519" z3="1.498144"/>
                <atom elementType="C" id="a3" x3="-0.564124" y3="2.670382" z3="0.097725"/>
                <atom elementType="C" id="a4" x3="0.730204" y3="1.954810" z3="-0.309581"/>
                <atom elementType="C" id="a5" x3="0.655804" y3="0.433103" z3="-0.132135"/>
                <atom elementType="C" id="a6" x3="-0.469551" y3="-0.264474" z3="-0.903415"/>
                <atom elementType="H" id="a7" x3="-0.334054" y3="-1.370706" z3="-0.872894"/>
                <atom elementType="C" id="a8" x3="-0.612044" y3="0.252243" z3="-2.347202"/>
                <atom elementType="O" id="a9" x3="-1.278276" y3="1.220387" z3="-2.618648"/>
                <atom elementType="C" id="a10" x3="0.147028" y3="-0.538562" z3="-3.368288"/>
                <atom elementType="H" id="a11" x3="-2.079050" y3="-0.311554" z3="1.490673"/>
                <atom elementType="H" id="a12" x3="-1.407361" y3="2.311564" z3="-0.556002"/>
                <atom elementType="H" id="a13" x3="-0.483910" y3="3.760823" z3="-0.105978"/>
                <atom elementType="H" id="a14" x3="0.961145" y3="2.208760" z3="-1.363741"/>
```

```xml
                        <atom     elementType="H"      id="a15"     x3="1.576590"     y3="2.344856"
z3="0.292116"/>
                        <atom    elementType="H"    id="a16"    x3="1.628505"    y3="-0.016675"    z3="-
0.412405"/>
                        <atom     elementType="H"     id="a17"     x3="0.516065"     y3="0.199930"
z3="0.947429"/>
                        <atom    elementType="H"    id="a18"    x3="-0.325803"    y3="-1.516215"    z3="-
3.546500"/>
                        <atom    elementType="H"    id="a19"    x3="0.165860"    y3="-0.018207"    z3="-
4.339440"/>
                        <atom    elementType="H"    id="a20"    x3="1.186865"    y3="-0.725941"    z3="-
3.073452"/>
                        <atom     elementType="O"     id="a21"     x3="-1.961162"     y3="-0.931384"
z3="0.706788"/>
                        <atom    elementType="O"    id="a22"    x3="-1.771362"    y3="0.032704"    z3="-
0.321731"/>
            </atomArray>
            <bondArray>
                        <bond atomRefs2="a19 a10" order="1"/>
                        <bond atomRefs2="a18 a10" order="1"/>
                        <bond atomRefs2="a10 a20" order="1"/>
                        <bond atomRefs2="a10 a8" order="1"/>
                        <bond atomRefs2="a9 a8" order="2"/>
                        <bond atomRefs2="a8 a6" order="1"/>
                        <bond atomRefs2="a14 a4" order="1"/>
                        <bond atomRefs2="a6 a7" order="1"/>
                        <bond atomRefs2="a6 a22" order="1"/>
                        <bond atomRefs2="a6 a5" order="1"/>
                        <bond atomRefs2="a12 a3" order="1"/>
                        <bond atomRefs2="a16 a5" order="1"/>
                        <bond atomRefs2="a22 a21" order="1"/>
                        <bond atomRefs2="a4 a5" order="1"/>
                        <bond atomRefs2="a4 a3" order="1"/>
                        <bond atomRefs2="a4 a15" order="1"/>
                        <bond atomRefs2="a5 a17" order="1"/>
```

```xml
                    <bond atomRefs2="a13 a3" order="1"/>

                    <bond atomRefs2="a3 a2" order="1"/>

                    <bond atomRefs2="a21 a11" order="1"/>

                    <bond atomRefs2="a2 a1" order="2"/>

            </bondArray>

            <propertyList>

                    <property dictRef="me:lumpedSpecies">

                            <array> </array>

                    </property>

                    <property dictRef="me:vibFreqs">

                            <array  units="cm-1">20.8063   31.7609   41.9773   47.0083   83.6901
89.8743 135.141 160.7815 169.1636 209.2794 238.5924 285.8088 312.9025 360.5493 410.1542 483.3314 515.4585
572.2681 652.2696 775.8425 869.5628 900.5139 915.4118 973.6399 996.4127 1011.3799 1016.057 1042.9407
1071.6191 1086.1089 1150.298 1154.5623 1175.1069 1197.5525 1210.3915 1216.189 1220.3812 1223.6068
1231.1112 1258.5588 1271.1906 1278.8894 1291.4384 1312.1359 1315.5512 1331.2168 1393.0055 1813.9551
2009.2692 2620.4048 2643.5273 2660.0591 2672.4147 2674.6935 2682.639 2691.2325 2716.0096 2736.3251
2738.0237 2786.6948</array>

                    </property>

<group>2</group>

                    <property dictRef="me:ZPE">

                            <scalar units="kJ/mol">-486.842</scalar>

                    </property>

                    <property dictRef="me:spinMultiplicity">

                            <scalar units="cm-1">2</scalar>

                    </property>

                    <property dictRef="me:epsilon">

                            <scalar>473.17</scalar>

                    </property>

                    <property dictRef="me:sigma">

                            <scalar>5.09</scalar>

                    </property>

            </propertyList>

            <me:energyTransferModel xsi:type="me:ExponentialDown">
```

```
                    <scalar units="cm-1">250</scalar>

            </me:energyTransferModel>

        </molecule>

        <molecule id="TS_O=CCCC[CH](C(=O)C)OO_[CH2]CC[CH](C(=O)C)OO" spinMultiplicity="2">

            <atomArray>

                    <atom    elementType="O"    id="a1"    x3="1.155529"    y3="1.843327"
z3="2.968748"/>

                    <atom    elementType="C"    id="a2"    x3="0.983026"    y3="1.527173"
z3="1.837875"/>

                    <atom    elementType="C"    id="a3"    x3="0.344743"    y3="2.242358"
z3="0.692521"/>

                    <atom    elementType="C"    id="a4"    x3="-1.052022"    y3="1.683085"
z3="0.390482"/>

                    <atom    elementType="C"    id="a5"    x3="-1.096277"    y3="0.163584"
z3="0.198252"/>

                    <atom   elementType="C"   id="a6"   x3="-0.192303"   y3="-0.353005"   z3="-
0.914999"/>

                    <atom   elementType="H"   id="a7"   x3="-0.223465"   y3="0.283034"   z3="-
1.830279"/>

                    <atom   elementType="C"   id="a8"   x3="-0.553404"   y3="-1.805227"   z3="-
1.277536"/>

                    <atom   elementType="O"   id="a9"   x3="-1.670178"   y3="-2.057333"   z3="-
1.668999"/>

                    <atom   elementType="C"   id="a10"   x3="0.524025"   y3="-2.823879"   z3="-
1.131200"/>

                    <atom    elementType="H"    id="a11"    x3="1.411442"    y3="0.348282"
z3="1.444727"/>

                    <atom    elementType="H"    id="a12"    x3="0.283609"    y3="3.330916"
z3="0.904817"/>

                    <atom   elementType="H"   id="a13"   x3="0.995302"   y3="2.134538"   z3="-
0.207183"/>

                    <atom    elementType="H"    id="a14"    x3="-1.746292"    y3="1.959608"
z3="1.212279"/>

                    <atom   elementType="H"   id="a15"   x3="-1.446512"   y3="2.188658"   z3="-
0.515462"/>

                    <atom    elementType="H"    id="a16"    x3="-0.826500"    y3="-0.353954"
z3="1.144897"/>

                    <atom   elementType="H"   id="a17"   x3="-2.145737"   y3="-0.144046"   z3="-
0.017801"/>
```

```
                    <atom    elementType="H"    id="a18"    x3="1.362469"    y3="-2.628601"    z3="-
1.816980"/>
                    <atom    elementType="H"    id="a19"    x3="0.962670"    y3="-2.801309"    z3="-
0.117287"/>
                    <atom    elementType="H"    id="a20"    x3="0.156010"    y3="-3.840752"    z3="-
1.325584"/>
                    <atom   elementType="O"   id="a21"   spinMultiplicity="2"   x3="1.533248"   y3="-
0.669327" z3="0.613427"/>
                    <atom    elementType="O"    id="a22"    x3="1.240618"    y3="-0.227129"    z3="-
0.584712"/>
            </atomArray>
            <bondArray>
                    <bond atomRefs2="a7 a6" order="1"/>
                    <bond atomRefs2="a18 a10" order="1"/>
                    <bond atomRefs2="a9 a8" order="2"/>
                    <bond atomRefs2="a20 a10" order="1"/>
                    <bond atomRefs2="a8 a10" order="1"/>
                    <bond atomRefs2="a8 a6" order="1"/>
                    <bond atomRefs2="a10 a19" order="1"/>
                    <bond atomRefs2="a6 a22" order="1"/>
                    <bond atomRefs2="a6 a5" order="1"/>
                    <bond atomRefs2="a22 a21" order="1"/>
                    <bond atomRefs2="a15 a4" order="1"/>
                    <bond atomRefs2="a13 a3" order="1"/>
                    <bond atomRefs2="a17 a5" order="1"/>
                    <bond atomRefs2="a5 a4" order="1"/>
                    <bond atomRefs2="a5 a16" order="1"/>
                    <bond atomRefs2="a4 a3" order="1"/>
                    <bond atomRefs2="a4 a14" order="1"/>
                    <bond atomRefs2="a3 a12" order="1"/>
                    <bond atomRefs2="a3 a2" order="1"/>
                    <bond atomRefs2="a11 a2" order="1"/>
                    <bond atomRefs2="a2 a1" order="2"/>
            </bondArray>
```

```xml
<propertyList>

    <property dictRef="me:vibFreqs">

                    <array      units="cm-1">25.13418572704809      43.78430644082872
87.50465467517697   105.38268034037192   142.0134080348396   191.06755994091435   210.0550632621102
225.07548192733668   276.09825780073396   314.0911375714748   343.7502648476827   408.67774100507665
412.20278329614746   454.5374068045882   516.4656713432407   551.1989536125178   572.122846251109
661.0329314469786   799.3401921327942   804.668724589122   852.644035477585   916.4671687195728
969.4454557669317   985.8195241279369   997.913563409009   1008.0228357319672   1021.888295720323
1046.441250494202   1088.6472282539326   1102.2690933577267   1130.6607218822778   1159.6970381047943
1178.7013687001133   1182.8439145687294   1194.5798669625397   1215.1411357908855   1220.914699512774
1245.8058783085253   1252.7953832773032   1257.9966610971915   1261.9824355495286   1263.4848628508646
1276.0106163401458   1309.3370720112725   1315.4708305598194   1333.8091930559588   1343.4769751508093
1801.0275964152236   1905.4602425864648   2647.677688615599   2653.6885928458764   2667.4095210809096
2668.198873367452   2686.1682556219366   2692.6035421000115   2726.4193201889566   2727.3053044328435
2735.7866279170457   2781.437478427309</array>

    </property>

    <property dictRef="me:imFreq">

                    <scalar units="cm-1">2608.6599232756794</scalar>

    </property>

<group>2</group>

    <property dictRef="me:spinMultiplicity">

                    <scalar units="cm-1">2</scalar>

    </property>

    <property dictRef="me:ZPE">

                    <scalar units="kJ/mol">-418.4450961</scalar>

    </property>

</propertyList>

</molecule>

<molecule id="[CH2]CC[CH](C(=O)C)OO" spinMultiplicity="2">

    <atomArray>

                    <atom elementType="C" id="a1" spinMultiplicity="2" x3="1.950510" y3="-0.674457" z3="0.547544"/>

                    <atom elementType="H" id="a2" x3="0.919030" y3="-0.429279" z3="0.683669"/>

                    <atom elementType="H" id="a3" x3="2.295348" y3="-1.607806" z3="0.935820"/>

                    <atom elementType="C" id="a4" x3="2.870616" y3="0.222298" z3="-0.167894"/>
```

```xml
                        <atom    elementType="C"    id="a5"    x3="2.255460"    y3="1.573815"    z3="-
0.552112"/>
                        <atom    elementType="C"    id="a6"    x3="1.924388"    y3="2.440754"
z3="0.664819"/>
                        <atom    elementType="H"    id="a7"    x3="1.725992"    y3="1.844168"
z3="1.584839"/>
                        <atom    elementType="C"    id="a8"    x3="2.953952"    y3="3.535018"
z3="0.976819"/>
                        <atom    elementType="O"    id="a9"    x3="2.610548"    y3="4.691045"
z3="1.088346"/>
                        <atom    elementType="C"    id="a10"    x3="4.370001"    y3="3.091755"
z3="1.178952"/>
                        <atom    elementType="O"    id="a11"    x3="0.677787"    y3="3.068585"
z3="0.243223"/>
                        <atom    elementType="O"    id="a12"    x3="0.121483"    y3="3.733109"
z3="1.355249"/>
                        <atom    elementType="H"    id="a13"    x3="3.793261"    y3="0.377980"
z3="0.434517"/>
                        <atom    elementType="H"    id="a14"    x3="3.220994"    y3="-0.290936"    z3="-
1.096038"/>
                        <atom    elementType="H"    id="a15"    x3="2.919373"    y3="2.115534"    z3="-
1.252050"/>
                        <atom    elementType="H"    id="a16"    x3="1.310247"    y3="1.408106"    z3="-
1.123410"/>
                        <atom    elementType="H"    id="a17"    x3="4.450520"    y3="2.289583"
z3="1.924211"/>
                        <atom    elementType="H"    id="a18"    x3="4.817172"    y3="2.720257"
z3="0.246749"/>
                        <atom    elementType="H"    id="a19"    x3="5.000325"    y3="3.926349"
z3="1.526219"/>
                        <atom    elementType="H"    id="a20"    x3="0.530822"    y3="4.658029"
z3="1.245181"/>
                </atomArray>
                <bondArray>
                        <bond atomRefs2="a15 a5" order="1"/>
                        <bond atomRefs2="a16 a5" order="1"/>
                        <bond atomRefs2="a14 a4" order="1"/>
                        <bond atomRefs2="a5 a4" order="1"/>
```

```xml
                    <bond atomRefs2="a5 a6" order="1"/>
                    <bond atomRefs2="a4 a13" order="1"/>
                    <bond atomRefs2="a4 a1" order="1"/>
                    <bond atomRefs2="a11 a6" order="1"/>
                    <bond atomRefs2="a11 a12" order="1"/>
                    <bond atomRefs2="a18 a10" order="1"/>
                    <bond atomRefs2="a1 a2" order="1"/>
                    <bond atomRefs2="a1 a3" order="1"/>
                    <bond atomRefs2="a6 a8" order="1"/>
                    <bond atomRefs2="a6 a7" order="1"/>
                    <bond atomRefs2="a8 a9" order="2"/>
                    <bond atomRefs2="a8 a10" order="1"/>
                    <bond atomRefs2="a10 a19" order="1"/>
                    <bond atomRefs2="a10 a17" order="1"/>
                    <bond atomRefs2="a20 a12" order="1"/>
            </bondArray>
            <propertyList>
                    <property dictRef="me:lumpedSpecies">
                            <array> </array>
                    </property>
                    <property dictRef="me:vibFreqs">
                            <array units="cm-1">41.8089  42.6383  77.1048  87.1645  123.7085
137.0871 182.9843 201.5935 276.3873 296.0194 385.92 440.9543 455.2806 476.8388 560.7724 620.1081 745.7295
818.1732 903.9345 908.545 935.8924 980.6297 1001.1639 1019.5839 1065.6029 1078.4061 1138.9352 1160.6362
1171.8998 1189.3988 1211.2466 1217.6392 1226.0833 1233.4969 1246.4209 1263.5802 1287.8377 1310.4524
1316.3122 1321.8994 1354.4061 1398.4677 1793.4825 2603.4492 2647.3598 2656.7778 2671.4678 2685.7488
2689.8255 2722.4228 2735.3725 2744.6126 2784.7864 2786.8878</array>
                    </property>
                    <property dictRef="me:ZPE">
                            <scalar units="kJ/mol">-471.8079</scalar>
                    </property>
                    <property dictRef="me:spinMultiplicity">
                            <scalar units="cm-1">2</scalar>
                    </property>
```

```xml
<group>2</group>
                                <property dictRef="me:epsilon">
                                        <scalar>473.17</scalar>
                                </property>
                                <property dictRef="me:sigma">
                                        <scalar>5.09</scalar>
                                </property>
                        </propertyList>
                        <me:energyTransferModel xsi:type="me:ExponentialDown">
                                <scalar units="cm-1">250</scalar>
                        </me:energyTransferModel>
                </molecule>
                <molecule id="[C]=O" spinMultiplicity="3">
                        <atomArray>
                                <atom     elementType="C"     id="a1"     spinMultiplicity="3"     x3="1.147696"
y3="0.016687" z3="0.054660"/>
                                <atom     elementType="O"     id="a2"     x3="2.287324"     y3="0.016687"
z3="0.054660"/>
                        </atomArray>
                        <bondArray>
                                <bond atomRefs2="a1 a2" order="2"/>
                        </bondArray>
                        <propertyList>
                                <property dictRef="me:lumpedSpecies">
                                        <array> </array>
                                </property>
                                <property dictRef="me:vibFreqs">
                                        <array units="cm-1">2052.1555</array>
                                </property>
                                <property dictRef="me:ZPE">
                                        <scalar units="kJ/mol">0.0</scalar>
                                </property>
                                <property dictRef="me:spinMultiplicity">
```

```xml
                              <scalar units="cm-1">1</scalar>
                      </property>
                      <property dictRef="me:epsilon">
                              <scalar>473.17</scalar>
                      </property>
                      <property dictRef="me:sigma">
                              <scalar>5.09</scalar>
                      </property>
              </propertyList>
              <me:energyTransferModel xsi:type="me:ExponentialDown">
                      <scalar units="cm-1">250</scalar>
              </me:energyTransferModel>
      </molecule>
      <molecule id="TS_O=CCCC[CH](C(=O)C)O[O]_O=CC[CH]C[CH](C(=O)C)OO" spinMultiplicity="2">
              <atomArray>
                              <atom    elementType="O"    id="a1"    x3="-1.336206"    y3="0.635008"
z3="3.371443"/>
                              <atom    elementType="C"    id="a2"    x3="-1.212631"    y3="1.015844"
z3="2.231264"/>
                              <atom    elementType="C"    id="a3"    x3="-0.054141"    y3="1.901857"
z3="1.809849"/>
                              <atom    elementType="C"    id="a4"    spinMultiplicity="2"    x3="0.416041"
y3="1.651619" z3="0.407032"/>
                              <atom    elementType="C"    id="a5"    x3="1.063715"    y3="0.326468"
z3="0.103481"/>
                              <atom    elementType="C"    id="a6"    x3="0.041132"    y3="-0.732092"    z3="-
0.351496"/>
                              <atom    elementType="H"    id="a7"    x3="-0.604211"    y3="-1.065479"
z3="0.488782"/>
                              <atom    elementType="C"    id="a8"    x3="0.701069"    y3="-1.942045"    z3="-
1.028483"/>
                              <atom    elementType="O"    id="a9"    x3="1.689995"    y3="-2.441184"    z3="-
0.548207"/>
                              <atom    elementType="C"    id="a10"    x3="0.026968"    y3="-2.438807"    z3="-
2.263281"/>
```

<atom elementType="H" id="a11" x3="-1.931271" y3="0.756869" z3="1.433139"/>

<atom elementType="H" id="a12" x3="-0.364842" y3="2.961845" z3="1.945006"/>

<atom elementType="H" id="a13" x3="0.790379" y3="1.774616" z3="2.527132"/>

<atom elementType="H" id="a14" x3="-0.709464" y3="1.694938" z3="-0.395686"/>

<atom elementType="H" id="a15" x3="0.891571" y3="2.512794" z3="-0.073881"/>

<atom elementType="H" id="a16" x3="1.837658" y3="0.451996" z3="-0.683298"/>

<atom elementType="H" id="a17" x3="1.609929" y3="-0.067458" z3="0.989951"/>

<atom elementType="H" id="a18" x3="-0.162230" y3="-1.627032" z3="-2.985799"/>

<atom elementType="H" id="a19" x3="0.623824" y3="-3.210845" z3="-2.771744"/>

<atom elementType="H" id="a20" x3="-0.956512" y3="-2.879408" z3="-2.036655"/>

<atom elementType="O" id="a21" x3="-1.555083" y3="0.826616" z3="-0.786433"/>

<atom elementType="O" id="a22" x3="-0.805689" y3="-0.106120" z3="-1.382116"/>

</atomArray>
<bondArray>
<bond atomRefs2="a18 a10" order="1"/>
<bond atomRefs2="a19 a10" order="1"/>
<bond atomRefs2="a10 a20" order="1"/>
<bond atomRefs2="a10 a8" order="1"/>
<bond atomRefs2="a22 a21" order="1"/>
<bond atomRefs2="a22 a6" order="1"/>
<bond atomRefs2="a8 a9" order="2"/>
<bond atomRefs2="a8 a6" order="1"/>
<bond atomRefs2="a21 a14" order="1"/>
<bond atomRefs2="a16 a5" order="1"/>
<bond atomRefs2="a6 a5" order="1"/>

```xml
            <bond atomRefs2="a6 a7" order="1"/>

            <bond atomRefs2="a15 a4" order="1"/>

            <bond atomRefs2="a5 a4" order="1"/>

            <bond atomRefs2="a5 a17" order="1"/>

            <bond atomRefs2="a4 a3" order="1"/>

            <bond atomRefs2="a11 a2" order="1"/>

            <bond atomRefs2="a3 a12" order="1"/>

            <bond atomRefs2="a3 a2" order="1"/>

            <bond atomRefs2="a3 a13" order="1"/>

            <bond atomRefs2="a2 a1" order="2"/>

        </bondArray>

        <propertyList>

            <property dictRef="me:vibFreqs">

                                    <array    units="cm-1">18.932020424276114    30.966548029154715
63.3672768841891    80.56572203226347    84.78239027882313    119.6466724557484    160.13105058300755
222.58143377635142    255.37068092528742    324.1213167126131    385.11825374332665    399.1083061982651
407.8336275916475    452.42305075281763    487.39518096519265    529.1551132187521    604.0056944935825
614.2664571537198    776.2403090553726    853.542866226261    872.2518762079669    921.6182809355927
982.172379824788    991.6647357545165    1005.683979703196    1024.180704165242    1033.2577127476645
1056.9708930012791    1078.4591260669558    1101.2970042420177    1134.2424544716973    1145.8906072256464
1175.824269176126    1178.6107468286295    1195.2132825003323    1203.024793093757    1207.790901561408
1221.256368003213    1223.4491077842606    1251.3140985076132    1260.6520439731892    1265.556658005857
1294.3879201727493    1307.355789619312    1316.7011437529297    1324.122896359976    1545.8311563462382
1791.7753555927368    1812.5248482527577    2618.4480146666106    2651.5089622492474    2654.944581737765
2672.64403923318    2680.603498737896    2686.6490535252788    2714.328159482877    2731.543014885078
2735.322238224648 2781.006330098668</array>

            </property>

<group>2</group>

            <property dictRef="me:imFreq">

                <scalar units="cm-1">2295.1079853126153</scalar>

            </property>

            <property dictRef="me:spinMultiplicity">

                <scalar units="cm-1">2</scalar>

            </property>

            <property dictRef="me:ZPE">

                <scalar units="kJ/mol">-427.1439</scalar>

            </property>
```

```
                    </propertyList>
            </molecule>
        <molecule id="O=CC[CH]C[CH](C(=O)C)OO" spinMultiplicity="2">
                <atomArray>
                                    <atom    elementType="O"    id="a1"    x3="-0.299758"    y3="0.066299"
z3="2.902538"/>
                                    <atom    elementType="C"    id="a2"    x3="0.146399"    y3="1.143885"
z3="3.229086"/>
                                    <atom    elementType="C"    id="a3"    x3="0.242061"    y3="2.346748"
z3="2.313232"/>
                                    <atom    elementType="C"    id="a4"    spinMultiplicity="2"    x3="-0.320761"
y3="2.105200" z3="0.966829"/>
                                    <atom    elementType="C"    id="a5"    x3="0.527860"    y3="1.599701"    z3="-
0.130909"/>
                                    <atom    elementType="C"    id="a6"    x3="-0.238132"    y3="0.784358"    z3="-
1.178889"/>
                                    <atom    elementType="H"    id="a7"    x3="0.462514"    y3="0.389751"    z3="-
1.963247"/>
                                    <atom    elementType="C"    id="a8"    x3="-1.330187"    y3="1.619927"    z3="-
1.875298"/>
                                    <atom    elementType="O"    id="a9"    x3="-1.038223"    y3="2.688316"    z3="-
2.364233"/>
                                    <atom    elementType="C"    id="a10"    x3="-2.695857"    y3="1.028642"    z3="-
1.931962"/>
                                    <atom    elementType="H"    id="a11"    x3="0.525475"    y3="1.314353"
z3="4.251181"/>
                                    <atom    elementType="H"    id="a12"    x3="-0.277109"    y3="3.209173"
z3="2.793329"/>
                                    <atom    elementType="H"    id="a13"    x3="1.308283"    y3="2.664845"
z3="2.248922"/>
                                    <atom    elementType="H"    id="a14"    x3="-1.077042"    y3="-0.349978"
z3="1.185629"/>
                                    <atom    elementType="H"    id="a15"    x3="-1.370715"    y3="2.276960"
z3="0.799001"/>
                                    <atom    elementType="H"    id="a16"    x3="1.007437"    y3="2.472794"    z3="-
0.641525"/>
                                    <atom    elementType="H"    id="a17"    x3="1.362145"    y3="0.979120"
z3="0.262493"/>
```

                                    <atom   elementType="H"   id="a18"   x3="-3.107723"   y3="0.893222"   z3="-0.914822"/>

                                    <atom   elementType="H"   id="a19"   x3="-3.392641"   y3="1.649002"   z3="-2.509359"/>

                                    <atom   elementType="H"   id="a20"   x3="-2.683056"   y3="0.015820"   z3="-2.364424"/>

                                    <atom   elementType="O"   id="a21"   x3="-1.666648"   y3="-0.286525"   z3="0.349549"/>

                                    <atom   elementType="O"   id="a22"   x3="-0.712536"   y3="-0.478921"   z3="-0.670752"/>

                    </atomArray>

                    <bondArray>

                                    <bond atomRefs2="a19 a10" order="1"/>

                                    <bond atomRefs2="a20 a10" order="1"/>

                                    <bond atomRefs2="a9 a8" order="2"/>

                                    <bond atomRefs2="a7 a6" order="1"/>

                                    <bond atomRefs2="a10 a8" order="1"/>

                                    <bond atomRefs2="a10 a18" order="1"/>

                                    <bond atomRefs2="a8 a6" order="1"/>

                                    <bond atomRefs2="a6 a22" order="1"/>

                                    <bond atomRefs2="a6 a5" order="1"/>

                                    <bond atomRefs2="a22 a21" order="1"/>

                                    <bond atomRefs2="a16 a5" order="1"/>

                                    <bond atomRefs2="a5 a17" order="1"/>

                                    <bond atomRefs2="a5 a4" order="1"/>

                                    <bond atomRefs2="a21 a14" order="1"/>

                                    <bond atomRefs2="a15 a4" order="1"/>

                                    <bond atomRefs2="a4 a3" order="1"/>

                                    <bond atomRefs2="a13 a3" order="1"/>

                                    <bond atomRefs2="a3 a12" order="1"/>

                                    <bond atomRefs2="a3 a2" order="1"/>

                                    <bond atomRefs2="a1 a2" order="2"/>

                                    <bond atomRefs2="a2 a11" order="1"/>

                    </bondArray>

```xml
<propertyList>
        <property dictRef="me:lumpedSpecies">
                <array> </array>
        </property>
        <property dictRef="me:vibFreqs">
                            <array  units="cm-1">30.078   34.4533   57.116   80.7878   100.0306
109.8882  148.9745  188.1364  199.7316  248.9114  297.2642  301.1578  338.7145  444.1771  468.9867  507.8073
552.9477  591.8392  667.0567  732.1896  757.6322  840.9908  879.1142  944.1629  956.2067  1006.2567  1011.6692
1022.0619  1042.6728  1073.3153  1110.6003  1125.2418  1182.0833  1183.8169  1191.9688  1208.1628  1222.1555
1223.6048  1239.9476  1245.5908  1260.9983  1262.6637  1280.5942  1317.4133  1320.7437  1387.5663  1399.1736
1781.4768  1799.0849  2535.2865  2642.8827  2650.7389  2654.4116  2665.9855  2671.3723  2697.8293  2722.2786
2732.1924  2759.5719  2782.445</array>
        </property>
        <property dictRef="me:ZPE">
                <scalar units="kJ/mol">-492.2359</scalar>
        </property>
<group>2</group>
        <property dictRef="me:spinMultiplicity">
                <scalar units="cm-1">2</scalar>
        </property>
        <property dictRef="me:epsilon">
                <scalar>473.17</scalar>
        </property>
        <property dictRef="me:sigma">
                <scalar>5.09</scalar>
        </property>
</propertyList>
<me:energyTransferModel xsi:type="me:ExponentialDown">
        <scalar units="cm-1">250</scalar>
</me:energyTransferModel>
</molecule>
<molecule id="TS_O=CCCC[CH](C(=O)C)O[O]_[O]O[CH]CCCC=O" spinMultiplicity="6">
        <atomArray>
                <atom     elementType="O"     id="a1"     x3="-2.184098"     y3="5.324316"
z3="0.847017"/>
```

<atom elementType="C" id="a2" x3="-0.994479" y3="5.332445" z3="0.626072"/>

<atom elementType="C" id="a3" x3="-0.372214" y3="4.924926" z3="-0.690252"/>

<atom elementType="C" id="a4" x3="-1.378044" y3="4.763843" z3="-1.833609"/>

<atom elementType="C" id="a5" x3="-2.349365" y3="3.588535" z3="-1.630084"/>

<atom elementType="C" id="a6" spinMultiplicity="2" x3="-1.687805" y3="2.284378" z3="-1.783501"/>

<atom elementType="H" id="a7" x3="-0.906054" y3="1.877116" z3="-1.185080"/>

<atom elementType="C" id="a8" spinMultiplicity="3" x3="-0.080334" y3="4.184307" z3="-5.269301"/>

<atom elementType="O" id="a9" x3="0.236172" y3="4.115438" z3="-6.360768"/>

<atom elementType="C" id="a10" spinMultiplicity="2" x3="-0.662206" y3="0.885604" z3="-6.168331"/>

<atom elementType="H" id="a11" x3="-0.270785" y3="5.639613" z3="1.400658"/>

<atom elementType="H" id="a12" x3="0.398234" y3="5.669861" z3="-0.980123"/>

<atom elementType="H" id="a13" x3="0.188626" y3="3.977334" z3="-0.539201"/>

<atom elementType="H" id="a14" x3="-1.971696" y3="5.696343" z3="-1.933270"/>

<atom elementType="H" id="a15" x3="-0.832092" y3="4.637264" z3="-2.791727"/>

<atom elementType="H" id="a16" x3="-2.813391" y3="3.663729" z3="-0.613403"/>

<atom elementType="H" id="a17" x3="-3.199740" y3="3.687096" z3="-2.341031"/>

<atom elementType="H" id="a18" x3="-1.293484" y3="1.247224" z3="-6.937056"/>

<atom elementType="H" id="a19" x3="0.369668" y3="0.740659" z3="-6.354419"/>

<atom elementType="H" id="a20" x3="-1.069722" y3="0.637487" z3="-5.216977"/>

<atom elementType="O" id="a21" spinMultiplicity="2" x3="-1.687571" y3="0.472479" z3="-3.079240"/>

```xml
                    <atom elementType="O" id="a22" x3="-2.190342" y3="1.565288" z3="-2.869497"/>
        </atomArray>
        <bondArray>
            <bond atomRefs2="a18 a10" order="1"/>
            <bond atomRefs2="a9 a8" order="2"/>
            <bond atomRefs2="a19 a10" order="1"/>
            <bond atomRefs2="a10 a20" order="1"/>
            <bond atomRefs2="a21 a22" order="1"/>
            <bond atomRefs2="a22 a6" order="1"/>
            <bond atomRefs2="a15 a4" order="1"/>
            <bond atomRefs2="a17 a5" order="1"/>
            <bond atomRefs2="a14 a4" order="1"/>
            <bond atomRefs2="a4 a5" order="1"/>
            <bond atomRefs2="a4 a3" order="1"/>
            <bond atomRefs2="a6 a5" order="1"/>
            <bond atomRefs2="a6 a7" order="1"/>
            <bond atomRefs2="a5 a16" order="1"/>
            <bond atomRefs2="a12 a3" order="1"/>
            <bond atomRefs2="a3 a13" order="1"/>
            <bond atomRefs2="a3 a2" order="1"/>
            <bond atomRefs2="a2 a1" order="2"/>
            <bond atomRefs2="a2 a11" order="1"/>
        </bondArray>
        <propertyList>
            <property dictRef="me:vibFreqs">
                <array units="cm-1">11.71360555933053    19.189758037728247
33.805057034104294    35.68584048680036    52.27609206498669    58.59929057274199    72.26356696696948
87.02061238229722    91.81438790362456    108.63948407350264    117.96706871061934    128.57015196643826
144.66547127152822    158.70658956397708    224.8746874981245    246.38270414061375    278.94985402872754
327.39238950647064    447.9013127205133    573.839734649091    616.7658242266406    755.5833123669297
836.4411503990024    866.8132307381262    940.7823591220523    942.0343826369561    959.4910376381922
1015.132113798345    1055.5446971191793    1094.3709070014024    1104.2076385772798    1123.309012650133
1162.9063203072305    1186.0051924573813    1194.9479400577725    1210.3391465442394    1233.2338898605915
1248.0404862891617    1255.7094415026156    1258.4859657688603    1262.6644637720592    1265.686820643351
1276.8414590314107    1304.7790881687622    1322.9403924736052    1366.220632738046    1792.8444344990535
```

2060.588885903238   2636.691567042242   2641.3574815347633   2659.2363362225096   2668.372461262945
2720.9861185384   2727.143119960301   2734.3785192985297   2739.3550690246584   2752.2679479055055
2758.3539694187134 2789.459711389141</array>

                                   </property>

                                   <property dictRef="me:imFreq">

                                              <scalar units="cm-1">36.32590051946986</scalar>

                                   </property>

<group>2</group>

                                   <property dictRef="me:spinMultiplicity">

                                              <scalar units="cm-1">2</scalar>

                                   </property>

                                   <property dictRef="me:ZPE">

                                              <scalar units="kJ/mol">-216.28381</scalar>

                                   </property>

                        </propertyList>

               </molecule>

               <molecule id="[O]O[CH]CCCC=O" spinMultiplicity="3">

                        <atomArray>
                                   <atom   elementType="O"   id="a1"   spinMultiplicity="2"   x3="0.655034"
y3="0.206253" z3="0.741465"/>

                                   <atom    elementType="O"    id="a2"    x3="1.876419"    y3="0.395494"
z3="0.764413"/>

                                   <atom  elementType="C"  id="a3"  spinMultiplicity="2"  x3="2.659467"  y3="-
0.602628" z3="1.141723"/>

                                   <atom    elementType="H"    id="a4"    x3="2.269746"    y3="-1.568884"
z3="1.418779"/>

                                   <atom    elementType="C"    id="a5"    x3="4.104585"    y3="-0.265749"
z3="1.139710"/>

                                   <atom    elementType="C"    id="a6"    x3="4.916290"    y3="-1.250192"
z3="0.282360"/>

                                   <atom  elementType="C"  id="a7"  x3="4.640467"  y3="-1.116473"  z3="-
1.213784"/>

                                   <atom  elementType="C"  id="a8"  x3="5.280747"  y3="0.099075"  z3="-
1.844949"/>

                                   <atom  elementType="O"  id="a9"  x3="5.427010"  y3="0.193294"  z3="-
3.041768"/>

```xml
                    <atom    elementType="H"    id="a10"    x3="4.269123"    y3="0.778168"
z3="0.793336"/>
                    <atom    elementType="H"    id="a11"    x3="4.480929"    y3="-0.286733"
z3="2.189393"/>
                    <atom    elementType="H"    id="a12"    x3="5.997252"    y3="-1.104276"
z3="0.485507"/>
                    <atom    elementType="H"    id="a13"    x3="4.697507"    y3="-2.290862"
z3="0.601951"/>
                    <atom  elementType="H"  id="a14"  x3="4.982361"  y3="-2.028503"  z3="-
1.751536"/>
                    <atom  elementType="H"  id="a15"  x3="3.546304"  y3="-1.086883"  z3="-
1.416706"/>
                    <atom  elementType="H"  id="a16"  x3="5.602990"  y3="0.895057"  z3="-
1.155470"/>
            </atomArray>
            <bondArray>
                    <bond atomRefs2="a9 a8" order="2"/>
                    <bond atomRefs2="a8 a7" order="1"/>
                    <bond atomRefs2="a8 a16" order="1"/>
                    <bond atomRefs2="a14 a7" order="1"/>
                    <bond atomRefs2="a15 a7" order="1"/>
                    <bond atomRefs2="a7 a6" order="1"/>
                    <bond atomRefs2="a6 a12" order="1"/>
                    <bond atomRefs2="a6 a13" order="1"/>
                    <bond atomRefs2="a6 a5" order="1"/>
                    <bond atomRefs2="a1 a2" order="1"/>
                    <bond atomRefs2="a2 a3" order="1"/>
                    <bond atomRefs2="a10 a5" order="1"/>
                    <bond atomRefs2="a5 a3" order="1"/>
                    <bond atomRefs2="a5 a11" order="1"/>
                    <bond atomRefs2="a3 a4" order="1"/>
            </bondArray>
            <propertyList>
                    <property dictRef="me:lumpedSpecies">
                            <array> </array>
```

```xml
              </property>
              <property dictRef="me:vibFreqs">
                          <array units="cm-1">22.4906  34.5956  44.593  96.8908  197.0665
252.5216  288.7109  351.6936  439.1894  515.8217  582.4808  747.1916  842.6327  888.7602  944.0026  973.1264
1033.0746  1057.524  1088.3907  1113.1996  1140.1877  1175.6145  1204.7225  1205.041  1218.6073  1253.6178
1266.8879  1276.5718  1279.6374  1299.3332  1317.0701  1335.591  1433.9945  1792.4074  2644.3144  2652.3582
2654.2342  2664.1935  2722.9946  2735.6517  2739.4406  2740.9604</array>
              </property>
              <property dictRef="me:ZPE">
                          <scalar units="kJ/mol">-287.66478</scalar>
              </property>
              <property dictRef="me:spinMultiplicity">
                          <scalar units="cm-1">1</scalar>
              </property>
<group>2</group>
              <property dictRef="me:epsilon">
                          <scalar>473.17</scalar>
              </property>
              <property dictRef="me:sigma">
                          <scalar>5.09</scalar>
              </property>
          </propertyList>
          <me:energyTransferModel xsi:type="me:ExponentialDown">
                          <scalar units="cm-1">250</scalar>
          </me:energyTransferModel>
      </molecule>
      <molecule id="C[C]=O" spinMultiplicity="2">
              <atomArray>
                          <atom     elementType="C"     id="a1"     x3="1.042219"     y3="-0.085020"
z3="0.053638"/>
                          <atom     elementType="C"     id="a2"     spinMultiplicity="2"     x3="0.860386"
y3="1.298096" z3="0.469816"/>
                          <atom     elementType="O"     id="a3"     x3="0.437868"     y3="1.959356"
z3="1.334402"/>
```

```xml
                              <atom elementType="H" id="a4" x3="0.535314" y3="-0.270438" z3="-0.908791"/>
                              <atom elementType="H" id="a5" x3="0.658002" y3="-0.812959" z3="0.784723"/>
                              <atom elementType="H" id="a6" x3="2.111086" y3="-0.302774" z3="-0.112829"/>
              </atomArray>
              <bondArray>
                      <bond atomRefs2="a4 a1" order="1"/>
                      <bond atomRefs2="a6 a1" order="1"/>
                      <bond atomRefs2="a1 a2" order="1"/>
                      <bond atomRefs2="a1 a5" order="1"/>
                      <bond atomRefs2="a2 a3" order="2"/>
              </bondArray>
              <propertyList>
                      <property dictRef="me:lumpedSpecies">
                              <array> </array>
                      </property>
                      <property dictRef="me:vibFreqs">
                              <array units="cm-1">73.8717 402.4786 933.8152 977.8873 1089.1815 1205.0768 1215.9683 1309.1112 2012.376 2681.2098 2682.7075 2780.0179</array>
                      </property>
                      <property dictRef="me:ZPE">
                              <scalar units="kJ/mol">0.0</scalar>
                      </property>
                      <property dictRef="me:spinMultiplicity">
                              <scalar units="cm-1">2</scalar>
                      </property>
                      <property dictRef="me:epsilon">
                              <scalar>473.17</scalar>
                      </property>
                      <property dictRef="me:sigma">
                              <scalar>5.09</scalar>
                      </property>
```

        </propertyList>

        <me:energyTransferModel xsi:type="me:ExponentialDown">

                <scalar units="cm-1">250</scalar>

        </me:energyTransferModel>

    </molecule>

    <molecule id="[O]OC(=O)CCC[CH](C(=O)C)OO" spinMultiplicity="2">

        <atomArray>
                                <atom    elementType="O"    id="a1"    x3="-2.487118"    y3="1.969056"
z3="4.144097"/>
                                <atom    elementType="C"    id="a2"    x3="-1.871952"    y3="1.937420"
z3="3.125975"/>
                                <atom    elementType="C"    id="a3"    x3="-2.149365"    y3="1.426280"
z3="1.760086"/>
                                <atom    elementType="C"    id="a4"    x3="-3.543873"    y3="1.827961"
z3="1.270659"/>
                                <atom    elementType="C"    id="a5"    x3="-3.622546"    y3="3.320832"
z3="0.928872"/>
                                <atom    elementType="C"    id="a6"    x3="-5.029455"    y3="3.743585"
z3="0.496375"/>
                                <atom    elementType="H"    id="a7"    x3="-5.101532"    y3="4.847539"
z3="0.351060"/>
                                <atom    elementType="C"    id="a8"    x3="-5.476960"    y3="3.047342"    z3="-
0.806190"/>
                                <atom    elementType="O"    id="a9"    x3="-4.773031"    y3="3.110493"    z3="-
1.787648"/>
                                <atom    elementType="C"    id="a10"    x3="-6.790722"    y3="2.344055"    z3="-
0.767323"/>
                                <atom    elementType="H"    id="a11"    x3="-6.951181"    y3="4.581400"
z3="2.525099"/>
                                <atom    elementType="H"    id="a12"    x3="-1.361083"    y3="1.780078"
z3="1.048284"/>
                                <atom    elementType="H"    id="a13"    x3="-2.049165"    y3="0.316217"
z3="1.763577"/>
                                <atom    elementType="H"    id="a14"    x3="-3.800178"    y3="1.221726"
z3="0.376834"/>
                                <atom    elementType="H"    id="a15"    x3="-4.306765"    y3="1.577105"
z3="2.040080"/>

```
                            <atom    elementType="H"    id="a16"    x3="-3.313502"    y3="3.930531"
z3="1.802113"/>

                            <atom    elementType="H"    id="a17"    x3="-2.913468"    y3="3.567098"
z3="0.107179"/>

                            <atom    elementType="H"    id="a18"    x3="-7.586779"    y3="2.999329"    z3="-
0.372411"/>

                            <atom    elementType="H"    id="a19"    x3="-7.095631"    y3="1.987102"    z3="-
1.760216"/>

                            <atom    elementType="H"    id="a20"    x3="-6.762248"    y3="1.481096"    z3="-
0.084739"/>

                            <atom    elementType="O"    id="a21"    x3="-5.882877"    y3="3.356055"
z3="1.605966"/>

                            <atom    elementType="O"    id="a22"    x3="-6.990953"    y3="4.252835"
z3="1.576667"/>

                            <atom    elementType="O"    id="a23"    spinMultiplicity="2"    x3="0.212680"
y3="2.619053" z3="2.338681"/>

                            <atom    elementType="O"    id="a24"    x3="-0.468846"    y3="2.564453"
z3="3.319230"/>

                    </atomArray>

                    <bondArray>

                            <bond atomRefs2="a9 a8" order="2"/>

                            <bond atomRefs2="a19 a10" order="1"/>

                            <bond atomRefs2="a8 a10" order="1"/>

                            <bond atomRefs2="a8 a6" order="1"/>

                            <bond atomRefs2="a10 a18" order="1"/>

                            <bond atomRefs2="a10 a20" order="1"/>

                            <bond atomRefs2="a17 a5" order="1"/>

                            <bond atomRefs2="a7 a6" order="1"/>

                            <bond atomRefs2="a14 a4" order="1"/>

                            <bond atomRefs2="a6 a5" order="1"/>

                            <bond atomRefs2="a6 a21" order="1"/>

                            <bond atomRefs2="a5 a4" order="1"/>

                            <bond atomRefs2="a5 a16" order="1"/>

                            <bond atomRefs2="a12 a3" order="1"/>

                            <bond atomRefs2="a4 a3" order="1"/>

                            <bond atomRefs2="a4 a15" order="1"/>
```

```xml
                    <bond atomRefs2="a22 a21" order="1"/>

                    <bond atomRefs2="a22 a11" order="1"/>

                    <bond atomRefs2="a3 a13" order="1"/>

                    <bond atomRefs2="a3 a2" order="1"/>

                    <bond atomRefs2="a23 a24" order="1"/>

                    <bond atomRefs2="a2 a24" order="1"/>

                    <bond atomRefs2="a2 a1" order="2"/>

            </bondArray>

            <propertyList>

                    <property dictRef="me:lumpedSpecies">

                            <array> </array>

                    </property>

                    <property dictRef="me:vibFreqs">

                            <array units="cm-1">9.4028 18.296 27.199 35.3317 57.1647 65.9436
92.3644 112.3471 118.0952 167.8792 185.4426 226.7847 254.761 274.7404 301.8656 353.8103 439.9769 448.009
479.2092 512.6071 517.5274 590.5531 675.109 723.1325 780.7531 859.6085 883.2911 959.2598 999.4521 1006.91
1009.8341 1021.7129 1064.6816 1081.7163 1119.0597 1151.6146 1158.0475 1167.8297 1185.9443 1194.2874
1201.1328 1210.7617 1221.9113 1223.2463 1232.0817 1256.4536 1267.12 1271.7764 1291.6263 1308.2173
1316.9151 1327.6754 1391.9792 1804.4739 1899.9596 2634.3632 2648.2814 2649.7654 2669.4233 2671.3437
2672.6419 2695.3 2720.0441 2725.2874 2737.8172 2783.6017</array>

                    </property>

                    <property dictRef="me:ZPE">

                            <scalar units="kJ/mol">-651.55285316894103</scalar>

                    </property>

<group>3</group>

                    <property dictRef="me:spinMultiplicity">

                            <scalar units="cm-1">2</scalar>

                    </property>

                    <property dictRef="me:epsilon">

                            <scalar>473.17</scalar>

                    </property>

                    <property dictRef="me:sigma">

                            <scalar>5.09</scalar>

                    </property>
```

```xml
        </propertyList>

        <me:energyTransferModel xsi:type="me:ExponentialDown">

                <scalar units="cm-1">250</scalar>

        </me:energyTransferModel>

    </molecule>

    <molecule              id="TS_[O]OC(=O)CCC[CH](C(=O)C)OO_OOC(=O)[CH]CC[CH](C(=O)C)OO"
spinMultiplicity="2">

        <atomArray>

                <atom    elementType="O"   id="a1"   x3="-1.366900"   y3="2.005814"   z3="-
0.840334"/>

                <atom    elementType="C"   id="a2"   x3="-1.421674"      y3="2.160623"
z3="0.349037"/>

                <atom    elementType="C"   id="a3"   x3="-0.659899"      y3="1.542687"
z3="1.484586"/>

                <atom    elementType="C"   id="a4"   x3="-0.774569"      y3="0.050377"
z3="1.634439"/>

                <atom    elementType="C"   id="a5"   x3="0.433569"      y3="-0.699850"
z3="1.060291"/>

                <atom    elementType="C"   id="a6"   x3="0.584615"   y3="-0.529161"   z3="-
0.453241"/>

                <atom    elementType="H"   id="a7"   x3="0.549278"   y3="0.538934"   z3="-
0.774895"/>

                <atom    elementType="C"   id="a8"   x3="1.902659"   y3="-1.153083"   z3="-
0.957255"/>

                <atom    elementType="O"   id="a9"   x3="2.932499"   y3="-0.926229"   z3="-
0.362859"/>

                <atom    elementType="C"   id="a10"   x3="1.811362"   y3="-1.990762"   z3="-
2.184107"/>

                <atom    elementType="H"   id="a11"   x3="-1.458918"   y3="0.074536"   z3="-
2.054836"/>

                <atom    elementType="H"   id="a12"      x3="0.348889"      y3="1.958832"
z3="1.629744"/>

                <atom    elementType="H"   id="a13"      x3="-1.293751"      y3="2.132201"
z3="2.493763"/>

                <atom    elementType="H"   id="a14"      x3="-0.881131"      y3="-0.211492"
z3="2.710904"/>

                <atom    elementType="H"   id="a15"      x3="-1.703232"      y3="-0.327868"
z3="1.148486"/>
```

<atom elementType="H" id="a16" x3="1.378954" y3="-0.375786" z3="1.550284"/>

<atom elementType="H" id="a17" x3="0.330738" y3="-1.781131" z3="1.298004"/>

<atom elementType="H" id="a18" x3="2.796495" y3="-2.334030" z3="-2.526249"/>

<atom elementType="H" id="a19" x3="1.174234" y3="-2.874827" z3="-2.015272"/>

<atom elementType="H" id="a20" x3="1.323638" y3="-1.444820" z3="-3.010624"/>

<atom elementType="O" id="a21" x3="-0.567506" y3="-1.224986" z3="-1.007517"/>

<atom elementType="O" id="a22" x3="-0.823927" y3="-0.681416" z3="-2.285010"/>

<atom elementType="O" id="a23" spinMultiplicity="2" x3="-2.292902" y3="2.978618" z3="2.250683"/>

<atom elementType="O" id="a24" x3="-2.322521" y3="3.112818" z3="0.861977"/>

</atomArray>

<bondArray>

<bond atomRefs2="a20 a10" order="1"/>

<bond atomRefs2="a18 a10" order="1"/>

<bond atomRefs2="a22 a11" order="1"/>

<bond atomRefs2="a22 a21" order="1"/>

<bond atomRefs2="a10 a19" order="1"/>

<bond atomRefs2="a10 a8" order="1"/>

<bond atomRefs2="a21 a6" order="1"/>

<bond atomRefs2="a8 a6" order="1"/>

<bond atomRefs2="a8 a9" order="2"/>

<bond atomRefs2="a1 a2" order="2"/>

<bond atomRefs2="a7 a6" order="1"/>

<bond atomRefs2="a6 a5" order="1"/>

<bond atomRefs2="a2 a24" order="1"/>

<bond atomRefs2="a2 a3" order="1"/>

<bond atomRefs2="a24 a23" order="1"/>

<bond atomRefs2="a5 a17" order="1"/>

```xml
        <bond atomRefs2="a5 a16" order="1"/>

        <bond atomRefs2="a5 a4" order="1"/>

        <bond atomRefs2="a15 a4" order="1"/>

        <bond atomRefs2="a3 a12" order="1"/>

        <bond atomRefs2="a3 a4" order="1"/>

        <bond atomRefs2="a3 a13" order="1"/>

        <bond atomRefs2="a4 a14" order="1"/>

    </bondArray>

    <propertyList>

        <property dictRef="me:vibFreqs">

                        <array    units="cm-1">29.076955868713966      31.71011044098436
36.30471015238132    69.65691461729841    77.94277333787912    83.3770758049076    112.44791578651773
122.34010459602686  150.6544937880492  194.7226023122538  245.24419035831895  264.49994672700564
268.9402205877795  365.23254255039205  377.50856989541734  421.4168912171439  446.65135534773935
495.58333844898567    511.57122472483    523.6104524582382    594.1031956521244    605.7754612439688
645.3311156736157  775.4172859543077  827.475146992905  870.7817238803749  908.5212758308667
931.2253954669277   973.3443813991003   979.4525056569224   1003.7298898040159   1021.9031568182393
1040.5642373312603  1048.2782347251425  1090.384486825502  1141.8816915889374  1158.1348253774947
1168.2239877283027  1177.1082260309045  1190.6566536158332  1197.6965017549217  1203.975540078149
1219.063074663718  1228.230279068644  1248.912259785416  1255.1136081269774  1259.5981768137585
1275.3781067411205  1294.449480937582  1315.4663395398136  1326.32232119549  1400.812319891632
1796.6334552939384  1802.37947182717  1919.2288933294506  2615.481619266483  2646.5753331679375
2654.9281404665535  2663.5403921214715  2668.7747779379556  2694.36368513395  2695.0160407900007
2728.9116483043726 2729.9323187717573 2782.064243040049</array>

        </property>

        <property dictRef="me:imFreq">

                        <scalar units="cm-1">1913.6612996668498</scalar>

        </property>

<group>3</group>

        <property dictRef="me:spinMultiplicity">

                        <scalar units="cm-1">2</scalar>

        </property>

        <property dictRef="me:ZPE">

                        <scalar units="kJ/mol">-585.12897855436168</scalar>

        </property>

    </propertyList>

</molecule>
```

```xml
<molecule id="OOC(=O)[CH]CC[CH](C(=O)C)OO" spinMultiplicity="2">
    <atomArray>
        <atom elementType="O" id="a1" x3="-6.666272" y3="3.603070" z3="1.521039"/>
        <atom elementType="C" id="a2" x3="-6.047622" y3="4.056275" z3="2.463939"/>
        <atom elementType="C" id="a3" spinMultiplicity="2" x3="-5.076608" y3="3.402989" z3="3.331403"/>
        <atom elementType="C" id="a4" x3="-5.033905" y3="1.931730" z3="3.411275"/>
        <atom elementType="C" id="a5" x3="-3.727054" y3="1.360366" z3="2.839388"/>
        <atom elementType="C" id="a6" x3="-3.624859" y3="1.585662" z3="1.328024"/>
        <atom elementType="H" id="a7" x3="-3.717843" y3="2.662925" z3="1.056001"/>
        <atom elementType="C" id="a8" x3="-2.295911" y3="1.051936" z3="0.755674"/>
        <atom elementType="O" id="a9" x3="-1.259101" y3="1.272229" z3="1.339383"/>
        <atom elementType="C" id="a10" x3="-2.388773" y3="0.302805" z3="-0.528946"/>
        <atom elementType="H" id="a11" x3="-6.047148" y3="2.080910" z3="0.174527"/>
        <atom elementType="H" id="a12" x3="-4.377176" y3="4.006932" z3="3.898330"/>
        <atom elementType="H" id="a13" x3="-5.773912" y3="6.867570" z3="3.672857"/>
        <atom elementType="H" id="a14" x3="-5.138600" y3="1.618794" z3="4.476549"/>
        <atom elementType="H" id="a15" x3="-5.894734" y3="1.465279" z3="2.874539"/>
        <atom elementType="H" id="a16" x3="-2.837659" y3="1.807995" z3="3.333112"/>
        <atom elementType="H" id="a17" x3="-3.680368" y3="0.273640" z3="3.059768"/>
        <atom elementType="H" id="a18" x3="-1.406933" y3="0.173777" z3="-1.003238"/>
```

```xml
                    <atom  elementType="H"  id="a19"  x3="-2.821679"  y3="-0.698041"  z3="-
0.368633"/>
                    <atom  elementType="H"  id="a20"  x3="-3.067023"  y3="0.798021"  z3="-
1.244822"/>
                    <atom    elementType="O"    id="a21"    x3="-4.736354"    y3="0.847465"
z3="0.746742"/>
                    <atom  elementType="O"  id="a22"  x3="-5.260999"  y3="1.640220"  z3="-
0.298658"/>
                    <atom    elementType="O"    id="a23"    x3="-5.608073"    y3="5.882467"
z3="3.810492"/>
                    <atom    elementType="O"    id="a24"    x3="-6.350975"    y3="5.417436"
z3="2.679812"/>
          </atomArray>
          <bondArray>
                    <bond atomRefs2="a20 a10" order="1"/>
                    <bond atomRefs2="a18 a10" order="1"/>
                    <bond atomRefs2="a10 a19" order="1"/>
                    <bond atomRefs2="a10 a8" order="1"/>
                    <bond atomRefs2="a22 a11" order="1"/>
                    <bond atomRefs2="a22 a21" order="1"/>
                    <bond atomRefs2="a21 a6" order="1"/>
                    <bond atomRefs2="a8 a6" order="1"/>
                    <bond atomRefs2="a8 a9" order="2"/>
                    <bond atomRefs2="a7 a6" order="1"/>
                    <bond atomRefs2="a6 a5" order="1"/>
                    <bond atomRefs2="a1 a2" order="2"/>
                    <bond atomRefs2="a2 a24" order="1"/>
                    <bond atomRefs2="a2 a3" order="1"/>
                    <bond atomRefs2="a24 a23" order="1"/>
                    <bond atomRefs2="a5 a17" order="1"/>
                    <bond atomRefs2="a5 a16" order="1"/>
                    <bond atomRefs2="a5 a4" order="1"/>
                    <bond atomRefs2="a15 a4" order="1"/>
                    <bond atomRefs2="a3 a4" order="1"/>
```

```xml
<bond atomRefs2="a3 a12" order="1"/>

<bond atomRefs2="a4 a14" order="1"/>

<bond atomRefs2="a13 a23" order="1"/>

</bondArray>

<propertyList>

<property dictRef="me:lumpedSpecies">

<array> </array>

</property>

<property dictRef="me:vibFreqs">

<array    units="cm-1">22.3142    25.0868    31.1405    52.7571    66.1908
68.9943 87.6872 101.9602 129.2579 152.7471 166.8147 248.8101 254.4823 281.3422 308.9802 397.7459 405.2117
434.1107 485.0958 490.8839 510.5962 540.2769 564.5857 607.5812 754.9215 831.9797 866.6344 910.1285
980.0713 983.531 1004.5982 1028.3566 1045.1051 1068.0809 1089.397 1120.0508 1158.851 1173.5108 1178.1719
1194.6166 1198.6292 1217.6738 1226.6924 1235.3365 1247.132 1253.9846 1264.2128 1279.9171 1305.8887
1317.0104 1378.4173 1390.9315 1425.1565 1783.9755 1803.195 2593.5623 2647.9947 2650.4707 2659.2261
2671.445 2675.0665 2694.8231 2719.7463 2732.9422 2737.016 2783.1887</array>

</property>

<property dictRef="me:ZPE">

<scalar units="kJ/mol">-694.65</scalar>

</property>

<property dictRef="me:spinMultiplicity">

<scalar units="cm-1">2</scalar>

</property>

<group>3</group>

<property dictRef="me:epsilon">

<scalar>473.17</scalar>

</property>

<property dictRef="me:sigma">

<scalar>5.09</scalar>

</property>

</propertyList>

<me:energyTransferModel xsi:type="me:ExponentialDown">

<scalar units="cm-1">250</scalar>

</me:energyTransferModel>
```

```xml
                    </molecule>
                    <molecule                id="TS_[O]OC(=O)CCC[CH](C(=O)C)OO_OOC(=O)C[CH]C[CH](C(=O)C)OO"
spinMultiplicity="2">
                         <atomArray>
                                        <atom      elementType="O"     id="a1"    x3="4.046312"      y3="-2.371176"
z3="0.319876"/>
                                        <atom      elementType="C"     id="a2"    x3="3.166836"      y3="-1.586119"
z3="0.544342"/>
                                        <atom    elementType="C"    id="a3"    x3="2.530066"   y3="-0.577712"    z3="-
0.375403"/>
                                        <atom    elementType="C"   id="a4"   spinMultiplicity="2"   x3="1.056351"   y3="-
0.400508" z3="-0.099302"/>
                                        <atom    elementType="C"    id="a5"    x3="0.502003"    y3="0.976666"     z3="-
0.364168"/>
                                        <atom    elementType="C"    id="a6"    x3="-1.010165"    y3="1.058367"     z3="-
0.129042"/>
                                        <atom      elementType="H"     id="a7"    x3="-1.342128"      y3="2.110139"
z3="0.046421"/>
                                        <atom    elementType="C"    id="a8"    x3="-1.825387"    y3="0.482326"     z3="-
1.311863"/>
                                        <atom    elementType="O"    id="a9"    x3="-1.387440"    y3="0.569431"     z3="-
2.435415"/>
                                        <atom    elementType="C"   id="a10"   x3="-3.149039"   y3="-0.112284"   z3="-
0.975764"/>
                                        <atom      elementType="H"     id="a11"    x3="-1.732801"      y3="1.019376"
z3="2.707756"/>
                                        <atom    elementType="H"   id="a12"   x3="3.081556"   y3="0.385871"    z3="-
0.267892"/>
                                        <atom    elementType="H"   id="a13"   x3="2.694692"   y3="-0.894700"   z3="-
1.432188"/>
                                        <atom    elementType="H"   id="a14"   x3="0.427986"   y3="-1.231326"   z3="-
0.450252"/>
                                        <atom      elementType="H"     id="a15"    x3="0.860859"      y3="-0.620156"
z3="1.200741"/>
                                        <atom      elementType="H"     id="a16"    x3="1.020156"      y3="1.730652"
z3="0.265400"/>
                                        <atom    elementType="H"   id="a17"   x3="0.709668"   y3="1.266584"    z3="-
1.424913"/>
```

            <atom   elementType="H"   id="a18"   x3="-3.418939"   y3="-0.920823"   z3="-1.671092"/>

            <atom   elementType="H"   id="a19"   x3="-3.183623"   y3="-0.513927"   z3="0.052534"/>

            <atom   elementType="H"   id="a20"   x3="-3.948847"   y3="0.641420"   z3="-1.044170"/>

            <atom   elementType="O"   id="a21"   x3="-1.276608"   y3="0.267579"   z3="1.059010"/>

            <atom   elementType="O"   id="a22"   x3="-2.215507"   y3="1.017403"   z3="1.824838"/>

            <atom   elementType="O"   id="a23"   x3="1.804786"   y3="-0.615654"   z3="2.112212"/>

            <atom   elementType="O"   id="a24"   x3="2.589211"   y3="-1.681429"   z3="1.848335"/>

        </atomArray>

        <bondArray>

            <bond atomRefs2="a9 a8" order="2"/>

            <bond atomRefs2="a18 a10" order="1"/>

            <bond atomRefs2="a13 a3" order="1"/>

            <bond atomRefs2="a17 a5" order="1"/>

            <bond atomRefs2="a8 a10" order="1"/>

            <bond atomRefs2="a8 a6" order="1"/>

            <bond atomRefs2="a20 a10" order="1"/>

            <bond atomRefs2="a10 a19" order="1"/>

            <bond atomRefs2="a14 a4" order="1"/>

            <bond atomRefs2="a3 a12" order="1"/>

            <bond atomRefs2="a3 a4" order="1"/>

            <bond atomRefs2="a3 a2" order="1"/>

            <bond atomRefs2="a5 a6" order="1"/>

            <bond atomRefs2="a5 a4" order="1"/>

            <bond atomRefs2="a5 a16" order="1"/>

            <bond atomRefs2="a6 a7" order="1"/>

            <bond atomRefs2="a6 a21" order="1"/>

            <bond atomRefs2="a1 a2" order="2"/>

            <bond atomRefs2="a2 a24" order="1"/>

```xml
        <bond atomRefs2="a21 a22" order="1"/>

        <bond atomRefs2="a15 a23" order="1"/>

        <bond atomRefs2="a22 a11" order="1"/>

        <bond atomRefs2="a24 a23" order="1"/>

      </bondArray>

      <propertyList>

        <property dictRef="me:vibFreqs">

                        <array    units="cm-1">18.88592992674915    25.511832511192086
31.54051174882022   50.366129664678034   57.433761909917436   59.73024264840239   92.76284841975375
145.74815569124993  184.71958654011016  212.20144015189365  228.02459178331114  286.98678983853586
298.460567293761    314.59575059777055   381.6700199693781   417.7892575416384   420.81811677923196
442.820848740065    490.94355436024875   533.752758678896    576.1495617374848   621.6360877137654
674.4479679403944   805.8663537643511    855.612544217139    884.8875547422351   907.3436339638487
961.0645958875959   989.8486717043401   1007.6851689449384  1021.9196864130471  1027.6325811307597
1049.392114089932   1094.3567779976163  1095.8642270550968  1144.9576782556926  1164.5888492698507
1177.498545528344   1184.3318256043085  1194.5908426883866  1201.0947710523133  1206.7445334388103
1214.3685384017551  1221.8659263767802  1226.55367013467    1242.7175528598511  1250.2629746189816
1255.3096505618132  1300.4242186620613  1321.1138553045134  1333.737084603934   1392.1772188091713
1630.365098084251   1806.330251858957   1844.4938306440145  2643.79102241605    2647.5436318165594
2648.353210603622   2664.5869423044487  2670.1555146001124  2684.208099410017   2700.4934869077456
2721.8271749536907  2729.168618736404  2779.946295024366</array>

        </property>

        <property dictRef="me:imFreq">

                        <scalar units="cm-1">1870.6885027366966</scalar>

        </property>

<group>3</group>

        <property dictRef="me:spinMultiplicity">

                        <scalar units="cm-1">2</scalar>

        </property>

        <property dictRef="me:ZPE">

                        <scalar units="kJ/mol">-586.78</scalar>

        </property>

      </propertyList>

    </molecule>

    <molecule id="OOC(=O)C[CH]C[CH](C(=O)C)OO" spinMultiplicity="2">

        <atomArray>

                        <atom    elementType="O"    id="a1"    x3="-4.117471"    y3="-0.763726"
z3="0.682858"/>
```

                              <atom    elementType="C"    id="a2"    x3="-3.593232"    y3="0.219046"
z3="1.151659"/>

                              <atom    elementType="C"    id="a3"    x3="-2.832012"    y3="1.300885"
z3="0.435683"/>

                              <atom   elementType="C"   id="a4"   spinMultiplicity="2"   x3="-3.757793"
y3="2.096392" z3="-0.399576"/>

                              <atom    elementType="C"    id="a5"    x3="-4.362234"    y3="3.354304"
z3="0.087642"/>

                              <atom    elementType="C"    id="a6"    x3="-5.825149"    y3="3.200781"
z3="0.529056"/>

                              <atom    elementType="H"    id="a7"    x3="-6.299980"    y3="4.202311"
z3="0.643195"/>

                              <atom   elementType="C"   id="a8"   x3="-6.650636"   y3="2.323675"   z3="-
0.443160"/>

                              <atom   elementType="O"   id="a9"   x3="-7.037834"   y3="2.814462"   z3="-
1.475383"/>

                              <atom   elementType="C"   id="a10"   x3="-6.905554"   y3="0.921077"   z3="-
0.020267"/>

                              <atom    elementType="H"    id="a11"    x3="-6.306864"    y3="3.427544"
z3="3.387757"/>

                              <atom    elementType="H"    id="a12"    x3="-2.278258"    y3="1.938022"
z3="1.168756"/>

                              <atom   elementType="H"   id="a13"   x3="-2.044374"   y3="0.823182"   z3="-
0.194999"/>

                              <atom   elementType="H"   id="a14"   x3="-4.003624"   y3="1.738863"   z3="-
1.387342"/>

                              <atom    elementType="H"    id="a15"    x3="-4.044184"    y3="2.086072"
z3="3.105634"/>

                              <atom    elementType="H"    id="a16"    x3="-3.791946"    y3="3.776611"
z3="0.946334"/>

                              <atom   elementType="H"   id="a17"   x3="-4.308813"   y3="4.125980"   z3="-
0.716152"/>

                              <atom   elementType="H"   id="a18"   x3="-7.719466"   y3="0.458113"   z3="-
0.599036"/>

                              <atom   elementType="H"   id="a19"   x3="-6.013637"   y3="0.280383"   z3="-
0.171127"/>

                              <atom    elementType="H"    id="a20"    x3="-7.158315"    y3="0.832483"
z3="1.048744"/>

                              <atom    elementType="O"    id="a21"    x3="-5.914985"    y3="2.505981"
z3="1.805286"/>

```xml
                              <atom    elementType="O"    id="a22"    x3="-5.492913"    y3="3.442377"
z3="2.796327"/>
                              <atom    elementType="O"    id="a23"    x3="-3.200173"    y3="1.503774"
z3="3.059780"/>
                              <atom    elementType="O"    id="a24"    x3="-3.752652"    y3="0.307854"
z3="2.550281"/>
       </atomArray>
       <bondArray>
              <bond atomRefs2="a9 a8" order="2"/>
              <bond atomRefs2="a14 a4" order="1"/>
              <bond atomRefs2="a17 a5" order="1"/>
              <bond atomRefs2="a18 a10" order="1"/>
              <bond atomRefs2="a8 a10" order="1"/>
              <bond atomRefs2="a8 a6" order="1"/>
              <bond atomRefs2="a4 a5" order="1"/>
              <bond atomRefs2="a4 a3" order="1"/>
              <bond atomRefs2="a13 a3" order="1"/>
              <bond atomRefs2="a19 a10" order="1"/>
              <bond atomRefs2="a10 a20" order="1"/>
              <bond atomRefs2="a5 a6" order="1"/>
              <bond atomRefs2="a5 a16" order="1"/>
              <bond atomRefs2="a3 a2" order="1"/>
              <bond atomRefs2="a3 a12" order="1"/>
              <bond atomRefs2="a6 a7" order="1"/>
              <bond atomRefs2="a6 a21" order="1"/>
              <bond atomRefs2="a1 a2" order="2"/>
              <bond atomRefs2="a2 a24" order="1"/>
              <bond atomRefs2="a21 a22" order="1"/>
              <bond atomRefs2="a24 a23" order="1"/>
              <bond atomRefs2="a22 a11" order="1"/>
              <bond atomRefs2="a23 a15" order="1"/>
       </bondArray>
       <propertyList>
```

```xml
            <property dictRef="me:lumpedSpecies">

                    <array> </array>

            </property>

            <property dictRef="me:vibFreqs">

                    <array units="cm-1">32.5886  40.0679  48.8161  67.4071  90.6683
113.1961  151.6025  163.2423  182.5695  192.3185  230.6372  261.8316  285.1035  288.9969  353.3318  375.3304
391.4909  445.6699  476.2086  499.7009  516.9181  544.681  609.2592  646.9162  706.8834  858.0662  884.8812
898.9776  942.5784  999.4785  1001.6742  1014.6654  1028.0368  1038.9888  1049.5375  1080.2318  1157.2432
1178.422  1192.6477  1195.7261  1211.9038  1213.7066  1217.0158  1224.3617  1238.6755  1247.1939  1260.7242
1271.4285  1304.6968  1317.4943  1378.0527  1385.8956  1396.2932  1811.0331  1815.7621  2551.5237  2641.5008
2643.047  2647.661  2656.1529  2673.3077  2682.479  2720.1694  2726.9231  2750.7917  2774.3074</array>

            </property>

            <property dictRef="me:ZPE">

                    <scalar units="kJ/mol">-646.99426</scalar>

            </property>

<group>3</group>

            <property dictRef="me:spinMultiplicity">

                    <scalar units="cm-1">2</scalar>

            </property>

            <property dictRef="me:epsilon">

                    <scalar>473.17</scalar>

            </property>

            <property dictRef="me:sigma">

                    <scalar>5.09</scalar>

            </property>

        </propertyList>

        <me:energyTransferModel xsi:type="me:ExponentialDown">

                <scalar units="cm-1">250</scalar>

        </me:energyTransferModel>

    </molecule>

    <molecule id="[O]O[CH](C[CH](C(=O)C)OO)CC(=O)OO" spinMultiplicity="2">

            <atomArray>

                    <atom    elementType="O"    id="a1"    x3="-5.002438"    y3="1.685995"
z3="2.728491"/>
```

```xml
<atom elementType="C" id="a2" x3="-4.237800" y3="2.034246" z3="1.853818"/>
<atom elementType="C" id="a3" x3="-4.128802" y3="1.523763" z3="0.450713"/>
<atom elementType="C" id="a4" x3="-5.332157" y3="1.821811" z3="-0.428489"/>
<atom elementType="C" id="a5" x3="-6.689278" y3="1.429463" z3="0.130650"/>
<atom elementType="C" id="a6" x3="-6.959509" y3="-0.074881" z3="0.113263"/>
<atom elementType="H" id="a7" x3="-7.962008" y3="-0.284501" z3="0.562061"/>
<atom elementType="C" id="a8" x3="-6.882505" y3="-0.730820" z3="-1.284431"/>
<atom elementType="O" id="a9" x3="-6.731918" y3="-0.064711" z3="-2.280980"/>
<atom elementType="C" id="a10" x3="-7.029360" y3="-2.215039" z3="-1.287909"/>
<atom elementType="H" id="a11" x3="-5.937423" y3="-0.076765" z3="2.638087"/>
<atom elementType="H" id="a12" x3="-3.199881" y3="1.910781" z3="-0.034516"/>
<atom elementType="H" id="a13" x3="-4.003327" y3="0.409575" z3="0.506197"/>
<atom elementType="H" id="a14" x3="-5.181902" y3="1.434012" z3="-1.465598"/>
<atom elementType="H" id="a15" x3="-2.730515" y3="4.346610" z3="1.155458"/>
<atom elementType="H" id="a16" x3="-6.798352" y3="1.806245" z3="1.174103"/>
<atom elementType="H" id="a17" x3="-7.480833" y3="1.949480" z3="-0.455059"/>
<atom elementType="H" id="a18" x3="-6.469897" y3="-2.683062" z3="-0.457933"/>
<atom elementType="H" id="a19" x3="-8.079366" y3="-2.519730" z3="-1.166199"/>
<atom elementType="H" id="a20" x3="-6.675179" y3="-2.657772" z3="-2.230899"/>
<atom elementType="O" id="a21" x3="-5.953868" y3="-0.790765" z3="0.887352"/>
```

```xml
                                <atom    elementType="O"    id="a22"    x3="-6.425904"    y3="-0.868432"
z3="2.214623"/>
                                <atom    elementType="O"    id="a23"    x3="-2.428892"    y3="3.389788"
z3="1.285441"/>
                                <atom    elementType="O"    id="a24"    x3="-3.323865"    y3="3.003209"
z3="2.313883"/>
                                <atom    elementType="O"    id="a25"    spinMultiplicity="2"    x3="-5.466648"
y3="4.084411" z3="0.128476"/>
                                <atom    elementType="O"    id="a26"    x3="-5.380572"    y3="3.319414"    z3="-
0.779276"/>
                    </atomArray>
                    <bondArray>
                                <bond atomRefs2="a9 a8" order="2"/>
                                <bond atomRefs2="a20 a10" order="1"/>
                                <bond atomRefs2="a14 a4" order="1"/>
                                <bond atomRefs2="a10 a8" order="1"/>
                                <bond atomRefs2="a10 a19" order="1"/>
                                <bond atomRefs2="a10 a18" order="1"/>
                                <bond atomRefs2="a8 a6" order="1"/>
                                <bond atomRefs2="a26 a4" order="1"/>
                                <bond atomRefs2="a26 a25" order="1"/>
                                <bond atomRefs2="a17 a5" order="1"/>
                                <bond atomRefs2="a4 a5" order="1"/>
                                <bond atomRefs2="a4 a3" order="1"/>
                                <bond atomRefs2="a12 a3" order="1"/>
                                <bond atomRefs2="a6 a5" order="1"/>
                                <bond atomRefs2="a6 a7" order="1"/>
                                <bond atomRefs2="a6 a21" order="1"/>
                                <bond atomRefs2="a5 a16" order="1"/>
                                <bond atomRefs2="a3 a13" order="1"/>
                                <bond atomRefs2="a3 a2" order="1"/>
                                <bond atomRefs2="a21 a22" order="1"/>
                                <bond atomRefs2="a15 a23" order="1"/>
                                <bond atomRefs2="a23 a24" order="1"/>
```

```xml
                    <bond atomRefs2="a2 a24" order="1"/>

                    <bond atomRefs2="a2 a1" order="2"/>

                    <bond atomRefs2="a22 a11" order="1"/>

            </bondArray>

            <propertyList>

                    <property dictRef="me:lumpedSpecies">

                            <array> </array>

                    </property>

                    <property dictRef="me:vibFreqs">

                            <array units="cm-1">19.608 30.2477 37.5115 42.046 74.5456 85.5875
93.4221 114.0227 135.2357 146.5351 163.1394 209.3709 223.1621 240.7453 260.2376 279.7499 314.8922
342.8385 384.4686 391.4069 430.7643 453.4727 501.2117 534.0857 548.6804 612.5751 661.3199 729.3278 797.19
858.5814 897.0035 901.9373 940.8142 978.9038 1002.5084 1014.1691 1034.8855 1044.5438 1072.3242 1097.0741
1120.913 1149.0755 1161.6495 1183.9004 1197.4522 1202.8955 1209.7969 1215.7586 1218.5216 1226.3495
1230.583 1260.3584 1266.4829 1282.7825 1314.5781 1317.1232 1326.9655 1391.3493 1396.6865 1794.8876
1811.584 2570.1484 2629.7798 2642.2798 2646.3849 2664.3304 2667.407 2670.9644 2686.0548 2704.7299
2725.4579 2780.7017</array>

                    </property>

                    <property dictRef="me:ZPE">

                            <scalar units="kJ/mol">-788.19077823796459</scalar>

                    </property>

<group>4</group>

                    <property dictRef="me:spinMultiplicity">

                            <scalar units="cm-1">2</scalar>

                    </property>

                    <property dictRef="me:epsilon">

                            <scalar>473.17</scalar>

                    </property>

                    <property dictRef="me:sigma">

                            <scalar>5.09</scalar>

                    </property>

            </propertyList>

            <me:energyTransferModel xsi:type="me:ExponentialDown">

                            <scalar units="cm-1">250</scalar>

            </me:energyTransferModel>
```

```xml
            </molecule>
            <molecule                    id="TS_[O]OC(=O)CCC[CH](C(=O)C)OO_OOC(=O)CCCC(=O)C(=O)C"
spinMultiplicity="2">
                <atomArray>
                    <atom    elementType="O"    id="a1"    x3="2.745083"    y3="1.797988"
z3="1.862665"/>
                    <atom    elementType="C"    id="a2"    x3="1.659245"    y3="1.667825"
z3="1.368729"/>
                    <atom    elementType="C"    id="a3"    x3="0.344593"    y3="1.312141"
z3="1.987090"/>
                    <atom    elementType="C"    id="a4"    x3="0.178361"    y3="-0.208494"
z3="2.109523"/>
                    <atom    elementType="C"    id="a5"    x3="0.504881"    y3="-0.991143"
z3="0.833275"/>
                    <atom elementType="C" id="a6" spinMultiplicity="2" x3="-0.307596" y3="-
0.596543" z3="-0.383412"/>
                    <atom    elementType="H"    id="a7"    x3="0.027694"    y3="0.650609"    z3="-
0.694844"/>
                    <atom    elementType="C"    id="a8"    x3="-0.010884"    y3="-1.396360"    z3="-
1.663369"/>
                    <atom    elementType="O"    id="a9"    x3="0.580697"    y3="-2.445731"    z3="-
1.581960"/>
                    <atom    elementType="C"    id="a10"    x3="-0.459786"    y3="-0.781030"    z3="-
2.942652"/>
                    <atom    elementType="H"    id="a11"    x3="-3.298920"    y3="0.022960"    z3="-
0.644719"/>
                    <atom    elementType="H"    id="a12"    x3="0.268367"    y3="1.783473"
z3="2.990933"/>
                    <atom    elementType="H"    id="a13"    x3="-0.490991"    y3="1.743133"
z3="1.384214"/>
                    <atom    elementType="H"    id="a14"    x3="0.825903"    y3="-0.585583"
z3="2.930257"/>
                    <atom    elementType="H"    id="a15"    x3="-0.866299"    y3="-0.422829"
z3="2.419955"/>
                    <atom    elementType="H"    id="a16"    x3="1.586049"    y3="-0.892350"
z3="0.579510"/>
                    <atom    elementType="H"    id="a17"    x3="0.364887"    y3="-2.082614"
z3="1.019720"/>
```

```xml
                    <atom  elementType="H"  id="a18"  x3="-0.294782"  y3="-1.455197"  z3="-
3.797045"/>
                    <atom  elementType="H"  id="a19"  x3="-1.534188"  y3="-0.524919"  z3="-
2.918422"/>
                    <atom  elementType="H"  id="a20"  x3="0.078427"  y3="0.155901"  z3="-
3.158331"/>
                    <atom   elementType="O"   id="a21"   x3="-1.651496"   y3="-0.560374"
z3="0.010273"/>
                    <atom  elementType="O"  id="a22"  x3="-2.382524"  y3="0.106973"  z3="-
1.042815"/>
                    <atom  elementType="O"  id="a23"  x3="0.458788"  y3="1.874466"  z3="-
0.595556"/>
                    <atom  elementType="O"  id="a24"  x3="1.674489"  y3="1.827701"  z3="-
0.073020"/>
         </atomArray>
         <bondArray>
                    <bond atomRefs2="a18 a10" order="1"/>
                    <bond atomRefs2="a20 a10" order="1"/>
                    <bond atomRefs2="a10 a19" order="1"/>
                    <bond atomRefs2="a10 a8" order="1"/>
                    <bond atomRefs2="a8 a9" order="2"/>
                    <bond atomRefs2="a8 a6" order="1"/>
                    <bond atomRefs2="a22 a11" order="1"/>
                    <bond atomRefs2="a22 a21" order="1"/>
                    <bond atomRefs2="a7 a23" order="1"/>
                    <bond atomRefs2="a23 a24" order="1"/>
                    <bond atomRefs2="a6 a21" order="1"/>
                    <bond atomRefs2="a6 a5" order="1"/>
                    <bond atomRefs2="a24 a2" order="1"/>
                    <bond atomRefs2="a16 a5" order="1"/>
                    <bond atomRefs2="a5 a17" order="1"/>
                    <bond atomRefs2="a5 a4" order="1"/>
                    <bond atomRefs2="a2 a1" order="2"/>
                    <bond atomRefs2="a2 a3" order="1"/>
                    <bond atomRefs2="a13 a3" order="1"/>
```

```xml
            <bond atomRefs2="a3 a4" order="1"/>

            <bond atomRefs2="a3 a12" order="1"/>

            <bond atomRefs2="a4 a15" order="1"/>

            <bond atomRefs2="a4 a14" order="1"/>

         </bondArray>

         <propertyList>

            <property dictRef="me:vibFreqs">

                              <array       units="cm-1">18.4999016294527       32.57552432251179
   69.89918073700194   72.1777538897478   104.95904296531452   121.68089721445745   138.63408113404458
   158.27327234654769   182.3321127098397   213.10224038568683   219.68117355701082   275.089593583438
   300.10705367742503   320.0554371404765   346.3027255369848   418.2988842237282   426.23403674459786
   450.87133687935136   526.369209431981   560.7129948990469   585.0688060337683   626.4706098234252
   681.1000579917255   771.6291137340131   842.3777061257857   867.079378364056   926.5925797354736
   952.6254149817471   990.0252264323916   1001.6729568861556   1003.1959131363265   1019.3243193416563
   1042.6725049571119   1054.7507456402711   1092.157904922273   1108.3947342409358   1150.1720132708435
   1163.4709676413383   1187.0358484761227   1193.6406305854894   1203.9911852695777   1213.2911110585428
   1221.6043335632326   1235.8006713876748   1243.3552374014691   1256.5310667819165   1260.39552902607
   1266.2388931007301   1275.0576155656822   1310.742862002011   1317.6826283647706   1331.2375913135897
   1402.3770597804673   1808.0018820931105   1853.7522185916926   2639.597636189945   2645.497087442013
   2649.317483922279   2663.0447354297357   2666.790238211517   2682.6470002965157   2722.8982264367314
   2726.366014875063   2733.318304311213   2775.915238985839</array>

            </property>

            <property dictRef="me:imFreq">

                              <scalar units="cm-1">2537.765617778737</scalar>

            </property>

<group>3</group>

            <property dictRef="me:spinMultiplicity">

                              <scalar units="cm-1">2</scalar>

            </property>

            <property dictRef="me:ZPE">

                              <scalar units="kJ/mol">-625.877196</scalar>

            </property>

         </propertyList>

      </molecule>

      <molecule id="OOC(=O)CCCC(=O)C(=O)C">

         <atomArray>

                              <atom       elementType="O"       id="a1"       x3="0.466015"       y3="-1.124284"
z3="1.347856"/>
```

z3="0.853729"/>
<atom elementType="O" id="a2" x3="0.316593" y3="0.189339" z3="0.910975"/>
<atom elementType="C" id="a3" x3="-1.029291" y3="0.593831" z3="1.128092"/>
<atom elementType="O" id="a4" x3="-1.911265" y3="-0.192328" z3="0.613860"/>
<atom elementType="C" id="a5" x3="-1.069400" y3="2.064664" z3="1.872457"/>
<atom elementType="C" id="a6" x3="-0.778471" y3="2.891931" z3="2.287056"/>
<atom elementType="C" id="a7" x3="0.691220" y3="2.846296" z3="1.552071"/>
<atom elementType="C" id="a8" x3="1.630072" y3="3.761227" z3="1.820156"/>
<atom elementType="O" id="a9" x3="2.802790" y3="3.853540" z3="0.407803"/>
<atom elementType="C" id="a10" x3="1.052635" y3="4.636519" z3="-0.643696"/>
<atom elementType="O" id="a11" x3="0.719082" y3="4.149176" z3="-0.643696"/>
<atom elementType="C" id="a12" x3="0.990908" y3="6.090057" z3="0.734790"/>
<atom elementType="H" id="a13" x3="0.091451" y3="-1.681374" z3="0.593732"/>
<atom elementType="H" id="a14" x3="-0.344681" y3="2.325752" z3="-0.199500"/>
<atom elementType="H" id="a15" x3="-2.072145" y3="2.319404" z3="0.204531"/>
<atom elementType="H" id="a16" x3="-1.109824" y3="3.936609" z3="1.703201"/>
<atom elementType="H" id="a17" x3="-1.406131" y3="2.522986" z3="2.713260"/>
<atom elementType="H" id="a18" x3="0.790330" y3="3.045188" z3="3.378313"/>
<atom elementType="H" id="a19" x3="1.086257" y3="1.798795" z3="2.179245"/>
<atom elementType="H" id="a20" x3="0.409353" y3="6.297577" z3="1.642649"/>
<atom elementType="H" id="a21" x3="2.003828" y3="6.495810" z3="0.906253"/>

```xml
                        <atom elementType="H" id="a22" x3="0.547208" y3="6.673543" z3="-0.087441"/>
            </atomArray>
            <bondArray>
                <bond atomRefs2="a11 a10" order="2"/>
                <bond atomRefs2="a14 a5" order="1"/>
                <bond atomRefs2="a22 a12" order="1"/>
                <bond atomRefs2="a15 a5" order="1"/>
                <bond atomRefs2="a10 a12" order="1"/>
                <bond atomRefs2="a10 a8" order="1"/>
                <bond atomRefs2="a13 a1" order="1"/>
                <bond atomRefs2="a5 a3" order="1"/>
                <bond atomRefs2="a5 a6" order="1"/>
                <bond atomRefs2="a12 a21" order="1"/>
                <bond atomRefs2="a12 a20" order="1"/>
                <bond atomRefs2="a2 a3" order="1"/>
                <bond atomRefs2="a2 a1" order="1"/>
                <bond atomRefs2="a3 a4" order="2"/>
                <bond atomRefs2="a8 a9" order="2"/>
                <bond atomRefs2="a8 a7" order="1"/>
                <bond atomRefs2="a16 a6" order="1"/>
                <bond atomRefs2="a6 a7" order="1"/>
                <bond atomRefs2="a6 a17" order="1"/>
                <bond atomRefs2="a19 a7" order="1"/>
                <bond atomRefs2="a7 a18" order="1"/>
            </bondArray>
            <propertyList>
                <property dictRef="me:lumpedSpecies">
                        <array> </array>
                </property>
                <property dictRef="me:vibFreqs">
                        <array units="cm-1">18.1726  31.7987  48.4573  55.4663  85.8103
92.2243 96.4233 136.3476 184.0153 213.5407 230.0248 252.2778 324.2327 333.3177 379.9426 428.4593 516.7087
```

581.3186 600.2516 616.2705 686.6523 759.8785 851.0803 897.8957 917.69 991.4859 1001.6838 1003.0406 1014.7976 1021.5251 1098.0028 1108.9038 1157.2017 1178.8451 1198.1925 1203.4249 1206.7276 1218.1028 1224.1261 1241.1149 1262.2598 1277.9659 1289.0504 1309.5723 1310.8454 1319.1842 1422.2152 1797.0151 1817.3148 1847.0385 2614.8423 2625.7775 2645.6911 2658.8627 2673.9937 2687.786 2717.606 2718.6197 2735.6557 2782.1964</array>

                                                </property>

                                                <property dictRef="me:ZPE">

                                                                <scalar units="kJ/mol">-817.877196</scalar>

                                                </property>

                                                <property dictRef="me:spinMultiplicity">

                                                                <scalar units="cm-1">1</scalar>

                                                </property>

                                                <property dictRef="me:epsilon">

                                                                <scalar>473.17</scalar>

                                                </property>

                                                <property dictRef="me:sigma">

                                                                <scalar>5.09</scalar>

                                                </property>

<group>3</group>

                                              </propertyList>

                                            <me:energyTransferModel xsi:type="me:ExponentialDown">

                                                <scalar units="cm-1">250</scalar>

                                            </me:energyTransferModel>

                                    </molecule>

                                    <molecule id="TS_OOC(=O)CCCC(=O)C(=O)C_[CH2]CCC(=O)C(=O)C" spinMultiplicity="3">

                                              <atomArray>

                                              <atom elementType="O" id="a1" x3="-1.212107" y3="0.409001" z3="-1.141102"/>

                                              <atom elementType="O" id="a2" x3="0.376651" y3="0.034303" z3="-1.406990"/>

                                              <atom elementType="C" id="a3" x3="0.759568" y3="0.706613" z3="-0.323200"/>

                                              <atom elementType="O" id="a4" spinMultiplicity="2" x3="1.900849" y3="0.887811" z3="0.031245"/>

                                              <atom elementType="C" id="a5" spinMultiplicity="2" x3="-0.471872" y3="1.223121" z3="0.451771"/>

```xml
                              <atom      elementType="C"      id="a6"      x3="-0.664986"      y3="2.707154"
z3="0.434883"/>
                              <atom      elementType="C"      id="a7"      x3="-1.784609"      y3="3.200259"
z3="1.354230"/>
                              <atom      elementType="C"      id="a8"      x3="-3.144879"      y3="2.657639"
z3="1.014394"/>
                              <atom      elementType="O"      id="a9"      x3="-3.356567"      y3="1.607913"
z3="0.452311"/>
                              <atom      elementType="C"      id="a10"     x3="-4.333324"      y3="3.542529"
z3="1.484382"/>
                              <atom      elementType="O"      id="a11"     x3="-4.515470"      y3="3.724308"
z3="2.659604"/>
                              <atom      elementType="C"      id="a12"     x3="-5.156087"      y3="4.091421"
z3="0.372951"/>
                              <atom      elementType="H"      id="a13"     x3="-1.937021"      y3="0.329006"      z3="-
1.836967"/>
                              <atom      elementType="H"      id="a14"     x3="-1.714718"      y3="0.622342"      z3="-
0.001294"/>
                              <atom      elementType="H"      id="a15"     x3="-0.656183"      y3="0.672774"
z3="1.380034"/>
                              <atom      elementType="H"      id="a16"     x3="0.298495"       y3="3.189640"
z3="0.741724"/>
                              <atom      elementType="H"      id="a17"     x3="-0.836739"      y3="3.058326"      z3="-
0.607602"/>
                              <atom      elementType="H"      id="a18"     x3="-1.564872"      y3="2.927159"
z3="2.416265"/>
                              <atom      elementType="H"      id="a19"     x3="-1.791653"      y3="4.312598"
z3="1.360887"/>
                              <atom      elementType="H"      id="a20"     x3="-4.576586"      y3="4.729721"      z3="-
0.309105"/>
                              <atom      elementType="H"      id="a21"     x3="-5.592665"      y3="3.282845"      z3="-
0.239374"/>
                              <atom      elementType="H"      id="a22"     x3="-5.996452"      y3="4.697245"
z3="0.751882"/>
                    </atomArray>
                    <bondArray>
                              <bond atomRefs2="a13 a1" order="1"/>
                              <bond atomRefs2="a2 a3" order="2"/>
                              <bond atomRefs2="a1 a14" order="1"/>
```

```
                <bond atomRefs2="a17 a6" order="1"/>

                <bond atomRefs2="a3 a4" order="1"/>

                <bond atomRefs2="a3 a5" order="1"/>

                <bond atomRefs2="a20 a12" order="1"/>

                <bond atomRefs2="a21 a12" order="1"/>

                <bond atomRefs2="a12 a22" order="1"/>

                <bond atomRefs2="a12 a10" order="1"/>

                <bond atomRefs2="a6 a5" order="1"/>

                <bond atomRefs2="a6 a16" order="1"/>

                <bond atomRefs2="a6 a7" order="1"/>

                <bond atomRefs2="a5 a15" order="1"/>

                <bond atomRefs2="a9 a8" order="2"/>

                <bond atomRefs2="a8 a7" order="1"/>

                <bond atomRefs2="a8 a10" order="1"/>

                <bond atomRefs2="a7 a19" order="1"/>

                <bond atomRefs2="a7 a18" order="1"/>

                <bond atomRefs2="a10 a11" order="2"/>

        </bondArray>

        <propertyList>

            <property dictRef="me:vibFreqs">

                        <array   units="cm-1">18.282370053916605      31.27869516883575
55.22068604657698   63.61215674326096   78.1720296342959   106.37744713140481   148.53342359871527
160.72483672735981   189.97612533173157   204.9096938512438   243.44396261896   275.6632787150323
325.10952908693434   399.3646687059436   486.75348916229564   519.2435467467219   572.6889144116867
581.701167685628   619.0667821018064   650.4428928422811   659.2321608466214   701.6091879791743
746.4182461185196   800.4338832535573   880.8251917267594   930.1188301289562   978.0406654982055
1003.9841326723466   1015.253578268939   1057.5868658300146   1077.4878540180327   1089.2720609107016
1145.09826530318   1152.3935007157806   1163.3866829600613   1164.1887909963975   1198.1036275026772
1214.7545647534916   1216.2256483458486   1226.6308962105174   1235.311901061497   1238.0287841404377
1254.2663165466674   1298.1251319490996   1307.9486993465828   1337.289213585871   1791.0813898400113
1819.168070842715   1828.81846509738   2058.8370309126776   2636.0216877744288   2646.129109615801
2651.137787686823   2672.6984815965575   2681.2825277824227   2696.256853943389   2725.1984088186505
2729.3194019304165   2779.778468376829</array>

            </property>

            <property dictRef="me:imFreq">

                        <scalar units="cm-1">1185.3337540579626</scalar>

            </property>
```

```xml
                    <property dictRef="me:spinMultiplicity">
                            <scalar units="cm-1">2</scalar>
                    </property>
<group>3</group>
                    <property dictRef="me:ZPE">
                            <scalar units="kJ/mol">-389.8090286075985</scalar>
                    </property>
            </propertyList>
        </molecule>
        <molecule id="[CH2]CCC(=O)C(=O)C" spinMultiplicity="2">
            <atomArray>
                    <atom elementType="C" id="a1" spinMultiplicity="2" x3="1.033187" y3="-0.239910" z3="0.119564"/>
                    <atom elementType="H" id="a2" x3="0.387866" y3="0.554376" z3="0.430077"/>
                    <atom elementType="H" id="a3" x3="0.590396" y3="-1.063033" z3="-0.398595"/>
                    <atom elementType="C" id="a4" x3="2.485968" y3="-0.190390" z3="0.334618"/>
                    <atom elementType="C" id="a5" x3="2.948854" y3="0.909024" z3="1.299639"/>
                    <atom elementType="C" id="a6" x3="2.625562" y3="0.567378" z3="2.731667"/>
                    <atom elementType="O" id="a7" x3="3.445883" y3="0.273196" z3="3.563847"/>
                    <atom elementType="C" id="a8" x3="1.119398" y3="0.632894" z3="3.100324"/>
                    <atom elementType="O" id="a9" x3="0.528803" y3="1.682702" z3="3.063479"/>
                    <atom elementType="C" id="a10" x3="0.526249" y3="-0.680080" z3="3.483589"/>
                    <atom elementType="H" id="a11" x3="2.866496" y3="-1.179538" z3="0.677230"/>
                    <atom elementType="H" id="a12" x3="2.983049" y3="-0.028565" z3="-0.654266"/>
                    <atom elementType="H" id="a13" x3="4.044802" y3="1.063526" z3="1.198589"/>
```

```xml
                                <atom     elementType="H"     id="a14"     x3="2.482234"     y3="1.883627"
z3="1.034812"/>
                                <atom     elementType="H"     id="a15"     x3="0.513619"     y3="-1.381009"
z3="2.635040"/>
                                <atom     elementType="H"     id="a16"     x3="1.099824"     y3="-1.160591"
z3="4.292945"/>
                                <atom     elementType="H"     id="a17"     x3="-0.511550"     y3="-0.570001"
z3="3.834076"/>
                </atomArray>
                <bondArray>
                        <bond atomRefs2="a12 a4" order="1"/>
                        <bond atomRefs2="a3 a1" order="1"/>
                        <bond atomRefs2="a1 a4" order="1"/>
                        <bond atomRefs2="a1 a2" order="1"/>
                        <bond atomRefs2="a4 a11" order="1"/>
                        <bond atomRefs2="a4 a5" order="1"/>
                        <bond atomRefs2="a14 a5" order="1"/>
                        <bond atomRefs2="a13 a5" order="1"/>
                        <bond atomRefs2="a5 a6" order="1"/>
                        <bond atomRefs2="a15 a10" order="1"/>
                        <bond atomRefs2="a6 a8" order="1"/>
                        <bond atomRefs2="a6 a7" order="2"/>
                        <bond atomRefs2="a9 a8" order="2"/>
                        <bond atomRefs2="a8 a10" order="1"/>
                        <bond atomRefs2="a10 a17" order="1"/>
                        <bond atomRefs2="a10 a16" order="1"/>
                </bondArray>
                <propertyList>
                        <property dictRef="me:lumpedSpecies">
                                <array> </array>
                        </property>
                        <property dictRef="me:vibFreqs">
                                <array  units="cm-1">31.3528  53.8648  94.2233  105.5611  171.7407
185.3044 202.9406 277.5776 354.5812 424.7409 490.3351 588.2284 708.2416 746.587 778.6213 878.832 912.8739
```

966.3751  1002.8235  1018.786  1048.4182  1091.8946  1151.9303  1166.3998  1205.2401  1216.3321  1221.1885  1234.9631  1257.0273  1260.1899  1286.6241  1311.6434  1320.1127  1350.4305  1806.886  1825.7532  2644.648  2661.7753  2678.1296  2685.2929  2720.3605  2734.5335  2742.0073  2780.6656  2782.6185</array>

                    </property>

                    <property dictRef="me:ZPE">

                            <scalar units="kJ/mol">-758.723571167156244</scalar>

                    </property>

                    <property dictRef="me:spinMultiplicity">

                            <scalar units="cm-1">2</scalar>

                    </property>

<group>3</group>

                    <property dictRef="me:epsilon">

                            <scalar>473.17</scalar>

                    </property>

                    <property dictRef="me:sigma">

                            <scalar>5.09</scalar>

                    </property>

            </propertyList>

            <me:energyTransferModel xsi:type="me:ExponentialDown">

                    <scalar units="cm-1">250</scalar>

            </me:energyTransferModel>

        </molecule>

        <molecule id="O=C=O_____[OH]" spinMultiplicity="2">

                <atomArray>

                            <atom     elementType="O"     id="a1"     x3="2.265879"     y3="0.124463" z3="0.204720"/>

                            <atom     elementType="C"     id="a2"     x3="3.238914"     y3="0.450251" z3="0.770878"/>

                            <atom     elementType="O"     id="a3"     x3="4.208805"     y3="0.774962" z3="1.335163"/>

                            <atom  elementType="O"  id="a4"  spinMultiplicity="2"  x3="-0.438780"  y3="-0.768491" z3="-1.347475"/>

                            <atom     elementType="H"     id="a5"     x3="0.382009"     y3="-0.498826"   z3="-0.878681"/>

                </atomArray>

```xml
<bondArray>
    <bond atomRefs2="a4 a5" order="1"/>
    <bond atomRefs2="a1 a2" order="2"/>
    <bond atomRefs2="a2 a3" order="2"/>
</bondArray>
<propertyList>
    <property dictRef="me:lumpedSpecies">
        <array> </array>
    </property>
    <property dictRef="me:vibFreqs">
        <array units="cm-1">6.4801  84.1536  163.2061  187.326  536.3297
536.4234 1378.7003 2321.3351 2630.5584</array>
    </property>
    <property dictRef="me:ZPE">
        <scalar units="kJ/mol">0.0</scalar>
    </property>
    <property dictRef="me:spinMultiplicity">
        <scalar units="cm-1">2</scalar>
    </property>
    <property dictRef="me:epsilon">
        <scalar>473.17</scalar>
    </property>
    <property dictRef="me:sigma">
        <scalar>5.09</scalar>
    </property>
</propertyList>
<me:energyTransferModel xsi:type="me:ExponentialDown">
    <scalar units="cm-1">250</scalar>
</me:energyTransferModel>
</molecule>
```

```
        </moleculeList>

        <reactionList>

                <reaction active="true" id="CC1=CCCCC1_C[C]12CCCC[CH]2OOO1">

                        <reactant>

                                <molecule me:type="modelled" ref="CC1=CCCCC1"/>

                        </reactant>

                        <reactant>

                                <molecule me:type="excessReactant" ref="[O]O[O]"/>

                        </reactant>

                        <product>

                                <molecule me:type="modelled" ref="C[C]12CCCC[CH]2OOO1"/>

                        </product>

<group>1</group>

                        <me:MCRCMethod xsi:type="MesmerILT">

                                <me:preExponential units="cm3 molecule-1 s-1">1E-10</me:preExponential>

                                <me:activationEnergy units="kJ/mol">0</me:activationEnergy>

                                <me:TInfinity>298.0</me:TInfinity>

                                <me:nInfinity>0.0</me:nInfinity>

                        </me:MCRCMethod>

                        <me:excessReactantConc>1E18</me:excessReactantConc>

                </reaction>

                <reaction active="true" id="C[C]12CCCC[CH]2OOO1_[O]O[CH]CCCCC(=O)C">

                        <reactant>

                                <molecule me:type="modelled" ref="C[C]12CCCC[CH]2OOO1"/>

                        </reactant>

                        <product>

                                <molecule me:type="modelled" ref="[O]O[CH]CCCCC(=O)C"/>

                        </product>

<group>1</group>

                        <me:transitionState>
```

```xml
                                <molecule                                    me:type="transitionState"
ref="TS2_C[C]12CCCC[CH]2OOO1_[O]O[CH]CCCCC(=O)C"/>

                    </me:transitionState>

                    <me:MCRCMethod name="SimpleRRKM"/>

            </reaction>

            <reaction active="true" id="C[C]12CCCC[CH]2OOO1_O=CCCCC[C](O[O])C">

                    <reactant>

                                <molecule me:type="modelled" ref="C[C]12CCCC[CH]2OOO1"/>

                    </reactant>

<group>1</group>

                    <product>

                                <molecule me:type="modelled" ref="O=CCCCC[C](O[O])C"/>

                    </product>

                    <me:transitionState>

                                <molecule                                    me:type="transitionState"
ref="TS_C[C]12CCCC[CH]2OOO1_O=CCCCC[C](O[O])C"/>

                    </me:transitionState>

                    <me:MCRCMethod name="SimpleRRKM"/>

            </reaction>

            <reaction active="true" id="O=CCCCC[C](O[O])C_CC(=CCCCC=O)OO">

                    <reactant>

                                <molecule me:type="modelled" ref="O=CCCCC[C](O[O])C"/>

                    </reactant>

<group>1</group>

                    <product>

                                <molecule me:type="modelled" ref="CC(=CCCCC=O)OO"/>

                    </product>

                    <me:transitionState>

                                <molecule                                    me:type="transitionState"
ref="TS_O=CCCCC[C](O[O])C_CC(=CCCCC=O)OO"/>

                    </me:transitionState>

                    <me:MCRCMethod name="SimpleRRKM"/>

            </reaction>
```

```xml
<reaction active="true" id="O=CCCC[CH]C(=O)C_CC(=CCCCC=O)OO">
    <reactant>
        <molecule me:type="modelled" ref="O=CCCC[CH]C(=O)C"/>
    </reactant>
<group>1</group>
    <reactant>
        <molecule me:type="excessReactant" ref="[OH]"/>
    </reactant>
    <product>
        <molecule me:type="modelled" ref="CC(=CCCCC=O)OO"/>
    </product>
    <me:MCRCMethod xsi:type="MesmerILT">
        <me:preExponential units="cm3 molecule-1 s-1">1E-10</me:preExponential>
        <me:activationEnergy units="kJ/mol">0.0</me:activationEnergy>
        <me:TInfinity>298.0</me:TInfinity>
        <me:nInfinity>0.0</me:nInfinity>
    </me:MCRCMethod>
    <me:excessReactantConc>1</me:excessReactantConc>
</reaction>
<reaction active="true" id="O=CCCC[CH]C(=O)C_O=[C]CCCCC(=O)C">
    <reactant>
        <molecule me:type="modelled" ref="O=CCCC[CH]C(=O)C"/>
    </reactant>
<group>1</group>
    <product>
        <molecule me:type="modelled" ref="O=[C]CCCCC(=O)C"/>
    </product>
    <me:transitionState>
        <molecule me:type="transitionState" ref="TS_O=CCCC[CH]C(=O)C_O=[C]CCCCC(=O)C"/>
    </me:transitionState>
    <me:MCRCMethod name="SimpleRRKM"/>
```

```
        </reaction>
        <reaction active="true" id="O=CCCC[CH]C(=O)C_O=CCCC[CH](C(=O)C)O[O]">
                <reactant>
                        <molecule me:type="modelled" ref="O=CCCC[CH]C(=O)C"/>
                </reactant>
<group>2</group>
                <reactant>
                        <molecule me:type="excessReactant" ref="O=O"/>
                </reactant>
                <product>
                        <molecule me:type="modelled" ref="O=CCCC[CH](C(=O)C)O[O]"/>
                </product>
                <me:MCRCMethod xsi:type="MesmerILT">
                        <me:preExponential units="cm3 molecule-1 s-1">1E-12</me:preExponential>
                        <me:activationEnergy units="kJ/mol">0</me:activationEnergy>
                        <me:TInfinity>298.0</me:TInfinity>
                        <me:nInfinity>0.0</me:nInfinity>
                </me:MCRCMethod>
                <me:excessReactantConc>1E18</me:excessReactantConc>
        </reaction>
        <reaction active="true" id="O=CCCC[CH](C(=O)C)O[O]_O=[C]CCC[CH](C(=O)C)OO">
                <reactant>
                        <molecule me:type="modelled" ref="O=CCCC[CH](C(=O)C)O[O]"/>
                </reactant>
<group>2</group>
                <product>
                        <molecule me:type="modelled" ref="O=[C]CCC[CH](C(=O)C)OO"/>
                </product>
                <me:transitionState>
                        <molecule                                    me:type="transitionState"
ref="TS_O=CCCC[CH](C(=O)C)O[O]_O=[C]CCC[CH](C(=O)C)OO"/>
                </me:transitionState>
```

```xml
                    <me:MCRCMethod name="SimpleRRKM"/>
            </reaction>
            <reaction active="true" id="O=CCCC[CH](C(=O)C)O[O]_O=CC[CH]C[CH](C(=O)C)OO">
                    <reactant>
                            <molecule me:type="modelled" ref="O=CCCC[CH](C(=O)C)O[O]"/>
                    </reactant>
<group>2</group>
                    <product>
                            <molecule me:type="modelled" ref="O=CC[CH]C[CH](C(=O)C)OO"/>
                    </product>
                    <me:transitionState>
                            <molecule                                    me:type="transitionState"
ref="TS_O=CCCC[CH](C(=O)C)O[O]_O=CC[CH]C[CH](C(=O)C)OO"/>
                    </me:transitionState>
                    <me:MCRCMethod name="SimpleRRKM"/>
            </reaction>-->
            <reaction active="true" id="O=[C]CCC[CH](C(=O)C)OO_[O]OC(=O)CCC[CH](C(=O)C)OO">
                    <reactant>
                            <molecule me:type="modelled" ref="O=[C]CCC[CH](C(=O)C)OO"/>
                    </reactant>
<group>3</group>
                    <reactant>
                            <molecule me:type="excessReactant" ref="O=O"/>
                    </reactant>
                    <product>
                            <molecule me:type="modelled" ref="[O]OC(=O)CCC[CH](C(=O)C)OO"/>
                    </product>
                    <me:MCRCMethod xsi:type="MesmerILT">
                            <me:preExponential units="cm3 molecule-1 s-1">1E-12</me:preExponential>
                            <me:activationEnergy units="kJ/mol">0</me:activationEnergy>
                            <me:TInfinity>298.0</me:TInfinity>
                            <me:nInfinity>0.0</me:nInfinity>
```

```xml
            </me:MCRCMethod>

            <me:excessReactantConc>1E18</me:excessReactantConc>

        </reaction>

        <reaction active="true" id="[CH2]CC[CH](C(=O)C)OO_O=[C]CCC[CH](C(=O)C)OO">

            <reactant>

                <molecule me:type="modelled" ref="[CH2]CC[CH](C(=O)C)OO"/>

            </reactant>

<group>2</group>

            <product>

                <molecule me:type="sink" ref="O=[C]CCC[CH](C(=O)C)OO"/>

            </product>

            <me:transitionState>

                <molecule                                        me:type="transitionState"
ref="TS_O=CCCC[CH](C(=O)C)OO_[CH2]CC[CH](C(=O)C)OO"/>

            </me:transitionState>

            <me:MCRCMethod name="SimpleRRKM"/>

        </reaction>

        <reaction active="true" id="[O]OC(=O)CCC[CH](C(=O)C)OO_OOC(=O)[CH]CC[CH](C(=O)C)OO">

            <reactant>

                <molecule me:type="modelled" ref="[O]OC(=O)CCC[CH](C(=O)C)OO"/>

            </reactant>

<group>3</group>

            <product>

                <molecule me:type="modelled" ref="OOC(=O)[CH]CC[CH](C(=O)C)OO"/>

            </product>

            <me:transitionState>

                <molecule                                        me:type="transitionState"
ref="TS_[O]OC(=O)CCC[CH](C(=O)C)OO_OOC(=O)[CH]CC[CH](C(=O)C)OO"/>

            </me:transitionState>

            <me:MCRCMethod name="SimpleRRKM"/>

        </reaction>

        <reaction active="true" id="[O]OC(=O)CCC[CH](C(=O)C)OO_OOC(=O)C[CH]C[CH](C(=O)C)OO">
```

```xml
                    <reactant>
                            <molecule me:type="modelled" ref="[O]OC(=O)CCC[CH](C(=O)C)OO"/>
                    </reactant>
<group>3</group>
                    <product>
                            <molecule me:type="modelled" ref="OOC(=O)C[CH]C[CH](C(=O)C)OO"/>
                    </product>
                    <me:transitionState>
                            <molecule                                    me:type="transitionState"
ref="TS_[O]OC(=O)CCC[CH](C(=O)C)OO_OOC(=O)C[CH]C[CH](C(=O)C)OO"/>
                    </me:transitionState>
                    <me:MCRCMethod name="SimpleRRKM"/>
            </reaction>
            <reaction active="true" id="[O]OC(=O)CCC[CH](C(=O)C)OO_OOC(=O)CCCC(=O)C(=O)C">
                    <reactant>
                            <molecule me:type="modelled" ref="[O]OC(=O)CCC[CH](C(=O)C)OO"/>
                    </reactant>
<group>3</group>
                    <product>
                            <molecule me:type="modelled" ref="OOC(=O)CCCC(=O)C(=O)C"/>
                    </product>
                    <me:transitionState>
                            <molecule                                    me:type="transitionState"
ref="TS_[O]OC(=O)CCC[CH](C(=O)C)OO_OOC(=O)CCCC(=O)C(=O)C"/>
                    </me:transitionState>
                    <me:MCRCMethod name="SimpleRRKM"/>
            </reaction>
            <reaction active="true" id="[CH2]CCC(=O)C(=O)C_OOC(=O)CCCC(=O)C(=O)C">
                    <reactant>
                            <molecule me:type="modelled" ref="OOC(=O)CCCC(=O)C(=O)C"/>
                    </reactant>
<group>3</group>
                    <product>
```

```xml
                        <molecule me:type="sink" ref="[CH2]CCC(=O)C(=O)C"/>
                </product>
                <me:transitionState>
                        <molecule                                    me:type="transitionState"
ref="TS_OOC(=O)CCCC(=O)C(=O)C_[CH2]CCC(=O)C(=O)C"/>
                </me:transitionState>
                <me:MCRCMethod name="SimpleRRKM"/>
        </reaction>

    </reactionList>
    <me:conditions>
        <me:bathGas>N2</me:bathGas>
        <me:PTs>
                <me:PTpair     me:P="750"     me:T="298"     me:precision="d"     me:units="Torr"
refReactionExcess="R2">       </me:PTpair>
        </me:PTs>
        <me:modelParameters>
                <me:grainSize units="cm-1">100</me:grainSize>
                <me:energyAboveTheTopHill>10.0</me:energyAboveTheTopHill>
        </me:modelParameters>
        <me:InitialPopulation>
                <me:molecule grain="1.0" population="1.0" ref="CC1=CCCCC1"/>
        </me:InitialPopulation>
    </me:conditions>
    <me:modelParameters>
        <me:numberStochasticTrials>100</me:numberStochasticTrials>
        <me:stochasticStartTime>1E-11</me:stochasticStartTime>
        <me:stochasticEndTime>1E-5</me:stochasticEndTime>
        <me:StochasticThermalThreshold>500000</me:StochasticThermalThreshold>
        <me:StochasticEqilThreshold>100000000</me:StochasticEqilThreshold>
        <me:StochasticAxdLimit>5</me:StochasticAxdLimit>
        <me:grainSize units="cm-1">100</me:grainSize>
        <me:energyAboveTheTopHill>10.0</me:energyAboveTheTopHill>
```

```
        </me:modelParameters>

        <me:control>

                <me:printSpeciesProfile/>

                <!--<me:stochasticOnePass/>-->

                <me:stochasticSimulation/>

        </me:control>

</me:mesmer>
```